\documentclass{elsart}
\usepackage{graphicx}
\usepackage{amssymb}
\usepackage{amsmath}
\usepackage{bm}
\usepackage{color}
\def\average#1{\left\langle {#1} \right\rangle}

\def\ddel#1#2{\frac{d^2 #1}{d #2 ^2}}

\def\delo#1{\frac{d}{d #1}}

\def\del#1#2{\frac{d #1}{d #2}}

\def\deld#1#2{\frac{\delta #1}{\delta #2}}
\def\vec2#1#2{\left(\begin{array}{c} #1 \\ #2 \end{array}\right)}
\def\vec3#1#2#3{\left(\begin{array}{c} #1 \\ #2 \\ #3 \end{array}\right)}
\def\Eqref#1{Eq. (\ref{#1})}
\def\Eqsref#1#2{Eqs. (\ref{#1})(\ref{#2})}

\newcommand{\lt}{\left}
\newcommand{\rt}{\right}

\newcommand{\adag}{{a^{\dagger}}}
\newcommand{\alphaz}{\alpha_0}
\newcommand{\alphasf}{\alpha_\spinflip}

\newcommand{\Av}{{\bm A}}
\newcommand{\Avem}{{\bm A}_{\rm em}}

\newcommand{\Ams}{\ {\rm A/m}^2}
\newcommand{\Aph}{A^{\phi}}
\newcommand{\Ath}{A^{\theta}}
\newcommand{\Aphv}{\Av^{\phi}}
\newcommand{\Athv}{\Av^{\theta}}
\newcommand{\Az}{{A^{z}}}

\newcommand{\Bv}{{\bm B}}

\newcommand{\Bc}{B_{\rm c}}
\newcommand{\Bvs}{{\bm B}_{S}}

\newcommand{\Bve}{{\bm B}_{\rm e}}
\newcommand{\betasf}{{\beta_\spinflip}}
\newcommand{\betana}{{\beta_{\rm na}}}
\newcommand{\betaw}{{\beta_{\rm w}}}
\newcommand{\cdag}{{c^{\dagger}}}
\newcommand{\chiz}{\chi^{(0)}}
\newcommand{\chio}{\chi^{(1)}}
\newcommand{\chitilo}{\tilde{\chi}^{(1)}}
\newcommand{\chitilz}{\tilde{\chi}^{(0)}}

\newcommand{\deltaS}{\delta S}

\newcommand{\dw}{{\rm w}}
\newcommand{\dx}{{d^3 x}}
\newcommand{\Deltatil}{\tilde{\Delta}}
\newcommand{\DOS}{{\nu}}
\newcommand{\DOSV}{{N(0)}}

\newcommand{\ef}{{\epsilon_F}}
\newcommand{\eF}{{\epsilon_F}}
\newcommand{\ekv}{\epsilon_{\kv}}
\newcommand{\ekvs}{\epsilon_{\kv\sigma}}
\newcommand{\Ev}{{\bm E}}
\newcommand{\ev}{{\bm e}}
\newcommand{\evth}{{\bm e}_{\theta}}
\newcommand{\evph}{{\bm e}_{\phi}}
\newcommand{\evs}{{\bm n}}
\newcommand{\evsz}{{\evs}_0}
\newcommand{\evsph}{(\evph\times\evz)}

\newcommand{\evy}{{\bm e}_{y}}
\newcommand{\evz}{{\bm e}_{z}}
\newcommand{\fl}{{\eta}}
\newcommand{\fltil}{\tilde{\eta}}

\newcommand{\fli}{{\fl_{\rm I}}}
\newcommand{\flr}{{\fl_{\rm R}}}

\newcommand{\fomega}{{\omega}}
\newcommand{\fbeta}{f^{\beta}}
\newcommand{\fpin}{{f_{\rm pin}}}
\newcommand{\Fpin}{{F_{\rm pin}}}

\newcommand{\fna}{{f_{\rm ref}}}
\newcommand{\Fe}{F}
\newcommand{\Fbeta}{F^{\beta}}

\newcommand{\Fbetafactor}{\mu}
\newcommand{\Fhall}{F^{\rm Hall}}
\newcommand{\Fhallv}{\Fv^{\rm Hall}}
\newcommand{\Fren}{F^{\rm ren}}
\newcommand{\Frenv}{\Fv^{\rm ren}}
\newcommand{\Fvna}{\Fv^{\rm ref}}

\newcommand{\Fna}{F^{\rm ref}}
\newcommand{\Fad}{\Fhall}

\newcommand{\Fv}{{\bm F}}
\newcommand{\Fo}{F^{(1)}}
\newcommand{\Fw}{F_{\rm w}}
\newcommand{\Fz}{F^{(0)}}
\newcommand{\Fzv}{\Fv^{(0)}}
\newcommand{\Fdelta}{\delta F}
\newcommand{\Fdeltav}{\delta \Fv}
\newcommand{\Fdel}{\delta \Fo}
\newcommand{\Fdelv}{\delta \Fv^{(1)}}
\newcommand{\Gv}{{\bm G}}

\newcommand{\gr}{g^{\rm r}}
\newcommand{\ga}{g^{\rm a}}
\newcommand{\gless}{g^{<}}

\newcommand{\gyro}{\gamma}
\newcommand{\gyroz}{\gyro_{0}}
\newcommand{\hf}{\frac{1}{2}}
\newcommand{\HA}{{H_{A}}}

\newcommand{\Hv}{\bm{H}}
\newcommand{\He}{H_{\rm e}}

\newcommand{\Hem}{H_{\rm em}}
\newcommand{\Hsd}{H_{sd}}

\newcommand{\Himp}{H_{\rm imp}}

\newcommand{\Hs}{{H_{\rm S}}}
\newcommand{\Hsf}{{H_\spinflip}}

\newcommand{\Hw}{H_{\dw}}
\newcommand{\hbarinv}{\frac{1}{\hbar}}
\renewcommand{\Im}{{\rm Im}}
\newcommand{\intinf}{\int_{-\infty}^{\infty}}
\newcommand{\intom}{\int \frac{d\omega}{2\pi}}
\newcommand{\intx}{\int {d^3x}}
\newcommand{\iv}{\bm{i}}

\newcommand{\Jsd}{J_{sd}}

\newcommand{\js}{j_{\rm s}}

\newcommand{\jsv}{\bm{j}_{\rm s}}

\newcommand{\JSv}{\bm{J}_{\rm S}}
\newcommand{\JS}{J_{\rm S}}
\newcommand{\JStotv}{\bm{J}_{S,{\rm tot}}}
\newcommand{\jc}{j_{\rm c}}
\newcommand{\jci}{{{j}_{\rm c}^{\rm i}}}

\newcommand{\jatil}{{\tilde{j}_{\rm a}}}
\newcommand{\jctil}{{\tilde{j}_{\rm c}}}
\newcommand{\jcitil}{{\tilde{j}_{\rm c}^{\rm i}}}
\newcommand{\jcetil}{{\tilde{j}_{\rm c}^{\rm e}}}

\newcommand{\jtil}{{\tilde{j}}}
\newcommand{\jv}{\bm{j}}
\newcommand{\kB}{{k_B}}
\newcommand{\kb}{{k_B}}
\newcommand{\kv}{{\bm k}}

\newcommand{\kvpq}{{\kv}+\frac{\qv}{2}}
\newcommand{\kvmq}{{\kv}-\frac{\qv}{2}}

\newcommand{\kf}{{k_F}}
\newcommand{\kF}{{k_F}}

\newcommand{\kfu}{k_{F+}}
\newcommand{\kfd}{k_{F-}}

\newcommand{\Kp}{{K_\perp}}
\newcommand{\ktil}{\tilde{k}}

\newcommand{\lamv}{{\lambda_{\rm v}}}
\newcommand{\lamso}{{\lambda_{\rm so}}}

\newcommand{\Le}{{L_{\rm e}}}
\newcommand{\Lez}{{L_{\rm e}^0}}
\newcommand{\Lb}{L_{\rm B}}
\newcommand{\Ldw}{L_{\dw}}
\newcommand{\Ls}{{L_{\rm S}}}
\newcommand{\Lsw}{L_{\rm sw}}

\newcommand{\lstil}{\tilde{l_\sigma}}
\newcommand{\mv}{{\bm m}}
\newcommand{\Mv}{{\bm M}}
\newcommand{\Mphi}{{M_{\phi}}}
\newcommand{\Mw}{{M_{\dw}}}
\newcommand{\Ms}{M_{\rm s}}

\newcommand{\mub}{\mu_B}
\newcommand{\muB}{\mu_B}

\newcommand{\Ne}{N_{\rm e}}
\newcommand{\nv}{{\bm n}}
\newcommand{\Nimp}{N_{\rm imp}}
\newcommand{\nimp}{n_{\rm imp}}
\newcommand{\Nw}{N_{\dw}}
\newcommand{\nvortex}{n_{\rm v}}
\newcommand{\nvz}{{\nv}_0}

\newcommand{\om}{{\omega}}
\newcommand{\omegap}{\omega'}
\newcommand{\Omegatil}{\tilde{\Omega}}

\newcommand{\Omegapin}{\Omega_{\rm pin}}
\newcommand{\Omz}{\Omega_0}
\newcommand{\ompOmz}{{\omega+\frac{\Omz}{2}}}
\newcommand{\ommOmz}{{\omega-\frac{\Omz}{2}}}
\newcommand{\phiz}{{\phi_0}}
\newcommand{\Phiv}{\bm{\Phi}}
\newcommand{\PhiB}{{\Phi_{\rm B}}}
\newcommand{\Ptil}{{\tilde{P}}}
\newcommand{\pv}{{\bm p}}
\newcommand{\qv}{{\bm q}}
\newcommand{\qtil}{{\tilde{q}}}
\newcommand{\ra}{\rightarrow}
\renewcommand{\Re}{{\rm Re}}
\newcommand{\rhow}{{\rho_{\dw}}}
\newcommand{\rhos}{{\rho_{\rm s}}}
\newcommand{\rhoS}{{\rho_{\rm s}}}
\newcommand{\rhoxy}{{\rho_{xy}}}
\newcommand{\RS}{{R_{\rm S}}}
\newcommand{\Rw}{{R_{\dw}}}

\newcommand{\Rv}{{\bm R}}
\newcommand{\sigmav}{{\bm \sigma}}
\newcommand{\se}{{s}}
\newcommand{\sev}{{\bm \se}}
\newcommand{\sevsf}{{\bm \se}_\spinflip}
\newcommand{\sgn}{{\rm sgn}}

\newcommand{\sv}{{{\bm s}}}
\newcommand{\seth}{{\se}_\theta}
\newcommand{\seph}{{\se}_\phi}
\newcommand{\sez}{{\se}_z}

\newcommand{\spol}{{M}}
\newcommand{\spinflip}{{\rm sr}}
\newcommand{\svtil}{\tilde{\bm s}}
\newcommand{\stil}{\tilde{\se}}
\newcommand{\stilz}{\stil_{z}}
\newcommand{\stilpm}{\stil^{\pm}}
\newcommand{\stilpmz}{\stil^{\pm(0)}}
\newcommand{\stilpma}{\stil^{\pm(1{\rm a})}}
\newcommand{\stilpmb}{\stil^{\pm(1{\rm b})}}
\newcommand{\stilpmo}{\stil^{\pm(1)}}

\newcommand{\Simpv}{{{\bm S}_{\rm imp}}}
\newcommand{\Simp}{{S_{\rm imp}}}
\newcommand{\Stot}{{S_{\rm tot}}}
\newcommand{\Stotv}{\bm{S}_{\rm tot}}

\newcommand{\Sv}{{{\bm S}}}

\newcommand{\sumx}{{\int \frac{d^3x}{a^3}}}

\newcommand{\sumkv}{{\sum_{\kv}}}
\newcommand{\sumom}{\int\frac{d\omega}{2\pi}}

\newcommand{\sumqv}{{\sum_{\qv}}}
\newcommand{\thickness}{{d}}
\newcommand{\thetaz}{{\theta_0}}
\newcommand{\tr}{{\rm tr}}
\newcommand{\Tc}{{T_{c}}}

\newcommand{\torque}{{\tau}}
\newcommand{\torquev}{{\bm \torque}}

\newcommand{\torqueve}{\torquev}
\newcommand{\torquee}{\torque}
\newcommand{\torquew}{{\torque_{\dw}}}
\newcommand{\tautil}{{\tilde{\tau}}}

\newcommand{\tauw}{\tau_{\dw}}

\newcommand{\thetast}{\theta_{\rm st}}
\newcommand{\ttil}{{\tilde{t}}}
\newcommand{\tz}{{t_0}}
\newcommand{\vc}{{v_{\rm c}}}

\newcommand{\vev}{{\vv_{\rm e}}}
\newcommand{\vv}{\bm{v}}
\newcommand{\vs}{{v_{\rm s}}}
\newcommand{\vsv}{{\vv_{\rm s}}}
\newcommand{\vw}{v_{\rm w}}
\newcommand{\vf}{{v_F}}
\newcommand{\vimp}{v_{\rm imp}}

\newcommand{\Vpin}{{V}_{\rm pin}}
\newcommand{\Vinv}{\frac{1}{V}}

\newcommand{\Vz}{{V_0}}
\newcommand{\Vztil}{{\tilde{V_0}}}
\newcommand{\Ws}{{W_{\rm S}}}
\newcommand{\Xtil}{{\tilde{X}}}
\newcommand{\xv}{{\bm x}}
\newcommand{\Xv}{{\bm X}}

\newcommand{\xw}{{z}}

\newcommand{\ztil}{u}

\begin{document}
\begin{frontmatter}


\title
{Microscopic approach to current-driven domain wall dynamics}


\author{Gen Tatara}
\ead{tatara@phys.metro-u.ac.jp}
\address{Graduate School of Science, Tokyo Metropolitan University,
Hachioji, Tokyo 192-0397,
Japan}
\address{
PRESTO, JST, 4-1-8 Honcho Kawaguchi, Saitama, Japan}

\author{Hiroshi Kohno}
\address{
Graduate School of Engineering Science, Osaka University,
Toyonaka, Osaka 560-8531, Japan}

\author{Junya Shibata}
\address{
Kanagawa Institute of Technology,
Shimo-Ogino, Atsugi, 243-0292, Japan
}
\today
\begin{abstract}
This review describes in detail the essential techniques used in microscopic theories on spintronics.
We have investigated the domain wall dynamics induced by electric current 
based on the $s$-$d$ exchange model.
The domain wall is treated as rigid and planar and is described by two collective coordinates: the position and angle of wall magnetization.
The effect of conduction electrons on the domain wall dynamics is calculated in the case of slowly varying spin structure (close to the adiabatic limit) by use of a gauge transformation.
The spin-transfer torque and force on the wall are expressed by Feynman diagrams and calculated systematically using non-equilibrium Green's functions, treating electrons fully quantum mechanically.
The wall dynamics is discussed based on two coupled equations of motion derived for two collective coordinates.
The force is related to electron transport properties, resistivity, and the Hall effect. 
Effect of conduction electron spin relaxation on the torque and wall dynamics is also studied. 
\end{abstract}

\begin{keyword}
spintronics \sep
spin transfer torque \sep
domain wall \sep
magnetoresistance \sep
Hall effect \sep
Keldysh Green's functions
\PACS
72.25.-b  \sep 
72.25.Pn  \sep 
72.25.Rb  \sep 
73.23.-b  \sep 
73.23.Ra  \sep 
75.47.De  \sep 
75.47.Jn  \sep 
75.60.Ch  \sep 
75.70.-i  \sep 
85.75.-d  
\end{keyword}
\end{frontmatter}

\tableofcontents
\newpage
\begin{table}[htb]
\begin{center}
\caption{List of important variables and their unit.
Equation or equation number indicates definition.
\label{variables}}

\begin{tabular}{cccc} \hline
  $\Sv$  & local spin vector & & {\rm -}\\ 
  $\evs$  & local spin direction & $ {\Sv}/{S}$ & {\rm -}\\ 
  $\evsz$  & local spin for domain wall configuration &  & {\rm -}\\ 
  $J$  & exchange interaction between local spins & & {\rm J/m$^2$}\\ 
  $K$  & easy axis energy gain of local spin per spin & & {\rm J}\\ 
  $\Kp$  & hard axis energy loss of local spin & & {\rm J}\\ 
  $\Jsd$   & $s$-$d$ exchange interaction & & {\rm J}\\ 
  $\alpha$ & Gilbert damping parameter of local spin & & - \\ 
  $\betasf$ & $\beta$ arising from electron spin relaxation & \Eqref{modLLG} & - \\ 
  $\betaw$ & effective $\beta$ or force acting on domain wall & \Eqref{betaeffective} & - \\ 
  $\lambda$   &  domain wall thickness & $\sqrt{{K}/{J}}$ & {\rm m}  \\ 
  $X$  & center position of domain wall & & {\rm m}\\ 
  $\phiz$  & collective angle out of easy plane of domain wall &\Eqref{phizdef} & {\rm -}\\ 
  $\Nw$  & number of spins in domain wall & 
     ${2\lambda A}/{a^3}$ & {\rm -}\\ 
  $\Mw$  & mass of domain wall & \Eqref{Mwdef}& {\rm kg}\\ 
  $\Rw$ & resistance due to domain wall &  &$\Omega$ \\ 
  $\rhos$ & resistivity due to spin structure & \Eqref{rhos}  &$\Omega$ m \\ 
  $\RS$ & resistance due to spin structure & $ \rhos {L}/{A}$ & $\Omega$ \\ 
\end{tabular}
\end{center}
\end{table}
\begin{table}[htb]
\begin{center}
\begin{tabular}{cccc} 
  $e$   & electron charge ($e<0$) & & C \\ 
  $j$   & current density & & {\rm A/m$^2$}\\ 
  $\js$   & spin current density (without spin length $\hf$) & $j_+ -j_-$ & {\rm A/m$^2$}\\ 
  $\jc$   & threshold current density & & {\rm A/m$^2$}\\ 
  $\jci$   & intrinsic threshold current density & \Eqref{jci} & {\rm A/m$^2$}\\ 
  $\DOS$  & electron density of states at Fermi energy per volume & ${mk_{F\sigma}}/({2\pi^2 \hbar^2})$ & {\rm 1/(J m$^3$)}\\ 
  $\se$  & electron spin density (without spin length $\hf$) & $n_+ - n_-$ & {\rm 1/m$^3$}\\ 
  $\sev$  & electron spin density vector & & {\rm 1/m$^3$}\\ 
  $\svtil$  & spin density vector in gauge transformed frame & & {\rm 1/m$^3$}\\ 
  $\tau$  & electron lifetime (spin-dependent in \S\ref{SEC:singlespin})& & {\rm s}\\ 
  $P$  & spin polarization of current &${\js}/{j}$ & - \\ 
  $\spol$  & spin polarization of conduction electron & $\Jsd S$  & J \\ 
  $\vs$  & drift velocity of electron spin & $Pj{a^3}/({2eS})$  & m/s \\ 
  $\vc$  & $\vs$ at intrinsic threshold & ${\Kp\lambda S}/({2\hbar})$  & m/s \\ 
  $k_{F\sigma}$  & Fermi wavelength of conduction electron with spin $\sigma=\pm$ & & 1/m \\ 
  $\epsilon_{F}$  & Fermi energy of conduction electron & & J \\ 
  $\epsilon_{F\sigma}$ & Fermi energy with spin splitting included ($\sigma=\pm$) & $\ef+\sigma\spol$ & J \\ 
  $a$   & lattice constant & & {\rm m}\\ 
  $L$   & system length along $\xw$ direction & & m\\ 
  $A$   & cross sectional area of system & & m$^2$\\ 
%
\hline
\end{tabular}
\end{center}
\end{table}

\section{Introduction}

\subsection{Magneto-electric effects and devices}

\begin{figure}[htb]\begin{center}
\includegraphics[width=0.3\linewidth]{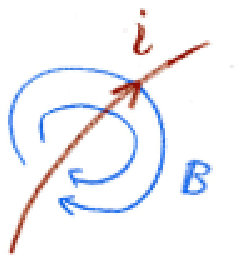}
\includegraphics[width=0.3\linewidth]{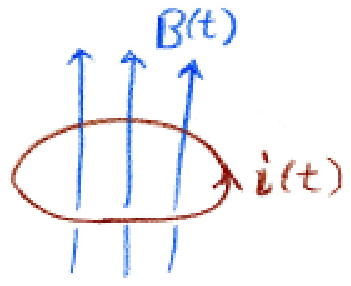}
\caption{
Amp\`ere and Faraday's laws.
\label{FIGmaxwell} }
\end{center}\end{figure}

\begin{figure}[htb]\begin{center}
\includegraphics[width=0.3\linewidth]{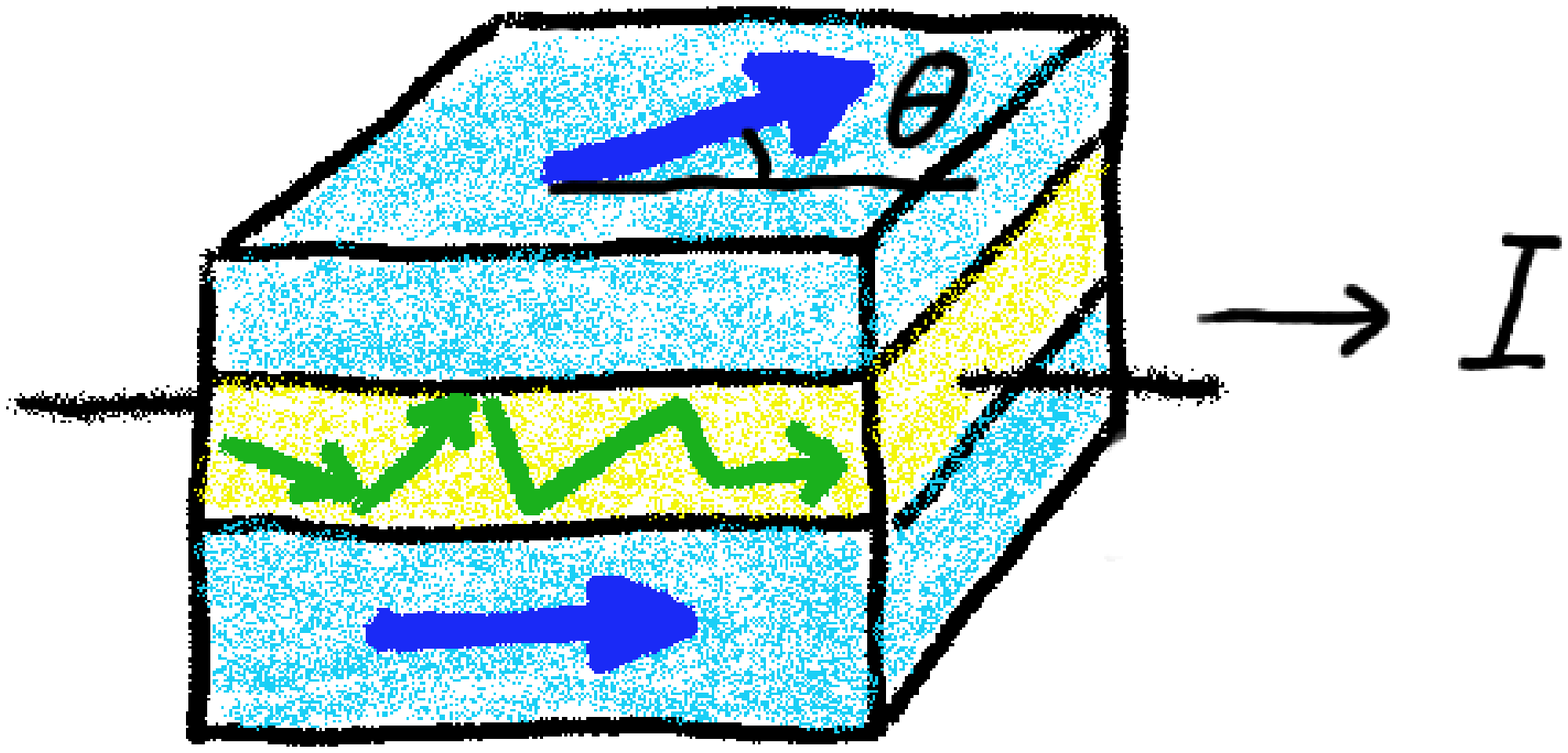}
\includegraphics[width=0.3\linewidth]{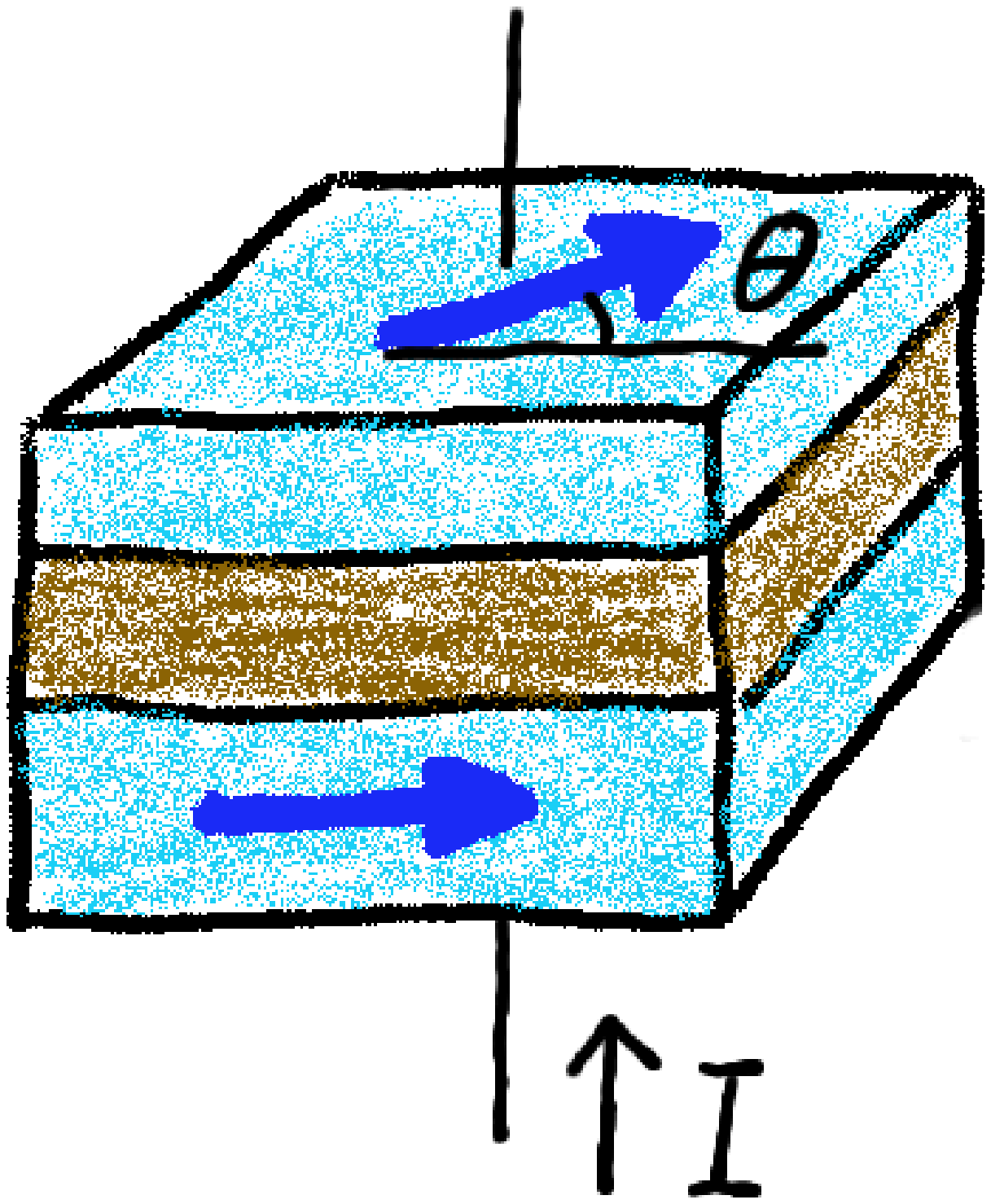}
\caption{
GMR and TMR systems that realize large magnetoresistance using $s$-$d$ interaction between localized spin and conduction electron.
\label{FIGgmr} }
\end{center}\end{figure}

Present information technology is based on electron transport and magnetism.
Magnetism has been most successful in high-density storages such as hard disks.
For integration of magnetic storages into electronic circuits, mechanisms are necessary to convert electric current/voltage into magnetic information and vice versa.
The most common and oldest electro-magnetic coupling is the one arising from Maxwell's equations.
Amp\`ere's law or Oersted's law, discovered in the early nineteenth century, describes the magnetic field created by an  electric current (Fig. \ref{FIGmaxwell}). 
This field can be applied to write information in magnetic information storages.
In fact, this mechanism is so far the only successful mechanism used in commercial high-density magnetic devices. 
On the other hand, Faraday's law provides us means to convert magnetic information into electric current or voltage, for instance, detecting magnetic information by scanning a read head (a coil) on the stored magnetic bits.
This mechanism is not, however, very useful in high density storages, and various magnetoresistive effects based on  solid-state systems have been discovered and applied in the late twentieth century, such as anisotropic magnetoresistance (AMR), giant magnetoresistance (GMR), and tunneling magnetoresistance (TMR)  effects (Fig. \ref{FIGgmr}). 
AMR is a resistivity dependent on the angle between the magnetization and the electric current, discovered in 1857\cite{Thomson1857}.
It arises from the coupling of magnetization and electrons' orbital motion due to spin-orbit interaction \cite{McGuire75}.
The resistivity change is of the order of only a few percent, but AMR is more efficient than using Faraday induction used in magnetic tape and hard disks in early days.
Magnetic heads with higher sensitivity were developed by use of the GMR effect in thin magnetic multilayers discovered in 1988 \cite{Baibich88,Binasch89}. 
In such multilayers,
a strong magnetization dependence of the resistivity arises from the spin-dependent scattering of electrons at the interface between a thin ferromagnetic layer and nonmagnetic metallic layers. 
A. Fert and P. Gr\"unberg were awarded the Nobel Prize in 2007 for the discovery of the GMR effect. 
Quite recently GMR heads are being replaced by even more efficient TMR heads, where the nonmagnetic layer is replaced by an insulating barrier \cite{Miyazaki95,Yuasa04,Parkin04}.
These rapid developments of read-out mechanisms by use of solid state systems have made possible so far the rapid increase of recording density. 
These magnetoresistances are due to the exchange interaction between localized spin and conduction electrons, arising from the overlap of electron wave functions and their correlation.
Present magnetic devices are therefore one of the most successful outcomes of material science.

\subsection{Magnetization switching by $s$-$d$ exchange interaction}

\begin{figure}[htb]\begin{center}
\includegraphics[width=0.3\linewidth]{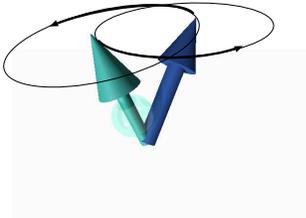}
\caption{The $s$-$d$ interaction induces precession of localized spin and electron spin around each other.
\label{FIGtwospin_rot} }
\end{center}\end{figure}

Electron transport in magnetic metals and semiconductors is modeled by the so-called $s$-$d$ model, where the conduction and magnetization degrees of freedom are separated from each other.
The conduction electrons we consider are non-interacting with each other, but are scattered by spin-independent impurities and also by spin-dependent impurities (resulting in spin relaxation).
The localized spin at position $\xv$ at time $t$ is described by a variable $\Sv(\xv,t)$.
The localized spin is related to magnetization as
\begin{equation}
\Mv(\xv)=  g \frac{e\hbar}{2m a^3}\Sv(\xv)
 =-\frac{g\muB}{a^3} \Sv(\xv)
 = -\frac{\hbar\gyro}{a^3}\Sv(\xv) , 
\end{equation}
 where 
$\muB\equiv \frac{|e|\hbar}{2m}$ 
is the Bohr magneton, $g=2$ is the g-factor, and
 $\gyro\equiv \frac{g\mub}{\hbar}(>0)$ is the gyromagnetic ratio. 
The electron charge $e$ is negative.
In this paper $\Sv$ is treated as a classical variable, since quantum fluctuation of $\Sv$ is blocked by the strong exchange interaction, $J$, among localized spins, and besides,
 we are interested in a semi-macroscopic object made of many spins, the domain wall. 
The localized spin interacts with the conduction electron by
an $s$-$d$ type exchange interaction (Fig. \ref{FIGtwospin_rot}),
\begin{equation}
\Hsd=-\Jsd \intx \Sv\cdot (c^\dagger \sigmav c).
\end{equation}
Here the conduction electron is represented by creation and annihilation operators $c^\dagger$ and $c$, and
$\sigmav\equiv(\sigma_x,\sigma_y,\sigma_z)$ where $\sigma_i$ are $2\times 2$ Pauli matrices satisfying the commutation relation 
$[\sigma_i,\sigma_i]=2i\epsilon_{ijk}\sigma_k$.
The description based on this $s$-$d$ exchange picture is an effective one, treating localized spin $\Sv$ as a variable independent from the conduction electrons, i.e., neglecting the hopping of $d$ electrons that form localized spin.
Still, we will take this effective $s$-$d$ model as the starting system for this investigation, and will not concern ourselves with the microscopic origin of the local moment.

The $s$-$d$ interaction is a coupling in spin space, which is decoupled from real space (as far as spin-orbit interaction is neglected).
Nevertheless, this spin coupling can affect charge transport if 
the localized spin has inhomogeneity, and various magnetoresistive effects such as GMR arise.
 
\begin{figure}[htb] \begin{center}
\includegraphics[width=0.6\linewidth]{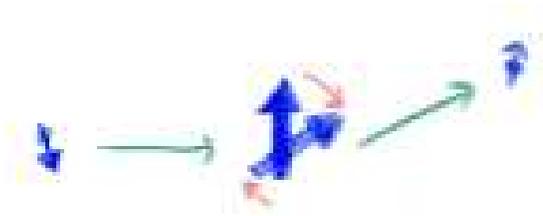}
\caption{$s$-$d$ exchange interaction describes the exchange of spin angular momentum between conduction electron and localized spin. 
The idea of spin flip using spin-polarized current is thus quite natural. 
\label{FIGexchange}
}
\end{center}\end{figure}

Since this exchange coupling describes the exchange of spin angular momentum, the idea of spin reversal by spin-polarized current arises naturally.
Namely, the injection of electron spin polarized in the opposite direction to a localized spin will cause flip of localized spin (Fig. \ref{FIGexchange}). 
This simple idea was integrated into realistic magnetization switching of thin film magnets by Slonczewski \cite{Slonczewski96} and Berger \cite{Berger96}.
The current-induced phenomena are expected to be applied to  memory devices like magnetoresistive random access memory (MRAM) that operates without magnetic field, and 
intensive studies on pillar systems and domain walls have then started at the end of the last century.
Domain-wall racetrack memory proposed by Parkin is one possibility of the high-density storage \cite{Parkin08,Hayashi08}.

Compared with switching by use of the   
Amp\`ere's field, the currend-induced magnetization switching 
has a great advantage in downsizing.
The field created by the Amp\`ere's law is proportional to the current, which decreases when the system size is reduced with a constant current density. 
Therefore, higher current density is necessary for the Amp\`ere's mechanism in smaller systems. 
In contrast, the current-induced switching rate is determined by the current density and material parameters, such as $s$-$d$ coupling, spin relaxation, and anisotropy energies, and the efficiency remains constant when the system size is reduced.
This is why current-induced switching becomes essential in high density devices.

In this paper, we review recent developments in the theory of current-driven domain wall motion.
In \S\ref{SEC:intro_st} and \S\ref{SEC:history}, the phenomenological argument and a brief history of current-induced domain wall dynamics are presented.
The theoretical study starts in \S\ref{SEC:spin} from the description of localized spin by the Lagrangian formalism.
Collective coordinates to describe domain wall are introduced in \S\ref{collective}.
We consider the case of a rigid one-dimensional (planar) domain wall, sometimes called the transverse wall. 
The conduction electrons and $s$-$d$ interaction are introduced in \S\ref{SEC:electron}.
The equation of motion of a domain wall coupled to the conduction electrons is derived in \S\ref{SEC:dweq0}.
The equation is expressed using the conduction electron spin density, which acts as the effective field on the localized spins.
Then the explicit equation of motion is obtained by calculating the conduction electron spin density in \S\ref{SEC:spindensity}.
The torque and force acting on the spin structure are obtained in \S\ref{SEC:torqueandforce}.
The adiabatic limit is briefly discussed in \S\ref{SEC:adiabaticspinlag}.
The full equation of motion of a domain wall is finally obtained in \S\ref{SEC:dweq} and is solved in \S\ref{SEC:sol}.
The case of a wall having vorticity, called the vortex wall, is considered briefly in \S\ref{SEC:vortex}.
The analysis in \S\ref{SEC:spin} to \S\ref{SEC:sol} is the main result of the paper, aiming at presenting our calculational method in a self-contained way.

Another approach to current-induced domain wall dynamics is to use the Landau-Lifshitz-Gilbert (LLG) equation taking account of the effect of current (as done in \S\ref{SEC:eqfromLLG}).
To do this, we need to calculate microscopically the torque induced by the electron. This is done in \S\ref{SEC:singlespin}.
In \S\ref{SEC:dwtransport}, electron transport properties in the presence of spin structures are discussed. 
The transport properties are shown to be the counter action of the current-induced forces.
Another counter action of current-induced magnetization dynamics, the pumping of current and spin current by magnetization dynamics, is briefly argued in \S\ref{SEC:ishe}.
Details of the Lagrangian formalism of spin and Green's functions are explained in the appendices.

\section{Current-driven domain wall dynamics}
\label{SEC:intro_st}

\subsection{Switching using domain wall motion}

\begin{figure}[tbh]
\label{DW}
\begin{center}
\includegraphics[width=0.4\linewidth]{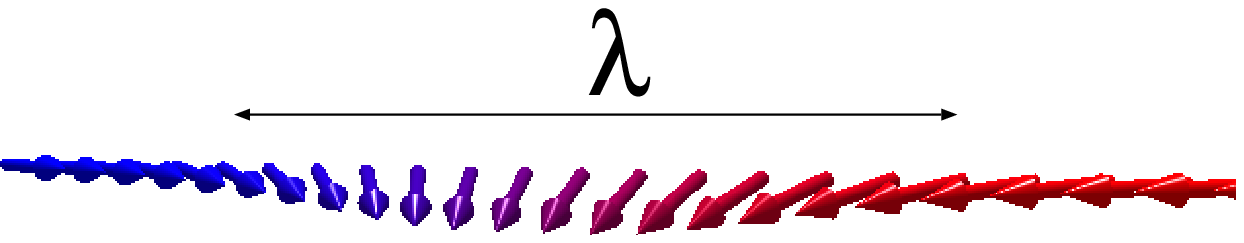}
\includegraphics[width=0.4\linewidth]{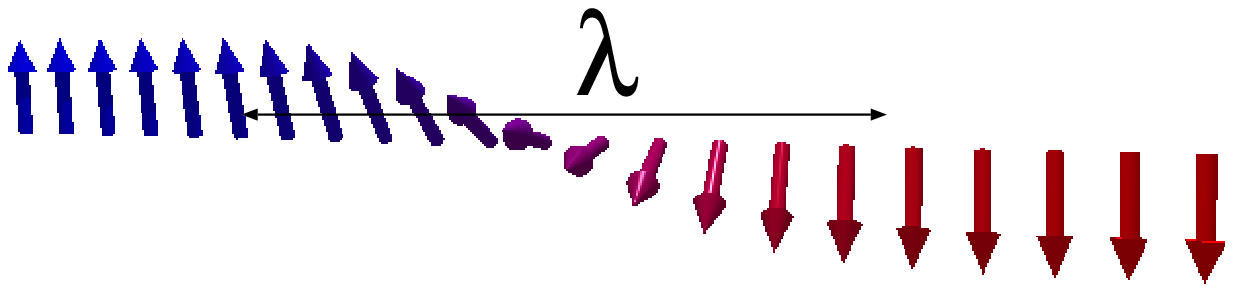}
\caption{ 
\label{FIGdw}
 Illustration of a N\'eel wall (left) and a Bloch wall (right), where the magnetic easy axis is along and perpendicular to the wall direction, respectively. 
 $\lambda$ is the thickness of the wall. 
}
\end{center}
\end{figure}

The idea of switching spin structure by electric current using the $s$-$d$ interaction was first discussed by Berger in 1978  \cite{Berger78}, much earlier than works by Slonczewski and Berger in 1996.
His idea was to push a domain wall by current.
A domain wall is a twisted spin structure where spins gradually rotate, which appears between two magnetic domains  \cite{Kittel49,Malozemoff79,Chikazumi97,Hubert98,Marrows05,Klaui08}
(Fig. \ref{FIGdw}).
The thickness of the wall, $\lambda$, is determined by the competition between the 
ferromagnetic exchange energy (between localized spins), $J$, which aligns the neighboring spins, and 
the 
magnetic anisotropy energy in the easy axis, $K$, which tends to reduce the wall thickness to minimize the deviation of spins from the easy axis,
as $\lambda =\sqrt{J/K}$.
(Here $J$ has dimensions of J/m$^2$.)
Thus $\lambda$ depends on the material and also on the sample shape 
since $K$ depends on the shape.
In the case of 3$d$ transition metals such as iron and nickel 
$\lambda\simeq 500\sim1000$ \AA \cite{Hong95,SMYT04}, and $\lambda\sim 150$ \AA\ 
in Co thin films \cite{Gregg96}.
This length scale is very large compared 
with the length scale of the electron, 
$k_{F}^{-1}\sim O(1 $ \AA$)$ ($k_{F}$ being the Fermi wave length of the 
electron).

\begin{figure}[htb] \begin{center}
\includegraphics[width=0.4\linewidth]{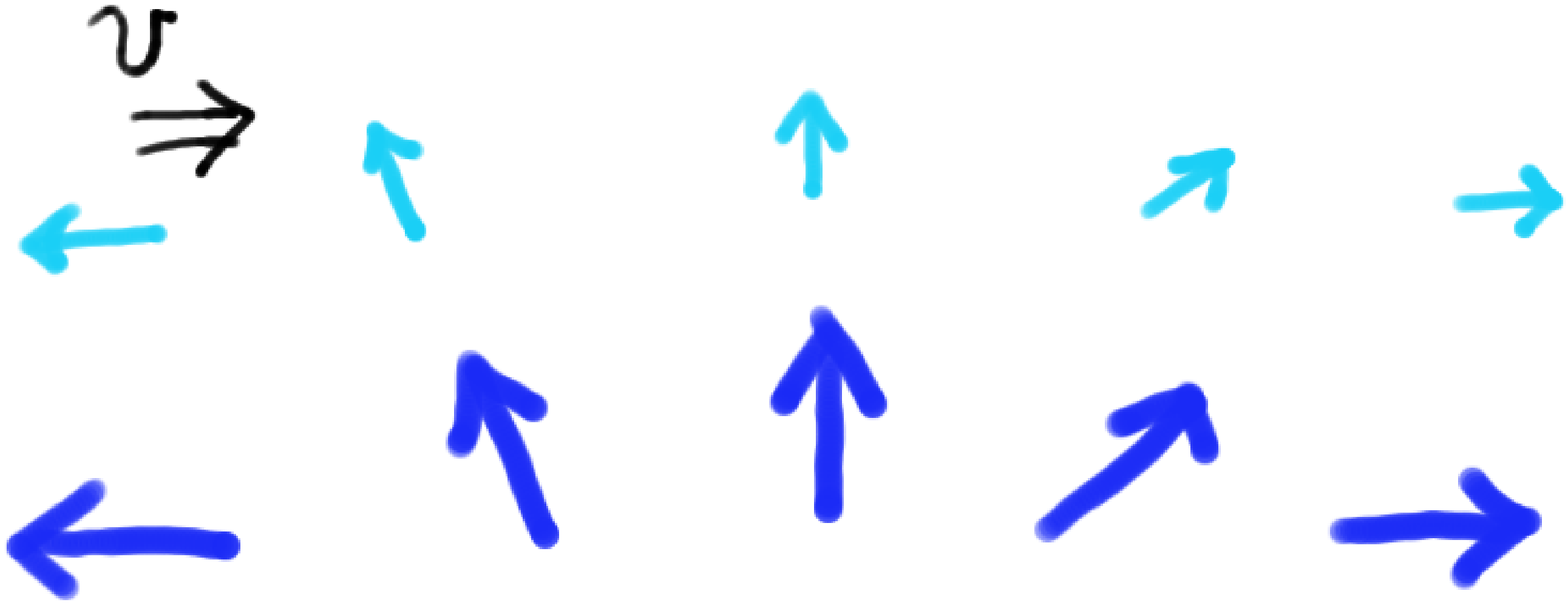}\\
\includegraphics[width=0.4\linewidth]{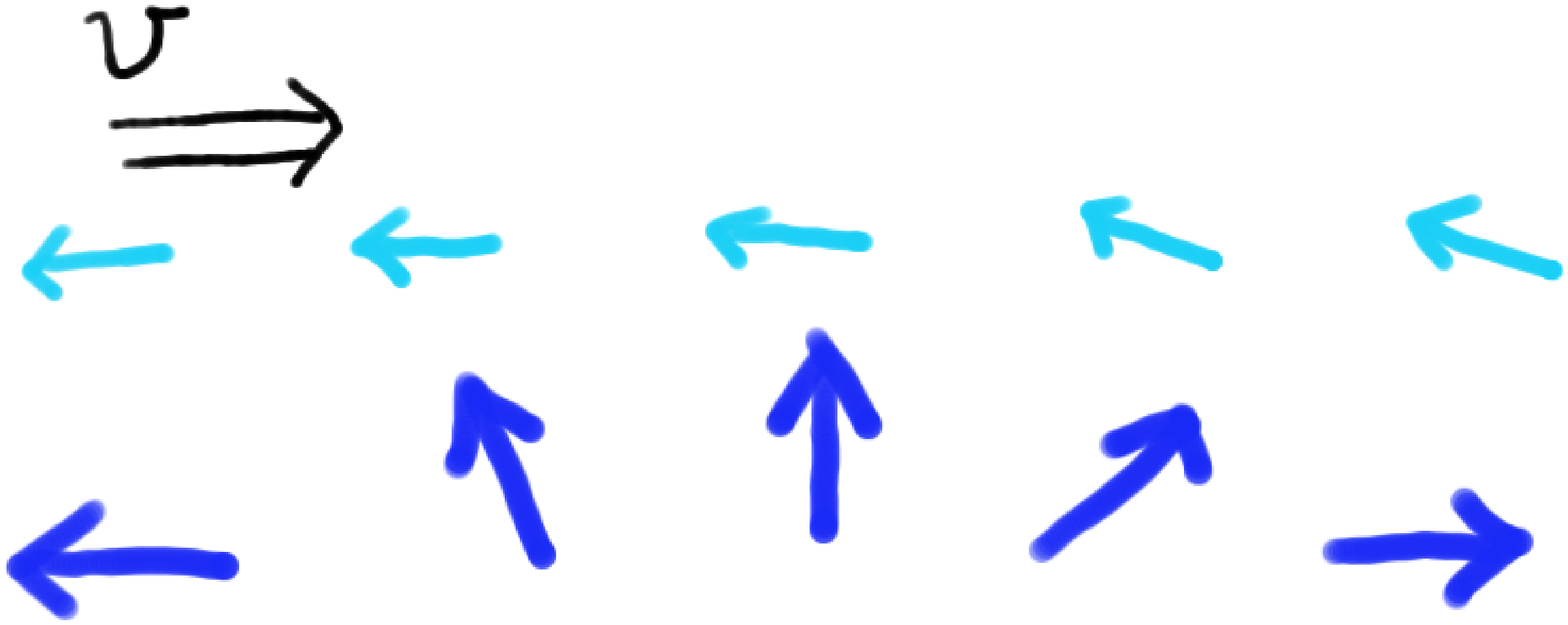}
\caption{
Schematic figure describing a conduction electron going through a domain wall. 
The electron spin is denoted by small arrows and localized spins are denoted by large arrows.
The upper figure is an adiabatic case, where the electron is slow and so can accommodate its spin as electron passes through the wall. ($v$ is velocity of the electron.)
The lower figure is the case of a fast electron, where the electron spin cannot follow the domain wall structure.
\label{FIGdwad}
}
\end{center}\end{figure}

Let us consider how the motion of the wall is induced by electric current.
When an electron is injected into the domain wall, there are basically two possibilities, reflection or transmission.
In the transmission process, there are again two possibilities as indicated in Fig. \ref{FIGdwad}, depending on the electron speed. 
If the electron is fast enough, it will pass through the wall without spin rotation, while the electron spin will be flipped by exchange coupling during the transmission if electron is slow.

Corresponding to the above possibilities, there are two different mechanisms of domain wall motion induced by electric current and exchange interaction.
The first one is due to reflection of the electron.
The exchange interaction 
describes a spin-dependent potential created by a localized spin $\Sv$, and so the electron is scattered if there is inhomogeneity, $\nabla \Sv$.
Namely, the electron feels a force from domain wall,
\begin{equation}
F\propto Rj,\label{Fv}
\end{equation}
where $R$ is the reflection probability for the electron.
(For correct expression, see Eq. (\ref{Fdef}) and \Eqref{Fandj}.)
From the conservation of linear momentum, electron scattering by a domain wall indicates that the wall must move. 
This process is an exchange of linear momentum, and is  sometimes called a momentum transfer process.
This force is strong if the domain wall is thin, since then the electron scattering is significant. 
In reality, in most experiments, domain walls are thick, and the exchange interaction is strong, and so most of the electrons do not get scattered
(except for systems with very thin walls \cite{Feigenson07} 
or in nano scale contacts \cite{Garcia99}). 
This case is called the adiabatic case, and is suggested in experiments by small resistivity due to domain walls \cite{Marrows05}. 
(For conditions of adiabaticity, see \S\ref{SEC:adiabatic}.)
Thus this force is not a major driving mechanism in most cases as discussed by Berger \cite{Berger92}.

\begin{figure}[htb] \begin{center}
\includegraphics[width=0.4\linewidth]{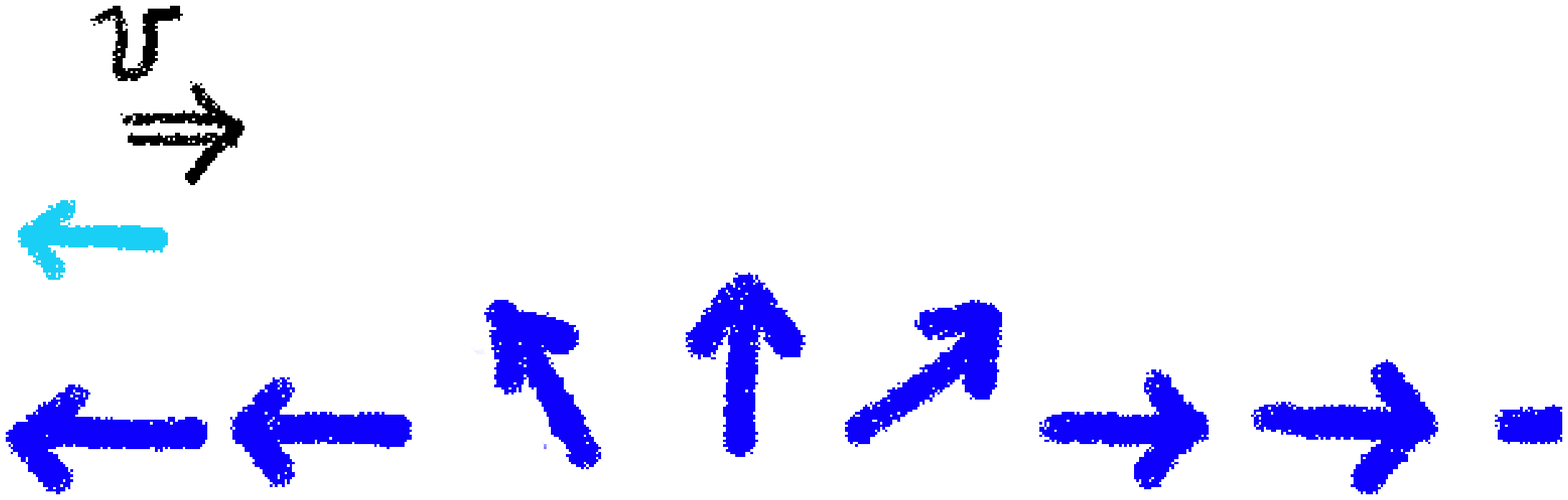}\\
\includegraphics[width=0.4\linewidth]{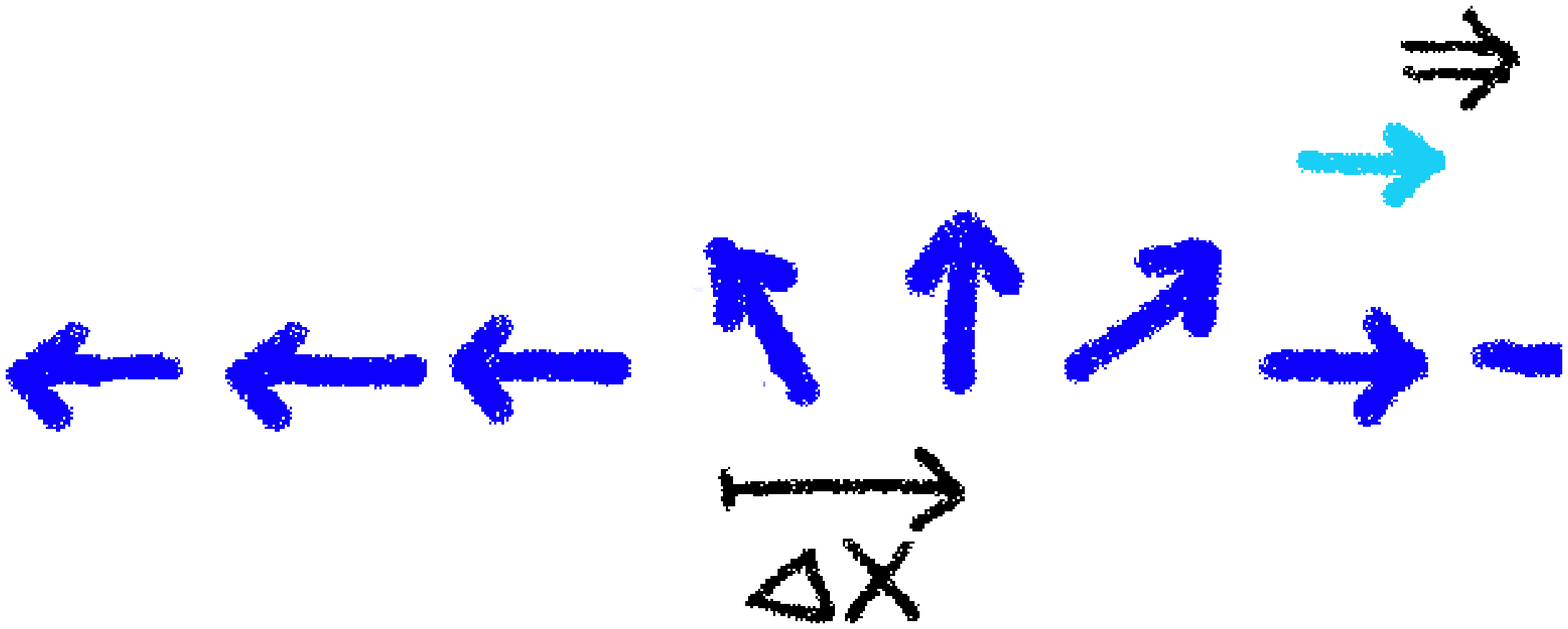}
\caption{Domain wall motion by injection of spin-polarized electron in the adiabatic limit (spin transfer effect).
\label{FIGdwmotion}
}
\end{center}\end{figure}

The other mechanism arises from  the adiabatic  electron transmission. 
As seen in Fig. \ref{FIGdwad}, the spins of slow electrons are flipped on transmission. 
The angular momentum of a conduction electron has changed by the amount $\hbar \times(\hf-(-\hf))=\hbar$ when one electron goes through.
From the conservation of angular momentum, the wall needs to shift by a distance of 
$\Delta X= \frac{a}{2S}$ ($a$ is the lattice constant) 
(Fig. \ref{FIGdwmotion}).
When a steady current density $j$ is injected, the wall then moves at speed of 
\begin{equation}
\vw=\Delta X \frac{Pja^2}{e}=\frac{Pja^3}{2eS}, \label{STv}
\end{equation}
where $P\equiv\frac{j_+-j_-}{j_++j_-}$ is the spin polarization of current ($j_\pm$ represents the current carried by the electron with spin $\pm$ and $j=j_+ +j_-$).  
This is so called spin-transfer mechanism of domain wall motion.
This argument applies to any adiabatic spin structure, and 
we can see that any spin structure tends to flow at the speed given by Eq. (\ref{STv}).
The direction of wall motion is the same as that of the electron, and so is opposite to the current (since the electron charge $e$ is negative).

\begin{figure}[htb]\begin{center}
\includegraphics[width=0.7\linewidth]{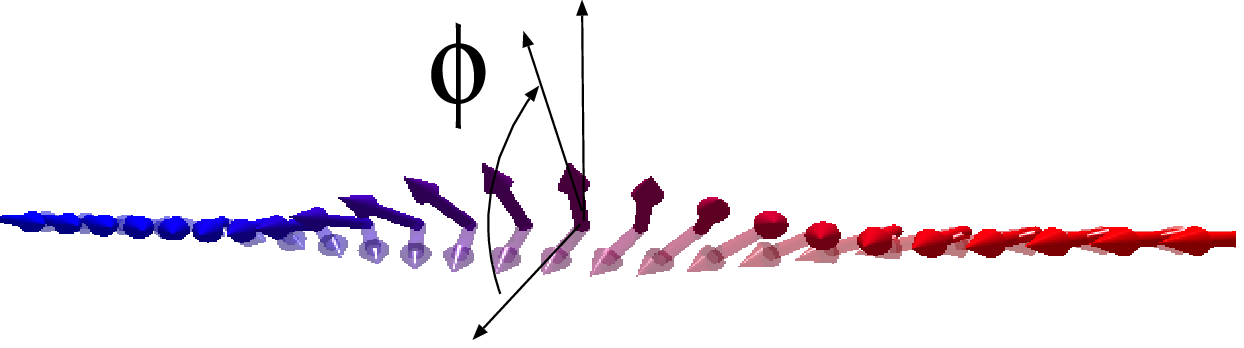}
\caption{
Domain wall configuration where the easy axis is along the wall direction, called the N\'eel wall. 
The angle out of the easy plane is $\phi$, which is a canonical momentum of the wall and which plays an essential role in dynamics. The dynamics of $\phi$ (called the chirality of the wall)
is discussed in Refs. \cite{Braun96,TT96}.
\label{FIGneelw} }
\includegraphics[width=0.7\linewidth]{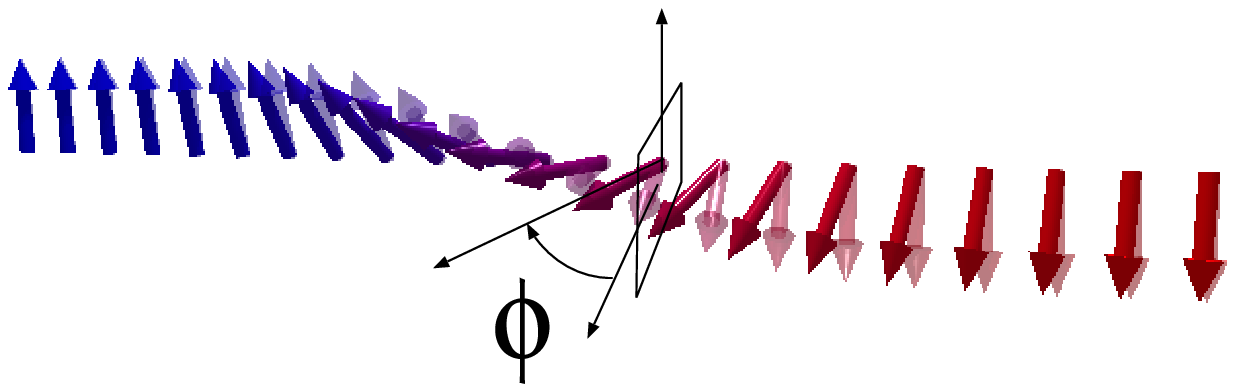}
\caption{
Domain wall configuration where the easy axis is perpendicular to the wall direction, called the Bloch wall. 
The easy plane is perpendicular to the wall direction as denote by a small square. The angle out of the plane, $\phi$, corresponds to the tilt of wall in the wall direction. 
\label{FIGblochw} }
\end{center}\end{figure}

In reality, two driving mechanisms exist and so the two equations, Eq.  (\ref{Fv}) and Eq. (\ref{STv}), need to be coupled.
One may naively guess simply that
$\Mw(\ddot{X}+\alpha'\dot{X})=F$ and $\dot{X}=\frac{Pja^3}{2eS}$ (where $\Mw$ is the wall mass and $\alpha'$ is the friction coefficient), but these are not correct, since the wall is not a simply a particle but has internal degrees of freedom. 
(One may notice also that these two equations for constant current have no solution.)
 Berger has proposed, based on phenomenological arguments, the  correct equation in his series of papers \cite{Berger78,Berger92}. 
Actually, a force on a domain wall induces not a simple acceleration ($\ddot{X}$) but an angle  out-of the easy plane, $\phi$ (see Figs. \ref{FIGneelw} and \ref{FIGblochw}).
This has been known in the case where a magnetic field $B$ is applied along the easy axis.
Let us visualize the motion in this case.
Under a magnetic field, each spin constituting the domain wall starts to precess around the field according to a torque equation of motion, 
\begin{equation}
\dot{\Sv}= \gyro \Sv\times \Bv, \label{precess}
\end{equation}
where $\gyro \ (>0)$ is the gyromagnetic ratio.
(We will use magnetic flux density $\Bv$ instead of magnetic field $\Hv$ ($\Bv=\mu(\Hv+\Mv)$), and $\Bv$ may be called the  "magnetic field", as in Kittel's textbook \cite{Kittel86}.)
The spin thus changes its direction perpendicular to $\Sv$ and $\Bv$.
The translational motion of the domain wall is therefore coupled with the out-of plane dynamics, and this is an essential and complicated feature of the wall dynamics.
The correct equation under a force is  given by \cite{Chikazumi97,TK04,KT06} (when friction is neglected)
\begin{equation}
\dot{\phi} = \eta F, \label{eq1}
\end{equation}
where $\eta$ is a numerical factor.
(It turns out that $\eta=\frac{\lambda}{\hbar \Nw S}$, where $\Nw$ is number of spins inside the wall.)
The effect of current in the adiabatic limit is to induce a wall velocity as we saw, but the wall velocity is also related to the hard-axis anisotropy energy (if it exists), since the translational motion of the wall needs to be induced by the effective magnetic  field perpendicular to the wall plane,  
again due to the precession equation (\ref{precess}). 
The other equation for the wall is therefore written as (without friction)
\begin{equation}
\dot{X} = \frac{P ja^3}{2eS}+\vc \sin2\phi, \label{eq2}
\end{equation}
where $\vc$ is a parameter that determines the hard axis anisotropy energy (\Eqref{vcdef}).
(Here, assuming hard axis anisotropy of the standard $\sin^2\phi$ type, the effective field perpendicular to the wall plane is given as  
$\propto \sin2\phi$.)
These two equations \Eqsref{eq1}{eq2} are not yet correct, since they do not include the effect of damping (friction), which is known to be quite essential in spin dynamics \cite{Chikazumi97}.
Damping can be incorporated phenomenologically by the prescription by Gilbert or derived from spin relaxation processes \cite{Tserkovnyak06,KTS06} (see \Eqsref{DWeq_a1}{DWeq_b1}).

Let us see how the wall motion changes when the sign of parameters changes.
If $\Jsd$ is negative, the spin-transfer torque gives a wall velocity opposite to the case of $\Jsd>0$, resulting in wall motion in the current direction (if carrier has negative charge). 
Mathematically, this is because $\spol$, $P$, and $\js$ change sign with $\Jsd$.
The forces due to non-adiabaticity and spin relaxation ($\Fna$ and $\Fbeta$) remain opposite to the current.
In the case of a hole in semi-conductors, the charge of the carrier is positive, and so the reflection force is along the current direction. 
The spin-transfer torque pushes the wall in the same direction  as in the electron case if $\Jsd$ has the same sign. For instance, in GaMnAs, exchange interaction between a $p$ hole and the $d$ localized spin is negative ($J_{pd}<0$) \cite{Dietl01} and so spin-transfer motion is opposite to the hole flow and the current.  

As is seen from the above arguments on spin transfer torque, what matters most is spin polarization of current, 
\begin{equation}
P=\frac{\sigma_+-\sigma_-}{\sigma_++\sigma_-}
 =\frac{n_+\tau_+ -n_-\tau_-}{n_+\tau_+ +n_-\tau_-},
\end{equation}
where $\sigma_\pm $, $n_\pm$, and $\tau_\pm$ are the Boltzmann conductivity, density, and lifetime of electrons with spin $\pm$, respectively. 
This polarization $P$ should not be confused with other definitions of spin polarization, such as 
polarization of electron density, 
$\frac{n_+ -n_-}{n_+ +n_-}$, or that of density of states, 
$\frac{\DOS_+-\DOS_-}{\DOS_++\DOS_-}=\frac{\kfu-\kfd}{\kfu+\kfd}$ (in three-dimensional free electron model).
It would be interesting to control the sign of $P$ in experiments and see how the wall motion changes.

When friction is included, these two equations work fine for qualitative argument close to the adiabatic limit. 
However, these phenomenological equations are too naive for quantitative study.
For instance, the definition of $\phi$ in \Eqref{eq1} is not clear for a domain wall whose $\phi$ can be position dependent, and the explicit form of $F$ is not given.
In addition, to include the effects of spin relaxation and non-adiabaticity correctly, 
one needs to use a many-body formalism treating the electron quantum mechanically.
This is what we are going to do in this paper.
Based on a Lagrangian formalism, the equation of motion is derived without worrying about complicated dynamics of each spin. 
(The result is \Eqref{DWeq3}.)
Mathematically, the coupling between the translational and out-of-plane dynamics is expressed by the fact that the canonical momentum of the wall is the average of $\phi$ (and not $\Mw\dot{X}$), a fact arising from SU(2) commutation relation of spins.

\section{Brief history}
\label{SEC:history}

\subsection{Berger's theories}

Berger considered a domain wall under an electric current, and saw that the $s$-$d$ exchange coupling between the localized spin and conduction electron spin is the dominant interaction that drives the wall under a current in the case of a thin film 
(e.g.,  thickness less than $\sim 0.1\ \mu$m), 
where the effect of an induced magnetic field can be neglected \cite{Berger78}.
In 1984 \cite{Berger84,Hung88}, he studied the effect of the force arising from the reflection of conduction electrons by the domain wall caused by this exchange coupling. 
This force was obtained as 
\begin{equation}
F=\Ms a_f(j-b_f \vw),
\end{equation}
 where 
$\Ms$ is the saturation magnetization, $a_f$ and $b_f$ are coefficients 
 introduced phenomenologically, $j$ is the current, and $\vw$ is the wall velocity.
The effect of the force was found to be small in most cases due to a very small reflection probability because the wall thickness is usually large compared with the Fermi wavelength.
In 1978 \cite{Berger78,Berger86}, he argued that the exchange interaction produces a torque,
\begin{equation}
\torque=-\frac{\hbar}{e}j,
\end{equation}
 (assuming full spin polarization of the conduction electron), 
 which tends to cant the wall magnetization out of the easy plane (angle $\phi$) and eventually induces a continuous rotation $\phi$ of a pinned wall under a large current \cite{Berger86}.
This torque was found to push the wall by a different mechanism from the exchange force, which turned out to be the dominant driving mechanism \cite{Berger92}.
The torque is nowadays called the spin transfer torque, after Slonczewski \cite{Slonczewski96}.
Based on the idea of torque-driven wall motion, an experimental study was carried out in 1993 \cite{Salhi93} on a thin film of Ni$_{81}$Fe$_{19}$. 
There, a domain wall velocity of 70 m/s
was reported at the current density of 
$1.35\times 10^{10}$ A/m$^2$ applied as a pulse of duration 0.14 $\mu$s.

There has been a renewal of interest on the current-induced domain wall motion for about a decade.
Recent experimental studies have been carried out on submicron-size wires, and the domain wall motion induced by current has been confirmed \cite{Grollier02,Grollier03,Tsoi03,Klaui03}.
The current density necessary for wall motion turned out to be rather high, of order of $10^{12}$ A/m$^2$. 
Measurement of the domain wall velocity was carried out by Yamaguchi et al.  \cite{Yamaguchi04} by observing wall displacement by use of magnetic force microscopy (MFM) after each current pulse of strength $1.2\times 10^{12}$ A/m$^2$ and duration of 5 $\mu$s.
The average velocity was found to be $2\sim 6$ m/s.
Other experiments also indicate a rather slow average wall velocity, of order of a few m/s under a steady current \cite{Klaui05}.

\subsection{Recent theories}

Those experiments motivated theoretical studies to look into the problem in more detail. 
Microscopic derivation of the equation of motion of the domain wall under current was carried out by Tatara and Kohno \cite{TK04,KT06,TKS08}.
They considered a planar (one-dimensional) wall and described the wall by the two collective coordinates, $X$ and $\phiz$, i.e. within Slonczewski's description \cite{Slonczewski72}. 
The variable $X$ represents the position of the wall, and $\phiz$ describes the tilt of the wall plane.  
Considering a small hard-axis anisotropy case, other deformation modes than $\phiz$ (such as change of wall width) were neglected (rigid wall approximation).  
The equation of motion with respect to $X$ and $\phiz$ was derived including the effect of conduction electrons via the $s$-$d$ exchange interaction.
The electron carrying a current was treated by the use of a non-equilibrium (Keldysh) Green's function.
Qualitatively, the equation of motion derived was indeed the same as that obtained by Berger long ago \cite{Berger84,Berger92}, namely, 
Eqs. (\ref{eq1}) (\ref{eq2}) with the damping term included. 
Berger's theory was thus confirmed by microscopic calculation.
The microscopic formalism made it possible for the first time to calculate the spin transfer torque and force systematically 
by representing these quantities by Green's functions and Feynman diagrams.
Based on the obtained equation of motion, the wall motion under steady current was studied. 
It was found that in the adiabatic limit, where the reflection force can be neglected, and in the absence of spin relaxation of conduction electrons ($\betasf$ term below), 
there is an threshold current determined by the hard-axis magnetic anisotropy energy, $\Kp$ as
\begin{equation}
 \jci =  \frac{e S^2}{a^3 \hbar P} \Kp \lambda.\label{jci}
\end{equation}
This is the intrinsic pinning of the wall arising from the "pinning" of $\phiz$ \cite{TK04}.
At larger currents, the wall gets depinned and its velocity becomes proportional to the spin current (spin polarization of the current flow), $\js$, as is required from the angular momentum conservation.

Of practical importance would be the wall motion below the intrinsic threshold.
Actually, the wall moves over quite a large distance 
(even in the absence of $\beta$ term) if there is no pinning.
According to the torque equation, when $j\ll\jci$ the current induces a tilt of the wall, $\phiz\sim \frac{j}{2 \jci}$ (see \Eqref{DWeq3nodim}).
This tilt is associated with wall translation (by the second equation of \Eqref{DWeq3nodim} without pinning and $\betaw$) over a distance of 
\begin{equation}
\Delta X \sim  \frac{\lambda}{2\alpha}\frac{j}{\jci} 
\;\;\;\;\; (j\ll\jci).  
\end{equation}
Since $\alpha$ is very small, e.g., $\alpha\sim 0.01$, this distance can be quite large compared with $\lambda$  even for a current 10\% of the intrinsic threshold.
Such motion at very low current would be enough for applications. 
One should note, however, that the wall below threshold goes back to the original position as soon as the current is cut.
One needs therefore a pinning site to maintain the wall displacement.

Numerical simulation was performed based on an equation of motion of each localized spin
by including the spin-transfer torque term in the adiabatic limit  \cite{Thiaville04}. 
The equation of motion is given by
\begin{equation}
\dot{\Sv}= \gamma \Bvs \times \Sv
-\frac{\alpha}{S} \Sv\times\dot\Sv
-\frac{a^3}{2eS}(\jsv\cdot\nabla)\Sv. \label{LLG1}
\end{equation}
Here $\Bvs$ is the effective field arising from the spin Hamiltonian, and $\alpha$ represents damping.
The equation 
$\dot{\Sv}=  \gyro \Bvs\times \Sv
-\frac{\alpha}{S} \Sv\times\dot\Sv$ has been well-known as the  Landau-Lifshitz-Gilbert equation describing magnetization dynamics in a magnetic field $\Bvs$. 
Spin-transfer torque from current is represented by the last term of \Eqref{LLG1}.
The simulation result was similar to the analytical (collective-coordinate) study, indicating the existence of an intrinsic threshold current. 
The motion of the domain wall under magnetic field and spin-transfer torque was solved in Ref.  \cite{Li04st}.

\subsubsection{The Landau-Lifshitz-Gilbert equation under current}
\label{SEC:LLGequation}

Later Zhang and Li  \cite{Zhang04} and Thiaville et al. \cite{Thiaville05} proposed to add a new torque term in the equation, which is perpendicular to the spin-transfer torque. 
 After Thiaville et al. \cite{Thiaville05}, 
we call this torque term the $\beta$ term.
Zhang and Li argued that the $\beta$ term 
arises from spin relaxation of conduction electrons \cite{Zhang04}. 
Thus the phenomenological equation of motion of localized spin under current becomes  
\begin{equation}
\dot{\Sv}= \gamma \Bvs \times \Sv
-\frac{\alpha}{S} \Sv\times\dot\Sv
-\frac{a^3}{2eS}(\jsv\cdot\nabla)\Sv
- \frac{a^3 \betasf}{2eS^2} [\Sv \times (\jsv\cdot\nabla)\Sv]
+\torquev_{\rm na}.
\label{modLLG}
\end{equation}
The fourth term is the new $\beta$ term. 
We explicitly wrote the coefficient $\betasf$ with a suffix $\spinflip$ to show that this term arises from spin relaxation.
The last term, $\torquev_{\rm na}$, denotes the non-adiabatic torque, which is spatially nonlocal \cite{Xiao06,TKSLL07}.
(So far, this nonlocal torque has not been taken account in numerical simulations.)
All these torques are derived in this paper 
based on \Eqref{LLGfull}.
Parameters $\alpha$ and $\betasf$ are calculated in \S\ref{SEC:singlespin}, and the nonlocal torque  $\torquev_{\rm na}$ is given by the nonlocal part of \Eqref{torqueresult}.

\begin{figure}[tbh]
  \begin{center}
  \includegraphics[scale=0.6]{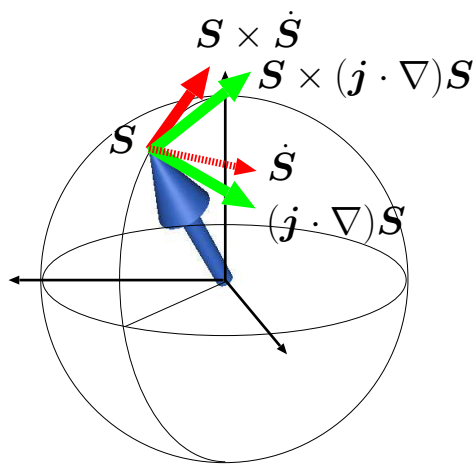}
  \end{center}
\caption{Torques acting on localized spin $\Sv$:
damping torque, $\Sv\times\dot{\Sv}$, 
spin transfer torque, $(\jv\cdot\nabla)\Sv$,
and $\beta$ torque, $\Sv\times(\jv\cdot\nabla)\Sv$.
\label{FIGspintorques}}
\end{figure}

The $\beta$ term turned out to modify the threshold current and the terminal velocity of the wall significantly if it is not small compared with damping parameter $\alpha$ \cite{Zhang04,Thiaville05,TTKSNF06}.
For a rigid domain wall, this $\betasf$ in the  Landau-Lifshitz-Gilbert equation has exactly the same effect as the force due to non-adiabaticity (electron reflection by wall) \cite{Thiaville05,TTKSNF06}.
In other words, the nonlocal torque $\torquev_{\rm na}$ for a  rigid wall is effectively represented by an additional $\beta$ term, $\betana$ \cite{TTKSNF06}.
Therefore, the effective $\beta$ or force 
that the rigid wall feels is given as
\begin{equation}
\betaw = \betasf+\betana,  \label{betawphenom}
\end{equation}
where $\betana$ denotes the contribution from non-adiabaticity 
(see \Eqref{betaeffective} and \Eqref{totalF}).
(For other spin structures such as vortices, an expression similar to \Eqref{betawphenom} holds.)
The Gilbert damping parameter originates from electron spin relaxation and other sources, so we can write 
$\alpha=\alphasf+\alpha_0$ (see \Eqref{eq:alpha1}).

It has been shown that when $\betaw/\alpha\gtrsim1$, the intrinsic threshold is smeared out and the true threshold current is determined by extrinsic pinning \cite{Thiaville05,TTKSNF06}.
The terminal wall velocity is also determined by $\betaw/\alpha$
 \cite{Zhang04,Thiaville05,TTKSNF06}.
The parameters $\betasf$ and $\betana$ are sensitive to the spin relaxation rate and wall thickness, respectively.
It would be therefore very interesting to experimentally identify the origin of $\betaw$ by changing material (spin-orbit interaction) and structure (wall thickness).

Microscopic derivation of the $\beta$-term has been carried out by several authors
 \cite{Tserkovnyak06,KTS06,Kohno07,Duine07}. 
Tserkovnyak et al. \cite{Tserkovnyak06} calculated $\betasf$ based on a one-band model considering spin-relaxation of conduction electrons semi-classically and assuming spin dynamics of small amplitude. 
They considered the limit of weak ferromagnetism and found that 
$\betasf=\alphasf$.
They also showed that in general 
$\betasf \neq \alphasf$ considering multiband effects or deviation from weak ferromagnetism.  
Their approach is, however, still phenomenological, treating the spin-flip process by a phenomenological spin-relaxation time in the equation of motion of spin. 
The relation $\betasf=\alphasf$ was suggested also by another  phenomenological argument \cite{Barnes05,Barnes06} (but see also Ref.  \cite{TK06}).
Fully microscopic calculation of $\beta$ and $\alpha$ due to spin relaxation was carried out on an $s$-$d$ model by Kohno et al. \cite{KTS06,Kohno07} using  standard diagrammatic perturbation theory, where the effect of spin relaxation are taken into account consistently and fully quantum mechanically. 
The result indicated $\betasf\neq \alphasf$.
The same result was obtained later in the functional Keldysh formalism by Duine et al. \cite{Duine07}.
 Determination of $\beta$ and $\alpha$ values needs a careful microscopic calculation, since they are quantities smaller by a factor of $1/(\ef\tau)$ compared with conventional transport coefficients. 
A phenomenological thermodynamic argument predicted $\beta=\alpha$  \cite{Barnes05,Barnes06}, but microscopic studies \cite{KTS06,Duine07,Kohno07} indicate that that is invalid. 
The error may arise because the argument of Refs.  \cite{Barnes05,Barnes06} lacks consistent consideration of the work done by the electric current \cite{TK06}.
So far, the effect of spin relaxation from spin-orbit interaction has not been calculated.
The effect is expected to be essentially the same as spin-flip impurities \cite{Kohno07}.
Such a term like $\beta$ was also theoretically found in ferromagnetic junctions \cite{Edwards05}.

Experimentally $\betaw$ can be determined from domain wall dynamics \cite{Thomas06,Heyne08}.
A very large ratio of $\frac{\beta}{\alpha}\sim 8$ was obtained from oscillating wall motion in a pinning potential \cite{Thomas06}, and the significant deformation of the domain wall observed by Heyne et al. \cite{Heyne08} clearly indicated $\betaw\neq\alpha$.
Recent systematic experimental study \cite{Oogane06} on the spin damping parameter in various ferromagnetic films revealed that $\alpha$ scales with $(g-2)^2$ ($g$ is the $g$-factor), indicating that damping arises mainly from spin-orbit interaction \cite{Elliott54}.

Sometimes the $\beta$ torque from spin relaxation is called the  non-adiabatic torque \cite{Zhang04}, extending the definition of non-adiabaticity to include spin deviation due to spin relaxation. 
The original meaning of adiabaticity in electron transport is the absence of scattering or the exchange of linear momentum.
In this paper, we use the term non-adiabatic in this original,  narrow sense. 
We therefore call non-adiabatic torques only those arising from momentum transfer (finite $\qv$ contributions in Eqs. (\ref{chiz1})(\ref{chiz2})(\ref{chio1})(\ref{chio2}) of torque given by  \Eqref{torqueuni}).
According to this definition, the $\betasf$ term from spin relaxation is still an adiabatic torque, since $\betasf$ is defined as the local torque in the Landau-Lifshitz-Gilbert equation (\Eqref{modLLG}). 
Correctly speaking,spin relaxation has additional truly non-adiabatic components that are nonlocal and oscillating (contributing to $\torquev_{\rm na}$ in \Eqref{modLLG}), which have not been discussed so far.

The Landau-Lifshitz-Gilbert equation is useful in numerical simulation of spin dynamics. 
However, one should pay attention to the fact that it is a phenomenological equation utilized only within a local approximation of torques, and is not suitable for studies including non-adiabatic effects (i.e., non-local torques).
The analysis in this paper is therefore not based on the  Landau-Lifshitz-Gilbert equation.
Instead, we will derive the equation of motion of the wall directly from the total Hamiltonian.

Spin damping torque in the magnetic field can also be expressed  using the Landau-Lifshitz prescription. The equation of motion then reads 
\begin{equation}
\dot{\Sv}= \gyro_{\rm LL} \Bvs \times \Sv
+\frac{\alpha}{S} 
\gyro_{\rm LL}  \Sv\times(\Sv\times\Bvs), 
\label{LLeq}
\end{equation}
where $\gyro_{\rm LL}$ is a constant.
This equation is equivallent to the Landau-Lifshitz-Gilbert (LLG) equation 
\begin{equation}
\dot{\Sv}= \gamma \Bvs \times \Sv
-\frac{\alpha}{S} \Sv\times\dot\Sv.
\end{equation}
In fact, \Eqref{LLeq} indicates 
$\gyro_{\rm LL}  (\Sv\times\Bvs)=
 -\dot{\Sv} +\frac{\alpha}{S} 
\gyro_{\rm LL}  \Sv\times(\Sv\times\Bvs)$, and thus
the equation reads 
\begin{equation}
\dot{\Sv}= \gyro_{\rm LL} \lt(1+\alpha^2 \rt) \Bvs \times \Sv
-\frac{\alpha}{S} (\Sv\times\dot{\Sv}).
\end{equation}
Therefore the two equations are equivalent either by defining $\gyro$ in the LLG equation as 
$\gyro \equiv \gyro_{\rm LL} \lt(1+\alpha^2 \rt)$, 
or by neglecting $O(\alpha^2)$. 
(Note that both equations are approximations neglecting higher derivatives such as $\ddot{\Sv}$.) 
From the microscopic theory point of view, the Landau-Lifshitz-Gilbert equation seems natural, since, as we will see in \S\ref{SEC:singlespin}, the damping torque of the Gilbert type,  $\frac{\alpha}{S} \Sv\times\dot\Sv$, is derived by a systematic gauge-field expansion.

\subsubsection{Non-adiabatic torque and spin-orbit interaction}

Waintal and Viret \cite{Waintal04} and  Xiao, Zangwill, and Stiles \cite{Xiao06} studied the spatial distribution of the current-induced torque around a domain wall by solving the Schr\"odinger equation and found a nonlocal oscillatory torque
($\tau_{\rm na}$ in Eq. (\ref{modLLG})). 
This torque is due to the non-adiabaticity arising from the finite domain wall width, or in other words,
from the fast-varying component of spin structure.
The oscillation period is $\sim \kf^{-1}$ ($\kf$ is the Fermi wavelength) and is of quantum origin similar to the Ruderman-Kittel-Kasuya-Yoshida (RKKY) oscillation.
This nonlocal torque looks complicated, but represents in fact a force acting on the wall, studied in Ref.  \cite{TK04}. 
In this paper, we demonstrate that this oscillating torque on each spin (\Eqref{tzphiz}) is indeed summed up to a force (\Eqref{Fna})  when looked collectively \cite{TKSLL07}.

Ohe and Kramer \cite{Ohe06} studied wall motion solving the torque due to the exchange interaction numerically, including
non-adiabaticity. 
Non-adiabaticity was studied further in Refs.  \cite{TKSLL07,Piechon07,Thorwart07}. 
Thorwart and Eggar  \cite{Thorwart07} applied a one-dimensional method to evaluate the conduction electrons, and carried out a gradient expansion with respect to slowly varying localized spins. The torque in the second order of the spatial derivative was derived and was shown to deform the wall significantly if the wall is very narrow, $\lambda\sim 5.5 a$.

Nonlocal oscillating torques were numerically studied by taking account of the strong spin-orbit interaction based on the Luttinger Hamiltonian  (i.e., in magnetic semiconductors) \cite{Nguyen07}.
It was shown there that due to the spin-orbit interaction, the oscillating torque becomes asymmetric around domain wall and that this feature results in high wall 
velocity.
Current-induced domain wall motion in the presence of Rashba spin-orbit interaction was recently studied \cite{Obata08}.
It was found that for one type of Bloch wall, a very large effective $\beta$ term is induced by the Rashba interaction, which acts as an effective magnetic field, and wall mobility is strongly enhanced.
A strong Rashba interaction is known to arise not only in semiconductors, but on the surface of metals doped with heavy ions \cite{Nakagawa07,Ast07}, and such systems would be very interesting in the context of domain wall dynamics.

\subsection{Recent experiments}

\subsubsection{Metallic systems}

So far experimental results on metallic samples all show threshold currents of the order of $10^{12}$ A/m$^2$.
If we use $\Kp/\kB\sim O(1\ {\rm K})$  estimated experimentally, \cite{Yamaguchi04,Yamaguchi05,Yamaguchi06err} 
the observed threshold is orders of magnitude ($10^{-2}-10^{-1}$ times) smaller than the intrinsic threshold, $\jci$.
For instance, a sample of Yamaguchi \cite{Yamaguchi04,Yamaguchi05,Yamaguchi06err} showed 
$\jc=1\times 10^{12}$ A/m$^2$. 
 The anisotropy energy is estimated to be $\Kp/\kB=2.4$ K, and using
$S\sim \frac{1}{2}$, $a\simeq 2.2$ \AA\ and $P\sim O(1)$, we obtain 
$\jci=5.8\times 10^{13}$ A/m$^2$, i.e., 
$\jc/\jci\sim 0.02$. 
 The observed low threshold currents in metals thus should be regarded as due to an extrinsic pinning.
 Actually, direct evidence excluding intrinsic pinning in permalloy wires so far was given by Yamaguchi et al. \cite{Yamaguchi06}.
 They prepared permalloy wires with different geometries, and realized different perpendicular anisotropy energies  
$\frac{S^2}{a^3}\Kp\simeq (0.1-7.6)\times 10^{5}$ J/m$^3$, 
which corresponds to $\Kp/\kB\simeq 0.03-2.4$ K
(per 1 spin).
The intrinsic pinning, Eq. (\ref{jci}),  predicts then
a threshold current of $5\times 10^{11}-4\times 10^{13}$ A/m$^2$.
In contrast to such a difference in the predicted values of the intrinsic threshold current, 
experimental values of the threshold for these samples do not vary so much,
$(3-8)\times 10^{11}$ A/m$^2$, and are smaller than the predicted intrinsic threshold by factors of 2 to 100.
Besides, data by Yamaguchi et al. indicate that these experimental values do not scale with $\Kp$, although there is a weak dependence on $\Kp$. 
Therefore the observed threshold in Ref.  \cite{Yamaguchi04,Yamaguchi06err} would be of some extrinsic origin.

The wall speed is another important quantity to determine the driving mechanism and efficiency.
Under long ($\gtrsim \mu$s) current pulses in metallic samples, the wall speed so far is very small, less than 1 m/s \cite{Klaui05} or $2\sim 6$ m/s at a current density of $10^{12}\Ams$ \cite{Yamaguchi06err}.
This is far below the perfect spin transfer limit, 
$\vw=\frac{a^3}{2eS}Pj\simeq 100$ m/s at $j=10^{12}\Ams$.
Processes involving strong dissipation of angular momentum or linear momentum during wall motion, such as deformation by extrinsic pinning centers or spin wave emission, might explain the discrepancy.
Measurements on clean samples are necessary.
(In semiconductors, in contrast, perfect spin transfer seems to be realized. See below.)

Direct observation of the spin structure indicates that the wall is considerably deformed upon motion \cite{Klaui05,Togawa06,Togawa06a,Biehler07}. 
It was shown \cite{Klaui05} that the initial state 
is not a planar wall but more like a vortex
 (called a vortex wall), 
which is the case in film or wide wires, the vortex wall moves by applying a current pulse of $2.2\times10^{12}$ A/m$^2$, and  the wall is deformed to become a transverse wall after some pulses.
It is interesting that although the vortex wall moves more easily, the transverse wall does not move at the same current density.
This would be explained by the absence of intrinsic pinning for the vortex wall (\Eqref{veq2}).
As far as the transverse wall is concerned, rigidity or non-rigidity does not seem essential since the analytical results for a rigid wall \cite{TK04,TKSLL07} and numerical simulation including deformation effect \cite{Thiaville05,Seo07} predict similar behaviors.

Heating effect in metallic samples has indeed been found to be crucially important \cite{Yamaguchi05,You07}.
For applications, heating may assist wall motion.
Sub ns pulses were reported to be quite efficient in driving the wall at a low current density of $\sim 10^{10}$ A/m$^2$ \cite{Lim04}.
This motion could be the motion due to $\beta$ term or below the intrinsic threshold.
Short pulses could be useful for applications.
Quite recently, Laufenberg et al. measured the temperature dependence of the threshold current and found that it decreases at low temperatures, for instance, from 
$2.4\times 10^{12} \Ams$  at $T\sim 170$ K to 
$1.9\times 10^{12} \Ams$  at $T\sim 100$ K
 \cite{Laufenberg06}.
Dissipation of spin-transfer torque by spin waves was suggested as a possible explanation, but a theoretical study is yet to be done.

For recent experimental results, see Ref.  \cite{Klaui08}.

\subsubsection{Thin wall}

Quite an interesting result was obtained recently by Feigenson et al. \cite{Feigenson07} in
SrRuO$_3$, an itinerant ferromagnet with perovskite structure.
The current density needed to drive the wall was 
$5.3\times 10^9\Ams$ at $T=140$ K and  $5.8\times 10^{10}\Ams$ at $T=40$ K.
A small threshold current at 140 K would be due to reduction of magnetization close to $\Tc=150$ K. 
The threshold current is about 2 orders of magnitude smaller than in other metals. 
This high efficiency would be due to a very narrow domain wall, $\lambda\sim 3$ nm, as a result of very strong uniaxial anisotropy energy ($K$) corresponding to a field of 10 T.
They defined a parameter determining the efficiency
as a ratio of the depinning field and depinning current density, 
$\Lambda\equiv \Bc/\jc$.
Their results were $\Lambda=10^{-12}$ T m$^2$/A.
They compared this value with the threshold current of extrinsic pinning \cite{TK04} 
(given by Eqs. (\ref{jcIb}) and (\ref{VzB})).
Using $\frac{\hbar a^3}{e\mub\lambda}\sim 0.5\times 10^{-11}$ T m$^2$/A  and $S\sim 3/2$, we see that $\Lambda=10^{-12}$ T m$^2$/A is realized if $\betaw\sim 0.5$.
This value would be too large if interpreted as due to spin relaxation.
Using the measured resistivity of the domain wall, the non-adiabatic force contribution to $\betaw$ was estimated and the result of $\jc$ was of similar order as the observed ones but with a discrepancy of a factor of  around 6 at low temperature 
(Fig. 4(a) of Ref.  \cite{Feigenson07}). 
This discrepancy seems not very crucial considering the crude rigid and planar approximation of the wall.
There is another extrinsic pinning threshold, 
$j_{\rm c}^{\rm Ia)}$ (\Eqref{jcIa}). 
If we use this expression,
$\Lambda=10^{-12}$ T m$^2$/A corresponds to 
$\Kp\sim \mub\Bc=4\times(10^{-3}-10^{-2})$ K (per site).
This number seems also possible although $\Kp$ was not measured.

Controlling magnetic anisotropy in metallic thin films is an interesting possibility. 
An easy axis perpendicular to the film is realized using a Pt layer, and in such case, the domain wall is a Bloch type and the thickness is known to be small, $\lambda\sim 12$ nm \cite{Ravelosona05}. 
Ravelosona et al.  \cite{Ravelosona05} found in such perpendicular systems that the threshold current density is reduced by a factor of $\sim \frac{1}{10}$ compared with other metallic systems,  $\jc \sim 10^{11}\Ams$.
The reduction could be explained by the intrinsic threshold, \Eqref{jci}, as due to reduction of $\lambda$ by roughly a factor of $\frac{1}{10}$ if $\Kp$ is the same order as in-plane anisotropy systems. 
It could be possible also that $\Kp$ is very small in perpendicular systems, since the system is quite symmetric within the plane perpendicular to the easy axis, resulting in reduced threshold current.
Although non-adiabatic effects are mentioned in Ref.  \cite{Ravelosona05}, a 12 nm wall would be still in the adiabatic regime, since $\kf \lambda \sim O(100)$.  
Reduction of threshold current in perpendicular anisotropy films was also confirmed in numerical simulation \cite{Fukami08,Suzuki08}.

\subsubsection{Magnetic Semiconductors}

Beautiful experiments were carried out at low current in ferromagnetic semiconductors by Yamanouchi et al \cite{Yamanouchi04,Yamanouchi06}.
They fabricated a wall structure of 20 $\mu$m width made of GaMnAs with different thicknesses, 
 which determines the ferromagnetic coupling and transition temperature, 
and trapped a domain wall. 
The wall position was measured optically after applying a current pulse, and the average velocity was estimated.
The current necessary was $\sim 4\times 10^{9}$ A/m$^2$, which is 2-3 orders of magnitude smaller than in metallic systems.
This is due to the small average magnetization, $S\sim 0.01$, carried by dilute Mn ions, and small hard-axis anisotropy $\Kp$  \cite{Yamanouchi06}.
The obtained velocity was rather high, $\sim 22$ m/s at $j=1.2\times 10^{10}$ A/m$^2$.
This velocity is consistent with the adiabatic spin-transfer mechanism, \Eqref{STv}, and the threshold appears to be consistent with the intrinsic pinning mechanism \cite{TK04} with anisotropy energy obtained from band calculations. 
 
However, there are some puzzles. 
 First, the theory of intrinsic pinning \cite{TK04} and adiabatic spin transfer does not take account of strong spin-orbit interaction in semiconductors. So the agreement with these theories might be a coincidence. 
The second puzzle is the validity of using purely adiabatic theory. In fact, quite a large momentum transfer (force) is expected from the wall resistance, $\Rw=1\ \Omega$ \cite{Chiba06}, corresponding to $\betaw\gg 1$ in terms of $\betaw$ \cite{Chiba06}.

Another puzzle, which was solved just recently, is the temperature dependence of the wall velocity. 
The observed velocity scaled as 
$\ln v \simeq -(\Tc-T)^2 j^{-1/2}$, similar to the creep behavior under a magnetic field \cite{Lemerle98}, but this fractional power of $j$ has not been explained in the current-driven case. 
A simple theory of thermal activation assuming a rigid wall under the spin-transfer torque predicts a different behavior, $\ln v \simeq j/T$ \cite{TVF05}, and thus creep motion would be essential in the experiment by Yamanouchi et al.
Successful explanation of creep behavior was just recently offered by Yamanouchi et al. \cite{Yamanouchi07}, by taking account of 
growth of $\phi$ at the pinning center.
The thermal effect and creep by current was studied in more detail by Duine et al. \cite{Duine07ta}.

Nguyen et al. studied theoretically the domain wall speed in magnetic semiconductors based on the 4-band Kohn-Luttinger Hamiltonian \cite{Nguyen07}.
It was shown there that the wall speed can be enhanced by the spin-orbit interaction by a factor of $10^{3}-10^{4}$ due to the increase of mistracking, hence reflection, of conduction electrons.
Rashba spin-orbit interaction is also expected to be  useful for efficient magnetization switching \cite{Obata08}.

\subsubsection{Excitation of wall} 
Time-resolved study of excitation of wall provides rich information on the wall character and driving mechanism.

Under alternating current (AC), the domain wall shows another aspect not seen in the direct current (DC) case.
AC can drive domain walls quite effectively at low current if the frequency is tuned close to the resonance with the pinning frequency. 
This resonance was realized in a recent experiment by Saitoh et al. \cite{SMYT04}.
 They applied a small AC (of amplitude of $10^{10}$ A/m$^2$) in a wire with a domain wall in a weak pinning potential controlled by a magnetic field.
Although the current is well below the threshold, the wall can shift slightly as we see below (distance is calculated to be around $\mu$m, which might be an overestimate).
Under a small current, $\phiz$ remains small, and the equation of motion reduces to that of a \lq\lq particle";
\begin{equation}
\Mw\ddot{X}+\frac{\Mw}{\tauw}\dot{X}+\Mw\Omegapin^2 X=F(t),
\end{equation}
where $\Mw$ is the wall mass, $\tauw\propto \alpha^{-1}$ is a damping time, $\Omegapin$ is the (extrinsic) pinning frequency, and $F(t)$ is a force due to current.
For AC, $I(t)=I_0 e^{i\omega t}$, where $\omega$ is the frequency, the force is given 
$F(t)=\frac{I(t)}{e}\lt[
\frac{2\hbar S}{\lambda}\betaw
-iP\hbar^2\frac{\omega}{\Kp\lambda}\rt]$, 
where $\betaw$, given by Eq. (\ref{betaeffective}), is the total force from momentum transfer and spin relaxation, and the last term, proportional to $\omega$, is from the spin-transfer torque.
The wall under a weak current thus shows a forced oscillation of a particle.
By measuring the energy dissipation (from complex resistance), a resonance peak would then appear when $\omega$ is tuned closely around $\Omegapin$.
From the resonance spectra, the mass and the friction constant were obtained as $\Mw=6.6\times 10^{-23}$ kg, $\tauw= 1.4\times 10^{-8}$ sec.
The experimental result seems to be well described by the rigid-wall picture, and this would be due to a low current density (by factor of $10^{-2}$ compared to DC experiments on metals), resulting in a small deformation.  
What is more, from the resonance line shape, the driving mechanism of the domain wall was identified to be the force ($\betaw$) rather than the spin-transfer torque.
This finding was surprising at that time, when the adiabatic spin-transfer torque was considered as the main driving mechanism. 
 The observed force corresponds to the value of $\betaw\sim 1.5$, which is too large if $\betaw$ arises from spin relaxation $\betasf$ ($\betasf$ is considered to be of the same order as $\alpha$, both arising from spin relaxation). 
If it comes purely from the momentum transfer, the wall resistance is estimated to be 
$\Rw=3\times 10^{-4}\ \Omega$, a quite reasonable value. 
 A striking point in this experiment is a significant enhancement of the effect of the force due to resonance, which made possible the low-current operation. 
 On the other hand, the spin-transfer torque is suppressed in the MHz range (as seen from the factor of $\omega$ in the spin-transfer torque term of $F(t)$).

Quite recently, Thomas et al. \cite{Thomas06,Thomas07} succeeded in detecting periodic oscillation of a wall in a confining potential by using a ns current pulse at $j=6.9\times10^{11}\Ams$.
The motion was consistent with the rigid wall description in terms of $X$ and $\phiz$.
Periodic variation of chirality, $\phiz$, of a wall was observed in the presence of magnetic field
and current pulse of 10ns at $1\times 10^{12}$ A/m$^2$ \cite{Hayashi07}.
 The results indicated that the chirality, $\phiz$, plays an important role in the wall propagation, as predicted theoretically  \cite{Slonczewski72,TK04,Thiaville05}.

\section{Localized spins }
\label{SEC:spin}

\subsection{Spin Lagrangian}

\begin{figure}[tbh]
  \begin{center}
  \includegraphics[scale=0.6]{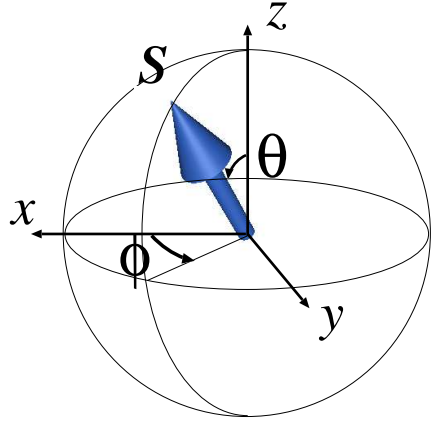}
  \end{center}
\caption{Polar coordinates $(\theta,\phi)$ of localized spin.
\label{FIGpolar2}
}
\end{figure}

Throughout this paper, we use the Lagrangian formalism, as it is useful in describing collective objects such as domain walls. 
The basics of the Lagrangian spin formalism are summarized in \S\ref{SEC:spinlag}.
The localized spin $\Sv(\xv,t)$ is a function of position $\xv$ and time $t$. 
We define polar coordinates as (Fig. \ref{FIGpolar2})
\begin{equation}
\Sv=S (\sin\theta\cos\phi,\sin\theta\sin\phi,\cos\theta).
\end{equation}

We will consider the effect of damping later.
The Lagrangian of the spin system with Hamiltonian $\Hs$ is given by
\begin{equation}
\Ls = \sumx \hbar S \dot{\phi}(\cos\theta -1) -\Hs,
\label{Ls}
\end{equation}
The first term, 
\begin{equation}
\Lb\equiv \sumx \hbar S \dot{\phi}(\cos\theta -1),
\end{equation}
is a term that describes the time development of the spin.
This term has the geometrical significance of a solid angle in spin space, or a spin Berry phase \cite{Auerbach94}.
As is easily seen, this Lagrangian (\ref{Ls}) results in the  correct Landau-Lifshitz equation of motion: 
\begin{equation}
 \hbar \dot{\Sv}=
 \hbar \gyro \Bvs \times \Sv,   \label{spin:LL}
\end{equation}
where 
\begin{equation}
\Bvs \equiv \frac{1}{\hbar \gyro}
 \deld{\Hs}{\Sv},
\end{equation}
 is the effective magnetic field acting on the spin 
(from the localized spin Hamiltonian $\Hs$).
(The energy density due to field is given as 
$-\Mv\cdot \Bvs = \frac{\hbar\gyro}{a^3} \Sv\cdot\Bvs$.)

\begin{figure}[tbh]
  \begin{center}
  \includegraphics[scale=0.6]{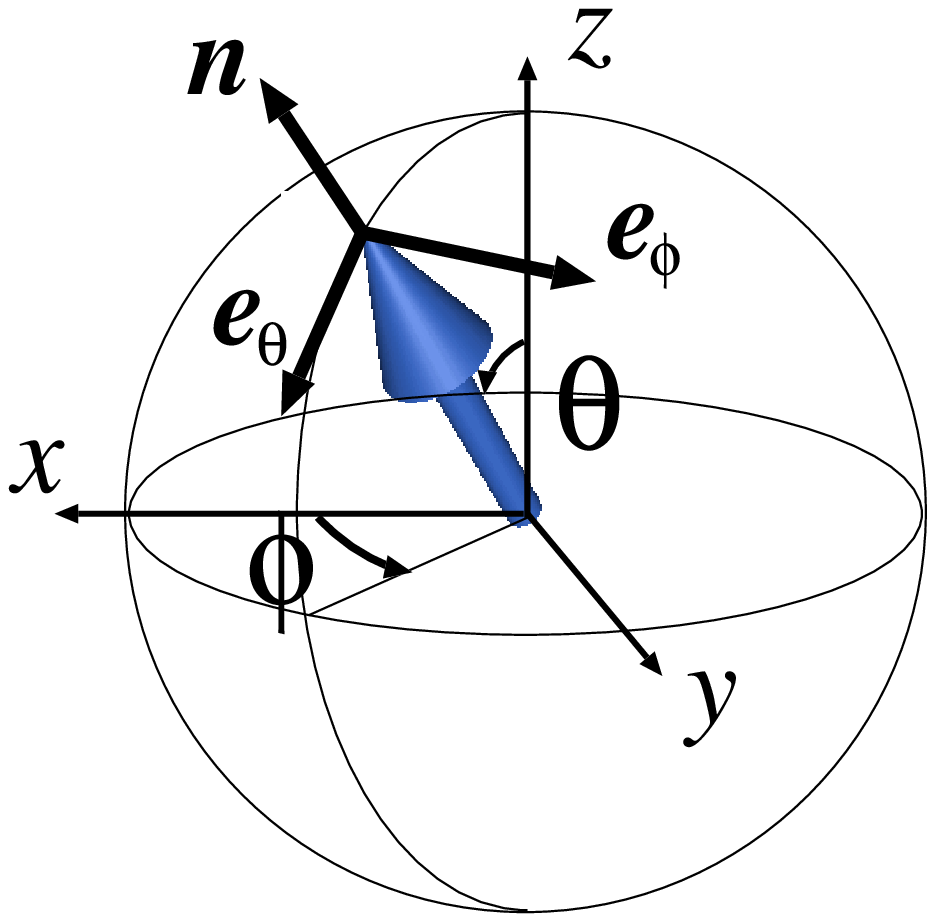}
  \end{center}
\caption{Unit vectors $\evs,\evth,\evph$.
They satisfy $\evs\times\evth=\evph$, $\evs\times\evph=-\evth$, and
$\evth\times\evph=\evs$.
\label{FIGpolar3}
}
\end{figure}

Let us demonstrate this. 
The equations of motion for the spin derived by taking derivatives of $\Ls$ with
respect to $\theta$ and $\phi$ are
\begin{eqnarray}
\delo{t} \deld{\Ls}{\dot{\theta}}
-\deld{\Ls}{\theta} &=& 0
\\
\delo{t} \deld{\Ls}{\dot{\phi}}
-\deld{\Ls}{\phi} &=& 0.
\end{eqnarray}
By use of the explicit form of $\Lb$, they become
\begin{eqnarray}
\hbar S \sin\theta \dot{\phi} &=&
-\deld{\Hs}{\theta} 
\label{spin:eqtheta0}\\
- \hbar S \sin\theta \dot{\theta} &=&
-\deld{\Hs}{\phi}    \label{spin:eqphi0}
\end{eqnarray}
Eqs. (\ref{spin:eqtheta0})(\ref{spin:eqphi0}) become more concise in terms of vector equations for $\Sv$.
The time-derivative of $\Sv$ is given by
\begin{equation}
\dot{\Sv}=S(\dot{\theta}\evth+\sin\theta\dot{\phi}\evph),
\label{spin:Sdot}
\end{equation}
where
\begin{eqnarray}
\evth &=& \left( \begin{array}{c}
  \cos\theta \cos\phi \\
  \cos\theta \sin\phi \\
   -\sin\theta  \end{array} \right)  \\
\evph &=& \left( \begin{array}{c}
   -\sin\phi \\
    \cos\phi \\
      0  \end{array} \right),
\end{eqnarray}
are unit vectors in the $\theta$- and $\phi$-directions, respectively (Fig. \ref{FIGpolar3}).
For any function $F(\theta,\phi)$ of $(\theta,\phi)$, the following identity holds: 
\begin{eqnarray}
\deld{F}{\theta} &=& S\left(\evth\cdot \deld{F}{\Sv} \right) \\
\deld{F}{\phi} &=& S\sin\theta \left(\evph\cdot \deld{F}{\Sv} \right).
\end{eqnarray}
Thus Eqs. (\ref{spin:eqtheta0})(\ref{spin:eqphi0})(\ref{spin:Sdot}) become
\begin{eqnarray}
\hbar \dot{\Sv} &=&
 S\left( \evth\left(\evph\cdot\deld{\Hs}{\Sv}\right) 
 -\evph\left(\evth\cdot\deld{\Hs}{\Sv}\right) \right)
\nonumber\\
&=&
S\left( \deld{\Hs}{\Sv} \times \evs \right)
\label{eqsdot},
\end{eqnarray}
which is Eq. (\ref{spin:LL}).
Here we used
\begin{equation}
\evth\times\evph=\evs,
\end{equation}
where
\begin{equation}
\evs \equiv \frac{\Sv}{S}= \left( \begin{array}{c}
  \sin\theta \cos\phi \\
  \sin\theta \sin\phi \\
   \cos\theta  \end{array} \right) ,
\end{equation}
is a unit vector along $\Sv$.

\subsection{Spin algebra}
The meaning of the "spin Berry phase" term can be understood if one notes that the canonical structure is contained in this kinematical term in the Lagrangian. 
Let us demonstrate this within classical mechanics.
The canonical momentum conjugate to $\phi$ is defined as 
\begin{equation}
 P_\phi \equiv \deld{\Ls}{\dot{\phi}} 
 = \hbar S_z -\hbar S .
\end{equation}
 Defining the Poisson bracket (times $\hbar$) by 
$\{ A , B \}_{\rm PB} 
= (\partial A/\partial \phi) (\partial B/\partial S_z) 
- (\partial B/\partial \phi) (\partial A/\partial S_z)$, 
we have $\{ \phi , S_z \}_{\rm PB} = 1$. 
 By using 
$S_x \pm iS_y = \sqrt{S^2 - S_z^2} \, e^{\pm i\phi}$,
we can derive the correct SU(2) algebra of the spin angular momentum as 
\begin{eqnarray}
 \{ S_i,S_j \}_{\rm PB} &= &\epsilon_{ijk} S_k .
\label{spin:Scom0}
\end{eqnarray}

\begin{figure}[t]
  \begin{center}
  \includegraphics[width=0.3\linewidth]{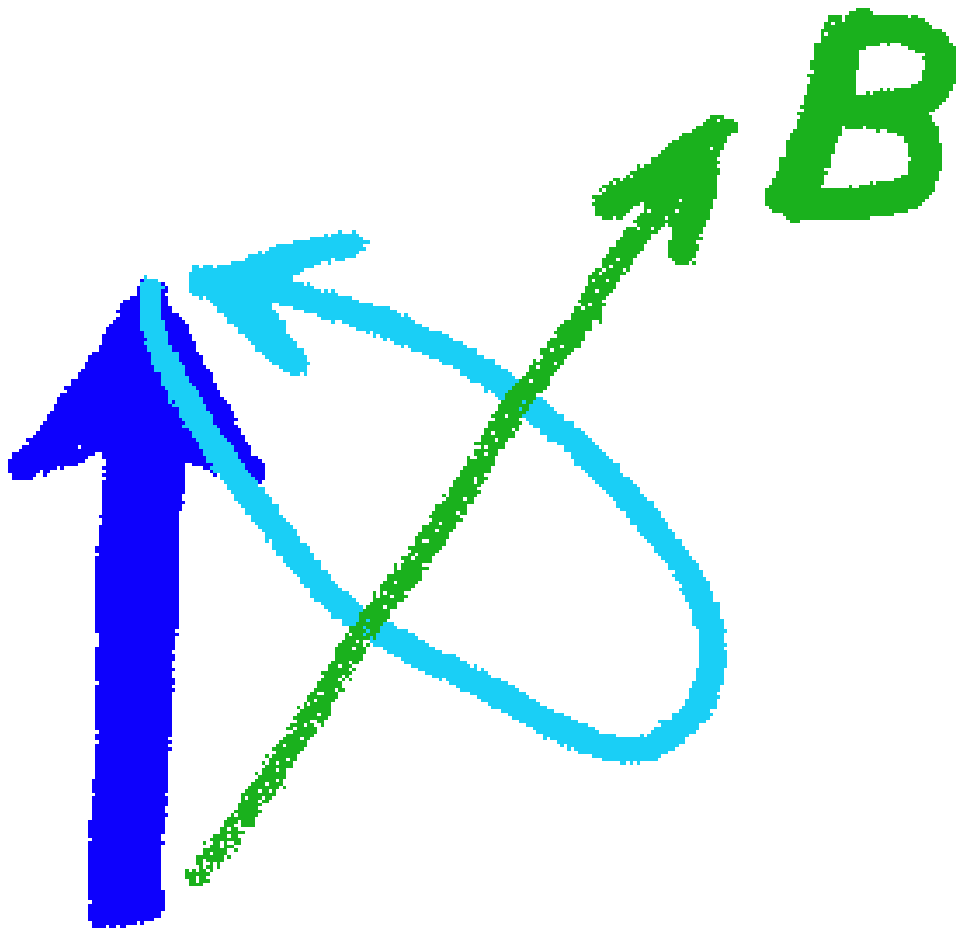}
  \includegraphics[width=0.3\linewidth]{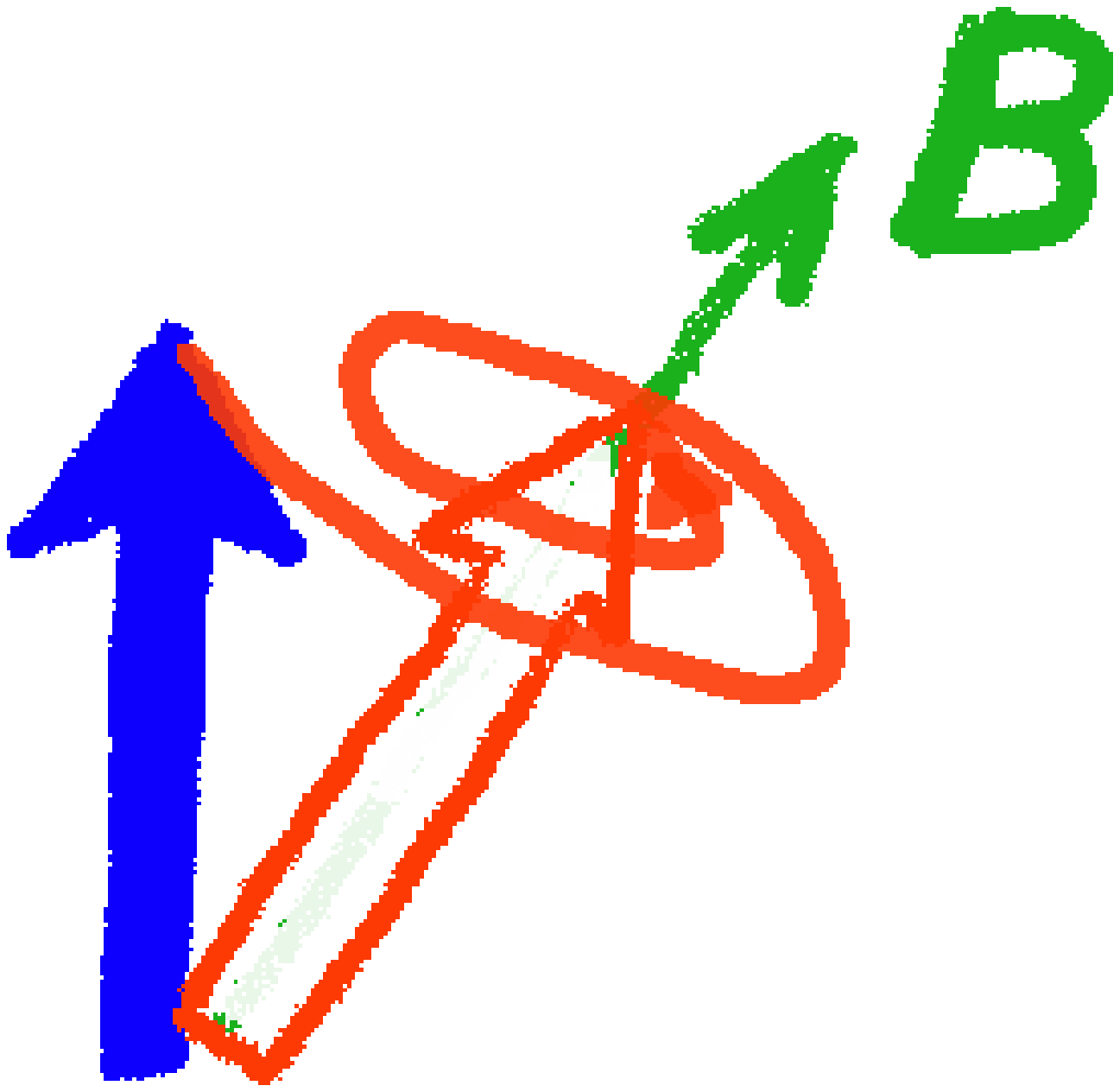}
\caption{Motion of localized spin under an effective field, 
$\Bv$, without and with damping $\alpha$. 
Without damping, the spin precesses forever, while with damping, the spin direction eventually relaxes to that of the field. 
\label{fig:precession}
}
  \end{center}
\end{figure}

\subsection{Dissipation}
In reality, the spins are subject to energy dissipation arising from various sources such as phonons.
This damping is rather essential in spin dynamics, for instance, to relax to the equilibrium configuration (Fig. \ref{fig:precession}).
The equation of motion with damping taken account is called the Landau-Lifshitz-Gilbert (LLG) equation \cite{Landau35}, which was discussed on a phenomenological basis. 
It reads 
\begin{equation}
  \del{\Sv}{t}=
   \gyro  \Bvs  \times \Sv - \frac{\alpha}{S}\Sv\times \del{\Sv}{t}.  
    \label{spin:LLG}
\end{equation}
 Here
$\alpha$ is a dimensionless damping constant, called the Gilbert damping parameter. 
This $\alpha$ here includes all the effects of conduction electrons and non-electron origins. 
The electron contribution to $\alpha$ is calculated in \S\ref{SEC:singlespin} (\Eqref{eq:alpha1}).
The parameter $\alpha$ is not therefore a phenomenological parameter, but can be derived \cite{Tserkovnyak06,KTS06}.

For the study of domain wall dynamics, the effect of damping is incorporated in the Lagrangian formalism 
 by use of Rayleigh's method as in classical mechanics \cite{Goldstein02} 
by introducing a dissipation function 
\begin{equation}
\Ws\equiv \sumx \frac{\hbar\alpha}{2S} \dot{\Sv}^2 
=\sumx  \frac{\alpha}{2} \hbar S \,(\dot\theta^2 + \sin^2\theta \, \dot\phi^2) ,
\end{equation}
where $\alpha$ is a dimensionless damping parameter.
The equation of motion with damping included is modified to be
\begin{equation}
\delo{t} \deld{\Ls}{\dot{q}}
-\deld{\Ls}{q} = -\deld{\Ws}{\dot{q}} \label{eqofmo0}
\end{equation}
where $q$ represents $\theta$ and $\phi$ and 
the last term describes the energy dissipated.
Explicitly, they read
\begin{eqnarray}
\hbar S \sin\theta \dot{\phi} &=&
-\deld{\Hs}{\theta} -\deld{\Ws}{\dot{\theta}}
\label{spin:eqtheta10}\\
- \hbar S \sin\theta \dot{\theta} &=&
-\deld{\Hs}{\phi} -\deld{\Ws}{\dot{\phi}}   \label{spin:eqphi10}
,
\end{eqnarray}
In terms of vector equations for $\Sv$
Eqs. (\ref{spin:eqtheta10})(\ref{spin:eqphi10}) read
\begin{equation}
 \hbar \dot{\Sv}=
  \left( \deld{\Hs}{\Sv} +\deld{\Ws}{\dot{\Sv}} \right) \times \Sv.   \label{spin:eqS0}
\end{equation}
Noting $\deld{\Ws}{\dot{\Sv}}=\frac{\hbar\alpha}{S}\dot{\Sv}$,
we see that the Landau-Lifshitz-Gilbert equation (\ref{spin:LLG}) is obtained.

Let us see that the damping term really corresponds to energy dissipation.
By multiplying $\Sv\times$ and then $\dot{\Sv}\cdot$ in Eq. (\ref{spin:LLG}), we obtain immediately the energy dissipation rate proportional to $\alpha\dot{\Sv}^2$; 
\begin{equation}
\del{\Hs}{t}=\dot{\Sv}\cdot\deld{\Hs}{\Sv}=- \frac{\alpha}{S}\dot{\Sv}^2.
\end{equation}

The above procedure of introducing dissipation seems artificial. 
We will later show that Gilbert damping can be derived microscopically
by considering a whole system including spin relaxation.
Actually, we will calculate the damping arising from the exchange coupling to conduction electron and spin relaxation \cite{KTS06}.

\subsection{Spin Hamiltonian}

\begin{figure}[tbh]
  \begin{center}
  \includegraphics[scale=0.45]{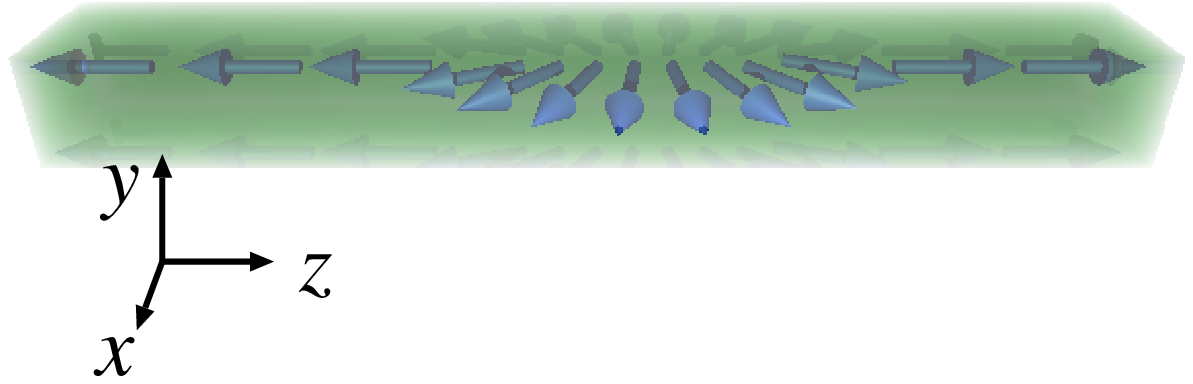}
  \includegraphics[scale=0.45]{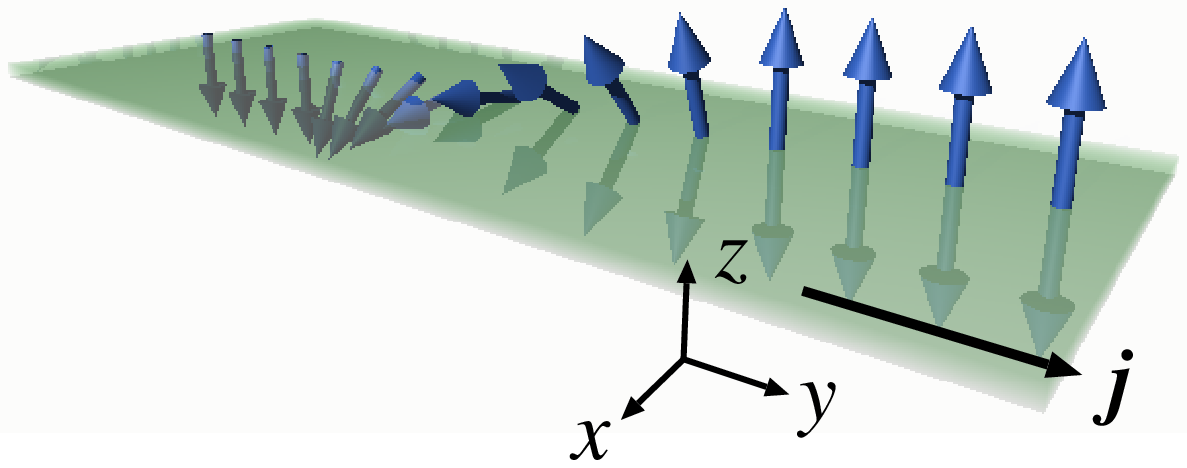}
\caption{Systems we consider: wires with easy axis along the wire and film with perpendicular easy axis.
The domain wall become the N\'eel type and the Bloch type, respectively (see Sec. \ref{SEC:NeelBloch}).
Coordinates in spin space are shown (which do not necessarily coincides with spatial coordinates).
The current is applied along the wall direction.
\label{FIGneelbloch}}
  \end{center}
\end{figure}

In this paper, we consider a spin system with an easy axis and a hard axis, chosen as the $z$ and $y$ directions, respectively. 
The energies of one spin pointing  in the easy and hard axes are denoted by $-\hf KS^2$ and  $\hf\Kp S^2$, respectively, where $K$ and $\Kp$ are positive.
With this two-axis anisotropy, the analysis here can be applied to both wires and films with perpendicular anisotropy (Fig. \ref{FIGneelbloch}).
We also include 
a pinning potential, $\Vpin$.
The explicit form of $\Vpin$ will be discussed later.
The Hamiltonian is thus
\begin{eqnarray}
\Hs &=& \sumx \left( \frac{J}{2}(\nabla \Sv)^2 -\frac{K}{2} (S_z)^2
  +\frac{\Kp}{2} (S_y)^2 
  \right) +\Vpin,\nonumber\\
&=& 
\sumx \left( 
 \frac{JS^2}{2}( (\nabla \theta)^2 +\sin^2\theta (\nabla \phi)^2 )
 +\frac{KS^2}{2} \sin^2\theta (1+\kappa \sin^2 \phi ) 
\right)+\Vpin, 
\nonumber\\&&
  \label{spin:Hs}
\end{eqnarray}
where $\kappa\equiv \Kp/K$.
The equation of motion \Eqref{spin:LLG} is then given by
\begin{eqnarray}
- \sin\theta \dot{\phi}-\alpha \dot{\theta}
  &=& \frac{KS}{\hbar}  \left(
  \lambda^2 \left( -\nabla^2 \theta +\frac{1}{2}\sin 2\theta(\nabla\phi)^2 \right)
+\frac{1}{2}\sin2\theta
(1+\kappa\sin^2\phi) \right)
\nonumber\\&&
+\frac{1}{\hbar S}\deld{\Vpin}{\theta}
\label{spin:eqtheta1}\\
\dot{\theta}-\alpha \sin\theta \dot{\phi} &=&
 \frac{KS}{\hbar}\frac{1}{\sin\theta}
  \left( -\lambda^2 \nabla(\sin^2\theta\nabla\phi)
   +\frac{\kappa}{2}\sin^2\theta \sin2\phi) \right)
\nonumber\\&&
     +\frac{1}{\hbar S \sin\theta}\deld{\Vpin}{\phi}
   .
  \label{spin:eqphi1}
\end{eqnarray}
Here we have introduced a length scale determined by 
\begin{equation}
\lambda\equiv \sqrt\frac{J}{K}. \label{thickness}
\end{equation}
This length governs the spatial scale of the magnetic structure and 
turns out to be the thickness of the domain wall.

\subsection{Static domain wall solution}
Throughout this paper, we consider a planar (i.e. one-dimensional) domain wall as realized in a narrow wire or film, where the spin configuration depends only on the coordinate $\xw$ in the wire direction.

We consider in this subsection the case without magnetic field and pinning.
Equations (\ref{spin:eqtheta1})(\ref{spin:eqphi1}) for a static configuration of a constant $\phi$
are then given by
\begin{equation}
\ddel{\theta}{\xw} =\frac{1}{\lambda^2}\sin\theta\cos\theta,
\end{equation}
which yields after integration 
\begin{equation}
\left(\del{\theta}{\xw}\right)^2 =\frac{1}{\lambda^2}\sin^2\theta+C,
\end{equation}
where $C$ is a constant.
For the configuration we consider,
$\sin\theta\rightarrow0$ at $\xw\rightarrow\pm\infty$, we see that $C=0$ and thus
\begin{equation}
\del{\theta}{\xw} = \mp \frac{1}{\lambda}\sin \theta.
\end{equation}
This equation is easily integrated to obtain
$\tan\frac{\theta}{2}=e^{\mp (\xw-X)/\lambda}$,
where a constant $X$ represent the wall position.
The solution for $\theta$ is thus obtained as 
$\cos\theta = \pm \tanh\frac{\xw-X}{\lambda}$.
The sign here corresponds to a "topological charge" of domain wall.
We consider in this paper domain wall with positive charge, i.e., spin texture changing from $-z$ at $\xw=-\infty$ to $+z$ at $\xw=\infty$. 
The wall solution we consider is thus given by 
\begin{eqnarray}
\cos\theta &\equiv&  \tanh\frac{\xw-X}{\lambda}  \\
\sin\theta &\equiv& \frac{1}{\cosh\frac{\xw-X}{\lambda}}
\label{spin:dw0}
\end{eqnarray}
with $\phi=0$.

\subsection{N\'eel wall and Bloch wall}
\label{SEC:NeelBloch}
We considered above the case of a domain wall where magnetization is changing in the spatial coordinate $\xw$, which coincides with the spin easy axis.
The wall structure looks in this case as in the left of Fig. \ref{FIGneelbloch}, and is called the N\'eel wall.
Another type of domain wall, called the Bloch wall as depicted in the right of Fig. \ref{FIGneelbloch}, is also possible. 
This corresponds to a case where the spatial coordinate along the wire is $y$, and the easy plane ($zx$-plane) is perpendicular to the wire direction.
These differences of structure do not affect the electron transport or the spin torque in the absence of an interaction that correlates the spin space and real space, such as spin-orbit interaction.

\subsection{Pinning potential}
Let us specify here the pinning potential.
Pinning arises from various origins.
Here we consider a simple case of a point defect that modifies the easy axis magnetic anisotropy, $K$.
Other cases can be treated in a similar manner, and the result of pinning potential for $X$ is essentially the same. 
A defect at $\xv=0$ is assumed to cause local enhancement of easy axis anisotropy of  $\delta\! K (>0)$, and then the pinning potential is given by
\begin{equation}
\Vpin = -\sumx \delta\! K \frac{ S^2 a^3}{2} \delta^3(\xv)\sin^2\theta(\xv),
\end{equation}
where $\delta^3(\xv)=\delta(x)\delta(y)\delta(z)$ represents a $\delta$-function in three dimensions. 
In terms of domain wall variables,
this potential reads
\begin{equation}
\Vpin(X) = -\frac{K S^2 }{2} \frac{1}{\cosh^2\frac{X}{\lambda}}.
\end{equation}
For generality, we model this potential by a harmonic one given by
\begin{equation}
\Vpin = \frac{1}{2}\Mw\Omegapin^2(X^2-\xi^2)\theta(\xi-|X|)
 =\frac{\Nw\Vz}{\xi^2} (X^2-\xi^2)\theta(\xi-|X|),
\label{Vpin}
\end{equation}
where $\theta(x)$ is a step function, 
$\Vz= \delta\! K S^2 \xi^2 /(2\Nw)$ is the pinning strength per spin, $\Mw\equiv \frac{2\hbar^2 A}{\Kp \lambda a^3}
=\frac{\hbar^2\Nw}{\Kp\lambda^2}$ 
is the wall mass, and 
$\Omegapin\equiv\sqrt{{2\Kp\Vz}} \frac{\lambda}{\hbar \xi}$ 
corresponds to the oscillation frequency at the potential minimum.
The range of the pinning $\xi$ is equal to $\lambda$ if the defect is point-like.

\section{Description of rigid planar wall}
\label{collective}

\subsection{Collective coordinates}

 To derive the equation of motion of a rigid domain wall, we here consider the collective coordinate description \cite{Rajaraman82,Sakita85}. 
 This treatment and the results are essentially the same as the one considered by  Slonczewski \cite{Slonczewski72,Hubert00} in the context of dynamics under a magnetic field.

 The idea is to consider the constant $X$ in Eq.(\ref{spin:dw0}) as a dynamical variable, $X(t)$. 
During wall motion,  the angle $\phi(\xw,t)$ can be excited, as suggested by the equation of motion,  Eq.(\ref{spin:LLG}) and \Eqsref{spin:eqtheta1}{spin:eqphi1}.
Another collective variable, $\phiz$, defined as the spatial average of $\phi(x)$ \cite{TT96}, is therefore required to describe the wall dynamics.
In fact, the features of spin dynamics are taken account of by the fact that $X$ and $\phiz$ are canonically conjugate to each other, as indicated by the fact that the first term of Eq.(\ref{Ls}) is written as $\propto \dot X \phiz$. 
Naively considering only the variable $X$ as dynamical results in the wrong answer in general.
These two variables are called collective coordinates since they describe the collective dynamics of many localized spins.

In the absence of sample inhomogeneity and a driving force, $X$ describes a gapless zero mode owing to the translational symmetry of the system. 
If the transverse anisotropy energy $\Kp$ is zero in addition, $\phi$ is also a gapless mode. 
We will in this section consider this case ($\Vz=0, \Kp=0$)  and show that the two variables $X$ and $\phiz$ indeed become dynamical variables when we take account of the fluctuation around the classical solution.

\subsection{Coherent representation of spin}

To describe fluctuations (spin waves) around the domain wall, coherent representation of spin is convenient.
We could instead define a fluctuation in rather a naive way as
$\theta=\thetaz+\delta\theta$ and $\phi=\phiz+\delta\phi$, as 
done in Ref.  \cite{TT96}, but this definition results in a rather complicated spin-wave Hamiltonian.
So let us here use a coherent representation and 
define a complex variable $\xi(\xv,t)$ as
\begin{equation}
\xi \equiv e^{i\phi}\tan\frac{\theta}{2}.
\label{coherent:xi}
\end{equation}
This variable is related to $\theta$, $\phi$ as
\begin{eqnarray}
\sin\theta\cos\phi &=& \frac{\xi+\bar{\xi}}{1+{|\xi|^2}}
  \nonumber\\
\sin\theta\sin\phi &=& -i\frac{\xi-\bar{\xi}}{1+{|\xi|^2}}
  \nonumber\\
\cos\theta &=& \frac{1-{|\xi|^2}}{1+{|\xi|^2}}
  \nonumber\\
 \partial_\mu\phi &=& 
  \frac{-i}{2{|\xi|^2}}
 ( \bar{\xi}\partial_\mu\xi -\partial_\mu\bar{\xi}\cdot \xi) 
.
\end{eqnarray}
Thus the spin Lagrangian is written as (without pinning and $\Kp$)
\begin{eqnarray}
\Ls &=& \sumx \frac{1}{1+|\xi|^2} 
 \left( i\hbar S(\bar{\xi}\dot\xi -\dot{\bar{\xi}}\xi) 
-\frac{2KS^2}{(1+|\xi|^2)^2} \left( \lambda^2|\nabla\xi|^2 +|\xi|^2 
 \right) \right).\label{LSzero}
\end{eqnarray}

We now study fluctuation around a static domain wall sotultion,
\begin{eqnarray}
\cos\theta &=& \tanh\frac{\xw-X}{\lambda}\nonumber\\
\phi &=& \phiz, \label{collective:Xphi}
\end{eqnarray}
where $X$ and $\phiz$ are arbitrary constants, since we here assume vanishing extrinsic pinning and hard-axis anisotropy.
We define a complex variable $\fl$ representing fluctuation as 
\begin{equation}
\xi \equiv e^{-\ztil(\xw,t)+i\phiz+\fl(\xw-X,t)},
\label{coherent:xidef}
\end{equation}
where $\ztil(\xw,t)\equiv \frac{\xw-X}{\lambda}$.
The fluctuation $\fl(\xw-X,t)$ here is thus defined with respect to the wall position.
The derivatives of $\xi$ are 
\begin{eqnarray}
\partial_i \xi  &=& (\frac{\delta_{i,\xw}}{\lambda}+\partial_i \fl)\xi \nonumber\\
\partial_t \xi  &=& \lt( \partial_t \fl \rt) \xi .
\end{eqnarray}
Writing $\fl\equiv \flr+i\fli$, we obtain
\begin{eqnarray}
\bar{\xi}\dot{\xi} -\dot{\bar{\xi}}\xi &=& 
 2i \dot{\fli} |\xi|^2 \nonumber\\
\lambda^2 |\nabla_i \xi|^2 
 &=& [\delta_{i,\xw}\left(1 -2\lambda\nabla_\xw\flr \right) 
    +\lambda^2 |\nabla_i \fl|^2] |\xi|^2 .
\end{eqnarray}
We expand the Lagrangian (\ref{LSzero}) up to the second order in $\fl$ by use of
\begin{eqnarray}
\frac{|\xi|^2}{1+|\xi|^2} &=& \frac{1}{1+e^{2\ztil}} +\frac{1}{2\cosh^2 \ztil}
\left(\flr+\flr^2\tanh \ztil  \right),\nonumber\\
\frac{|\xi|^2}{(1+|\xi|^2)^2} &=& \frac{1}{4\cosh^2 \ztil}
 \left(1+2\flr\tanh \ztil 
 +\left(2-\frac{3}{\cosh^2\ztil}\right) \flr^2 \right),
\end{eqnarray}
and obtain
\begin{eqnarray}
\Ls &=& -\hbar S \sumx \frac{1}{\cosh^2\ztil}
 \left( \flr \partial_t \fli \right)
  -KS^2 \sumx  \frac{1}{\cosh^2\ztil} 
  \frac{\lambda^2}{2}|\nabla\fl|^2 -\Hw, \label{Lssw1}
\end{eqnarray}
where $\Hw=\Nw KS^2$ is the domain wall energy. Since $\Hw$ is a constant, we neglect it below.
In \Eqref{Lssw1}, linear terms in $\fl$ vanish since $\fl$ is defined around a classical solution, Eq. (\ref{collective:Xphi}).
The fluctuation $\fl$ is not a proper mode since the cubic terms have a 
weight factor $ \frac{1}{\cosh^2\ztil} $  in the integration. 
This is removed by redefining the fluctuation as
\begin{equation}
\fl \equiv 2 \fltil \cosh\ztil .
\end{equation}
Then we have, for example,
\begin{eqnarray}
\partial_t \fl &=& 2\cosh\ztil \partial_t \fltil 
 \nonumber\\
\partial_i \fl &=& 2\cosh\ztil 
 \left(\partial_i \fltil 
 +\tanh \ztil \frac{1}{\lambda}\delta_{i,\xw} \fltil \right).
\end{eqnarray}
The Lagrangian is then written as
(after symmetrization of the time-derivative term)
\begin{eqnarray}
 \Ls &=& 
\sumx \left[ 
  {i\hbar S} (\bar{\fltil} \dot{\fltil}-\dot{\bar{\fltil}} \fltil)
 -2KS^2  
 \left( {\lambda^2}|\nabla\fltil|^2 
 +\left(1-\frac{2}{\cosh^2\ztil}\right)|\fltil|^2 \right) 
\rt].
\end{eqnarray}
The dispersion of fluctuation is determined by the quadratic term.

\subsection{Spin-wave dispersion}
The spin-wave part is written conveniently by use of eigenfunctions satisfying
\begin{equation}
\left( -\lambda^2 \nabla_{\xw}^2
 +\left( 1- \frac{2}{\cosh^2\frac{\xw}{\lambda}} \right) \right)
\varphi_\omega =\omega \varphi_\omega,
\end{equation}
where $\omega$ is the eigenvalue. 
The eigenfunctions are well known \cite{Landau77}.
There is a single bound state with $\omega=0$,
\begin{equation}
\varphi_0(\xw) = \frac{1}{\cosh\frac{\xw}{\lambda}},
\end{equation}
and continuum states labeled by $k$:
\begin{equation}
\omega = 1+k^2\lambda^2 \equiv \omega_k,
\end{equation}
with
\begin{equation}
\varphi_k(\xw) = \frac{1}{\sqrt{2\pi\omega_k }} 
 \left(-ik\lambda+\tanh\frac{\xw}{\lambda}\right)e^{ik\xw}.
\end{equation}
The wave function $\varphi_0$ is called a zero mode after its zero eigenvalue.
The most important feature of the fluctuation is that 
the bound state function is a derivative of domain wall solution;
$\varphi_0(\xw) = -\lambda \nabla_{\xw}\thetaz$.
Other wave functions are orthogonal to zero-mode wave function:
\begin{equation}
\int_{-\infty}^{\infty} d\xw {\varphi}_{k}(\xw)\varphi_0(\xw)
 =0,
\label{coherent:orthognal1}
\end{equation}
and 
form an orthogonal base (dropping an oscillating term at $|\xw|\rightarrow\infty$),
\begin{eqnarray}
\int_{-\infty}^{\infty} d\xw \bar{\varphi}_{k'}(\xw) {\varphi}_{k}(\xw)
 &=& \frac{1}{2\pi\sqrt{\omega_k\omega_k'}} \int_{-\infty}^{\infty} d\xw
\left(e^{i(k-k')\xw}(1+\lambda^2 kk')
-\lambda \frac{d}{d\xw}\left(e^{i(k-k')\xw}\tanh\frac{\xw}{\lambda}\right) \right) \nonumber\\
&=&
\delta({k-k'}).
\label{coherent:orthogonal2}
\end{eqnarray}

\subsection{Zero mode}
By use of these eigenfunctions, we can expand
$\fltil$ as
\begin{equation}
\fltil(\xv,t) = 
\left( \frac{1}{2}\eta_0(t)\varphi_0(\xw) 
 +\sum_{k}\eta_k(t)\varphi_k(\xw)\right) ,
\label{coherent:wrongexpansion}
\end{equation}
(Here we neglect dependences of $\eta_0(t)$ and $\eta_k(t)$ on the wave vector perpendicular to the wire direction, considering a narrow wire.)
The Lagrangian is then written as
\begin{eqnarray}
\Ls &=& \Nw \left[ 
  \frac{i\hbar S}{4} (\bar{\eta_0} \dot{\eta_0}-\dot{\bar{\eta_0}} \eta_0)
  + \sum_{k} \left(  {i\hbar S} 
 (\bar{\eta}_k\dot{\eta}_k-\dot{\bar{\eta}}_k \eta_k)
 -2{KS^2}\omega_k |\eta_k|^2 \right) 
\right].
\end{eqnarray}
The zero mode, $\eta_0$, has an important role.
This can be seen by rewriting Eq. (\ref{coherent:xidef}) 
by use of fluctuation modes $\eta_0$ and $\eta_k$'s;
\begin{eqnarray}
\xi &=& e^{-\frac{\xw-X}{\lambda}+i\phiz} 
 \exp \left[
2\cosh \frac{\xw}{\lambda}
 (\frac{1}{2}\eta_0 \varphi_0 +\sum_k \eta_k \varphi_k)  \right]
\nonumber\\
&\simeq& e^{-\frac{\xw-X-\lambda \Re \eta_0(t)}{\lambda}}
  e^{i(\phiz+\Im \eta_0(t))}
 \exp \left[
2\cosh \frac{\xw}{\lambda} \sum_k \eta_k \varphi_k  \right]
. \label{coherent:xizero}
\end{eqnarray}
Thus the real part of $\eta_0$ corresponds to a translation of domain wall, $X$ and the imaginary part to $\phiz$.
Therefore, taking account of zero-mode fluctuations is equivalent to treating $X$ and $\phiz$ as dynamical variables, $X(t)$ and $\phiz(t)$, i.e., 
\begin{eqnarray}
X(t)&=& X+\lambda\Re \eta_0(t) \nonumber\\
\phiz(t)&=& \phiz+\Im \eta_0(t).
\end{eqnarray}
The spin Lagrangian is now written as the sum of the domain wall part and the spin wave part as
\begin{equation}
\Ls=\Ldw^{(0)}+\Lsw,
\end{equation}
where
\begin{eqnarray}
\Ldw^{(0)} &=& \frac{\hbar \Nw S}{2\lambda} (\dot{X}\phiz-X\dot{\phiz})  \nonumber\\
\Lsw &=& 
\Nw \sum_{k} \left(  {i\hbar S} 
 (\bar{\eta}_k\dot{\eta}_k-\dot{\bar{\eta}}_k \eta_k)
 -2{KS^2}\omega_k |\eta_k|^2 \right) .
\end{eqnarray}


\subsection{Condition of rigid wall}

When $\Kp$ is finite, the definition of $\phiz$ is determined by the condition of vanishing linear coupling to fluctuations  
 \cite{TT96}:
\begin{equation} 
 \phiz (t) \equiv \int \frac{d\xw}{2\lambda}\sin^2 \thetaz \, \phi (z,t)   \label{phizdef}
\end{equation}
where  
$\sin\thetaz = \left[ \cosh\frac{z-X(t)}{\lambda} \right]^{-1}$.

 If a pinning potential $V_0$ is present, the energy scale of $X$ motion is $V_0$. 
 Similarly, the energy scale of the $\phiz$-mode is given by $\Kp$. 
 Since the energy gap of the spin-wave mode is $\sim \sqrt{KK_\perp}$, the modes described by $X$ and $\phiz$ are at low energy compared to others if the following condition is satisfied: 
\begin{equation}
 \Vz \ll \sqrt{KK_\perp}, \ \ \  \Kp \ll K. 
\label{rigidity}
\end{equation}
 In this case, the low-energy wall dynamics is described by the two variables, $X$ and $\phiz$. 
 Otherwise, the pinning and/or $\Kp$ leads to deformation of the wall, whose description requires other variables than $X$ and $\phiz$.
 The condition (\ref{rigidity}) gives a criterion that such deformations can be neglected.

 Precisely speaking, we need one more condition for justifying collective description, namely, 
vanishing linear coupling of spin-wave modes to $X$ or $\phiz$. 
 In reality, when $V_0$ and $\Kp$ are finite,  such linear couplings arise, and the wall dynamics is not closed in $X$ and $\phiz$ in a strict sense.
This is quite natural since the pinning and $\Kp$ result in a deformation of the wall whose description requires other variables than $X$ and $\phiz$.
 However, the condition (\ref{rigidity}) also assures that such linear couplings are small. 
 We assume the condition (\ref{rigidity}) in this paper.

The wall solution we start from is thus given by $\theta\equiv\thetaz(\xv,t)$, where
\begin{eqnarray}
\cos\thetaz &\equiv&  \tanh\frac{\xw-X(t)}{\lambda}  \nonumber\\
\sin\thetaz &\equiv& \frac{1}{\cosh\frac{\xw-X(t)}{\lambda}},
\label{spin:dw1}
\end{eqnarray}
and 
\begin{equation}
\phi=\phiz(t). \label{spin:dw2}
\end{equation}

\subsection{Rigid domain wall Lagrangian}

From these considerations, the Lagrangian for the low-energy dynamics of a rigid wall is given by using ${\bm n}_0 = (\theta_0, \phi_0)$.
As a result,  $\Ls$ describing many spins 
(with Hamiltonian given by Eq. (\ref{spin:Hs}))
reduces to $\Ldw^{(0)}$ of two dynamical variables
($^{(0)}$ denotes without electrons):
\begin{eqnarray}
\Ldw^{(0)} &=& \hbar \Nw S \left( \frac{\dot{X}}{\lambda}\phiz 
   -\frac{\Kp}{2\hbar}S \sin^2\phiz \right) 
-\Vpin[\nvz],
  \label{Ldw0}
\end{eqnarray}
where $\Nw \equiv 2\lambda A/a^3$ is the number of spins in the wall. ($A$ is the cross-sectional area of the system.) 

The equation of motion of the wall is obtained simply by taking variations  with respect to $X$ and $\phiz$, including the  dissipation function,
\begin{equation}
\Ws=\frac{\alpha \Nw \hbar S}{2}
  \left( \frac{\dot{X}^2}{\lambda^2}+\dot{\phiz}^2 \right).
\end{equation}
In the absence of electrons and a magnetic field, the equation becomes (using Eq. (\ref{eqofmo0}) and Eq. (\ref{Vpin}) for pinning)
\begin{eqnarray}
\dot\phiz+\alpha \frac{\dot{X}}{\lambda}
 &=& -\frac{2\lambda\Vz}{\hbar S\xi^2}X\theta(\xi-|X|)
\label{DWeq_1sta}
\\ 
\dot{X}-\alpha\lambda\dot{\phiz} 
&=& \frac{\Kp\lambda}{2\hbar}S \sin 2\phiz  \label{DWeq_1stb}
.
\end{eqnarray}
The solution without any driving force is of course trivial: $\phiz=0$ and $X=0$.

\subsection{Domain wall particle ?}

The equation of motion (\ref{DWeq_1sta})(\ref{DWeq_1stb}) is one example of the Hamilton equation of motion for a classical object.
For a classical particle with mass $M$, the  
Hamilton equation for the position $X$ and momentum $P$ is given by
\begin{eqnarray}
\dot{P} &=& F \label{Pdot} \\
\dot{X} &=& \frac{P}{M} \label{Xdot}
\end{eqnarray}
where $F$ is the force.
For a domain wall, $\phiz$ plays the role of momentum $P$, and so the relation between $\dot{X}$ and momentum is not simply Eq. (\ref{Xdot}) but is non-linear (Eq. (\ref{DWeq_b})).
In this sense, the domain wall is not a classical particle.
 This is natural, since the domain wall (even in the rigid case) has an internal degree of freedom, $\phiz$.

Let us consider the equation of motion with harmonic pinning, 
\begin{eqnarray}
\dot\phiz+\alpha \frac{\dot{X}}{\lambda}
 &=& \frac{\lambda}{\hbar NS} \lt( F-2\Nw \frac{\Vz}{\xi^2} X \rt), 
\label{DWeq_a1}
\\ 
\dot{X}-\alpha\lambda\dot{\phiz} 
&=& \frac{\Kp\lambda}{2\hbar}S \sin 2\phiz  
   + \frac{\lambda}{\hbar NS} \torque.
\label{DWeq_b1}
\end{eqnarray}
Here we included force $F$ and torque $\torque$ (from magnetic field or current), which are constant in time here.

There is a limiting case, where the domain wall behaves like a particle described by one linear equation of motion. 
This occurs if $\phiz$ remains small, i.e, if the hard-axis anisotropy is strong.

\subsubsection{Weak extrinsic pinning}
Let us first consider the case of large $\Kp$.
In this case $\phiz\sim 0$ and so we can linearize the shape anisotropy term as $\sin 2\phiz \sim 2\phiz$.
Then we obtain from (\ref{DWeq_b1})  (neglecting $\alpha^2\ll1$)
\begin{equation}
\phiz= \frac{\hbar}{\Kp\lambda S}
 \lt( \dot{X}-\alpha \frac{\lambda^2}{\hbar \Nw S} 
 \lt( F-2\Nw \frac{\Vz}{\xi^2} X \rt)
  -\frac{\lambda}{\hbar \Nw S} \torque \rt) .
\end{equation}
Substituting $\phiz$ in Eq. (\ref{DWeq_a1}), we obtain the equation for $X$ as
\begin{equation}
\Mw\ddot{X}+\Mw\frac{1}{\tauw}\dot{X}+\Mw{\Omegapin}^2 X=F,
\end{equation}
where 
\begin{eqnarray}
\Mw &=& \frac{\hbar^2 \Nw}{\Kp \lambda^2} \label{Mwdef}\\
\Omegapin &=& \sqrt{\frac{2\Vz\Kp \lambda^2}{\hbar^2\xi^2} } \\
\frac{1}{\tauw} &=& \frac{\alpha}{\hbar} \lt(\Kp+2S\Vz\frac{\lambda^2}{\xi^2} \rt),
\end{eqnarray}
are mass, pinning frequency, and reciplocal decay time of the wall, respectively.
In this limit, the domain wall becomes a simple particle.
The existence of mass of the wall was first discussed by D\"oring \cite{Doring48}.

\subsubsection{Strong pinning or large $\phiz$}
When the pinning potential is harmonic, we can always delete $X$ from the equations of motion (\ref{DWeq_a1})(\ref{DWeq_b1}) and obtain one equation for $\phiz$ as
\begin{equation}
\Mphi\ddot{\phiz}
+\alpha \hbar S \Nw \dot{\phiz} 
  \lt(1+\frac{\Mphi \Kp}{\hbar^2 \Nw}\cos2\phiz \rt)
+\Nw\Kp S^2 \frac{1}{2}\sin 2\phiz = -\torque,
\end{equation}
where 
\begin{equation}
\Mphi = \Nw \frac{\hbar^2 S^2 \xi^2}{2\Vz \lambda^2},
\end{equation}
is the mass of the $\phi$-particle.
The $\phi$ variable thus is a particle moving in a sine or cosine potential \cite{Braun96,TT96}.

Comparing two masses $\Mw$ and $\Mphi$, we see that $\Mw$ is lighter than $\Mphi$ when pinning is weak, $\Vz \ll \Kp$. 
This indicates that $X$ is a better variable to describe
the weak pinning case.
In contrast, $\Mw$ is heavier if pinning is strong, $\Vz \gg \Kp$, and in this case, the system is better described in terms of the dynamics of $\phiz$ \cite{Braun96,TT96}.
This applies at the quantum level, too. 
In a path integral representation, the vacuum transition amplitude for the Lagrangian Eq. (\ref{Ldw0}) is given as (approximating $\phiz\ll1$)
\begin{equation}
Z_{\dw} =\int {\cal D}X{\cal D}\phiz 
e^{-\frac{\Nw}{\hbar} \int dt\lt(\hbar S\frac{\dot{X}}{\lambda}\phiz
-\frac{\Kp S^2}{2}\sin^2 \phiz -\frac{\Vz}{\xi^2}X^2  \rt)}.
\end{equation}
When the hard-axis anisotropy is strong, $\phiz$ is a high-energy mode, and should be integrated out first, resulting in
\begin{equation}
Z_{\dw} =\int {\cal D}X 
e^{-\frac{1}{\hbar} \int dt\lt( 
\frac{\Mw}{2} \dot{X}^2 -\frac{\Mw}{2}\Omegapin^2 X^2  \rt)}.
\end{equation}
Thus the wall behaves like a standard particle.
When pinning is strong, we can integrate out $X$ and obtain 
\begin{equation}
Z_{\dw} =\int {\cal D}\phiz 
e^{-\frac{1}{\hbar} \int dt\lt(
\frac{\Mphi}{2} \dot{\phiz}^2
-\Nw \frac{\Kp S^2}{2}\sin^2\phiz \rt)}.
\end{equation}
In this case, $\phiz$ (chirality of the wall) becomes a good variable, and tunneling of chirality can occur as discussed in Refs.  \cite{Braun96,TT96}.

\section{Conduction electrons}
\label{SEC:electron}

In this section, we consider conduction electrons.
We use the second quantized representation, and write
annihilation and creation operators for the electron with spin $\sigma=\pm$ at site $\xv$ and at time $t$  by $c_\sigma(\xv,t)$ and $\cdag_\sigma (\xv,t)$, respectively. 
The electron density is $\cdag(\xv) c(\xv)$.
We sometimes suppress spin indices when obvious, {\it e.g.},
$\cdag c$ denotes $\sum_{\sigma} \cdag_\sigma c_\sigma$.
The Hamiltonian for the electron we consider is given by a free part $H_0$, an impurity scattering $\Himp$, a spin flip part $\Hsf$, and an interaction with the electric field $\Hem$ which drives the current.
(The interaction with localized spin is explained later in \S \ref{SEC:adiabatic}).
The electron Hamiltonian is thus 
\begin{equation}
\He \equiv \intx
  \left( \frac{\hbar^2}{2m}|\nabla c|^2 -\eF c^\dagger c \right)+\Himp+\Hsf+\Hem,   \label{He}
\end{equation}
with $\eF$ being Fermi energy.

\subsection{Free electron}
The free part, $H_0$, is given in Fourier space as
$H_0=\sumkv \ekv c^\dagger_{\kv} c_{\kv}$ where 
$\ekv\equiv \frac{\hbar^2k^2}{2m}-\eF$ and the Fourier transform is defined as 
$c(\xv)=\frac{1}{\sqrt{V}}\sumkv e^{i\kv\cdot\xv} c_{\kv} $.
Let us here briefly examine the behavior of the free electron Green's function. 
Conventional Green's functions such as retarded and advanced Green's functions are not physical quantities, but are useful in calculation.
Let us consider the behavior of a function 
$\rho(\xv,t,\xv',t')\equiv \average{c^\dagger(\xv',t')c(\xv,t)}_{H_0}$, which is the electron density if at equal times and equal positions. 
The average $\average{\cdots}_{H_0}$ is for free states determined by $H_0$.
This function satisfies an "equation of motion"
\begin{eqnarray}
\partial_t \rho(\xv,t\xv',t') &=& 
-\frac{i}{\hbar} \average{c^\dagger(\xv',t')[H_0,c(\xv,t)]}_{H_0}
\nonumber\\
 &=& i\lt(\frac{\hbar}{2m}\nabla_{x}^2+\eF/\hbar\rt) \rho(\xv,t,\xv',t').
\end{eqnarray}
This equation is in fact a continuity equation, but is not convenient for calculating non-free cases.
A more useful function can be defined if one includes anti-commutator and step function in time as
\begin{equation}
\gr(\xv,t,\xv',t')\equiv -\frac{i}{\hbar}
\theta(t-t')\average{\{c^\dagger(\xv',t'),c(\xv,t)\}}_{H_0}.
\end{equation}
The equation of motion is then 
\begin{eqnarray}
\lt( i\hbar \partial_t +\lt(\frac{\hbar^2}{2m}\nabla_{x}^2+\eF/\hbar\rt) \rt) \gr(\xv,t,\xv',t') 
&=& \delta(t-t')\delta^3(x-x'),\label{dysonfree}
\end{eqnarray}
which is a Green's function with a source term. 
This type of equation is mathematically useful allowing systematic calculation.
This function is called a retarded Green's function due to a factor $\theta(t-t')$.
The equation (\ref{dysonfree}) is easily solved using the  Fourier transform, 
$\gr(\xv,t,\xv',t)=\frac{1}{V}\sumom\sumkv e^{i(\kv\cdot\xv-\omega t)} g^{\rm r}_\kv(\omega)$ 
as
\begin{equation}
g^{\rm r}_\kv(\omega) =\frac{1}{\omega-\ekv+i0},
\end{equation}
where the small imaginary part denoted by $i0$ is to reproduce 
the retardation factor, $\theta(t-t')$.

\subsection{Impurity scattering (spin-independent)}

We include here the effect of electron scattering by impurities.
The scattering is treated as elastic and spin-independent, and is then described by a term 
\begin{eqnarray}
\Himp&=& \sum_{i=1}^{\Nimp} \intx v(\xv-\Rv_i) \cdag(\xv)c(\xv) \nonumber\\
 &=& \sum_{i=1}^{\Nimp}\sum_{\kv\kv'} 
 v(\qv)e^{i(\kv-\kv')\cdot\Rv_i} \cdag_{\kv'}c_{\kv},
\end{eqnarray}
where $v$ represents the potential due to an impurity, $\Rv_i$ represents the position of random impurities, $\Nimp$ is the number of impurities, and 
$v(\qv)\equiv \frac{1}{V}\intx e^{i\qv\cdot\xv}v(\xv)$.
We approximate the potential as an on-site type, 
$v(\xv-\Rv_i)=\vimp a^3 \delta^3(\xv-\Rv_i)$ ($\vimp$ is a constant), i.e., $v(\qv)=\frac{\vimp}{N}$ ($N\equiv V/a^3$ is number of sites).
To estimate physical quantities, we need to take the random average over impurity positions as 
\begin{eqnarray}
\sum_i \average{e^{-i\qv\cdot\Rv_i}} &\equiv& 
\sum_i \int \frac{d^3R_i}{V} e^{-i\qv\cdot\Rv_i}
=\Nimp \delta_{\qv,0}=0\;\;\;(\mbox{\rm for } q\neq0) \nonumber\\
\sum_{ij} \average{e^{-i\qv_1\cdot\Rv_i}e^{-i\qv_2\cdot\Rv_j}}
& =& \sum_{ij} \int \frac{d^3R_i}{V} \int \frac{d^3R_j}{V} e^{-i\qv_1\cdot\Rv_i}e^{-i\qv_2\cdot\Rv_j}\nonumber\\
&=&\Nimp \delta_{\qv_1+\qv_2,0}
 +\Nimp^2 \delta_{\qv_1,0} \delta_{\qv_2,0} 
=\Nimp \delta_{\qv_1+\qv_2,0},
\end{eqnarray}
where the term linear in $\Nimp$ is a contribution at $i=j$.
Taking account of successive impurity scattering by ladders, the electron Green's function 
in the presence of random impurities is given by e.g.,
$\gr_{\kv}(\omega)=\frac{1}{\omega-\ekv+\frac{i}{2\tau}}$, 
where the inverse lifetime is given as
\begin{equation}
\frac{1}{\tau} =-\frac{2}{\hbar} 
 \lt(\frac{\vimp}{N}\rt)^2\Im\sum_{ij}
\sum_{\qv_1\qv_2}
\average{e^{-i\qv_1\cdot\Rv_i}e^{-i\qv_2\cdot\Rv_j}
}g^{\rm r}_{\kv+\qv_1}(\omega) 
\simeq 
\frac{2\pi}{\hbar}{\nimp} \vimp^2 \DOS a^3.
\end{equation}
Here $\DOS$ is the density of states per volume ($\DOS\equiv N(0)/Na^3$, $N(0)$ is the density of states) and 
$\nimp\equiv \frac{\Nimp}{N}$ is the impurity concentration.
In the derivation we used 
$\sumkv \gr_{\kv}(\omega)
\simeq N(0)\int_{-\infty}^{\infty}d\epsilon \frac{1}{-\epsilon+\frac{i\hbar}{2\tau}}$
$\simeq -i\pi N(0)$.
We consider a conducting system, and so
\begin{equation}
\frac{\ef\tau}{\hbar} \gg 1.
\end{equation}
When spin polarization due to (uniform component of) localized spin is taken into account, the lifetime becomes spin-dependent.
For the most part of this paper, we neglect this spin-dependence, to avoid unnecessary complication.
The spin-dependence is correctly included for evaluation of delicate quantities such as $\alpha$ and $\beta$ below in \S\ref{SEC:singlespin}.

\subsection{Spin relaxation}

We also include spin flip interaction due to random magnetic impurities to take account of the spin relaxation effect.
The interaction is represented by
\begin{equation}
\Hsf = u_s \intx \sum_i \Simpv_i \delta(\xv-\Rv'_i)
(\cdag\sigmav c)_\xv,
\end{equation}
where $\Simpv_i$ represents the impurity spin at site $\Rv'_i$.
(The spin-orbit interaction leads to essentially the same results as spin flip case here.)
A quenched average for the impurity spin was taken as 
\begin{eqnarray}
  \overline{\Simp_i^\alpha \Simp_j^\beta} 
&=& \delta_{ij} \delta_{\alpha\beta} \times \left\{ \begin{array}{cc} 
    \overline{\Simp_\perp^2} & (\alpha, \beta = x,y) \\ 
    \overline{\Simp_z^2}     & (\alpha, \beta = z) 
    \end{array} \right. .
\end{eqnarray}
The effect of spin relaxation is considered in \S\ref{SEC:singlespin}.

\subsection{Electric field}

In this study, the applied electric current has the most important role of inducing spin dynamics. 
The effect of current is calculated as a response to the applied electric field. 
The interaction is expressed by use of uniform charge current density $\jv$ and electromagnetic gauge field $\Avem$ as
\begin{equation}
\Hem = - \int \dx \Avem \cdot \jv, \label{Hemdef}
\end{equation}
The gauge field is given by use of $\Ev\equiv -\dot{\Av}$ as 
\begin{equation}
\lt. 
\Avem= i\frac{\Ev}{\Omz}e^{i\Omz t}\rt|_{\Omz\ra 0},
\end{equation}
where $\Ev$ is the applied electric field assumed to be spatially homogeneous.
For calculation purpose, it is treated as having finite frequency, $\Omz$, which is chosen as $\Omz\rightarrow0$ at the last stage of the calculation (this is a standard technique in linear response calculation \cite{Rammer86}).
The current density is given in the presence of the gauge field by
($e<0$ is electron charge)
\begin{eqnarray}
j_\nu &\equiv& -\Vinv \int \dx \frac{e\hbar}{m}\frac{i}{2} 
(c^{\dagger}\stackrel{\leftrightarrow}{\nabla}_{\nu}c)
-\frac{e^2\hbar}{m} A_{{\rm em},\nu} \sum_{\kv}
c^{\dagger}_{\kv}c_{\kv}
\nonumber\\
&=& \frac{e}{m}\sum_{\kv}
\lt(\hbar k_\nu- eA_{{\rm em},\nu}\rt) c^{\dagger}_{\kv}c_{\kv}.
\label{jemdef}
\end{eqnarray}
Within the linear response to $\Ev$, the last term is neglected. 
We used the coupling $(\Av\cdot\jv)$ between gauge field and current, and not the one between electric charge and scalar potential $(\Phi\rho)$, since the $\Av\cdot\jv$-description is known to be convenient in describing the system as spatially uniform as in the Kubo formula.
(In contrast, $\Phi\rho$ description is useful in a Laudauer type description treating the spatial difference of chemical potential explicitly.)

The electron part of the Lagrangian
is given by use of the Hamiltonian $\He$ as 
\begin{equation}
\Le =  i\hbar \intx  \cdag \dot{c} 
-  \He
,\label{Le}
\end{equation}
where the first term is a dynamical term that correctly reproduces the Schr\"odinger equation.

\subsection{$s$-$d$ exchange interaction and adiabatic condition}
\label{SEC:adiabatic}
The most important interaction for us is the exchange interaction between electron spin and localized spin, which is a source of all current-driven spin dynamics and 
magnetic transport properties.
This term is given by
\begin{equation}
\Hsd \equiv -\Jsd \intx \Sv \cdot (c^\dagger \sigmav c)
= -{\spol}\intx
  \evs \cdot (c^\dagger \sigmav c), \label{Hsd}
\end{equation}
where $\Jsd$ is the strength of $s$-$d$ exchange interaction and $\spol\equiv \Jsd S$ is half the exchange splitting.
 An important point is that $J_{sd}$ is rather strong in 3-d ferromagnets: 
$J_{sd}/\eF \gtrsim O(0.1 -1)$. 
 These values are indicated from experimental observations of large magnetoresistances such as the GMR.

The most non-trivial part of the theory is the treatment of this strong exchange interaction when the localized spin has a spatial structure and/or is dynamical. 
 Fortunately, spin structures in 3-d ferromagnets are slowly varying compared to the scale of conduction electrons. 
 This is a consequence of the strong exchange interaction, $J$, between localized spins, which is of order of 1000 K as indicated by the high critical temperature of 3-d ferromagnets (For Fe, $\Tc\sim 1043$ K). 
(Correctly, the typical length scale, $\lambda$, is determined by the ratio of exchange energy and magnetic anisotropy 
(Eq. (\ref{thickness})).) 
Since many localized spins within the scale of $\lambda$ are coupled, the spin structure is (semi-) macroscopic and its time scale is slow compared to that of electrons. 
From these considerations, the electron can go through the spin structures adiabatically.

In our study, we carry out the expansion with respect to the gauge field representing the non-adiabaticity.
The gauge field contains the space and time derivatives of localized spins. 
The expansion parameters are given by
\begin{eqnarray}
\frac{\ef}{\spol} \frac{1}{\kf\lambda}\frac{\ell}{\lambda}
&=& \frac{1}{(\kf\lambda)^2}\frac{\ef^2 \tau}{\hbar \spol} \ll 1  \nonumber\\
\frac{\hbar  \omega_{\dw}^2 \tau}{\spol} &\ll& 1,
\label{adiabaticconditon1}
\end{eqnarray}
where $\omega_{\dw}$ is the frequency scale of the domain wall motion.
This is understood by noting that the electron spin density lowest order in the gauge field is given by Figs. \ref{FIGse0} and \ref{FIGsevc}, and the correction to these processes contains a factor of
$\frac{1}{m^2} (\kv\cdot\Av^+) (\kv\cdot\Av^-) g_{\kv,+}g_{\kv,-}$ or
$A_0^+ A_0^- g_{\kv,+}g_{\kv,-}$ in the adiabatic limit, where $g$ is either the retarded or advanced Green's function.
Noting $A_i \sim |\nabla_i \Sv|\sim {\lambda}^{-1}$,
$A_0 \sim \omega_{\dw}$, and $g_{\kv,+}g_{\kv,-}\sim \frac{\tau}{\hbar\spol}$ when $\spol\tau/\hbar \gg1$, 
we have identified the expansion condition as given by \Eqref{adiabaticconditon1}.
The gauge field expansion is thus justified if either the spin splitting is large or the spin structure is slowly varying, and it results in a different series expansion from the simple gradient expansion assuming $(\kf\lambda)^{-1}\ll1$ and $\omega_{\dw}\tau \ll1$.
(In Ref.  \cite{Thorwart07}, the gradient expansion condition was argued to be 
$\lambda_{\rm st} /\lambda \ll1$ and 
$\lambda_{\rm st}\omega_{\dw}/\vf \ll1$, where they introduced  a phenomenological parameter of spin transport length scale $\lambda_{\rm st}$ \cite{Zhang04}.
However, the result (Eq. (39) of Ref. \cite{Thorwart07}) appears to be
that of a simple gradient expansion assuming  
$(\kf\lambda)^{-1}, \omega_{\dw}\tau \ll1$. )

In actual 3-d ferromagnets, $\spol/\ef \lesssim 1$ and therefore the gauge field expansion would become essentially the same as a gradient expansion. 
Nevertheless, it would be formally useful to carry out the gauge field expansion first and then consider a slowly varying limit as we will do in \S\ref{SEC:force} in estimating the reflection force.

The adiabatic condition obtained above has an extra factor of $\ell/\lambda$ due to disorder scattering if we compare with the condition proposed by Waintal and Viret \cite{Waintal04}.
They obtained in the ballistic case the adiabaticity condition of 
\begin{equation}
\frac{1}{\kF \lambda}\frac{\eF}{\spol} \simeq \frac{1}{(\kfu-\kfd)\lambda} \ll 1 , \label{adiabatic2}
\end{equation}
where the left-hand side is a ratio of the precession time of conduction electron due to the exchange interaction, $\hbar/\spol$, to the time needed for the electron to pass through the spin structure,  $\lambda/\vf$.

In the context of quantum electron transport, Stern \cite{Stern92}  introduced a different condition in the disordered case,
\begin{equation}
\hbar/(\spol \tau) \ll 1 . \label{adiabatic1}
\end{equation}
This would be satisfied in 3-d ferromagnets, but is not a necessary condition in our  calculation.
For quantum transport, other conditions have been proposed \cite{Popp03}, which appear 
to depend on the system considered.

\section{Equation of motion of domain wall under current}
\label{SEC:dweq0}
\subsection{Effective Lagrangian}

As we have discussed, the total system we consider is described by the Lagrangian $L\equiv \Ls+\Le-\Hsd$ (eqs.(\ref{Ls})(\ref{spin:Hs})(\ref{He})(\ref{Le})(\ref{Hsd})).
We are interested in localized spin dynamics, and for this purpose, we derive the effective Lagrangian for localized spin by integrating out the electron.
"Integrate out" here means taking the trace over the quantum mechanical states of the electron. 
In the path-integral formalism \cite{Feynman65}, this process corresponds indeed to an integration in the following way.
The partition function of the system is represented as
\begin{equation}
Z=\int {\cal D}\theta {\cal D}\phi \sin\theta {\cal D}\bar{c}{\cal D} c
e^{i\int dt \lt(\Ls[\Sv]+\Le[\bar{c},c]-\Hsd[\Sv,\bar{c},c]\rt)}.
\end{equation}
Here ${\cal D}\theta $ denotes integration over the field variable
(${\cal D}\theta =\Pi_{\xv,t}d\theta(\xv,t)$), and 
$\bar{c}$ and $c$ are Grassmann numbers corresponding to creation and annihilation operators for the electron. 
The time integration is on the real axis, but the argument here applies also to the case of the Keldysh contour $C$. 
By integration over the electron, $Z$ reduces to 
\begin{equation}
Z=\int {\cal D}\theta {\cal D}\phi \sin\theta
e^{i\int dt L^{\rm eff}_S[\Sv]},
\end{equation}
where 
$L^{\rm eff}_S[\Sv] = \Ls+\Delta L_S$ is the effective Lagrangian for localized spin and 
\begin{equation}
\int dt \Delta L_S[\Sv]\equiv -i\ln 
\int {\cal D}\bar{c} {\cal D}c 
e^{i\int dt \lt(\Le[\bar{c},c]-\Hsd[\Sv,\bar{c},c]\rt)}
\equiv -i\ln Z_e, \label{delLs}
\end{equation}
is the contribution from electrons, which includes formally everything from the electrons exactly.
The equation of motion of spin with all the effects from electrons included is then written as
(neglecting dissipation)
\begin{equation}
\deld{L^{\rm eff}_S}{\Sv}=0. \label{eqfull}
\end{equation}
Let us look into this equation in more detail.
The electron contribution is written as
\begin{eqnarray}
\deld{\Delta L_S}{\Sv} &=&
\frac{1}{Z_e}\int {\cal D}\bar{c} {\cal D}c 
\lt(-\Jsd (\bar{c} \sigmav c) \rt)
e^{i\int dt \lt(\Le[\bar{c},c]-\Hsd[\Sv,\bar{c},c]\rt)}
\nonumber\\
&=& -\Jsd \sev 
\end{eqnarray}
where we noted that the right-hand side of the first line is the definition of electron spin density, 
\begin{equation}
\sev(\xv,t)\equiv \average{c^\dagger(\xv,t) \sigmav c(\xv,t)}.
\label{selectrondef}
\end{equation}
(We define spin density without the factor of $\hf$ representing electron spin magnitude.)
The average here is taken using the full electron Hamiltonian, $\He+\Hsd$, namely taking account of background localized spin structure, electric field, impurity scattering, and spin relaxation.
Let us define the effective field from the electron, $\Bve$, as
\begin{equation}
\Bve\equiv - \frac{1}{\hbar \gamma} \Jsd \sev .
\label{Bvedef}
\end{equation}
We now understand that the effect of the electron is taken into account simply by adding the effective field from the electron, $\Bve$, to the equation of motion (\ref{spin:eqS0}). 
The full equation of motion (\ref{eqfull}) thus reduces to 
(now including dissipation)
\begin{equation}
\del{\Sv}{t}=
\gyro \Bvs\times \Sv  -\gyro \Sv\times \Bve  
- \frac{\alpha}{S}\Sv\times \del{\Sv}{t}.
    \label{LLGfull}
\end{equation}
The effective Lagrangian for localized spin is therefore given by 
\begin{equation}
\Ls^{\rm eff} = \Ls -\Jsd \intx \Sv\cdot\sev.
\label{Leffdef}
\end{equation}
Note carefully, however, that if we solve for the electron spin density $\sev$ (\Eqref{svdef}) and put the result in \Eqref{Leffdef}, we obtain the trivial answer of no effect from the current.
To discuss spin dynamics, we have to first derive the equation of motion regarding $\sev$ as independent variable as $\Sv$, and then apply the result of $\sev$. 
This is what we do below.

\subsection{Equation of motion}

From the considerations above, the effective Lagrangian of the domain wall in the presence of electrons is given by
$\Ls+\average{\Hsd}$. 
Replacing the localized spin direction $\evs$ by the domain wall configuration $\evsz$ (whose polar coordinates are $(\theta_0, \phi_0)$ (Eq. (\ref{spin:dw1})))
yields
\begin{eqnarray}
\Ldw &=& \hbar NS \left( \frac{\dot{X}}{\lambda}\phiz 
   -\frac{\Kp}{2\hbar}S \sin^2\phiz \right) 
-\Vpin[\evsz]
 +{\spol} \intx \, \evsz \cdot \sev.
  \label{Ldw}
\end{eqnarray}
Noting 
\begin{eqnarray} 
\deld{\evsz}{X} &=& -\nabla_{\xw} \evsz \nonumber\\
\deld{\evsz}{\phiz} &=& \sin\theta_0 \evph \cdot \sev  
 = (\evsz \times \sev)_z,
\end{eqnarray}
the equation of motion of the domain wall under current is given by
\begin{eqnarray}
\dot\phiz+\alphaz \frac{\dot{X}}{\lambda}
 &=& \frac{\lambda}{\hbar NS} (\Fe+\Fpin) , 
\label{DWeq_a}
\\ 
\dot{X}-\alphaz\lambda\dot{\phiz} 
&=& \frac{\Kp\lambda}{2\hbar}S \sin 2\phiz  
   + \frac{\lambda}{\hbar NS} \torquee_{z}.
\label{DWeq_b}
\end{eqnarray}
Here $\Fpin\equiv -\del{\Vpin[\evsz]}{X}$, and 
the force and torque due to electrons are defined as
\begin{eqnarray}
\Fe & \equiv & - \average{\deld{\Hsd}{X}} 
  = -{\spol}\intx \, \nabla_{\xw}\evsz \cdot\sev , \label{Fdef}
\\
\torqueve & \equiv & -\intx \average{\deld{\Hsd}{\Sv}}  \times \Sv
 = -{\spol}\intx  (\evsz \times\sev).\label{Tdef}
\end{eqnarray}
We stress here again that these equations contain all the effects of the electron without any approximation so far. 
We note also that this set of equations, (\ref{DWeq_a}) and (\ref{DWeq_b}), is essentially the same as those obtained by Berger \cite{Berger84,Berger92}.
What is new and essential in the present theory is that we have formal but exact expressions of force and torque, which we can evaluate by a systematic diagrammatic method.

\subsection{Equation of motion from the Landau-Lifshitz-Gilbert equation}
\label{SEC:eqfromLLG}
The equation of motion of the domain wall can be derived from the LLG equation, \Eqref{modLLG}, by using the domain wall solution including collective coordinates, \Eqsref{spin:dw1}{spin:dw2}.
Here we neglect nonlocal terms for simplicity. 
(These terms are calculated in \S\ref{SEC:force}.)
Using
$\partial_\mu\Sv=S( (\partial_\mu\theta_0)\evth
 +\sin\theta_0(\partial_\mu\phiz)\evph)$, 
and $\dot{\theta_0}=\frac{\dot{X}}{\lambda}\sin\theta_0$,
$\nabla_\xw{\theta_0}=-\frac{1}{\lambda}\sin\theta_0$, the LLG equation reduces to (see also \Eqsref{spin:eqtheta1}{spin:eqphi1})
\begin{eqnarray}
\dot{\theta_0}-\alpha \sin\theta_0 \dot{\phiz} &=&
\sin\theta_0\lt(\frac{\dot{X}}{\lambda}-\alpha\dot{\phiz}\rt) \nonumber\\
&=&
\frac{a^3}{2eS\lambda}\js \sin\theta_0
+
 \frac{\Kp S}{2\hbar} \sin\theta_0 \sin2\phiz
\nonumber\\
 \sin\theta_0 \dot{\phiz}+\alpha \dot{\theta_0}
 &=&
 \sin\theta_0\lt(\dot{\phiz}+\alpha\frac{\dot{X}}{\lambda}\rt)
   \nonumber\\ 
 &=& 
\frac{a^3}{2eS\lambda}\betasf \js \sin\theta_0
-\frac{KS}{2\hbar} \sin2\theta_0
 (1+\kappa\sin^2\phiz)   .
\end{eqnarray}
Integrating over position $\xv$, we obtain 
\begin{eqnarray}
\frac{\dot{X}}{\lambda}-\alpha\dot{\phiz}
&=&
\frac{a^3}{2eS\lambda}\js 
+
 \frac{\Kp S}{2\hbar} \sin2\phiz
\nonumber\\
\dot{\phiz}+\alpha\frac{\dot{X}}{\lambda}
 &=& 
\frac{a^3}{2eS\lambda}\betasf \js ,
\end{eqnarray}
which is \Eqsref{DWeq_a}{DWeq_b} with force from the $\betasf$ term included (see \S\ref{SEC:force}).

This derivation of the domain wall equation from the LLG equation is useful as far as only local torque is concerned, but is wrong when non-adiabaticity is to be considered.
In this paper, we proceed based on equations \ref{DWeq_a}
 to \ref{Tdef}, which include all the torques without local approximation.

\subsection{Spin conservation law}
Let us briefly look into the conservation law of spin.
The equation of motion of localized spin is given by \Eqref{LLGfull}.
The spin part of the effective field, $\Bvs$, is described by \Eqref{spin:Hs} as (neglecting pinning)
\begin{eqnarray}
\hbar\gyro\Bvs &=&
-J\nabla^2\Sv-K\evz S_z+\Kp\evy S_y  ,
\end{eqnarray}
and so the LLG equation \Eqref{LLGfull} is written as
\begin{eqnarray}
\dot{\Sv}+\nabla\cdot \JSv &=&
 - \frac{\Jsd}{\hbar a^3} \Sv\times \sev + \bm{\tau}_{S}
,  \label{Scont}
\end{eqnarray}
where the spin current associated with localized spin is given by
\begin{equation}
{\JS} _\mu ^\alpha \equiv \frac{J}{\hbar} 
 ((\nabla_\mu \Sv)\times \Sv)^\alpha,
\end{equation}
and the spin source or sink is given by anisotropy and Gilbert damping as
\begin{equation}
\bm{\tau}_{S} =\frac{\alpha}{S}\Sv\times\dot{\Sv}
  - K \lt( \begin{array}{c} - S_y S_z \\
             S_x S_z \\
          0  \end{array} \rt)
  - \Kp \lt( \begin{array}{c}  S_y S_z \\
             0 \\
           -S_x S_y  \end{array} \rt).
\end{equation}

The equation for the spin density of the conduction electron, defined by \Eqref{selectrondef}, can be derived by considering its time derivative,
\begin{eqnarray}
\del{\sev}{t} &=& \frac{i}{\hbar}
  \average{c^\dagger \sigmav [H,c] -[H,c^\dagger]\sigmav c },
\end{eqnarray}
and evaluating the commutation relation with the total Hamiltonian $H$.
Using, e.g., $[\cdag_{\xv'}c_{\xv'},c_{\xv}]=-c_{\xv}\delta(\xv-\xv')$,
we obtain
\begin{eqnarray}
\hf\del{\sev}{t} &=& -\hf\nabla\cdot \jsv 
 + \frac{\Jsd}{\hbar a^3} \Sv\times \sev +\bm{\tau}_{s}.  \label{scont}
\end{eqnarray}
Here $\jsv$ is the spin current density 
(divergence here is with respect to spatial coordinate),
defined as
\begin{equation}
{\js} _{\mu}^\alpha =\frac{-i\hbar}{2m}  
  \average{\cdag \sigma^\alpha 
\overleftrightarrow{\nabla}_\mu c },
\end{equation}
where 
$A \overleftrightarrow{\nabla}_\mu B 
\equiv A (\nabla_\mu B) -(\nabla_\mu A) B$,
and $\bm{\tau}_{s}$ represents relaxation of electron spin.

Combining \Eqref{Scont} and \Eqref{scont}, we see that $s$-$d$ exchange torques cancel each other in the equation of motion for the total spin, $\Stotv\equiv \Sv+\frac{a^3}{2}\sev$, and total spin current, $\JStotv\equiv \JSv+\frac{a^3}{2}\jsv$. 
This is natural since exchange interaction is the internal exchange of angular momentum, which does not change total spin dynamics.
The continuity equation thus becomes
\begin{eqnarray}
\dot{\Stotv}+\nabla\cdot \JStotv &=&
\bm{\tau}_{S}+\bm{\tau}_{s}
.  \label{Stotcont}
\end{eqnarray}
This continuity equation is another representation of current-induced torques, where the torque due to current is included in the $\nabla\cdot\jsv$ term \cite{TE08}. 

\section{Calculation of electron spin density}
\label{SEC:spindensity}


In this section, the electron spin density is calculated.
The calculation in this section is done for general spin structures, not restricted to domain walls.

\subsection{Gauge transformation}
Our task now is to calculate the electron spin density $\sev(\xv,t)$.
This is non-trivial, since the electrons are interacting with the background spin, which is spatially and temporally non-uniform.
For estimating $\sev$, we consider first the free part of the electron with the exchange coupling. 
The corresponding Lagrangian (we call $\Lez$) is 
\begin{eqnarray}
\Lez\equiv \hbarinv \left[ \intx [i\hbar \cdag \dot{c} 
-  \left( \frac{\hbar^2}{2m}|\nabla c|^2 -\eF c^\dagger c \right)\right]
+ {\spol} \intx
  \evs \cdot (c^\dagger \sigmav c).
  \label{Le0}
\end{eqnarray}
As we discussed in \S\ref{SEC:adiabatic}, we are interested in the adiabatic regime, and the treatment using a gauge transform becomes useful in this case.

The idea of a local gauge transformation is simply to diagonalize locally the $s$-$d$ exchange interaction. 
This is always possible by choosing an appropriate $2\times 2$ unitary matrix $U(\xv,t)$ such as
\begin{equation}
U^\dagger(\xv,t) ( \evs(\xv,t) \cdot \sigmav )U(\xv,t) =\sigma_z.
\end{equation}
(Since the localized spin direction depends on position and time, the matrix $U$ also is, i.e., $U(\xv,t)$.)
This transformation is implemented by choosing 
\begin{equation}
U(\xv,t)= \mv(\xv,t)\cdot\sigmav,
\end{equation}
where $\mv$ is a real three-component unit vector given as
\begin{equation}
\mv\equiv\left(
\sin\frac{\theta}{2}\cos\phi,\sin\frac{\theta}{2}\sin\phi,\cos\frac{\theta}{2} \right).
\label{mdef}
\end{equation}
The matrix $U$ satisfies $U^2=1$, or
\begin{equation}
U(\xv,t)^{-1}=U(\xv,t).
\end{equation}
This unitary transformation
corresponds to defining a new electron operator 
$a\equiv ({a_+},{a_-})^{\rm t}$ (t represents transpose) as
\begin{equation}
c(\xv,t)\equiv U(\xv,t) a(\xv,t).
\end{equation}
Now the exchange interaction is diagonalized, but this redefinition affects the kinetic term of the electron. In fact, we immediately see from the identity
\begin{equation}
\partial_\mu c(\xv,t)=U(\xv,t)(\partial_\mu+U(\xv,t)^{-1}\partial_\mu U(\xv,t))a
  = U(\xv,t)(\partial_\mu+iA_\mu)a ,
\end{equation}
that a gauge field appears, defined as
\begin{equation}
A_\mu\equiv -iU(\xv,t)^{-1}\partial_\mu U(\xv,t).
\end{equation}

By use of (\ref{mdef}), $A_\mu$ is written as
\begin{equation}
A_\mu=(\mv\times\partial_\mu \mv)\cdot \sigmav \equiv A_\mu^\alpha \sigma_\alpha  ,
\end{equation}
where summation over $\alpha=x,y,z$ is suppressed.
The gauge field is explicitly obtained
in a vector notation with respect to the spin index ($\alpha$) as
\begin{equation}
\Av_\mu= \hf 
\vec3{
-\partial_\mu \theta \sin \phi -\sin\theta \cos\phi \partial_\mu \phi }{
\partial_\mu \theta \cos \phi -\sin\theta \sin\phi \partial_\mu \phi }{
  (1-\cos\theta)\partial_\mu \phi  }
\equiv \Ath_\mu\evth+\Aph_\mu\evph-\Az_\mu\evs.
\end{equation}
Here the components of gauge fields are defined as
\begin{eqnarray}
\Ath_\mu &\equiv & \evth\cdot\Av_\mu
= \hf \sum_{\pm} e^{\mp i\phi} A_\mu^\pm 
  = -\hf\sin\theta \partial_\mu{\phi}
\nonumber\\
\Aph_\mu &\equiv& \evph\cdot\Av_\mu
=\frac{i}{2} \sum_{\pm} \mp e^{\mp i\phi} A_\mu^\pm 
  = \hf\partial_\mu{\theta},
\end{eqnarray}
where
\begin{eqnarray}
A_\mu^\pm &\equiv& A_\mu^x \pm  i A_\mu^y
\nonumber\\
&=&
 \hf e^{\pm i \phi}
 \left( \pm i \partial_\mu \theta  -\sin\theta \partial_\mu \phi \right)
=e^{\pm i\phi}(\Ath_\mu\pm i\Aph_\mu).
\end{eqnarray}
Some useful relations between gauge field and spin vector are
\begin{eqnarray}
 \partial_\mu \evs &=& 2(-\Ath_\mu \evph+\Aph_\mu \evth)
  =2 \Av_\mu\times \evs \nonumber\\
 (\partial_\mu \evs \times \evs) &=& 
   2(\Ath_\mu \evth+\Aph_\mu \evph)
\nonumber\\
\partial_\mu \evs \times \partial_\nu \evs 
&=& 4\evs 
  (\Ath_\mu \Aph_\nu -\Ath_\nu \Aph_\mu) \nonumber\\
\Av_\mu &=& \hf\evs\times \partial_\mu \evs -\PhiB_\mu\evs 
=\hf(\partial_\mu\theta\evph-\partial_\mu\phi \evs+\partial_\phi \evz)
,
\end{eqnarray}
where $\PhiB_\mu\equiv \hf (1-\cos\theta)\partial_\mu \phi =-\Av_\mu\cdot\evs$.

\begin{figure}[tbh]
\begin{center}
\includegraphics[scale=0.4]{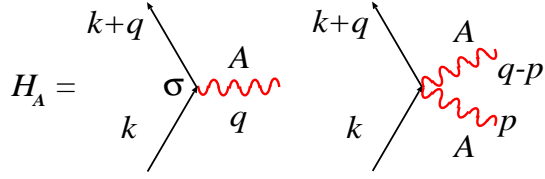}
\caption{Diagrammatic representation of interaction between domain wall and conduction electron.
Solid lines are electron Green's functions and wavy lines denote interaction with the magnetization configuration (e.g. domain wall). 
\label{electron:FIGinteractions}
}
\end{center}
\end{figure}

After gauge transformation, 
the electron Lagrangian is written in terms of the $a$-electron as
\begin{eqnarray}
\Lez &=&   \intx \left[
i\hbar \adag \dot{a} -\frac{\hbar^2}{2m}|\nabla a|^2 +\eF \adag a
 -\spol \adag \sigma_z a
  \right.\nonumber\\
 && \left.+i\frac{\hbar^2}{2m}\sum_{i}
 (\adag A_i \nabla_i a - (\nabla_i\adag) A_i a )
  -\frac{\hbar^2}{2m}A^2 \adag a -\hbar \adag A_0 a
 \right].
\end{eqnarray}
Defining a Fourier transform as
$a(\xv)\equiv \frac{1}{\sqrt{V}}\sumkv a_{\kv}e^{i\kv\cdot\xv}$ and
$A_\mu^\alpha(\qv)\equiv \Vinv\intx e^{i\qv\cdot\xv} A_\mu^\alpha(\xv)$, the expression becomes 
\begin{eqnarray}
\Lez 
&=&
\sumkv 
i\hbar 
a^\dagger_{\kv\sigma} \lt( \partial_t -\epsilon_{\kv\sigma} \rt) a_{\kv\sigma} 
-\HA, \label{Lgauge}
\end{eqnarray}
where
\begin{equation}
\epsilon_{\kv\sigma} \equiv \frac{\hbar^2 k^2}{2m}-\eF-\sigma\spol,  \label{ekvs}
\end{equation}
is the electron polarized uniformly along the $z$-axis
and
\begin{eqnarray}
\HA &\equiv&  \sum_{\kv\qv\alpha} \lt[
\sum_{\mu}
  \hbar\left(J_\mu\left(\kv+\frac{\qv}{2}\right)\cdot A_\mu^\alpha(-\qv) \right)
  \adag_{\kv+\qv}\sigma_\alpha a_{\kv} \right.  \nonumber\\ && \left.
 + \frac{\hbar^2}{2m}\sum_{\pv i}A_i^\alpha(-\qv-\pv) A_i^\alpha(\pv)
 \adag_{\kv+\qv} a_{\kv}
 \right], \label{HAdef}
\end{eqnarray}
is the interaction with the gauge field
(the Greek suffix $\mu$ runs over $x,y,z,t$ and $i$ runs over $x,y,z$).
Here 
\begin{equation}
J_\mu(\kv)\equiv \left( \frac{\hbar}{m}\kv, 1  \right),
\end{equation}
for $\mu=x,y,z,t$.
The interaction with the gauge field is shown diagramatically in Fig. \ref{electron:FIGinteractions}.

\begin{figure}[tbh]
\begin{center}
\includegraphics[width=0.3\linewidth]{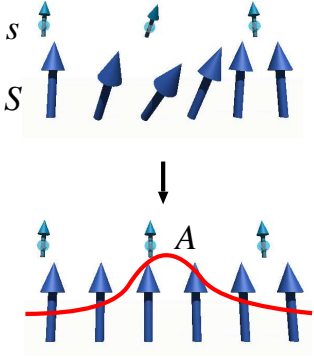}
\caption{By gauge transformation, electrons (denoted by their spin $\se$) interacting with a non-uniform localized spin structure ($\Sv$) are transformed to electrons uniformly polarized and interacting with the SU(2) gauge field, $\Av$, localized around the inhomegeneity.
Thus the non-adiabatic component causing scattering is separated from the adiabatic component.
\label{FIGgaugetr}
}
\end{center}
\end{figure}

The Lagrangian (\ref{Lgauge}) is very useful for our purpose, since the perfect adiabaticity (represented by uniform polarization in Eq. (\ref{ekvs})) and non-adiabaticity (gauge field) are clearly separated (Fig. \ref{FIGgaugetr}). 

%

\begin{figure}[tbh]
\begin{center}
\includegraphics[scale=0.4]{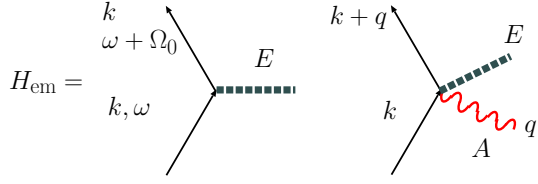}
\caption{Diagrammatic representation of interaction with the applied electric field, $\Ev$, denoted by dashed line.
\label{electron:FIGHem}
}
\end{center}
\end{figure}

The  electric current density, Eq. (\ref{jemdef}),  is modified by the gauge transformation and the electromagnetic gauge field as
\begin{eqnarray}
j_i &=& -\int \dx \frac{e\hbar}{m}\frac{i}{2} 
(c^{\dagger}\stackrel{\leftrightarrow}{\nabla}_{i}c)
-\frac{e^2\hbar}{m} A_{{\rm em},i} \sum_{\kv}
c^{\dagger}_{\kv}c_{\kv}
\nonumber\\&=&       
\frac{e\hbar}{m} \sum_{\kv} \lt[
k_i a^{\dagger}_{\kv} 
a_{\kv}
+ \sum_{\qv\alpha}
A_i^\alpha (\qv) a^{\dagger}_{\kv+\qv}\sigma^{\alpha}a_{\kv} 
 -e A_{{\rm em},i} a^{\dagger}_{\kv}a_{\kv} \rt]
 , \label{jdef}
\end{eqnarray}
Thus the interaction with external electric field, Eq. (\ref{Hemdef}), is given by (Fig. \ref{electron:FIGHem})
\begin{eqnarray}
\Hem &=& \lt.
\sum_i \frac{ie\hbar E_i}{m^2\Omz}e^{i\Omz t}
\sumkv \lt[ k_i  a^{\dagger}_{{\kv}} a_{{\kv}}
+\sum_{\alpha\qv} A^\alpha _{i}(\qv) a^{\dagger}_{\kv+\qv}\sigma^{\alpha}a_{\kv}
\rt]\rt|_{\Omz\ra0} +O(E^2). \label{hemmod}
\end{eqnarray}
 
\subsection{Electron spin density}

Calculation of electron spin density is carried out in the gauge transformed frame by the standard diagrammatic method. 
What we obtain there is the spin density in the gauge transformed frame, $\svtil$, defined by
\begin{equation}
\svtil(\xv,t) \equiv  \average{\adag\sigmav a}.
\end{equation}
It is related to the physical spin density, $\sev$, 
as
\begin{equation}
\sev(\xv,t)\equiv \average{\cdag\sigmav c}
  = \average{\adag U^{\dagger} \sigmav U a} 
  =2\mv(\mv\cdot\svtil)-\svtil. \label{svdef0}
\end{equation}
Let us define 
\begin{equation}
\sv \equiv \seth\evth +\sez \evs +\seph \evph.\label{svdef}
\end{equation}
Each component is written in terms of components of $\svtil$ as
\begin{eqnarray}
\seth &=& 
 -\hf \sum_{\pm} e^{\mp i\phi} \stil^{\pm} \nonumber \\
\seph &=& 
-\hf \sum_{\pm} (\mp)ie^{\mp i\phi} \stil^{\pm},
    \label{sethph}
\end{eqnarray}
where $\stil^{\pm}\equiv \stil_{x}\pm i \stil_{y}$.
Eq. (\ref{sethph})  can easily be checked by use of 
$\mv= -\sin\frac{\theta}{2} \evth+\cos\frac{\theta}{2}\evs$
and $\cos\theta\evth+\sin\theta\evs=(\cos\phi,\sin\phi,0)=(\evph\times\evz)$.

Let us see how the force and torque on the spin structure (Eqs. (\ref{Fdef})(\ref{Tdef}))
are represented in terms of $\seth$ and $\seph$.
Using 
$\nabla \evs=\evth\nabla\theta+\evph \sin\theta\nabla\phi$
and $\evs\times\evth=\evph$, $\evs\times\evph=-\evth$,
we obtain
\begin{eqnarray}
\Fv 
 & = &
-\spol \intx
   \left( \seth\nabla\theta +\seph \sin\theta\nabla\phi \right)
= 2\spol \intx \left(-\seth \Aphv +\seph \Athv\right)
   \label{force2}\\
\torquev 
  &=&  \spol \intx \left( \seph \evth - \seth \evph \right).
 \label{torque2}
\end{eqnarray}
We see here that $\stilz$, the perfectly adiabatic component, does not contribute to the force or torque.
For the rigid wall we are considering, $\nabla\phi=0$, 
only components $F_{\xw}$ and $\torque_z$ affects its dynamics,
and we obtain
\begin{eqnarray}
F_{\xw} 
  =  - \spol \intx  \seth \nabla_\xw \theta
   \nonumber\\
\torque_z  = -\spol \intx \sin\thetaz \seph
\label{fandtorquedw}
\end{eqnarray}
Thus the force and spin-transfer torque on a rigid wall arises from 
$\seth$ and $\seph$, respectively.

\subsection{Electron Green's functions}
\begin{figure}[tbh]
  \begin{center}
  \includegraphics[scale=0.3]{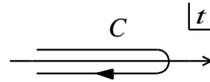}
\caption{Keldysh contour in the complex time plane used in non-equilibrium Green's functions.
\label{FIGC}
}
  \end{center}
\end{figure}

The spin density $\svtil(\xv,t)$ is defined in terms of an $a$-electron as
\begin{eqnarray}
 \stil^{\alpha}(\xv,t)
  & = &  \average{\adag(\xv,t)\sigma_\alpha a(\xv,t)}
 	   \nonumber\\
 &=& -i\hbar\tr [\sigma_\alpha G^{<}(\xv,t,\xv,t)],
\end{eqnarray}
where
\begin{equation}
G_{\sigma,\sigma'}^< (\xv,t,\xv',t') \equiv
  \frac{i}{\hbar} \average{ \adag_{\sigma'}(\xv',t') a_{\sigma}(\xv,t)},
\end{equation}
is a lesser component of the Keldysh Green's function \cite{Haug07,Mahan90} and 
the trace ($\tr$) is over the spin index.
This Green's function is an extension of the standard equilibrium Green's functions (retarded and advanced Green's functions) to a contour-ordered Green's function defined on the complex time plane.
(For more details see \S \ref{SEC:Keldysh}.)
The contour-ordered Green's function is defined as
\begin{equation}
G_{\sigma,\sigma'}(\xv,t,\xv',t') \equiv
 -\frac{i}{\hbar} \average{T_C a_{\sigma}(\xv,t) \adag_{\sigma'}(\xv',t')},
\end{equation}
where $t, t'$ are defined on a contour on the complex plane and
$T_C$ denotes ordering on the Keldysh contour $C$ (Fig. \ref{FIGC}).
The average here denotes both quantum expectation value and averaging over random normal impurities. 
This Green's function contains information on retarded, advanced, and lesser Green's functions.

In the free case, the lesser component is defined as 
\begin{equation}
g_{\sigma,\sigma'}^< (\xv,t,\xv',t') \equiv
  \frac{i}{\hbar} \average{ \adag_{\sigma'}(\xv',t') a_{\sigma}(\xv,t)}_0 \equiv g_{\sigma,\sigma'}^< (\xv-\xv',t-t'),
\end{equation}
where $\average{\ }_0$ denotes the expectation value by use of the free Hamiltonian, $H_0$, but in the presence of impurities ($\Himp$).
By standard ladder summation over multiple impurity scattering, the free lesser component 
is obtained in the Fourier space as 
\begin{equation}
g_{\kv}^< (\omega) =
 f(\omega)(\ga_{\kv}(\omega)-\gr_{\kv}(\omega)),
\label{freelesser}
\end{equation}
where the advanced and retarded Green's functions are 
\begin{eqnarray}
\ga_{\kv} (\omega) &=& \frac{1}{\omega-\ekv-\frac{i}{2\tau}},
\label{freeranda}
\end{eqnarray}
and  $\gr=(\ga)^*$.

Calculation of $\svtil$ is done by standard perturbation theory with respect to the gauge field $A_\mu$, up to the lowest, i.e., linear, order.
We define the Fourier transform of a Green's function as
\begin{eqnarray}
G_{\kv\sigma,\kv+\qv,\sigma'}(t,t')
 &\equiv&
 -i\average{T_C a_{\kv\sigma}(t)\adag_{\kv+\qv,\sigma'}(t')}
\nonumber\\
&=& \Vinv \intx\intx' e^{-i(\kv\cdot\xv-\kv'\cdot\xv')}
G_{\sigma,\sigma'}(\xv,t,\xv',t') .
\end{eqnarray}
The full Keldysh Green's function defined on a complex contour satisfies the Dyson equation similar to the conventional time-ordered and retarded Green's functions:
\begin{eqnarray}
i\hbar \partial_t G_{\kv\kv'}(t,t')
  &=& \delta(t-t')\average{ \{ a_{\kv}(t),\adag_{\kv'}(t') \} }
  + \frac{i}{\hbar} \average{ T_C [H,a_{\kv}(t)]\adag_{\kv'}(t') }.
\end{eqnarray}
Here $H=H_0+\Hem+\HA$ ($\Himp$ is already included in $\tau$ of free Green's functions, e.g., in Eq. (\ref{freelesser})).
Evaluation of the anti-commutation relation can be done 
using $[AB,C]=A\{B,C\}-\{A,C\}B$ for any three operators $A, B$, and $C$ as 
\begin{eqnarray}
[H,a_{\kv}(t)] &=&
-\sum_{\qv\alpha} \lt[
  \sum_{\mu} \hbar J_{\mu}\lt(\kvpq\rt)A_{\mu}^{\alpha}(\qv,t)
   \sigma_{\alpha} a_{\kv+\qv}(t) \rt. \nonumber\\
&& \lt.
+ \frac{\hbar^2}{2m}\sum_{\pv,i}
   A_i^\alpha(\qv-\pv,t) A_i^\alpha(\pv,t)
   a_{\kv+\qv}(t) \rt].
\end{eqnarray}
We then obtain  
\begin{eqnarray}
i\hbar \partial_t G_{\kv\kv'}(t,t') &=&   \delta(t-t')\delta_{\kv,\kv'} + \epsilon_{\kv} G_{\kv\kv'}(t,t')
 +\hbar\sumqv\sum_{\mu\alpha} J_{\mu}\lt(\kvpq\rt)A_{\mu}^{\alpha}(\qv,t)
   \sigma_{\alpha} G_{\kv+\qv,\kv'}(t,t')   \nonumber\\
&&
+\sum_{i}\sumqv \frac{ieE_i}{m\Omz} e^{i\Omz t_1}
\lt(k_i\delta_{\qv,0} 
  + \sum_\alpha A_{i}^{\alpha}(\qv,t_1)\sigma^\alpha \rt)
 \sigma_{\alpha} G_{\kv+\qv,\kv'}(t,t')  
\nonumber\\
&&
+\frac{\hbar^2}{2m}\sum_{\qv\pv}\sum_{\alpha i}
   A_i^\alpha(\qv-\pv,t) A_i^\alpha(\pv,t)
   G_{\kv+\qv,\kv'}(t,t')
+O(E^2).\nonumber\\
&& 
\end{eqnarray}
By multiplying 
$[i\hbar\partial_t -\epsilon_{\kv}]^{-1} =g_{\kv}$, we
obtain the Dyson equation on contour $C$,
\begin{eqnarray}
\lefteqn{ G_{\kv\kv'}(t,t')
  = g_{\kv}(t-t')\delta_{\kv,\kv'}   
 +
 \hbar\sumqv \int_{C}\! dt_1 
  g_{\kv}(t-t_1) }
\nonumber\\
&&  
\times 
\lt[ \sum_{\alpha\mu} J_{\mu}\lt(\kvpq\rt) A_{\mu}^{\alpha}(\qv,t_1) \sigma_{\alpha}
+ \sum_i \frac{ieE_i}{m\Omz} e^{i\Omz t_1}
\lt(k_i\delta_{\qv,0} 
  + \sum_\alpha A_{i}^{\alpha}(\qv,t_1)\sigma^\alpha \rt)\rt]
  G_{\kv+\qv,\kv'}(t_1,t') 
\nonumber\\&& 
 +O(E^2,A^2)\nonumber\\
&=&  g_{\kv}(t-t')\delta_{\kv,\kv'}   
   + \hbar\sum_{\alpha\mu} J_{\mu}\lt(\frac{\kv+\kv'}{2}\rt)
 \int_{C}\! dt_1 g_{\kv}(t-t_1) A_{\mu}^{\alpha}(\kv'-\kv,t_1)
   \sigma_{\alpha}
  g_{\kv'}(t_1-t') \nonumber\\
&&  
+\frac{ie\hbar}{m\Omz} 
\sum_{\alpha i} E_i  
\int_{C}\! dt_1  A_{i}^{\alpha}(\kv'-\kv,t_1) e^{i\Omz t_1}
 g_{\kv}(t-t_1) \sigma_{\alpha} g_{\kv'}(t_1-t')
\nonumber\\
&&+\frac{ie\hbar}{m\Omz} 
\sum_{\alpha\mu i} E_i J_{\mu}\lt(\frac{\kv+\kv'}{2}\rt) 
\int_{C}\! dt_1 \! \int_{C}\! dt_2 
\nonumber\\
&&\times
\lt[k'_i e^{i\Omz t_2} A_{\mu}^{\alpha}(\kv'-\kv,t_1)
   g_{\kv}(t-t_1) \sigma_{\alpha} g_{\kv'}(t_1-t_2) g_{\kv'}(t_2-t')  \rt.  \nonumber\\
&& \lt.
+ k_i e^{i\Omz t_1} A_{\mu}^{\alpha}(\kv'-\kv,t_2) 
g_{\kv}(t-t_1)  g_{\kv}(t_1-t_2) 
   \sigma_{\alpha} g_{\kv'}(t_2-t') \rt]
\nonumber\\
&& +O(E^2,A^2)
\label{dysonfull}.
\end{eqnarray}
Here we neglected a trivial term which does not contain $A$.
Green's functions are written by use of matrix representation in spin space,
\begin{equation}
G_{\kv\kv'} = \left(\begin{array}{cc} 
        G_{\kv+,\kv'+} & G_{\kv+,\kv'-} \\
	G_{\kv-,\kv'+} & G_{\kv-,\kv'-} \end{array}  \right),
\end{equation}
and free Green's function is diagonal,
\begin{equation}
g_{\kv} = \left(\begin{array}{cc} 
        g_{\kv+} & 0 \\
	0 & g_{\kv-} \end{array}  \right).
\end{equation}
The Equation (\ref{dysonfull}) is for Green's function defined on complex time contour, and is not physical.
To evaluate spin density, we need to take a lesser component,
$G^<$, of the contour ordered Green's functions. 
As described in \S\ref{SEC:Keldysh}, 
lesser component of product is mixed with a lesser component and retarded and advanced components as
\begin{equation}
\lt[\int_C dt_1 A(t,t_1) B (t_1,t')\rt]^<
=\intinf dt_1 (A^{\rm r}(t,t_1) B^<(t_1,t')+A^{<}(t,t_1) B^{\rm a}(t_1,t')),\label{lesserrule}
\end{equation}
while retarded and advanced components are the products:
\begin{equation}
\lt[\int_C dt_1 A(t,t_1) B (t_1,t')\rt]^{\rm r}
=\intinf dt_1 A^{\rm r}(t,t_1) B^{\rm r}(t_1,t').
\end{equation}
We define Fourier transform of gauge fields as
\begin{equation}
A_\mu^\alpha(\qv,\Omega) \equiv
\intinf dt e^{-i\Omega t} A_\mu^\alpha(\qv,t).
\end{equation}
and electron Green's function
as ($\tau$ denotes $<$, r, a)
\begin{equation}
G^\tau_{\kv,\kv'}(t,t') \equiv
\intom \int\frac{d\omega'}{2\pi} e^{-i\omega t} e^{i\omega' t'}
G^\tau_{\kv,\kv'}(\omega,\omega'),
\end{equation}
or 
$G^\tau_{\kv,\kv'}(\omega,\omega') \equiv
\int dt \int dt' e^{i\omega t} e^{-i\omega' t'}
G^\tau_{\kv,\kv'}(t,t')
$.
In general $\omega\neq\omega'$, electron absorbs energy (frequency) from the electric field (and from spin structures if dynamical). 
Taking the lesser component, Fourier representation of Eq. (\ref{dysonfull}) is written as
\begin{eqnarray}
\lefteqn{ 
G^<_{\kv,\kv+\qv}(\omega,\omega+\Omega)
  = g_{\kv\omega}^< \delta_{\qv,0} \delta_{\Omega,0}  
+ \hbar \sum_{\alpha\mu} J_{\mu}\lt(\kvpq\rt) A_{\mu}^{\alpha}(\qv,\Omega) 
[g_{\kv\omega} \sigma_{\alpha} g_{\kv+\qv,\omega+\Omega}]^<
}
 \nonumber\\
&& 
+\frac{ie\hbar}{m\Omz} 
\sum_{\alpha i} E_i A_{i}^{\alpha}(\qv,\Omega-\Omz)
\lt[g_{\kv\omega}\sigma_{\alpha} g_{\kv+\qv,\omega+\Omega} 
\rt]^<
 \nonumber\\
&& + 
\frac{ie\hbar}{m\Omz} 
\sum_{\alpha\mu i} E_i J_{\mu}\lt(\kvpq\rt) 
A_{\mu}^{\alpha}(\qv,\Omega-\Omz)
\nonumber\\
&&\times
\lt[(k+q)_i g_{\kv\omega}\sigma_{\alpha} g_{\kv+\qv,\omega+\Omega-\Omz} g_{\kv+\qv,\omega+\Omega}
+ k_i  g_{\kv,\omega} g_{\kv,\omega+\Omz} \sigma_{\alpha} g_{\kv+\qv,\omega+\Omega}\rt]^<
\nonumber\\
&& +O(E^2,A^2)
\label{dysonless}.
\end{eqnarray}
We consider a slow spin dynamics and assume 
that the spin gauge field has only a zero frequency component, i.e., 
$A_{\mu}^{\alpha}(\qv,\Omega) = \delta_{\Omega,0} A_{\mu}^{\alpha}(\qv)$.
This is justified when $\Omega \tau \ll 1$.
We then obtain
\begin{eqnarray}
\lefteqn{ 
G^<_{\kv,\kv+\qv}(\omega,\omega+\Omz)
  = g_{\kv\omega}^< \delta_{\qv,0} \delta_{\Omz,0} 
+ \hbar \sum_{\alpha\mu} J_{\mu}\lt(\kvpq\rt) A_{\mu}^{\alpha}(\qv) 
[g_{\kv\omega} \sigma_{\alpha} g_{\kv+\qv,\omega+\Omz}]^<
}
 \nonumber\\
&& 
+\frac{ie\hbar}{m\Omz} 
\sum_{\alpha i} E_i A_{i}^{\alpha}(\qv)
\lt[g_{\kv\omega}\sigma_{\alpha} g_{\kv+\qv,\omega+\Omz} 
\rt]^<
 \nonumber\\
&& + 
\frac{ie\hbar}{m\Omz} 
\sum_{\alpha\mu i} E_i J_{\mu}\lt(\kvpq\rt) 
A_{\mu}^{\alpha}(\qv)
\nonumber\\
&&\times
\lt[(k+q)_i g_{\kv\omega}\sigma_{\alpha} g_{\kv+\qv,\omega} g_{\kv+\qv,\omega+\Omz}
+ k_i  g_{\kv,\omega} g_{\kv,\omega+\Omz} \sigma_{\alpha} g_{\kv+\qv,\omega+\Omz}\rt]^<
\nonumber\\
&& +O(E^2,A^2)
\label{dysonless2}.
\end{eqnarray}

\subsection{Spin densities}

\begin{figure}[tbh]
  \begin{center}
  \includegraphics[scale=0.3]{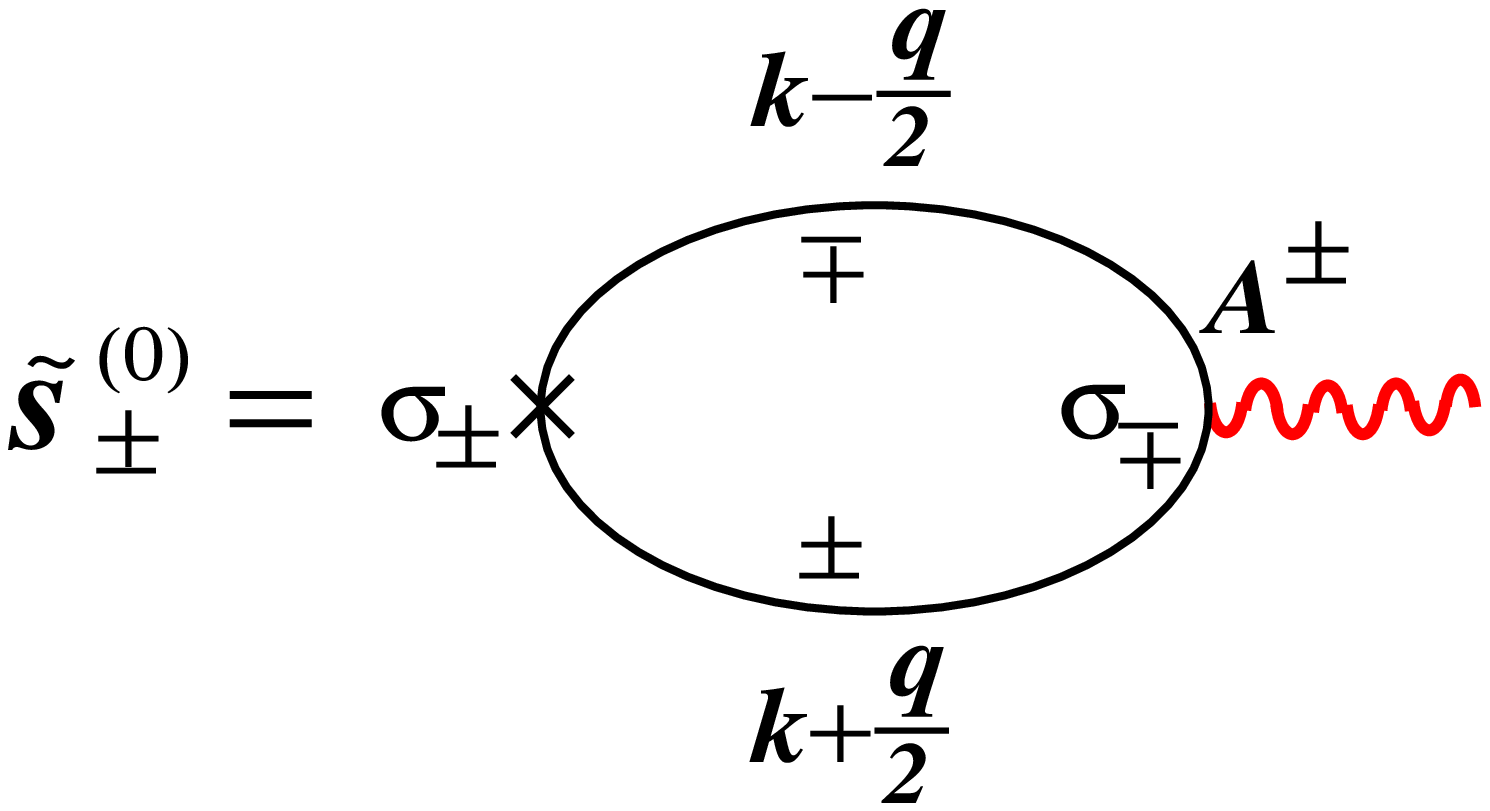}
  \end{center}
\caption{
Electron spin density without current at the lowest order in gauge field (represented by wavy lines).
\label{FIGse0}
}
\end{figure}

\begin{figure}[tbh]
  \begin{center}
  \includegraphics[scale=0.3]{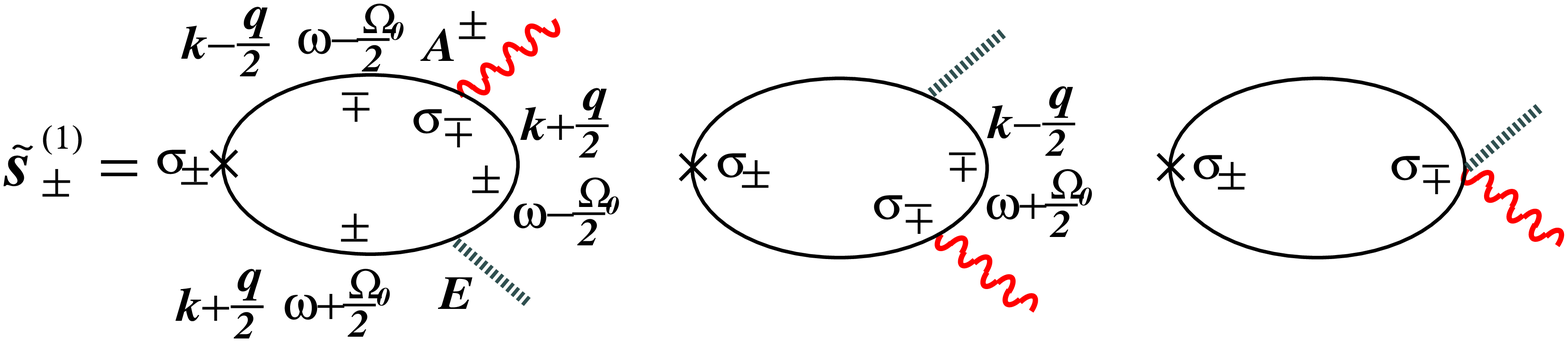}
\caption{
Diagrams representing the current-driven part of the electron spin density at the linear order in both gauge field (represented by wavy lines) and  applied electric field (dotted lines).
The first two processes are self-energy corrections , $\stilpma$, and the last process is from the correction of the current vertex, $\stilpmb$.
\label{FIGsevc}
}
  \end{center}
\end{figure}

The spin density in the rotated frame is calculated by multiplying the lesser Green's function by a Pauli matrix and taking a trace.
Only $\pm$ components are necessary, whose Fourier transforms are given as
\begin{eqnarray}
\stilpm_\qv &=& -2i \lim_{\Omz\ra0}\sumom\sumkv
 \tr[\sigma^\pm G^<_{\kv,\kv+\qv}(\omega,\omega+\Omz)]
\nonumber\\
&\equiv& \stilpmz_\qv+\stilpmo_\qv,
\end{eqnarray}
where $\sigma_x \pm i\sigma_y=2\sigma_\pm$,
$\stilpmz_\qv$ and $\stilpmo_\qv\equiv \stilpma_\qv+\stilpmb_\qv$ are the equilibrium and current-driven parts, respectively. 
Each contribution is given as 
\begin{eqnarray}
\stilpmz_\qv & = &
-2i \frac{\hbar^2}{V} \lim_{\Omz\ra0} \sumom \sum_{\kv} 
\sum_{\mu} J_\mu(\kv)A_\mu^\pm (\qv) 
\lt[g_{\kvmq,\mp\ommOmz} g_{\kvpq,\pm,\ompOmz}\rt]^< 
\nonumber\\
\stilpma_\qv & = &
\lim_{\Omz\ra0}
\sum_{\omega\kv} \sum_i \frac{B_i^\pm}{\Omz} 
\lt[
\lt( k+\frac{q}{2} \rt)_i 
\lt[ g_{\kvmq,\mp,\ommOmz} g_{\kvpq,\pm,\ommOmz} g_{\kvpq,\pm,\ompOmz}\rt]^<
\right.\nonumber\\
&& \left.
+
\lt(k-\frac{q}{2}\right)_i 
\lt[g_{\kvmq,\mp,\ommOmz} g_{\kvmq,\mp,\ompOmz} g_{\kvpq,\pm,\ompOmz}\rt]^<
\rt]
\label{sSE}\\
\stilpmb_\qv & = &
\lim_{\Omz\ra0} \sum_{\omega\kv} \frac{B_0^\pm}{\Omz} 
\lt[
g_{\kvmq,\mp,\ommOmz}g_{\kvpq,\pm,\ompOmz} 
\rt]^<,
\label{sV}
\end{eqnarray}
where
\begin{eqnarray}
B_i^\pm(\kv,\qv) &\equiv&
  2 \frac{eE_i}{m V} \sum_{\mu}
 J_\mu(\kv)A_\mu^\pm (\qv) \nonumber\\
B_0^\pm(\kv,\qv)  &\equiv&
  2 \frac{eE_i}{m V} A_i^\pm (\qv) .
\end{eqnarray}
These contributions are shown in Figs. \ref{FIGse0} and 
\ref{FIGsevc}.

In this section, we will derive the leading contribution in the clean limit, $\tau\rightarrow\infty$. 
(When spin relaxation is concerned, we need to estimate the next-order contribution of order of $(\ef\tau)^{-1}$. See \S\ref{SEC:singlespin}.)
Using Eq. (\ref{lesserrule}), 
the equilibrium part is obtained as
\begin{eqnarray}
\stilpmz_\qv & = &
-2i \frac{\hbar^2}{V} \lim_{\Omz\ra0} \sumom \sum_{\kv\mu} 
 J_\mu(\kv)A_\mu^\pm (\qv) 
\lt[\gr_{\kvmq,\mp,\ommOmz} \gless_{\kvpq,\pm,\ompOmz}
+\gless_{\kvmq,\mp,\ommOmz} \ga_{\kvpq,\pm,\ompOmz}\rt] .
\nonumber\\&&
\end{eqnarray}
By use of 
\begin{equation}
g^<_{\kv\sigma}(\omega)
 \simeq f_{\kv\sigma}\delta(\omega-\ekvs),
\label{freelesser2}
\end{equation} 
we obtain the equilibrium contribution as
\begin{eqnarray}
\stilpmz_\qv & = &
 \frac{2\hbar^2}{V} \sum_{\kv} A_0^\pm (\qv)
\frac{f_{\kv+\frac{\qv}{2},\pm}-f_{\kv-\frac{\qv}{2},\mp}}
{\epsilon_{\kvpq,\pm}-\epsilon_{\kvmq,\mp}+\frac{i}{\tau}} .
\nonumber\\
&&\label{sequil}
\end{eqnarray}
The first contribution to current-driven part is obtained by use of Eqs. (\ref{freeranda})(\ref{lesserrule}) as
\begin{eqnarray}
\lefteqn{
\stilpma_\qv  = 
\sum_{\omega\kv} \sum_i \frac{B_i^\pm}{\Omz} 
\lt[
\lt( k+\frac{q}{2} \rt)_i \rt.
} \nonumber\\
&& \times 
\lt[ 
f\lt(\ompOmz\rt) \gr_{\kvmq,\mp,\ommOmz} \gr_{\kvpq,\pm,\ommOmz} 
\lt(\ga_{\kvpq,\pm,\ompOmz}-\gr_{\kvpq,\pm,\ompOmz}\rt) \rt.
\nonumber\\
&&  \lt.
+f\lt(\ommOmz\rt) 
\lt( \ga_{\kvmq,\mp,\ommOmz} \ga_{\kvpq,\pm,\ommOmz} -\gr_{\kvmq,\mp,\ommOmz} \gr_{\kvpq,\pm,\ommOmz} \rt)
\ga_{\kvpq,\pm,\ompOmz} \rt]
\nonumber\\
&& 
 + \lt( k-\frac{q}{2} \rt)_i  \nonumber\\
&& \times 
\lt[ 
f\lt(\ompOmz\rt) \gr_{\kvmq,\mp,\ommOmz} 
\lt( \ga_{\kvmq,\mp,\ompOmz}\ga_{\kvpq,\pm,\ompOmz}
-\gr_{\kvmq,\mp,\ompOmz}\gr_{\kvpq,\pm,\ompOmz}\rt) \rt.
\nonumber\\
&& \left.
+f\lt(\ommOmz\rt) 
\lt(\ga_{\kvmq,\mp,\ommOmz}-\gr_{\kvmq,\mp,\ommOmz}\rt) \ga_{\kvmq,\mp,\ompOmz} \ga_{\kvpq,\pm,\ompOmz}\rt]
.
\end{eqnarray}
The dominant term arises from terms containing both $\gr$ and $\ga$, since the contribution from $\gr$ only is higher order of $(\ef\tau)^{-1}$ after summation over the wave vector ($\kv$), as is well-known in electron transport phenomena.
This is easily understood from following example.
In the case of $(\ef\tau)^{-1} \ll 1$, the summation over the wave vector can be replaced by an energy integral as
\begin{equation}
\Vinv\sumkv \gr_{\kv\sigma} \ga_{\kv\sigma}
=\int_{-\ef}^{\infty} d\epsilon \DOS(\epsilon) 
\frac{1}{(\epsilon-\sigma\spol)^2+\lt(\frac{1}{2\tau}\rt)^2} 
\simeq \DOS_\sigma \int^\infty_{-\infty}d\epsilon \frac{1}{\epsilon^2+\lt(\frac{1}{2\tau}\rt)^2}=2\pi\DOS_\sigma \tau,
\end{equation}
where $\DOS(\epsilon)$ is the energy-dependent density of states and $\DOS_\sigma \equiv \DOS(\sigma\spol)$.
Within this approximation, the product of $\gr$ results in
$\sumkv (\gr_{\kv\sigma})^2 
\simeq \DOS_\sigma \int^\infty_{-\infty}d\epsilon (\epsilon-\frac{i}{2\tau})^{-2}=0$. 
We therefore see that the dominant term is simply proportional to
$f\lt(\ompOmz\rt) -f\lt(\ommOmz\rt) =\Omz f'(\omega)$ ($\Omz\ra0$), as is familiar in low energy transport properties.
Therefore, we obtain 
\begin{eqnarray}
\stilpma_\qv & \simeq &
\sum_{\omega\kv} \sum_i B_i^\pm f'(\omega) 
\nonumber\\ && \times 
\left[ \left(k+\frac{q}{2}\right)_i 
 \gr_{\kv-\frac{q}{2},\mp,\omega} 
 \gr_{\kv+\frac{q}{2},\pm,\omega} 
 \ga_{\kv+\frac{q}{2},\pm,\omega}
+\left( k-\frac{q}{2}\right)_i 
\gr_{\kv-\frac{q}{2},\mp,\omega}
\ga_{\kv-\frac{q}{2},\mp,\omega} \ga_{\kv+\frac{q}{2},\pm,\omega}
\right]\nonumber\\
&& \label{stilpma}
\end{eqnarray}
The contribution $\stilpmb_\qv$ is of higher order in $(\ef\tau)^{-1}$ than 
$\stilpma_\qv $, since it has a smaller number of Green's functions \cite{TKSLL07}.
Thus the current-induced part is given by
$\stilpmo_\qv \sim \stilpma_\qv $. 
Since we are interested in low temperatures, $\kb T/\eF \ll 1$, we can replace
$f'(\omega)\simeq -\delta(\omega)$ in 
\Eqref{stilpma},
and hence
\begin{eqnarray}
\stilpmo_\qv  &=&
-\frac{1}{2\pi}\sum_{\kv} \sum_i B_i^\pm 
\nonumber\\ && \times 
\left[ \left(k+\frac{q}{2}\right)_i 
 \gr_{\kv-\frac{q}{2},\mp} 
 \gr_{\kv+\frac{q}{2},\pm} 
 \ga_{\kv+\frac{q}{2},\pm}
+\left( k-\frac{q}{2}\right)_i 
\gr_{\kv-\frac{q}{2},\mp}
\ga_{\kv-\frac{q}{2},\mp} 
\ga_{\kv+\frac{q}{2},\pm}
\right],
\nonumber\\ &&
\end{eqnarray}
where $\gr_{\kv-\frac{q}{2},\mp}\equiv \gr_{\kv-\frac{q}{2},\mp,\omega=0}$.
Using the identities
\begin{eqnarray}
 \gr_{\kv,\sigma} \ga_{\kv,\sigma} &=&  
 i\tau (\gr_{\kv,\sigma} - \ga_{\kv,\sigma})
\\
 \gr_{\kv-\frac{q}{2},\mp} \ga_{\kv+\frac{q}{2},\pm} &=&  
-\frac{1}
{\epsilon_{\kv+\frac{q}{2}}-\epsilon_{\kv-\frac{q}{2}} \mp2\spol +i\gamma} (\gr_{\kv-\frac{q}{2},\mp} - \ga_{\kv+\frac{q}{2},\pm})
\\
 \gr_{\kv-\frac{q}{2},\mp} \gr_{\kv+\frac{q}{2},\pm} &=&  
-\frac{1}
{\epsilon_{\kv+\frac{q}{2}}-\epsilon_{\kv-\frac{q}{2}} \mp2\spol } (\gr_{\kv-\frac{q}{2},\mp} - \gr_{\kv+\frac{q}{2},\pm}),
\end{eqnarray}
where $\gamma\equiv \frac{1}{\tau}$,
we obtain, in the limit of $\tau\rightarrow\infty$, 
\begin{eqnarray}
\lefteqn{
\stilpmo_\qv  =
-\frac{i}{\pi}\sum_{\kv} \sum_i  \frac{eE_i}{m V} \sum_{\mu}
 A_\mu^\pm (\qv) \tau
}
\nonumber\\ && \times 
\left[ 
 k_i  J_\mu\lt(-\lt(\kvpq\rt)\rt) 
 \frac{\gr_{\kv,\pm} -\ga_{\kv,\pm} }
{\epsilon_{\kv+q}-\epsilon_{\kv} \pm2\spol} 
-  k_i J_\mu\lt(\kvpq\rt)
 \frac{\gr_{\kv,\mp} - \ga_{\kv,\mp}}
{\epsilon_{\kv+\qv}-\epsilon_{\kv} \mp2\spol}
\right.\nonumber\\
&& \left.
-i\pi q_i \left(
\gr_{\kv,\mp} J_\mu\lt(\kvpq\rt) 
\delta({\epsilon_{\kv+\qv}-\epsilon_{\kv} \mp2\spol})
-\ga_{\kv,\pm}  J_\mu\lt(-\lt(\kvpq\rt)\rt) 
\delta({\epsilon_{\kv+\qv}-\epsilon_{\kv} \pm2\spol})
\right)
\right].
\nonumber\\ \label{S1_2}
\end{eqnarray}
The first line in square brackets arises from the real part of 
$(\epsilon_{\kv+q}-\epsilon_{\kv} \pm2\spol+i\gamma)^{-1}$ etc., while the second line arises from the imaginary part.

The adiabatic limit can be obtained easily from the results \Eqsref{sequil}{S1_2} by setting 
$\epsilon_{\kv+\qv}\simeq\epsilon_{\kv}$.
The result is 
(${\stil}^{\pm{\rm (ad)}}_\qv \equiv (\stilpmz_\qv+\stilpmo_\qv)|^{\rm (ad)}$)
\begin{eqnarray}
{\stil}^{\pm{\rm (ad)}}_\qv &=&
-\frac{1}{\spol V}
\sumkv \lt[ A_0^\pm (\qv) (f_{\kv+}-f_{\kv-})
+i \frac{e E_i A_i^\pm (\qv) \tau}{d \pi m^2  }
k^2(\gr_{\kv+} - \gr_{\kv-}) \rt] \nonumber\\
&=& -\frac{1}{\spol} 
\lt( \se A_0^\pm (\qv) +\frac{\jsv}{e} \cdot \Av^\pm (\qv) \rt).
\label{sad}
\end{eqnarray}
Here $d$ is the dimensionality, 
\begin{eqnarray}
\se &\equiv& n_+-n_-\nonumber\\
\jsv&\equiv& \frac{1}{m} (n_+-n_-)e^2\tau \Ev =\jv_+-\jv_-,
\end{eqnarray}
 are the electron spin density and spin current density
(both defined without spin length of $\hf$), respectively, where 
$n_\sigma \equiv \frac{k_{F\sigma}^3}{6\pi^2}$ is the spin-resolved electron density.

Going back to the general case, 
spin polarization in real space, $\stilpm(\xv)$, is obtained by Fourier transform as
$\stilpm(\xv)=\sumqv e^{-i\qv\cdot\xv}\stilpm_\qv$.
After some calculation, the equilibrium contributions, $\seth^{(0)}(\xv)$ and $\seph^{(0)}(\xv)$, are obtained as
\begin{eqnarray}
\seth^{(0)}(\xv) 
 &=&
- \frac{1}{\spol}\sumqv e^{-i\qv\cdot\xv}
\left[
\lt( \evsph_\xv\cdot \Av_0(\qv) \rt) 
\chiz_1(\qv)+ \lt(\evph(\xv)\cdot \Av_0(\qv) \rt) \chiz_2(\qv)
 \right]
\nonumber\\
\seph^{(0)}(\xv) 
 &=&
-\frac{1}{\spol}\sumqv e^{-i\qv\cdot\xv}
\left[
\lt( \evph(\xv)\cdot \Av_0(\qv) \rt) \chiz_1(\qv)- \lt( \evsph_\xv\cdot \Av_0(\qv) \rt) \chiz_2(\qv)
 \right],\nonumber\\
&&\label{sephi2}
\end{eqnarray}
where we used
$\half\sum_\pm e^{\mp i\phi(\xv)} A_\mu^\pm(\qv)
=(\evph(\xv)\times \evz) \cdot \Av_\mu(\qv)$ and
$ \half\sum_\pm (\mp i)e^{\mp i\phi(\xv)}A_\mu^\pm(\qv)
=\evph(\xv)\cdot \Av_\mu(\qv)$.
The correlation functions are given by
\begin{eqnarray}
\chiz_1(\qv) &=& \frac{2\hbar\spol}{V} \sum_{\kv\pm} 
 {\rm P}
\frac{f_{\kv\pm}}{\pm2\spol+\frac{2\kv\cdot\qv+q^2}{2m}} 
\nonumber\\
\chiz_2(\qv) &=& \frac{2\hbar\spol}{V} \sum_{\kv\pm} 
\frac{\pi}{2}(f_{\kv+}-f_{\kv-})
\delta\left(\pm2\spol+\frac{2\kv\cdot\qv+q^2}{2m}\right).
\label{chizdef}
\end{eqnarray}
Similarly, the current-induced contribution is calculated as
\begin{eqnarray}
\seth^{(1)}(\xv) 
 &=&
-\frac{\sum_i E_i }{\spol}\sumqv e^{-i\qv\cdot\xv}
\left[
\lt( \evsph_\xv\cdot\Av_i(\qv) \rt) \chio_1(\qv)+ \lt( \evph(\xv)\cdot \Av_i(\qv) \rt) \chio_2(\qv)
 \right]
\nonumber\\
\seph^{(1)}(\xv) 
 &=&
-\frac{\sum_i E_i }{\spol} \sumqv e^{-i\qv\cdot\xv}
\left[
\lt( \evph(\xv)\cdot \Av_i(\qv) \rt) \chio_1(\qv) - \lt( \evsph_\xv\cdot \Av_i(\qv) \rt) \chio_2(\qv)
 \right],  \nonumber\\
&& \label{stils}
\end{eqnarray}
where
\begin{eqnarray}
\chio_1(\qv) &=& \frac{e\tau\spol}{3\pi m^2 V} \sum_{\kv\pm} 
\left( \kv\cdot\left(\kv+\frac{\qv}{2}\right)\right) 
i\frac{\gr_{\kv\pm}-\ga_{\kv\pm}}
 {\epsilon_{\kv+\qv}-\epsilon_{\kv}\pm2\spol} 
\nonumber\\
\chio_2(\qv) &=& \frac{e\tau\spol}{3\pi m^2 V} \sum_{\kv\pm} 
\left( \pm\frac{\pi}{2} \right) 
\left( \qv\cdot\left(\kv+\frac{\qv}{2}\right)\right) 
\delta( {\epsilon_{\kv+\qv}-\epsilon_{\kv}\pm2\spol} )
i(\gr_{\kv\pm}-\ga_{\kv\pm}).\nonumber\\
&&\label{chi2}
\end{eqnarray}

\begin{figure}[tbh]
  \begin{center}
  \includegraphics[width=5cm]{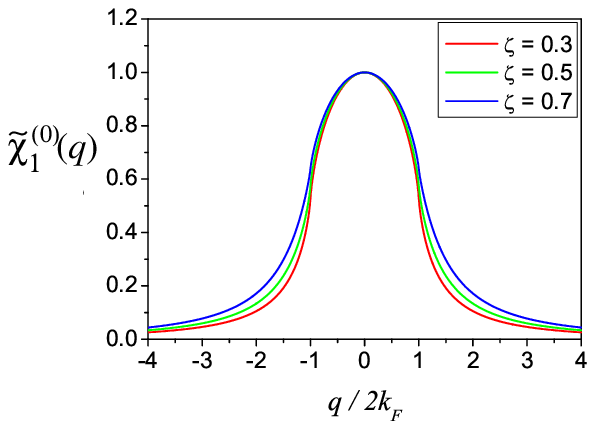}
  \includegraphics[width=5cm]{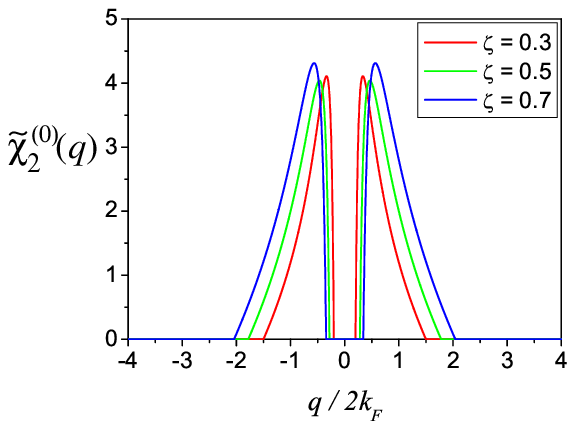}
  \includegraphics[width=5cm]{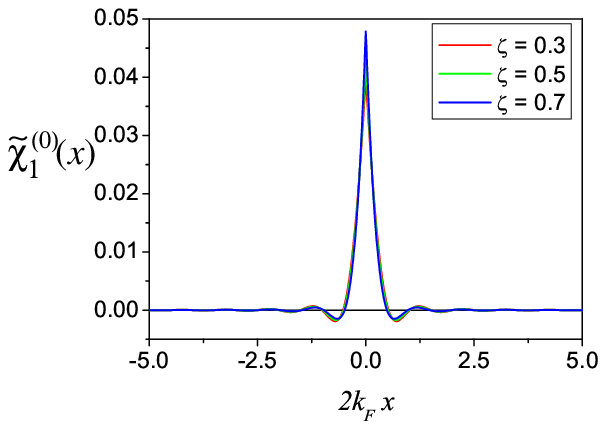}
  \includegraphics[width=5cm]{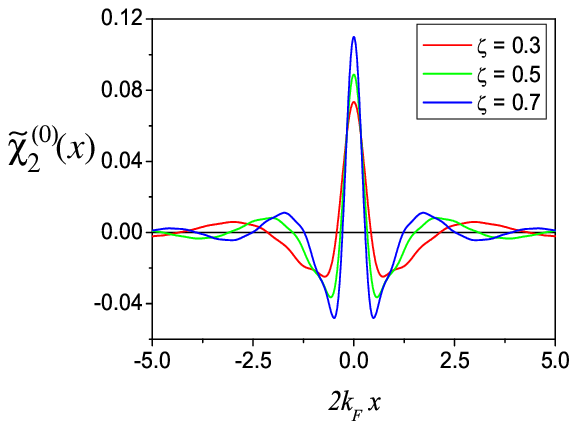}
  \end{center}
\caption{ 
Plot of correlation functions, $\chitilz_1(q)$ and $\chitilz_2(q)$, and
their Fourier transforms, $\chitilz_1(x)$ and $\chitilz_2(x)$,  describing the nonlocal component of spin density (and torque) in the absence of current.
$\chitilz_1$ has a finite adiabatic component ($q=0$, i.e., local component), while $\chitilz_2$ does not.
\label{FIGnlt1}
}
\end{figure}

Integration over $\kv$ in Eq. (\ref{chizdef}) is carried out to obtain
\begin{eqnarray}
\chiz_1(\qv) &=& \se \chitilz_1(\qv)
\nonumber\\
\chitilz_1(\qv) & \equiv &
 \frac{3}{4}\frac{1}{(3+\zeta^2)}\frac{1}{\qtil^2}
\sum_{\pm} \left[
-\frac{1}{2\qtil}(\qtil^2-1)(\qtil^2-\zeta^2)
\ln\left|\frac{(\qtil+1)(\qtil\pm\zeta)}{(\qtil-1)(\qtil\mp\zeta)}\right|
\pm (1\pm\zeta)(\zeta\pm\qtil^2) \right],\nonumber\\
\label{chiz1}
\end{eqnarray}
where $\qtil=|\qv|/(2\kf)$, 
$\se=n_+-n_-=\frac{\kf^3}{3\pi^2}\zeta(3+\zeta^2)$ is the electron spin density,
$\kf\equiv \half(\kf_++\kf_-)$, 
$\zeta\equiv \frac{\kf_+-\kf_-}{\kf_++\kf_-}$, 
and $\chitilz_1(q)$ is normalized to be
$\chitilz_1(q)=1+O(q^2)$.
Similarly, we obtain
\begin{eqnarray}
\chiz_2(\qv) &=& 
\frac{\kf^3 \zeta}{16\pi}\frac{1}{\qtil}
\sum_{\pm}\int_{1-\zeta}^{1+\zeta} d\ktil \ktil 
\theta(\ktil\qtil -|\qtil^2\pm\zeta)|)
\equiv \kf^3 \chitilz_2(\qv),
\label{chiz2}
\end{eqnarray}
where the integral is to be taken only in the regime
${\ktil>|\frac{1}{\qtil}(\qtil^2\pm\zeta)|}$.
Correlation functions, $\chitilz_1$ and $\chitilz_2$, are plotted in 
momentum space and real space in Fig. \ref{FIGnlt1}.

\begin{figure}[tbh]
  \begin{center}
  \includegraphics[width=5cm]{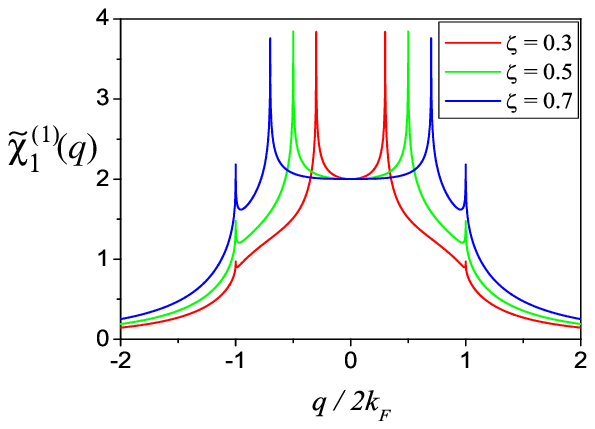}
  \includegraphics[width=5cm]{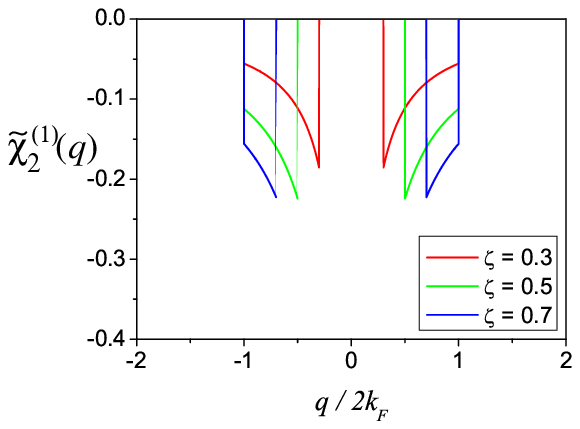}
  \includegraphics[width=5cm]{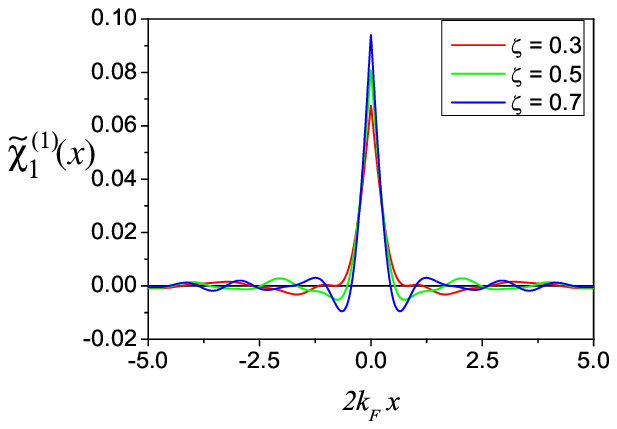}
  \includegraphics[width=5cm]{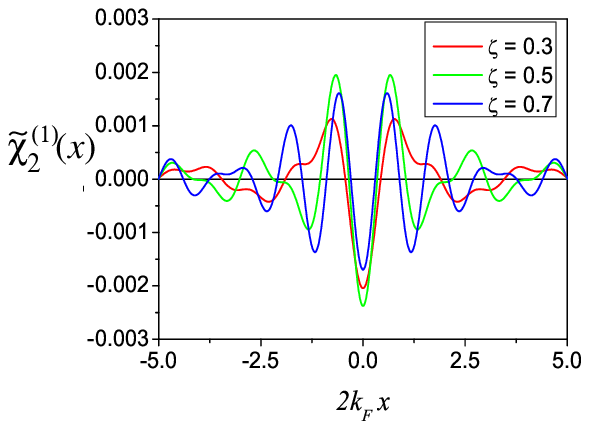}
  \end{center}
\caption{
Plot of correlation functions $\chitilo_1(q)$, $\chitilo_2(q)$ and
their Fourier transforms, $\chitilo_1(x)$ and $\chitilo_2(x)$,   describing nonlocal components of spin density and torque in the presence of current.
$\chitilo_1$ has a finite adiabatic component ($q=0$ i.e., local component) while $\chitilo_2$ does not.
\label{FIGnlt2}
}
\end{figure}

The summation over $\kv$ in Eq. (\ref{chi2}) is carried out using
$\Vinv\sum_{\kv}=\int d\epsilon \DOS(\epsilon) \int_{-1}^{1}\frac{d\cos\theta_k}{2}$,
where $\theta_k$ is the angle of $\kv$ measured from the $\qv$ direction.
We carry out the energy integration first, assuming that the dominant pole arises from the Green's function and the residue contribution from $\frac{1}{\epsilon_{\kv+q}-\epsilon_{\kv} \pm2\spol} $ is neglected.
After some calculation, we obtain
\begin{eqnarray}
\chio_1(\qv) &=& \frac{e\tau}{m}(n_+-n_-) \chitilo_1(\qv)
\nonumber\\
\chitilo_1 &\equiv& \frac{1}{3+\zeta^2} \sum_{\pm}
(1\pm\zeta)\left[1+\frac{1+\zeta^2\pm\zeta-\qtil^2}{2(1\pm\zeta)\qtil}
\ln \left| \frac{(1+\qtil)(\qtil\pm\zeta)}{(1-\qtil)(\qtil\mp\zeta)} \right|
\right],\label{chio1}
\end{eqnarray}
where $n_\pm=\frac{\kf_{\pm}^3}{6\pi^2}$ is the spin-resolved electron density.
$\chitilo_1$ is normalized to be $\chitilo_1(q)=1+O(\qtil^2)$, and 
thus $\chio_1(\qv) \Ev =\frac{1}{e}\chitilo_1(\qtil)\jsv $.
We similarly obtain
\begin{equation}
\chio_2(q) = -\frac{m^2\spol^2}{6 \pi n\kf}
\frac{\sigma_0}{e}\frac{\thetast(q)}{\qtil} 
\equiv \frac{\sigma_0}{e} \chitilo_2(q),
\end{equation}
where
\begin{equation}
\chitilo_2(q) = -\frac{\pi}{2}
\frac{\zeta^2}{1+3\zeta^2}\frac{\thetast(q)}{\qtil} ,
\label{chio2}
\end{equation}
and
\begin{equation}
\thetast(q) \equiv 
\left\{ 
 \begin{array}{cc}
1 \;\;\;  & (\kf_+-\kf_- \leq |q| \leq \kf_++\kf_-) \\
0 & {\rm otherwise} \\
\end{array} \right. \label{thetastdef}
\end{equation}
represents the regime of Stoner excitation, 
$\sigma_0=e^2n\tau/m$ is the Boltzmann conductivity, and $n=n_++n_-$ is the total electron density.
As is obvious, for small $\qtil$, $\chitilo_2=0$.
Correlation functions, $\chitilo_1$ and $\chitilo_2$, are plotted in 
momentum space and real space in Fig. \ref{FIGnlt2}.
Note that the $\tilde{\chi}_1$- and $\tilde{\chi}_2$-terms are proportional to spin current and charge current, respectively, only in the adiabatic limit ($\qtil=0$), but are not necessarily so when nonadiabaticity sets in, since $\tilde{\chi}_1(\qtil)$ and $\tilde{\chi}_2(\qtil)$ can depend on polarization $\zeta$ in a complicated manner.


In the real space representation, the spin densities are given as
\begin{eqnarray}
\seth^{(0)}(\xv) 
 &=&
- \frac{1}{\spol}\intx' 
\left[ 
\se \chitilz_1(\xv-\xv')\lt( \evsph_\xv\cdot \Av_0(\xv') \rt) 
+ {\kf^3}\chitilz_2(\xv-\xv')\lt(\evph(\xv)\cdot \Av_0(\xv') \rt)  \right]
\nonumber\\
\seph^{(0)}(\xv) 
 &=&
-\frac{1}{\spol}\intx' 
\left[
\se \chitilz_1(\xv-\xv') \lt( \evph(\xv)\cdot \Av_0(\xv') \rt) 
- \kf^3 \chitilz_2(\xv-\xv')\lt( \evsph_\xv\cdot \Av_0(\xv') \rt)
 \right],\nonumber\\
&& \label{sezresult}
\end{eqnarray}
and
\begin{eqnarray}
\seth^{(1)} (\xv) 
&=&
-\frac{1}{e\spol}  \sum_{i}\intx'  \left[
{\js^i}\;
\chitilo_1(\xv-\xv') \lt(\evsph_\xv\cdot\Av_i(\xv')\rt) 
\rt.\nonumber\\
&& \lt. 
+{j^i}
\chitilo_2(\xv-\xv') (\evph(\xv)\cdot\Av_i(\xv') )
\right]\nonumber\\
\seph^{(1)} (\xv) 
&=& 
-\frac{1}{e\spol}  \sum_{i}\intx'
\left[
{\js^i}  
\chitilo_1(\xv-\xv') (\evph(\xv)\cdot\Av_i(\xv'))
\rt.\nonumber\\
&& \lt. 
-{j^i}\chitilo_2(\xv-\xv') (\evsph_\xv \cdot\Av_i(\xv') )
\right].\label{seoresult}
\end{eqnarray}

\begin{figure}[tbh]
\begin{center}
\includegraphics[width=0.5\linewidth]{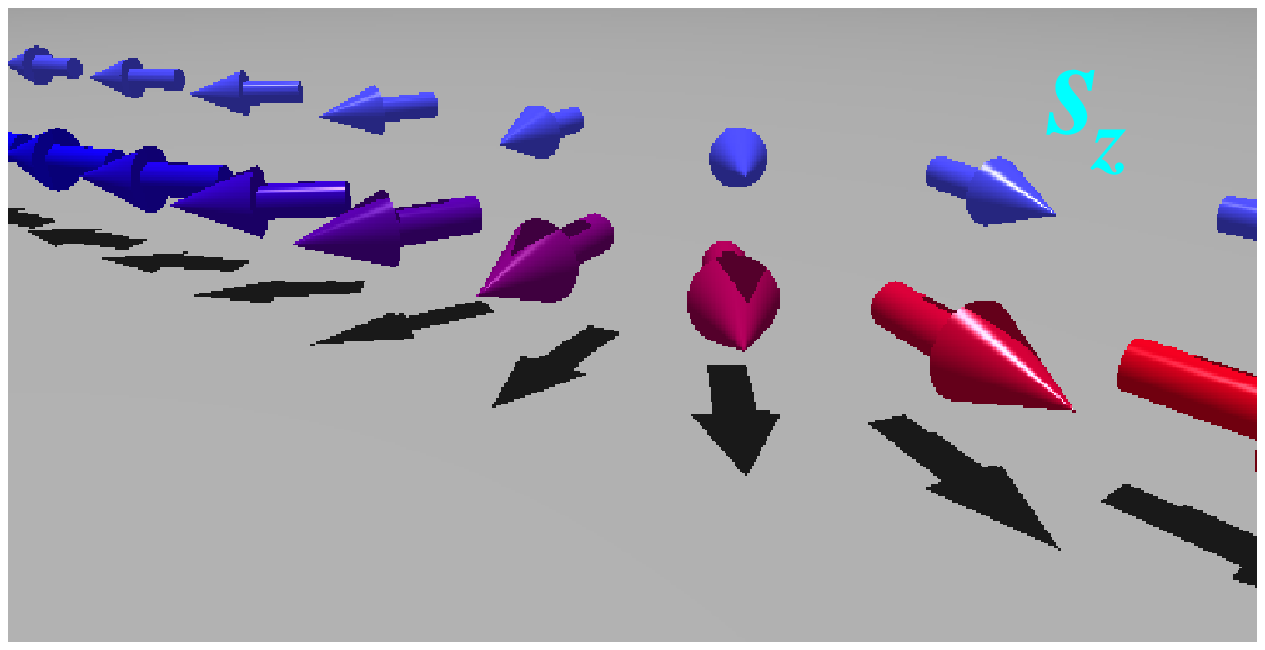}\\
\includegraphics[width=0.5\linewidth]{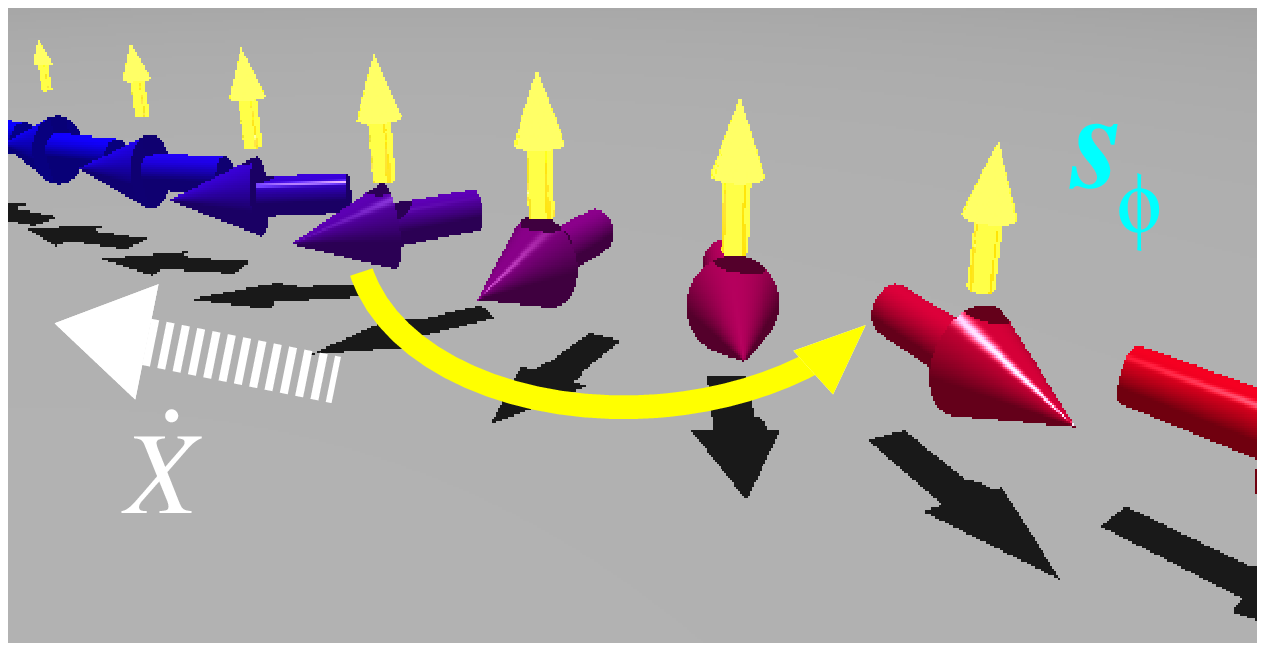}\\
\includegraphics[width=0.5\linewidth]{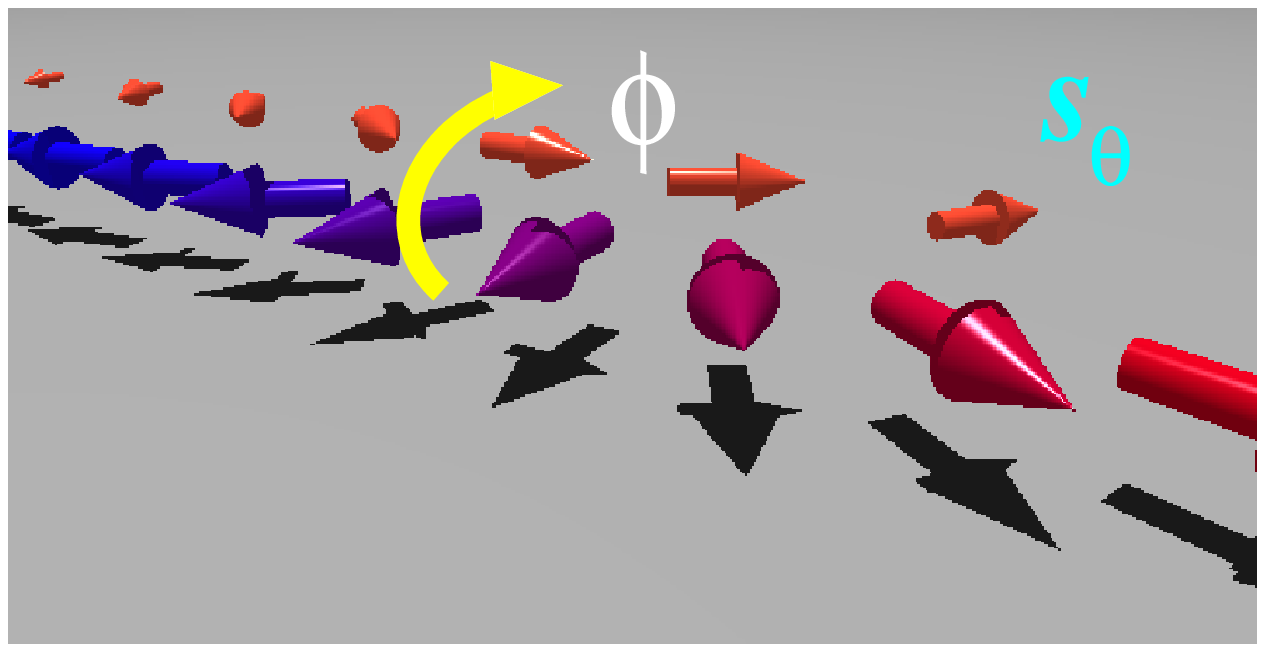}\\
\caption{ \label{FIGelectronspin}
Spin polarization of conduction electrons (denoted by small arrows) around a domain wall (large arrows), and its effect on the domain-wall dynamics.  Top: The adiabatic component, $\se_z$,  along the localized spin does not affect the dynamics. 
Middle: Under current, the spin-transfer torque is induced by the polarization $\seph$ out of the wall plane, which induces rotation of localized spins within the plane, resulting in translational motion, $\dot{X}$.
Bottom: $\dot{\phiz}$ is induced by the component $\seth$  in the $\theta$-direction in the wall plane.  Such polarization arises from spin relaxation processes \cite{Zhang04}. 
Non-adiabaticity also induces $\dot{\phiz}$ \cite{TK04} but the spin density profile is nonlocal and is different from the one due to spin relaxation. }
\end{center}
\end{figure}

The electron spin density induced around domain wall is summarized in Fig. \ref{FIGelectronspin}

\section{Torque and force from electron}
\label{SEC:torqueandforce}
\subsection{Torque}
\label{SEC:torque}

The torque acting on a localized spin at $\xv$ is given by
\begin{eqnarray}
\torquev(\xv)&=&\spol\left( \seph \evth - \seth \evph \right)
\nonumber\\
&=&
- \intx' 
\left[ 
\se \chitilz_1(\xv-\xv') 
\lt[\evth(\xv)\lt( \evph(\xv)\cdot \Av_0(\xv') \rt)
-\evph(\xv) \lt( \evsph_\xv\cdot \Av_0(\xv') \rt) \rt]
\rt.\nonumber\\
&& - {\kf^3}\chitilz_2(\xv-\xv')
\lt[ \evth(\xv)\lt(\evsph_\xv\cdot \Av_0(\xv') \rt)+ \evph(\xv)\lt(\evph(\xv)\cdot \Av_0(\xv') \rt) \rt]
\nonumber\\
&& 
+\sum_{i} \frac{\js^i}{e}  
\chitilo_1(\xv-\xv')
\lt[ \evth(\xv) (\evph(\xv)\cdot\Av_i(\xv'))
-\evph(\xv) (\evsph_\xv\cdot\Av_i(\xv')) \rt]
\nonumber\\
&& \lt. 
-\sum_i \frac{j^i}{e} \chitilo_2(\xv-\xv') 
\lt[ \evth(\xv)(\evsph_\xv \cdot\Av_i(\xv') )
+\evph(\xv)(\evph(\xv) \cdot\Av_i(\xv') ) \rt] \rt].
\nonumber\\&&
\label{torqueresult}
\end{eqnarray}
This torque is the sum of an adiabatic part and a non-adiabatic part,
\begin{equation}
\torquev(\xv)=\torquev^{\rm ad} (\xv)+\torquev_{\rm na} (\xv),
\end{equation}
where the adiabatic part $\torquev^{\rm ad}$ is a purely local part, given later in 
\Eqref{torqueadiabaticres}, 
and the nonlocal part is defined as
\begin{equation}
\torquev_{\rm na} (\xv)\equiv \torquev (\xv)-\torquev^{\rm ad} (\xv). \end{equation}

It is in general non-local due to the non-adiabatic correction.
The total torque on the whole structure is given by
its integral, $\torquev =\intx \torquev(\xv)$.

In the case of uniform polarization, $\phi(\xv)=\phiz$, Eq. (\ref{torqueresult}) becomes somewhat simpler using 
$\evsph_\xv\cdot \Av_\mu(\xv')= \evsph_{\xv'}\cdot \Av_\mu(\xv')
=\Ath_\mu(\xv')$:
\begin{eqnarray}
\lefteqn{
\torquev(\xv)|_{\phi(\xv)=\phiz}
=
\intx' 
\left[ 
\se \chitilz_1(\xv-\xv') 
\lt( \evs(\xv)\times\Av_0(\xv') \rt) \rt.}
\nonumber\\
&&
+\sum_{i} \frac{\js^i}{e}  
\chitilo_1(\xv-\xv')
\lt(  \evs(\xv)\times\Av_i(\xv') \rt)
 \nonumber\\ 
&& 
+ {\kf^3}\chitilz_2(\xv-\xv')
\lt[ \evth(\xv)\lt(\evth(\xv)\cdot \Av_0(\xv') \rt)+ \evph\lt(\evph\cdot \Av_0(\xv') \rt) \rt]
\nonumber\\
&& \lt. 
+\sum_i \frac{j^i}{e} \chitilo_2(\xv-\xv') 
\lt[ \evth(\xv)\lt(\evth(\xv)\cdot \Av_i(\xv') \rt)+ \evph\lt(\evph\cdot \Av_i(\xv') \rt) \rt] \rt],
\label{torqueuni}
\end{eqnarray}
where we used $\nv=\evth\times\evph$.
The $z$-component is given as 
\begin{eqnarray}
\lefteqn{
\torque_z(\xv)|_{\phi(\xv)=\phiz}
=
- \sin\thetaz(\xv) \intx' 
\left[ 
\se \chitilz_1(\xv-\xv') 
\Aph_0(\xv')
+\sum_{i} \frac{\js^i}{e}  
\chitilo_1(\xv-\xv')
\Aph_i(\xv')  \rt.}
 \nonumber\\ 
&&   \lt. 
+ {\kf^3}\chitilz_2(\xv-\xv')
\lt(\evth(\xv)\cdot \Av_0(\xv') \rt)
+\sum_i \frac{j^i}{e} \chitilo_2(\xv-\xv') 
\lt(\evth(\xv)\cdot \Av_i(\xv') \rt)\rt],\nonumber\\
&&  \label{tzphiz}
\end{eqnarray}

\subsection{Adiabatic limit}

The expression for the spin density takes the form of an integral such that the spin density at a given point is determined by all other points in space.
The adiabatic limit is the only exception.
In this limit, the momentum transfer $\qv$ from the gauge field is 
negligibly small compared with the electron momentum $\kv$, and so we can estimate all correlation functions $\tilde\chi^{(n)}_m(\qv)$ at $\qv=0$. 
Thus, $\tilde\chi^{(0)}_1(\qv)=\tilde\chi^{(1)}_1(\qv)=1$ in Eqs. (\ref{sephi2})(\ref{stils}), 
and $\tilde\chi^{(0)}_2(\qv)=\tilde\chi^{(1)}_2(\qv)=0$.
Namely,  
\begin{eqnarray}
\chitilz_1(\xv-\xv')&=&\chitilo_1(\xv-\xv')=\delta(\xv-\xv')\nonumber\\
\chitilz_2(\xv-\xv')&=&\chitilo_2(\xv-\xv')=0.
\end{eqnarray} 
We therefore obtain, by noting 
$  \lt( \evsph_\xv \cdot \Av_\mu(\xv) \rt) 
=  \lt( \evth(\xv) \cdot \Av_\mu(\xv) \rt)$,  
\begin{eqnarray}
\seth^{\rm (ad)}(\xv) 
 &=&
- \frac{1}{\spol}\left[ 
\se \Ath_0(\xv) 
+\frac{\js^i}{e} \Ath_i(\xv) 
\right]
\nonumber\\
\seph^{\rm (ad)}(\xv) 
 &=&
-\frac{1}{\spol}
\left[
\se \Aph_0(\xv)  
+ \frac{\js^i}{e} \Aph_i(\xv)
 \right]\label{sezad}
\end{eqnarray}
The local torque in the adiabatic limit, given using eq. (\ref{torque2}) as
$\torquev^{\rm ad}=\spol \intx 
\left( \seph^{\rm (ad)} \evth - \seth^{\rm (ad)} \evph \right)$,
is then given as
\begin{eqnarray}
\torquev^{\rm ad} (\xv)
 &=&
\evs\times
\lt(\se \Av_0+\frac{\js^i}{e} \Av_i \rt)
\nonumber\\
&=&
\frac{1}{2} 
\lt( \se \partial_0 +\frac{1}{e}\jsv\cdot \nabla \rt) \nv.
\label{torqueadiabaticres}
\end{eqnarray}
The second term is a spin transfer torque, and the first term represents the renormalization of the magnitude of the spin.

\subsection{Force}
\label{SEC:force}

Let us look into the force exerted by the electron.
The total force is given as a sum of equilibrium and current-driven parts, 
$F_i=\Fo_i+\Fz_i$, 
arising from $ \stilpmz$ and $\stilpmo$, respectively.
We first look into the force from the current, $\Fo$.
Using Eq. (\ref{S1_2}) and noting 
$\gr_{\kv\sigma} \simeq -i\pi\delta(\epsilon_{\kv\sigma})$
on $\kv$-integration, we obtain 
\begin{eqnarray}
\Fo_i &=&
-\sum_{\pm}\sum_{\kv\qv} \sum_{jk}  \frac{eE_k\tau\Delta}{m^2V}
 (\pm)\delta(\epsilon_{\kv\pm})
\nonumber\\ && \times 
\left[ 
 k_k \left(k+\frac{q}{2}\right)_j \frac{-i}
{\epsilon_{\kv+q}-\epsilon_{\kv} \pm2\Delta} 
\left( 
 A_j^\pm(\qv)  A_i^\mp(-\qv)
 - A_j^\mp(\qv) A_i^\pm(-\qv)
\right)
\right. \nonumber\\
&& \left.
-\pi q_k \left(\kv+\frac{q}{2}\right)_j 
\delta({\epsilon_{\kv+q}-\epsilon_{\kv} \pm2\Delta})
\left( 
 A_j^\pm(\qv) A_i^\mp(-\qv) + A_j^\mp(\qv) A_i^\pm(-\qv)
\right)
\right].\nonumber\\
&\equiv &  \Fna_i+\Fdel_i,
\end{eqnarray}
where
the first and second terms are defined as
\begin{eqnarray}
\Fna_i &=& 
\sum_{\pm}\sum_{\kv\qv} \sum_{jk}  \frac{eE_k\tau\Delta}{m^2V}
 (\pm)\delta(\epsilon_{\kv\pm})
\nonumber\\ && \times 
\pi q_k \left(\kv+\frac{q}{2}\right)_j 
\delta({\epsilon_{\kv+q}-\epsilon_{\kv} \pm2\Delta})
\left( 
 A_j^\pm(\qv) A_i^\mp(-\qv) + A_j^\mp(\qv) A_i^\pm(-\qv)
\right)\nonumber\\
\Fdel_i &=&
\sum_{\pm}\sum_{\kv\qv} \sum_{jk}  \frac{eE_k\tau\Delta}{m^2V}
 (\pm)\delta(\epsilon_{\kv\pm})
\nonumber\\ && \times 
 k_k \left(k+\frac{q}{2}\right)_j 
\frac{i}
{\epsilon_{\kv+q}-\epsilon_{\kv} \pm2\Delta} 
\left( 
 A_j^\pm(\qv)  A_i^\mp(-\qv)
 - A_j^\mp(\qv) A_i^\pm(-\qv)
\right).
 \nonumber\\&&
\label{FnaFdel}
\end{eqnarray}
The first term $\Fna$ vanishes in the adiabatic limit ($q=0$) while the second term $\Fdel$ remains finite.
We now demonstrate that the term $\Fna$  represents the reflection of the electron due to the spin structure.
In fact,  we find, by using
$\epsilon_{\kv+\qv}-\epsilon_{\kv}=\frac{1}{m}\qv\cdot(\kv+\frac{\qv}{2})$,
that  $\Fna$ is 
\begin{eqnarray}
\Fna_i
&=&
-\sum_{\pm}\sum_{\kv\qv} \sum_{j}  \frac{4\pi eE_j\tau\Delta^2}{mV}
\delta(\epsilon_{\kv\pm}) 
\delta({\epsilon_{\kv+q}-\epsilon_{\kv} \pm2\Delta})
 A_j^\pm(\qv)
A_i^\mp(-\qv). \nonumber\\&& \label{Fna}
\end{eqnarray}
This force is proportional to the resistivity due to the spin structure, which will be discussed in \S \ref{SECmori} on the basis of the Mori formula.
The resistivity is calculated as (in the case of current along $\xw$-direction)
\begin{equation}
\rhos= \frac{4\pi \Delta^{2}}{e^{2}n^{2}}\frac{1}{V}
 \sum_{\kv q \sigma} |A_\xw^\sigma(\qv)|^{2} 
\delta(\epsilon_{\kv+q,-\sigma} -\epsilon_{\kv,\sigma} ) \delta(\epsilon_{\kv,\sigma} ).
\label{rhos}
\end{equation}
We therefore see the relation between force and resistivity
\begin{equation}
\Fvna= \frac{e^3 \Ev \tau}{m}\rhos n^2 V 
  =e\Ne\rhos \jv, \label{Fandj}
\end{equation}
where $\Ne=nV$ is the total electron number.
This is the result in Ref.  \cite{TK04} extended to a general spin structure.

The resistance
due to spin structure, $\RS\equiv \frac{L}{A}\rhoS $,
is related to the reflection probability $R$, according to the four-terminal Landauer-B\"uttiker formula \cite{Buttiker85}, as 
$\RS=\frac{\pi\hbar}{e^2} \frac{R}{1-R}$, and hence Eq. (\ref{Fandj}) 
indeed relates the force to the reflection of the electron.
Equation (\ref{Fandj}) can also be written, using the density of states 
$\DOSV=\frac{m \kf V}{2\pi^2 \hbar^2}$ (neglecting spin splitting), as
\begin{equation}
\Fna= \frac{1}{3} eV_{\rm S} \DOSV {2\kf\hbar}\frac{\kf\hbar}{mL},
\end{equation}
where $V_{\rm S}\equiv \RS I$ is a voltage drop due to spin structure and  $I\equiv Aj$ is the current ($A$ being the cross section).
This equation clearly indicates that the force is due to the momentum transfer
(${2\kf\hbar}$ per electron, with frequency $\frac{\kf\hbar}{mL}$)  
multiplied by the number of electrons that contribute to the resistance
($\frac{1}{3}eV_{\rm S} \DOSV$).

What is the origin of the other term $\Fdel$ 
(Eq. (\ref{FnaFdel})) ?
Its meaning becomes clear in  the adiabatic limit.
In fact, in this limit, $\Fdel_i$ reduces to
$\Fhall$, where 
\begin{eqnarray}
\Fhall_i 
&=&
\sum_{\pm}\sum_{\kv\qv} \sum_{j}  \frac{eE_j\tau}{2m}
n_{\pm} i \delta(\epsilon_{\kv\pm})
k^2 \left(
A_j^\pm(\qv) A_i^\mp(-\qv)
+ A_j^\mp(\qv) A_i^\pm(-\qv)
\right) 
\nonumber\\
&=&
i \sum_j \frac{{\js}_j}{e} \intx \left(
A_j^+(\xv) A_i^-(\xv)
- A_j^-(\xv) A_i^+(\xv)
\right) .
\end{eqnarray}
Using
\begin{equation}
\left(
A_i^+(\xv) A_j^-(\xv)
- A_i^-(\xv) A_j^+(\xv)
\right) 
= -i \half \sin\theta 
(\partial_i\theta \partial_j\phi-\partial_j\theta \partial_i\phi),
\end{equation}
and
\begin{equation}
\partial_i\Sv \times \partial_j\Sv =\evs S^2 \sin\theta 
(\partial_i\theta \partial_j\phi
  -\partial_j\theta \partial_i\phi),
\end{equation}
we see that
\begin{eqnarray}
\Fhall_i 
&=& -\frac{1}{2S^3} \sum_j \frac{{\js}_j}{e} \intx 
\Sv\cdot(\partial_i\Sv \times \partial_j\Sv )
=-2\pi \sum_j \frac{{\js}_j}{e} \Phi_{ij},
\label{Fad0}
\end{eqnarray}
where
\begin{equation}
\Phi_{ij}\equiv  \frac{1}{4\pi S^3}  \intx
\Sv\cdot(\partial_i\Sv \times \partial_j\Sv ) \label{topcharge}
\end{equation}
is a vortex number in a plane perpendicular to $i$ and $j$ directions.
In the case of a thin system (with thickness $\thickness$), this reduces  to  
\begin{equation}
\Phi_{xy}=\nvortex \thickness,
\end{equation}
where
\begin{equation}
\nvortex \equiv  \frac{1}{4\pi S^3}  \int d^2x
\Sv\cdot(\partial_x\Sv \times \partial_y\Sv ),
\end{equation}
is a topological number in two dimensions. 
This force is, in fact, a back reaction of the Hall effect due to spin chirality \cite{Ye99,TK02,OTN04}, and was derived by Thiele \cite{Thiele73} and by Berger \cite{Berger86}, then rigorously in Ref.  \cite{KTSS07}, and also
assuming a vortex structure in Ref.  \cite{SNTKO06}.
The Hall effect due to vorticity arises from the spin Berry phase \cite{Ye99} or spin chirality \cite{TK02}.
For a domain wall, the vorticity is zero and so $\Fhall=0$.
Defining the gyrovector of the vortex, $\Gv$, as
\begin{equation}
G_i\equiv \frac{2\pi S}{a^3}\epsilon_{ijk}\Phi_{jk}, \label{Gvdef}
\end{equation}
we can write the Hall force as
\begin{eqnarray}
\Fhallv 
&=& =-\frac{P a^3}{2eS}(\jv\times\Gv),
\end{eqnarray}
where $P\jv$ represents direction of spin current flow.

\begin{figure}[tbh]
  \begin{center}
  \includegraphics[scale=0.3]{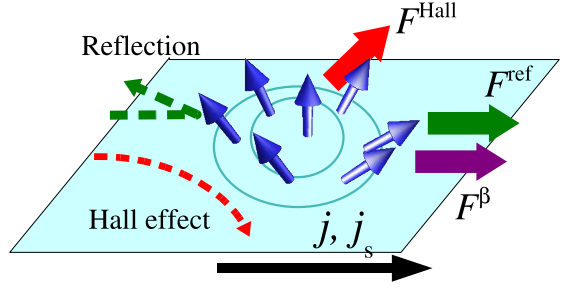}
\caption{
Schematic summary of three forces acting on spin structures. 
Electron reflection pushes the structure along the current flow ($\Fna$), and the Hall effect due to spin chirality results in a force in the perpendicular direction ($\Fhall$). 
For a domain wall, $\Fhall=0$.
Spin relaxation results in a force along the direction of the  current ($\Fbeta$).
\label{FIGvortexwall}}
  \end{center}
\end{figure}

There is another force arising from spin relaxation, the $\betasf$ term. 
Comparing the definition of the $\beta$ term, \Eqref{modLLG}, with \Eqsref{LLGfull}{Bvedef}, we see that the $\beta$ term corresponds to spin polarization of
\begin{equation}
\sevsf = -\frac{\hbar P}{2eS\spol}(\jv\cdot\nabla)\nv. 
\end{equation}
This spin polarization exerts a force, according to \Eqref{Fdef}, of
 \cite{Thiaville05,Zhang04,KTSS07,TKSLL07}
\begin{equation}
F^{\beta}=\frac{\hbar}{2eS} \Fbetafactor \betasf Pj,
\end{equation}
where $\Fbetafactor \equiv  \intx \, (\nabla_\xw \nv)^2$.
For a planar wall,
\begin{equation}
\Fbetafactor=\frac{a^3}{\lambda^2}\Nw.
\end{equation}
These three forces are schematically illustrated in Fig. \ref{FIGvortexwall}.

One may think this strange, but there is a force arising from the equilibrium component of spin polarization. 
This force, given by (see \Eqref{force2})
\begin{equation}
\Fz_i \equiv 2\spol \intx (-\seth^{(0)}\Aph_i +\seph^{(0)}\Ath_i) ,
\end{equation}
with  $\seth^{(0)}$ and $\seph^{(0)}$ given by \Eqref{sezresult}, 
 arises when the spin structure is moving, and represents the effect of relative velocity between the spin structure and 
the current-carrying electron.
Let us consider again the adiabatic limit for simplicity.
Using Eq. (\ref{sezad}), the force in this limit is given by
\begin{eqnarray}
\Fren_i 
&\equiv &
\Fz_i|_{\rm adiabatic} 
 =
2\se\intx   \evs \cdot\lt(\Av_0\times \Av_i \rt) 
\nonumber\\
&=&
\frac{\se}{2} \intx 
 \nv\cdot(\partial_0\nv\times \partial_i \nv) \equiv  (\Phiv_0)_i,
\end{eqnarray}
where the vector  $\Phiv_0$ is defined as 
$(\Phiv_0)_i\equiv \Phi_{0i}$ of \Eqref{topcharge}.
This equilibrium contribution to the adiabatic force 
thus has again a topological meaning but in two dimensions including time and space.
The total force due to current is summarized as
\begin{equation}
\Fv=
\left( e\Ne \rhoS + P\Fbetafactor \frac{1}{2eS}\betasf \right)
\jv
+P \frac{a^3}{2eS} \jv\times \Gv 
+2\pi\se \Phiv_0 +\Fdeltav, \label{totalF}
\end{equation}
where 
\begin{equation}
\Fdeltav  \equiv   \Fdelv-\Fhallv + \Fzv -\Frenv,
\end{equation}
is an additional non-adiabatic correction to the force (besides $\Fna$).
In the adiabatic limit it reduces to 
\begin{eqnarray}
\Fv^{\rm (ad)} &=& 
P\Fbetafactor \frac{\hbar}{2eS}\betasf \jv 
+P \frac{\hbar a^3}{2eS} \jv\times \Gv 
+2\pi\hbar \se \Phiv_0 
.\label{Fad}
\end{eqnarray}

\section{Effective Lagrangian (adiabatic limit)}
\label{SEC:adiabaticspinlag}

\subsection{Spin transfer torque and Berry phase}

In the gauge transformed frame, the effective Lagrangian of the  localized spin is obtained simply by taking the expectation value of $\HA$, \Eqref{HAdef} \cite{Bazaliy98}.
In the adiabatic limit, this value is given by
\begin{eqnarray}
\average{\HA} &=& 
  \hbar \sum_{\kv\qv\alpha} 
\lt( A_0^\alpha(\qv)
+\frac{\hbar}{m}  \kv\cdot \Av^\alpha(\qv) \rt)
\average{  \adag_{\kv}\sigma_\alpha a_{\kv} }
\nonumber\\ 
&& =  \intx 
\lt(\hbar \se A_0^z(\xv)
+\frac{P}{e} \jv\cdot \Av^z(\xv) \rt),
\end{eqnarray}
(only $z$ spin components are finite at the lowest order in $A$).
Therefore the effective spin Lagrangian reads (neglecting dissipation)
\begin{eqnarray}
\Ls^{\rm eff} &=& \Ls-\average{\HA} \nonumber\\
 &=& \hbar \sumx \lt[\Stot \dot{\phi}(\cos\theta-1)
  +\frac{Pa^3}{2e}[(\jv\cdot\nabla)\phi](\cos\theta-1) \rt]
-\Hs,
\end{eqnarray}
where $\Stot=S+\frac{\se a^3}{2}$ is the total spin including electron spin polarization.

This Lagrangian for the adiabatic case can be written as
\begin{eqnarray}
\Ls^{\rm eff}
 &=& \sumx \lt[ \hbar \Stot
[(\partial_t+\vsv\cdot\nabla)\phi](\cos\theta-1) \rt]
-\Hs,  \label{LeffST}
\end{eqnarray}
where
\begin{equation}
\vsv\equiv \frac{Pa^3}{2e\Stot}\jv,
\end{equation}
is the drift velocity of the electron spin.
The solution has the form 
$f(\xv-\vs t)$ ($f$ is any function), which clearly indicates that the adiabatic spin transfer torque induces a stream motion of localized spin structure with velocity $\vev$. 
This applies to any spin structure.
We can also say from \Eqref{LeffST} that the spin transfer torque induces a spatial spin Berry phase.

\subsection{Nucleation of domain wall by spin current}

 As seen from \Eqref{LeffST}, the spin current favors a magnetic configuration with 
spatial gradient, or more precisely, with finite Berry-phase curvature. 
 It is thus expected that a large spin current destabilizes a uniform 
ferromagnetic state.
 This is indeed seen from the spin-wave energy around a uniform ferromagnetic state  \cite{Bazaliy98,Rossier04,Li04}, 
\begin{eqnarray}
\label{Omega-uni}
\Omega_{\bm k}^{\rm (F)}
=\frac{KS}{\hbar}\sqrt{(k^2\lambda^2+1)(k^2\lambda^2+1+\kappa)} 
+ {\bm k}\cdot \vsv .
\end{eqnarray}
The first term is the well-known spin-wave dispersion with an anisotropy gap. 
The effect of spin current appears in the second term as the Doppler shift due to the drift velocity of electron spins
 \cite{Rossier04}.
 For sufficiently large $\vsv$, 
$\Omega_{\bm k}^{\rm (F)}$ 
becomes negative for a range of ${\bm k}$. 
 This means that there exist states with negative excitation energy, 
indicating the instability of the assumed uniformly-magnetized state 
 \cite{Bazaliy98,Rossier04,Li04}. 
 The wavelength of the unstable mode around the uniform magnetization is 
$k=-(K(K+K_\perp)/J^2)^{1/4}$ \cite{Li04}. 
The critical current, $j_{\rm nuc}$, is given by \cite{Rossier04,Li04} 
\begin{equation}
j_{\rm nuc}=\frac{2eS^2}{\hbar a^{3}}K\lambda(1+\sqrt{1+\kappa}).
\end{equation} 
 The true ground state under a large spin current was found to be a multi-domain state \cite{STK05}. 
 Namely, the energy of a domain wall becomes lower than the uniform ferromagnetic state 
 when a spin current exceeds a critical value 
$j_{\rm nuc}$. 
Nucleated domains flow at a  velocity $\vsv$ if $\Kp \ll K$. 

When $K \ll \Kp$, as is usually the case of a film or wide strip, vortex nucleation occurs at a lower current than $j_{\rm nuc}$, as was shown in Ref.  \cite{NSTKTM08}.

\section{Torque and force on domain wall}
\label{SEC:dweq}

In this section, the torque and force on a rigid domain wall are estimated. 

The gauge fields for a domain wall is given by
\begin{eqnarray}
\Ath_0 &=& -\hf\sin\thetaz \dot{\phiz} \nonumber\\
\Aph_0 &=& \hf\frac{\dot{X}}{\lambda}\sin\thetaz \nonumber\\
\Ath_i &=& 0 \nonumber\\
\Aph_i &=& -\delta_{i,\xw} \frac{1}{2\lambda}\sin\thetaz,
\end{eqnarray}
where $\sin\thetaz=[\cosh \frac{x-X(t)}{\lambda}]^{-1}$.
Noting that $\sin\thetaz(\xv)$ is odd in $\xw-X$ and 
$\intx \chitilz_2(\xv-\xv')=\chitilz_2(\qv=0)=0$, we see that the total torque on a wall is obtained from Eq. (\ref{tzphiz}) as
\begin{eqnarray}
\torquew
=
-\hf \intx\intx' \sin\thetaz(\xv) \sin\thetaz(\xv') 
\lt[ 
\se \frac{\dot{X}}{\lambda} \chitilz_1(\xv-\xv') 
- \sum_{i} \frac{\js^i}{e\lambda}  
\chitilo_1(\xv-\xv')
\rt].\nonumber\\
&&
\end{eqnarray}
This is the full expression of the torque acting on the wall including non-adiabaticity (but without spin relaxation).
Separating the adiabatic contribution, it can be written 
as
\begin{equation}
\torquew=\frac{\Nw}{2\lambda}\lt(\frac{Pj}{e}-\se\dot{X}\rt)
+\delta\torque,
\end{equation}
where $\delta\torque$ represents non-adiabatic contributions.

For domain walls, the adiabatic Hall force term $\Fhall$ vanishes since the domain wall has no vortex charge. 
The total force on the domain wall is given by
\begin{eqnarray}
\Fw&=& \Fna+\Fren+\Fbeta+\Fdelta \nonumber\\
&=& e\Ne\rhow j + \frac{\Nw}{2\lambda} \se\dot{\phiz}
  +\Fbeta+\Fdelta ,
\end{eqnarray}
where $\rhow$ is the resistivity due to the domain wall.

From the above argument, the equation of motion of a rigid wall is therefore obtained as (Eqs. (\ref{DWeq_a})(\ref{DWeq_b}))
\begin{eqnarray}
\dot\phiz+\alpha \frac{\dot{X}}{\lambda}
 &=& \frac{\lambda}{\hbar \Nw S} (\Fw+F_{\rm pin})
\nonumber \\ 
\dot{X}-\alpha\lambda\dot{\phiz} 
&=& \frac{\Kp\lambda}{2\hbar}S \sin 2\phiz  
   + \frac{\lambda}{\hbar \Nw S} \torquew,
\end{eqnarray}
or
\begin{eqnarray}
\dot\phiz+\alpha \frac{\dot{X}}{\lambda}
 &=& 
 \fna + \fbeta + \fpin -\frac{\se a^3}{2S}\dot{\phiz}+\delta f
\nonumber\\
\dot{X}-\alpha\lambda\dot{\phiz} &=&  \vc \sin 2\phiz  
+\frac{a^3}{2S}\lt( P\frac{j}{e}-\se\dot{X} \rt)
+\frac{\lambda}{\hbar \Nw S} \delta\torque.
\label{dweqsimple}
\end{eqnarray}
where $f^{({\rm ref, pin},\beta)}\equiv \frac{\lambda}{\hbar\Nw S} F^{({\rm ref, pin},\beta)}$.
We see that the contribution to the $\dot\phiz$ term from $\Fren$ 
represents the dressing effect of localized spin due to conduction electron polarization. Namely, the spin constituting the domain wall is now given by the total spin
\begin{equation}
\Stot \equiv S+\frac{\se a^3}{2} \equiv S+\deltaS,
\label{stotdef}
\end{equation}
and the equations become
\begin{eqnarray}
\Stot \dot\phiz+\alpha S \frac{\dot{X}}{\lambda}
 &=& \frac{\lambda}{\hbar \Nw} (\Fna+\Fbeta+\Fdelta+F_{\rm pin})
\nonumber \\ 
\Stot \dot{X}-\alpha S\lambda\dot{\phiz} 
&=& \frac{\Kp\lambda}{2\hbar}S^2 \sin 2\phiz   
   +\frac{P a^3}{2e} j
   + \frac{\lambda}{\hbar \Nw }  \delta\torque.
\end{eqnarray}
The spin renormalization factor appears also in the equation for $\dot{X}$, and this contribution guarantees correctly that the spin-transfer effect arises proportional to the relative velocity $\lt( \frac{P}{e}j-\se\dot{X} \rt)$ between the wall and conduction electron spin.
These terms were considered phenomenologically in Refs.  \cite{Berger86,Berger92} and  \cite{Barnes05}, but these effects naturally appear in our formulation.

\section{Domain wall dynamics}
\label{SEC:sol}

In the following calculation, we simply write $\Stot$ as $S$.
In non-adiabatic contributions, we include $\Fna$, which is qualitatively important in dynamics, while we neglect  $\Fdelta$ and $\delta\torque$, since these affect the dynamics only quantitatively.
The equation of motion of the wall we consider is given by 
\begin{eqnarray}
\dot\phiz+\alpha \frac{\dot{X}}{\lambda}
 &=& \frac{a^3}{2eS\lambda} \betaw j +\fpin
\nonumber\\
\dot{X}-\alpha\lambda\dot{\phiz} &=&  \vc \sin 2\phiz  
+\frac{a^3}{2eS}P{j},
\label{DWeq3}
\end{eqnarray}
where 
\begin{eqnarray}
\betaw 
&\equiv&
 \frac{e^2}{\hbar}n A\lambda \Rw + P\betasf,  \label{betaeffective}
\end{eqnarray}
represents the total force acting on the wall
($\Rw=\rhow L/A$ is the wall resistance),
\begin{equation}
\vc\equiv 
\frac{\Kp\lambda S}{2\hbar},\label{vcdef}
\end{equation}
 and
\begin{eqnarray}
\fpin &=& 
-\frac{2\lambda\Vz}{\hbar S \xi^2} X \theta(\xi-|X|).
\end{eqnarray}
The coefficient $\betaw$ is the effective $\beta$ that acts on a rigid wall. 
In other words, a rigid wall feels electron spin relaxation and non-adiabaticity (reflection) exactly the same. 
The equation of motion in terms of dimensionless parameters is written as
\begin{eqnarray}
{\partial_\ttil} \left( {\Xtil} -\alpha {\phiz} \right)
 &=& \sin 2\phiz + \Ptil\jtil \nonumber\\
{\partial_\ttil} \left( {\phiz} +\alpha{\Xtil} \right)  
       &=&
  -\Vztil \Xtil\theta(\xi/\lambda-|\Xtil|) +\betaw \jtil,
\label{DWeq3nodim}
\end{eqnarray}
where
$\ttil\equiv t \vc/\lambda$, $\Xtil\equiv X/\lambda$, 
$\Omegatil\equiv \Omegapin \lambda/\vc 
 = \frac{2\sqrt{2}}{S}\frac{\lambda}{\xi}\sqrt{\frac{\Vz}{\Kp}}$, 
$\Ptil\equiv {P}$, 
$\jtil\equiv \frac{a^3}{2eS\vc}j$,  
$\Vztil\equiv \frac{1}{2}\Omegatil^2
 = \Vz \frac{2}{\hbar S} \frac{\lambda^3}{\vc \xi^2}
 = \frac{4}{S^2} \frac{\Vz}{\Kp}\left(\frac{\lambda}{\xi}\right)^2$, and
$\Vz\equiv \frac{\Mw}{2\Nw}\Omegapin^2\xi^2
 = 
\frac{\hbar^2}{2}\frac{\Omegapin^2}{\Kp}\left(\frac{\xi}{\lambda}\right)^2
=\frac{\hbar S}{4}\frac{\Omegapin^2}{\vc}\frac{\xi^2}{\lambda}
$.
The equation of motion can be written as
\begin{eqnarray}
{\partial_\ttil} {\Xtil} &=& \frac{1}{1+\alpha^2}  
 \left( \sin 2\phiz + (\Ptil+\alpha\betaw)\jtil 
 -\alpha \Vztil \Xtil\theta(\xi/\lambda-|\Xtil|) \right) \label{DWvelocity}\\
{\partial_\ttil} {\phiz} &=& \frac{1}{1+\alpha^2}  
 \left( (\betaw-\Ptil\alpha)\jtil-\alpha \sin 2\phiz 
-\Vztil \Xtil\theta(\xi/\lambda-|\Xtil|)  \right) .
\label{DWeq4}
\end{eqnarray}
We will 
solve this equation of motion below for the case of steady current.
A moving domain wall, free from the extrinsic pinning effect, 
is described by the above equation with $\Vztil=0$.
There is an analytical solution in this case as we see just below.

\subsection{Solution without extrinsic pinning}
\label{SECnopin}
Let us consider first the case without pinning potential of extrinsic origin. 
In this case the analytical solution is easily obtained.
In fact Eq. (\ref{DWeq4}) has a form of
\begin{equation}
{\partial_\ttil} {\phiz} = B-A \sin 2\phiz,
\label{eqst}
\end{equation}
where
\begin{eqnarray}
 A &=& \frac{\alpha}{1+\alpha^2} \nonumber\\
 B &=& \frac{(\betaw-\Ptil\alpha)\jtil}{1+\alpha^2}.
\end{eqnarray}
The solution to this differential equation is given by
\begin{equation}
\tan \phiz =\frac{1}{B} \left(A -\om \cot(\om(t-\tz)) \right),
\label{tansol}
\end{equation}
where $\cot\theta\equiv\frac{\cos\theta}{\sin\theta}$, 
\begin{equation}
\om\equiv \sqrt{B^2-A^2}= \frac{\alpha}{1+\alpha^2}
\sqrt{\lt(\frac{\jtil}{\jatil}\rt)^2-1},
\end{equation} 
and
\begin{equation}
\jatil \equiv \frac{1}{\Ptil-\frac{\betaw}{\alpha}}.
\label{jatildef}
\end{equation}
In the case of imaginary $\om$ (i.e., $A^2 >B^2$), 
Eq. (\ref{tansol}) is still applicable by replacing
$\om \rightarrow i|\om|$ and 
$\tan(\om(t-\tz)) \rightarrow i \tanh(|\om|(t-\tz))$.
We require the initial condition $\phiz(0)=0$ at $t=0$. 
Then we find the solution as 
\begin{equation}
\tan \phiz = \frac{ B \tan\om t}{\om+ A\tan \om t},
\label{tansol2}
\end{equation}
and thus $\sin 2\phiz$, which appears in the wall velocity, 
is obtained as
\begin{eqnarray}
\sin 2\phiz 
&=&
 \frac{A+ B \sin (2\om t -\vartheta)}
  {B +A \sin (2\om t -\vartheta)} \nonumber\\
&=&
\left( \frac{\betaw}{\alpha}-\Ptil \right) \jtil
-
\frac{ \left[\left(\Ptil-\frac{\betaw}{\alpha}\right)\jtil\right]^2-1}
{\lt(\frac{\betaw}{\alpha}-\Ptil\rt)\jtil 
   +\sin(2\om t-\vartheta)}  \nonumber\\
&=&
-\frac{\jtil}{\jatil}
+\frac{\lt(\frac{\jtil}{\jatil}\rt)^2-1}
  {\frac{\jtil}{\jatil}-\sin(2\om t-\vartheta)},
\label{sin2phi}
\end{eqnarray}
where 
$\sin \vartheta\equiv \frac{A}{B}=\frac{1}{(\frac{\betaw}{\alpha}-\Ptil)\jtil}$, 
$\cos \vartheta\equiv \frac{\om}{B}$.
We see that an anomaly appears when $\om$ switches from real to imaginary, i.e., at $|\jtil| = |\jatil|$.
Above and below $|\jatil|$, the wall dynamics is quite different. 
For  $|\jtil|>|\jatil|$, the wall velocity (\ref{DWvelocity}) has an oscillating component.
This oscillating dynamics is similar to the behavior in a high magnetic field, known as Walker's breakdown \cite{Hubert00}.
In contrast, for $|\jtil| < |\jatil|$, $\phiz$ reaches its stable angle (given by \Eqref{phizterm}), and simple sliding motion of wall with constant velocity (\Eqref{vwallslide}) is realized. 

The average velocity for $|\jtil| \geq |\jatil|$ is obtained by use of 
Eqs. (\ref{DWvelocity})(\ref{sin2phi}) and 
\begin{equation}
\frac{1}{T} \int_0^T dt \frac{1}{C-\sin \left(2\pi \frac{t}{T}\right) }
 =\frac{1}{\sqrt{C^2-1}}\frac{C}{|C|},
\end{equation}
(for $|C|>1$) to be ($T=\frac{\pi}{\omega}$)
\begin{eqnarray}
\average{\dot{X}} &=&  \frac{\betaw}{\alpha} \jtil
 +\frac{1}{1+\alpha^2}\sgn\lt[\frac{\jtil}{\jatil}\rt]
 \sqrt{ \lt( \frac{\jtil}{\jatil} \rt)^2-1  }
\nonumber\\
&=&
\frac{\betaw}{\alpha} \jtil
 +\frac{1}{1+\alpha^2}\frac{1}{\jatil}
 \sqrt{\jtil^2-\jatil^2}
.\label{averagewallvelocity}
\end{eqnarray}
When $\jtil \gg |\jatil|$, the velocity becomes 
\begin{equation}
\average{\dot{X}} \rightarrow  \frac{\Ptil+\alpha\betaw}{1+\alpha^2} \jtil  \;\;\;\;\;
  (\jtil\gg |\jatil|)
 .\label{vwallST}
\end{equation}

Below $\jatil$, $\omega$ is pure imaginary and $\sin2\phiz$ approaches at $t\rightarrow\infty$ a steady value of 
\begin{equation}
\sin 2\phiz \rightarrow  \frac{B}{A}
 =\left(\frac{\betaw}{\alpha}-\Ptil\right) \jtil, \label{phizterm}
\end{equation}
and so the terminal velocity is obtained as
\begin{equation}
\average{\dot{X}} = \frac{\betaw}{\alpha} \jtil. \label{vwallslide}
\end{equation}
This speed can be larger than the spin-transfer limit, 
$\average{\dot{X}} = \Ptil\jtil$, if $\frac{\betaw}{\alpha}$ is large, but this high speed is realized only at a small current,
$\jtil < |\jatil|$, i.e.,  
$j\lesssim \frac{\alpha}{\betaw}\frac{2eS}{a^3}\vc$.

A special case arise when $\frac{\betaw}{\alpha}=\Ptil$, where simple sliding with $\phiz=0$ and $\dot{\Xtil}=\Ptil \jtil$ is realized \cite{Barnes05}.

\begin{figure}[tbp]
\begin{center}
\includegraphics[scale=1]{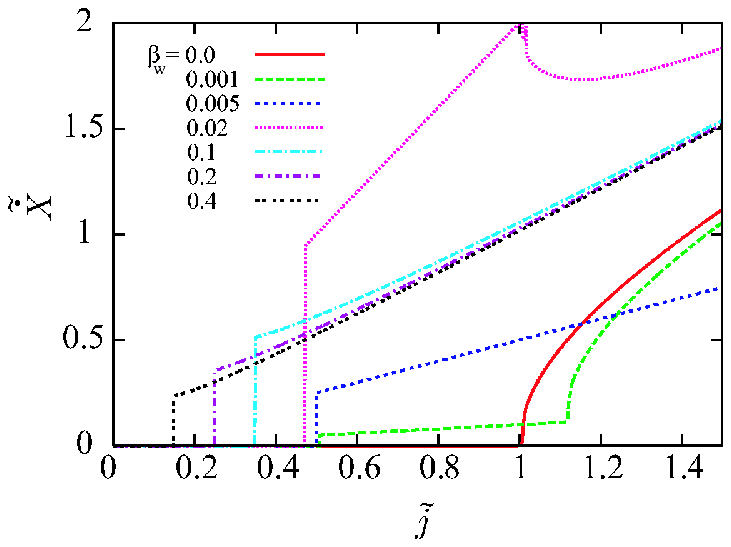}
\caption{ Wall velocity as function of current for $\Omegatil=0.5$ and $\alpha=0.01$.
A jump in wall velocity is seen at threshold, $\jtil=\jctil$.
A crossover from I-a)($\jctil\sim \Omegatil/\betaw$) to 
I-b)($\jctil \sim \Omegatil$) regime is seen at 
$\betaw\simeq \frac{\Omegatil}{4}\sim 0.1$. 
Intrinsic pinning occurs only for $\betaw=0$ in the present case of $\Omegatil \lesssim1$ (see Table \ref{TABLE}).
Linear behavior of velocity seen for large $\betaw(\geq0.1)$ is the perfect adiabatic limit realized for $\jtil\gg\jatil$ 
(\Eqref{vwallST}).
\label{FIGvj} }
\end{center}
\end{figure}

These behaviors of wall speed are seen in Fig. \ref{FIGvj}
(in the presence of extrinsic pinning potential).
The most significant effect of $\betaw$ is to shift $\jatil$.
When $\betaw\lesssim\alpha\Ptil$, $\jatil$ remains 
close to the intrinsic threshold, but when 
$\betaw$ exceeds $\alpha\Ptil$, $\jatil$ becomes negative, meaning that no anomaly appears in the velocity curve.

\subsection{Pure spin-transfer case ($\betaw=0$)}
In this subsection, we look into the case of $\betaw=0$, i.e., the dynamics driven by purely spin torque. 
As we see, a significant feature of the domain wall, intrinsic pinning, appears in this regime.
\subsubsection{Intrinsic pinning}
\begin{figure}[tb]
  \begin{center}
 \includegraphics[width=6cm]{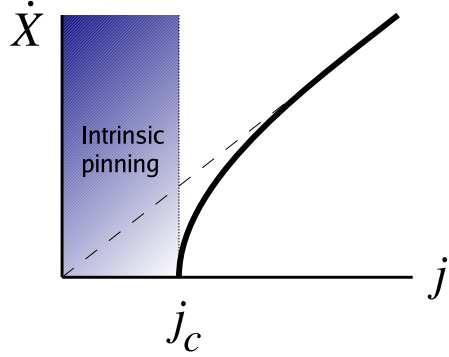}
  \end{center}
\caption{Domain wall velocity in the case of $\betaw=0$ with not too strong extrinsic pinning, $\Vztil \lesssim \alpha^{-1}$.
A finite threshold appears due to the intrinsic pinning due to perpendicular magnetic anisotropy energy.  \label{FIG:v-j}}
\end{figure}

In the case of $\betaw=0$ and $\Vztil=0$, the wall velocity becomes 
\begin{equation}
\average{\dot{X}} =
\left\{ \begin{array}{lrr} 
  0  &  \;\;\;\;\; & (\jtil < \jcitil) \\
 \frac{|\Ptil|}{1+\alpha^2}\sqrt{\jtil^2-(\jcitil)^2} 
  & \;\;\;\;\; & (\jtil \geq \jcitil )
\end{array}\right.
\end{equation}
and $\jcitil \equiv \frac{1}{\Ptil}$ becomes a threshold current density \cite{TK04} (Fig. \ref{FIG:v-j}).
This effect, which could be called an intrinsic pinning, arises from the fact that a domain wall is a collective object that can deform (i.e., can develop $\phiz$) and absorb spin torque.
In dimensioned quantities, the intrinsic threshold is at 
\begin{equation}
 \jci =  \frac{e S^2}{a^3 \hbar P} \Kp \lambda.
\end{equation}

The intrinsic pinning effect is robust against extrinsic pinning \cite{TK04}.
In fact, the dynamics near threshold is studied based on the equation of $\phiz$ derived by eliminating $X$ from Eq. (\ref{DWeq3nodim});
\begin{equation}
(1+\alpha^2) \partial_{\ttil}^2 \phiz
+ \alpha\partial_{\ttil}{\phiz}
\left(2\cos2\phiz +\Vztil\right)
+\Vztil\sin2\phiz + \jtil\Vztil\Ptil=0.
\label{phieq0}
\end{equation}
This equation indicates that $\phiz$ is like a classical particle moving in a potential
\begin{equation}
V_\phi=\frac{\Vztil}{1+\alpha^2}
\lt(\frac{1}{2}\sin2\phiz + \jtil\Ptil\phiz\rt).
\label{phipot}
\end{equation}
From Eq. (\ref{phieq0}), we see that the energy barrier for $\phiz$ vanishes when
\begin{equation}
\jctil\sim \Ptil^{-1}, \label{jcbetazero}
\end{equation}
irrespective of extrinsic pinning strength.
Once $\phiz$ escapes from a local minimum, its average velocity is given by Eq. (\ref{phieq0}) as
\begin{equation}
\partial_{\ttil}\phiz \simeq \frac{\jtil\Ptil}{\alpha}.
\end{equation} 
This velocity corresponds, via Eq. (\ref{DWeq4}), 
to a maxmum displacement of the wall inside the pinning potential,  
\begin{equation}
\Xtil_{\rm max} \simeq - \frac{1}{\Vztil} \partial_{\ttil}\phiz
 \simeq \frac{\jtil\Ptil}{\alpha\Vztil}.
\label{XII}
\end{equation}
Since $\alpha$ is small, $|\Xtil_{\rm max}|$ easily exceeds $\xi$ (unless $\Vztil \gtrsim \alpha^{-1}$), and the wall is depinned as soon as $\phiz$ runs.
Thus the threshold current is given by $\jctil\sim \Ptil^{-1}$, even in the presence of extrinsic pinning if
\begin{equation}
\Vztil \lesssim \frac{1}{\alpha}.
\end{equation}

Let us note that the domain wall moves even below the intrinsic threshold. Actually, the shift of the domain wall can be quite large.
Here we consider no extrinsic pinning and $\betaw=0$.
For $\jtil <\jatil$
($\jatil$ coinsides with intrinsic threshold), $\phiz$ reaches the value given by \Eqref{phizterm}.
Equation (\ref{DWeq3nodim}) then indicates that the shift of $\phiz $ is associated with the shift of the wall itself as
\begin{equation}
\Delta X= \frac{\lambda}{\alpha} \Delta\phiz 
= \frac{\lambda}{2\alpha}\sin^{-1} \Ptil\jtil.  \label{delXbelow}
\end{equation}
Therefore, even for a current 2\% of the intrinsic threshold,
the domain wall can move over a distance $\lambda$ (if $\alpha=0.01$).
Such motion at very low current would be enough for applications. 
For this, a very clean sample is required, since 
 even weak extrinsic pinning can result in creep motion in the low current regime \cite{Yamanouchi07}.


\subsubsection{Extremely strong pinning case}
\begin{figure}[tb]
  \begin{center}
  \includegraphics[width=6cm]{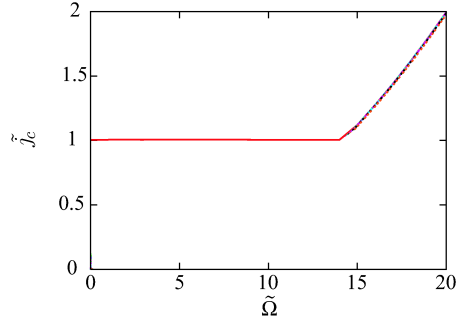}
  \end{center}
\caption{Threshold current $\jctil$ plotted as function of pinning frequency $\Omegatil$ for $\betaw=0$ with $\alpha=0.01$ and $\Ptil=1$. 
We see that intrinsic pinning persists if $\Omegatil \lesssim \sqrt{\frac{2}{\alpha}}$.  }
\label{FIG:jcbz}
\end{figure}

When $\Vztil \gtrsim \frac{1}{\alpha}$, the threshold  current 
is governed by the extrinsic pinning. 
In fact, Eqs. (\ref{DWvelocity})(\ref{DWeq4}) in this strong-pinned case (and $\betaw=0$) reduce to (within the pinning potential ($|X|<\xi$) and at large current)
\begin{eqnarray}
{\partial_\ttil} {\Xtil} & \simeq & \frac{1}{1+\alpha^2}  
 \left( \Ptil\jtil -\alpha\Vztil \Xtil \right) \nonumber\\
{\partial_\ttil} {\phiz} &\simeq & - \frac{1}{1+\alpha^2}\Vztil \Xtil
.
\end{eqnarray}
We thus see that stream motion of the wall occurs if $\jctil > \jcetil$, where
\begin{equation}
\jcetil \equiv \frac{\alpha}{\Ptil} \Vztil \xi.
\end{equation}
The behavior of the numerically determined $\jctil$ in the case of $\betaw=0$ is plotted in Fig. \ref{FIG:jcbz}.

\subsection{Effect of force ($\betaw$)}

Now we look into the case of non-vanishing $\betaw$.
We will see that there are three different behaviors depending on the extrinsic pinning strength.
The analysis here is based on Ref.  \cite{TTKSNF06}.
\subsubsection{Weak extrinsic pinning regime : I)}

Under weak current, $\jtil\lesssim 1$, $\phiz$ remains small and the wall dynamics is well described by $X$ only. 
This is the first regime I).
Linearizing the sine term in Eq. (\ref{DWeq3nodim}) as $\sin 2\phiz \simeq 2\phiz$, $\phiz$ can be eliminated to obtain a simple equation for $X$ as \cite{SMYT04,TSIK05}
\begin{equation}
(1+\alpha^2)\partial_{\ttil}^2 {\Xtil}
 +\frac{1}{\tautil}\partial_{\ttil}{\Xtil}
+\Omegatil^2 \Xtil = \tilde{F},
\label{DWeq2weakpin}
\end{equation}
where 
$1/\tautil=2\alpha\left(1+\frac{1}{2}\Vztil\right)$
and 
\begin{equation}
\tilde{F}\equiv 2\betaw \jtil,
\end{equation}
is a dimensionless force due to current.
We consider a case of steady current and weak damping; $2\Omegatil\tautil > 1$
A general solution to Eq. (\ref{DWvelocity})  
(inside the pinning potential) is given as
\begin{equation}
\Xtil(t)=  \frac{\betaw \jtil}{\Omegatil^2}
+e^{-\frac{\ttil}{2\tautil}}
 (A\cos\Omegatil' \ttil 
 + B\sin \Omegatil' \ttil )  ,
\end{equation}
where 
$\Omegatil' \equiv \sqrt{\Omegatil^2-\frac{1}{4\tautil^2}}$
and $A$, $B$ are constants.
The initial condition required is $\Xtil(0)=0$
and $\partial_{\ttil}\Xtil(0)=\Ptil \jtil$.
The second condition on the wall speed comes from the first equation in Eq. (\ref{DWeq3nodim}) (with $\phiz(0)=0$ and $\alpha\ll1$), and is the most important
consequence of the spin-transfer torque; namely, spin-transfer torque gives initial speed to the wall. 
With these initial conditions, we obtain a solution as 
\begin{equation}
\Xtil(t)=  \frac{2\betaw \jtil}{\Omegatil^2}
\left(1-e^{-\frac{\ttil}{2\tautil}}
 \left(\cos\Omegatil' \ttil 
 +\frac{1}{2\Omegatil' \tautil} \sin \Omegatil' \ttil \right) \right) 
+\frac{\Ptil \jtil}{\Omegatil'} 
e^{-\frac{\ttil}{2\tautil}} \sin \Omegatil' \ttil.
\end{equation}
The first part is governed by a force from $\betaw$ and the second is driven by a spin-transfer torque term.
In the case of small $\betaw$, the spin torque contribution leads to 
a maximum displacement 
\begin{equation}
|{\Xtil}_{\rm max}|\simeq \frac{\Ptil}{\Omegatil}\jtil,
\end{equation}
while
\begin{equation}
|{\Xtil}_{\rm max}|\simeq \frac{4\betaw}{\Omegatil^2}\jtil,
\end{equation}
if $\betaw$ is large. 
(We assumed here that damping is weak ($\Omegatil'\tau\gg1$).) 
The first regime corresponds to regime I-a) and the second to I-b).
The threshold current in each case is given as
\begin{equation}
{\jctil}^{\rm Ia)} \sim \frac{\Omegatil}{\Ptil} \frac{\xi}{\lambda},
\end{equation}
and
\begin{equation}
{\jctil}^{\rm Ib)} \sim \frac{\Omegatil^2}{4\betaw} \frac{\xi}{\lambda}.
\end{equation}
The crossover between regimes Ia) and Ib) occurs at
\begin{equation}
\betaw^{\rm c}\simeq \frac{\Ptil}{2}\Omegatil.
\end{equation}
In terms of dimensioned quantities, 
\begin{equation}
{\jc}^{\rm Ia)} \simeq  \frac{2\sqrt{2}S}{P} \frac{e}{a^3}\frac{\sqrt{\Kp\Vz}}{\hbar} \lambda
=\frac{2\sqrt{2}}{S}\sqrt{\frac{\Vz}{\Kp}} \jci,\label{jcIa}
\end{equation}
and
\begin{equation}
{\jc}^{\rm Ib)}= 
\frac{e\lambda}{4 \vc a^3}
\frac{\Omega^2}{|\betaw|}\xi 
 = \frac{Se\Vz}{\hbar a^3} 
\frac{1}{|\betaw|} \frac{\lambda^2}{\xi}
 =\frac{P}{S} \frac{1}{|\betaw|}\frac{\Vz}{\Kp} \frac{\lambda}{\xi} \jci.
 \label{jcIb}
\end{equation}
Note that simple comparison of the pinning force and $\tilde{F}$ in Eq. (\ref{DWeq2weakpin}) gives a result correct up to a numerical factor, 
${\jc}^{\rm Ib)}\sim \frac{1}{2}\frac{Se\Vz}{\hbar a^3} 
\frac{1}{|\betaw|} \frac{\lambda^2}{\xi}$.

The pinning strength $\Vz$ is experimentally accessible by driving the wall by a magnetic field.
A magnetic field $B$ along the easy axis adds a term in Eq. (\ref{dweqsimple})
\begin{equation}
f_{B} = \frac{g \muB}{\hbar}B, \label{fb}
\end{equation}
where $g=2$.
By a simple comparison of pinning force and magnetic field, $\Vz$ is written in terms of the  depinning magnetic field $\Bc$ as
\begin{equation}
\Vz= \frac{S}{2}g\muB \Bc \frac{\xi}{\lambda}, \label{VzB}
\end{equation}
 and so ${\jc}^{\rm Ib)}$ is simplified to be
\begin{equation}
{\jc}^{\rm Ib)} = \frac{e }{\hbar a^3}\frac{S^2}{2}g \muB \Bc \lambda
   \frac{1}{|\betaw|}  .
 \label{JcIb}
\end{equation}

\subsubsection{Intermediate regime (intrinsic pinning): II}
When the extrinsic pinning is not weak, the threshold current becomes $\jctil\gtrsim O(1)$. 
For such a case, $\phiz$ no longer remains small, and the wall dynamics is governed by this angle variable. 
In fact,  the effective mass of a $\phiz$-"particle" is given by $1/\Vz$ \cite{TT96}(see Eq. (\ref{phieq})), and it becomes lighter than the corresponding mass of an $X$-"particle" given by $1/\Kp$ when extrinsic pinning is strong, $\Omegatil\gtrsim1$.
By eliminating $X$ from Eqs. (\ref{DWeq3nodim}), we obtain
\begin{equation}
(1+\alpha^2) \partial_{\ttil}^2 \phiz
+ \alpha\partial_{\ttil}{\phiz}
\left(2\cos2\phiz +\Vztil\right)
+\Vztil\sin2\phiz + \jtil\Vztil\Ptil=0.
\label{phieq}
\end{equation}
Thus $\betaw$ does not affect the dynamics of $\phiz$.
(Correctly, this feature is specific to a harmonic pinning potential, and unharmonicity results in the appearance of $\betaw$.
In fact, the $\beta$ term is eliminated from the equation of motion if one replaces $X$ in Eq. (\ref{DWeq3nodim}) by 
$X'\equiv X-\frac{2\betaw}{\Omegatil^2}\jtil$ (i.e., shift of stable point of $X$).
Even in the unharmonic case, nevertheless, we have numerically checked that the $\betaw$ does not lead to an important contribution in this regime.)

From Eqs. (\ref{phieq})(\ref{XII}), 
the threshold is roughly given by $\jctil\sim \Ptil^{-1}$,
and is actually found numerically as
\begin{equation}
\jctil\sim 0.7\times \Ptil^{-1}.\label{jcII}
\end{equation}

This story is the same as the case of $\betaw=0$, \Eqref{jcbetazero}, but
with a different numerical factor ($0.7$ here instead of $1$ for $\betaw=0$).
This difference comes from a different definition of threshold current with and without $\betaw$.
In the analysis of the $\betaw=0$ case, 
even if $X$ escapes from the pinning center at current 
$\jtil > 0.7\Ptil^{-1}$,
the terminal velocity vanishes if $\jtil <\Ptil^{-1}$, since the motion stops
due to the intrinsic pinning effect (i.e., $\phiz$ reaches a steady value and $\dot{X}$ becomes zero).
On the other hand, if $\betaw\neq 0$, steady motion of $X$ is possible as soon as $X$ escapes from the pinning.
This is the reason the threshold value in the intermediate regime is different for $\betaw=0$ and $\betaw\neq0$
(Fig. \ref{FIGjc}).
If $\betaw < 0$, or more precisely, if the relative sign between 
$\betaw$ and $\tilde P$ in Eq.(\ref{DWeq3nodim}) is negative, the $\betaw$-term 
will drive the depinned wall back to the pinning center, and the 
threshold in this regime is given by the intrinsic value $\jctil=1$.

\subsubsection{Strong pinning regime : III}

Equation (\ref{XII}) indicates that for extremely strong pinning, 
$\Vztil \gtrsim \alpha^{-1}$, the wall is not always depinned even after $\phiz$ escapes from the potential minimum. 
Depinning occurs at
\begin{equation}
\jctil\sim \frac{\alpha\Vztil}{\Ptil}\frac{\xi}{\lambda},
\label{jcIII}
\end{equation}
as in the case of $\betaw=0$.

\begin{figure}[tbp]
\begin{center}
\includegraphics[width=6cm]{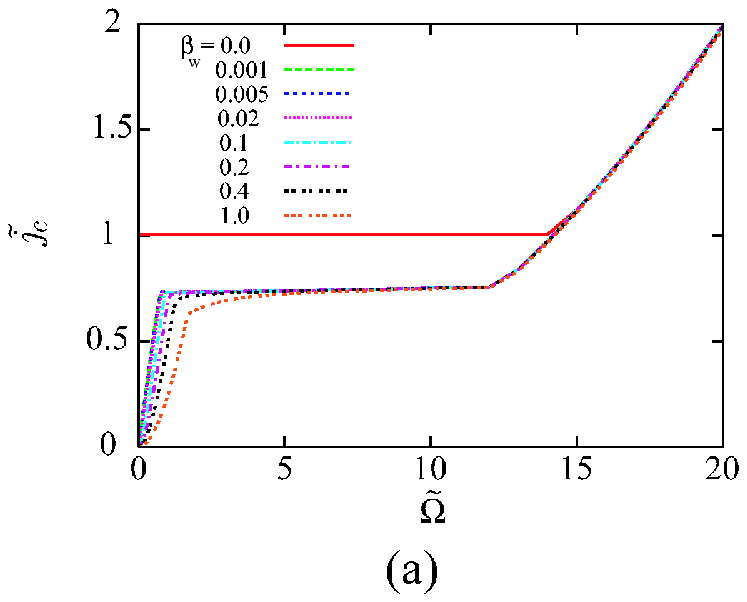}
\includegraphics[width=6cm]{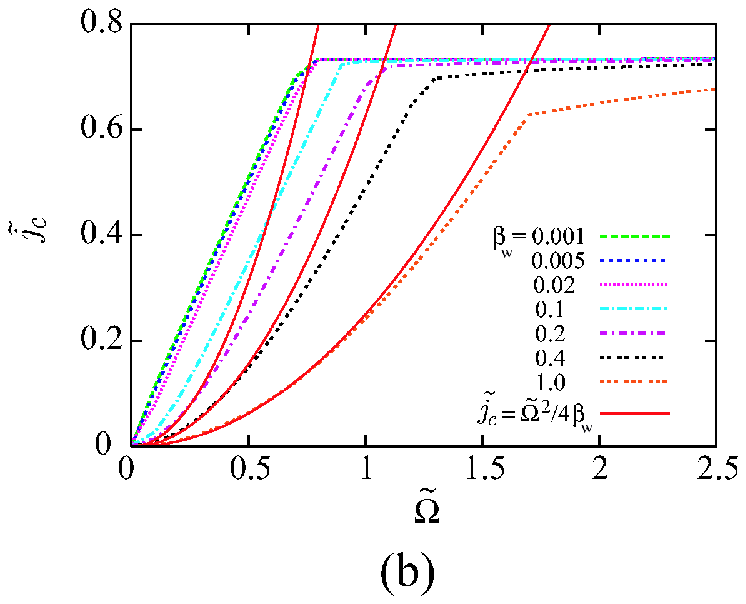}
\caption{(a): Threshold current $\jctil$ plotted as function of pinning frequency $\Omegatil\equiv\sqrt{2\Vztil}$ for several values of $\betaw$ with $\alpha=0.01$ and $\Ptil=1$. 
(b): Threshold current in weak pinning regime. 
Fitted curves for $\betaw\gtrsim 0.1$ are 
$\jctil\propto\Omegatil^2$.
For small $\betaw(\lesssim0.02)$, $\jctil$ is linear in $\Omegatil$.
\label{FIGjc} }
\end{center}
\end{figure}

\begin{figure}[tbp]
\begin{center}
\includegraphics[width=6cm]{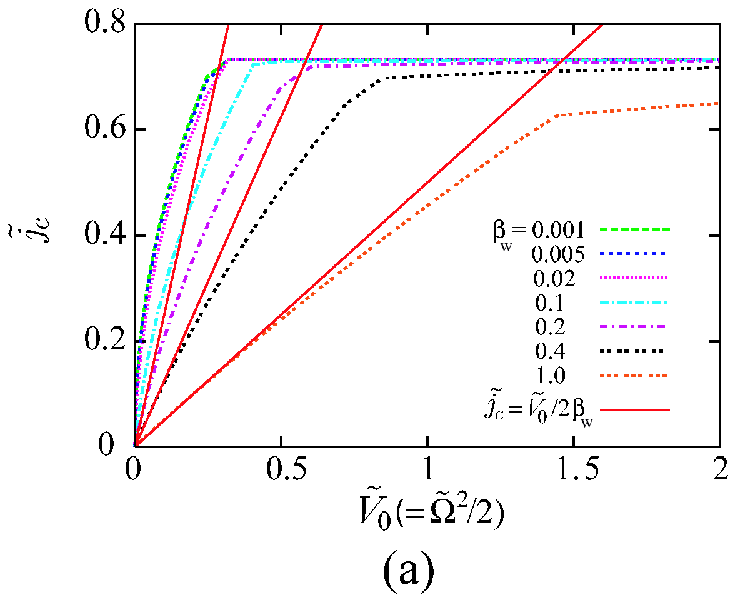}
\includegraphics[width=6cm]{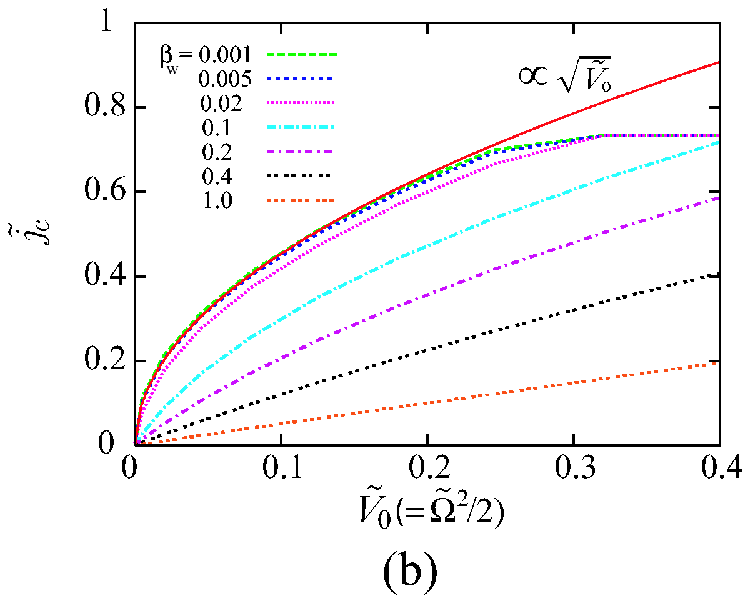}
\caption{(a): Threshold current $\jctil$ in weak pinning regime plotted as function of pinning strength $\Vztil$ for several values of $\betaw$ with $\alpha=0.01$ and $\Ptil=1$. 
For $\betaw\gtrsim0.1$, $\jctil\simeq \frac{\Vztil}{2\betaw}$ is linear in $\Vztil$.
(b): Weaker pinning regime. 
For small $\betaw(\lesssim0.02)$, $\jctil\propto\sqrt{\Vztil}$.
\label{FIGjcv} }
\end{center}
\end{figure}

Figure \ref{FIGjcv} shows a numerically determined threshold.
It is clearly seen in Fig. \ref{FIGjc}(a) that behaviors for $\betaw=0$ and 
$\betaw\neq 0$ are quite different except for the extremely strong pinning regime ($\Vztil \gtrsim 100\simeq 1/\alpha$).

\begin{table}[tb]
\caption{Summary of threshold current for strength of extrinsic pinning. 
The third, fourth, and fifth columns indicate the origin of threshold (extrinsic (E) or intrinsic (I)), critical current,  
and mechanism of depinning (either spin transfer (ST) or force ($\betaw$)), respectively.
The last column is a "good" variable to describe depinning.
\label{TABLE}}
\begin{center}
\begin{tabular}{c|c|c|c|c|c}
    &  condition   & threshold & $\jc$ & depinning &   \\ \hline
I-a  & $\Omegatil \lesssim 1$, $\betaw\lesssim \Omegatil$ &  
E & $\propto \sqrt{\Kp\Vz}$ & ST  & $X$ \  ($\phiz\sim0$) \\ 
I-b  & $\Omegatil \lesssim 1$, $\betaw\gtrsim \Omegatil$ & E & $\propto {\Vz}/\betaw$ & $\betaw$ &  $X$ \  ($\phiz\sim0$) \\ 
II & $1 \lesssim \Omegatil \lesssim \alpha^{-1}$ & I & $\propto \Kp$ & ST &  $\phiz$ \\ 
III &  $\Omegatil \gtrsim \alpha^{-1}$  & E & $\propto {\Vz}/\alpha$ & ST & $\phiz$  \\ \hline
\end{tabular}
\end{center}
\end{table}
Results of threshold current is summarize in Table \ref{TABLE}.
It is interesting that such a simple set of equation of motion results in so rich behaviors.

\subsection{Wall speed}

The wall speed after depinning is obtained as function of current based on Eq. (\ref{averagewallvelocity}) 
as in Fig. \ref{FIGvj}.
The anomaly at $\jatil$ are seen in Fig. \ref{FIGvj} in the case of 
$\betaw=0.001$ (at $\jtil=\frac{1}{0.9}\sim 1.1$)
and 
$\betaw=0.02$ (at $\jtil=1$), and the behavior is in agreement with 
numerical result of Ref.  \cite{Thiaville05}, where 
the anomaly of velocity under current was first reported.
This anomaly is essentially the same as the anomaly under magnetic field, known as Walker breakdown \cite{Schryer74,Hubert00}.

\subsection{Effect of external magnetic field}
Let us here briefly consider the effect of external magnetic field along easy axis.
From \Eqsref{dweqsimple}{fb}, the field replaces
$\betaw \jtil$ by $\betaw \jtil +b$, where
$b\equiv -\frac{g\muB B_z\lambda }{\hbar \vc} = 
-\frac{2g \muB B_z}{S\Kp}$ ($g=2$).
The average wall velocity with $\Vztil=0$ is then given 
if $\jtil \geq \jatil$ by
\begin{equation}
\average{\dot{X}} \rightarrow \frac{\betaw\jtil+b}{\alpha} 
+\frac{\sgn[[(\Ptil-\frac{\betaw}{\alpha})\jtil]- \frac{b}{\alpha}]}
{1+\alpha^2}
\sqrt{\left[\left(\Ptil-\frac{\betaw}{\alpha}\right)\jtil- \frac{b}{\alpha}\right]^2-1},
\end{equation}
 and by
\begin{equation}
\average{\dot{X}} \rightarrow \frac{\betaw \jtil+b}{\alpha} ,
\end{equation}
if $\jtil < \jatil$,
where the anomaly now occurs at
\begin{equation}
\jatil \rightarrow \frac{b\pm \alpha}
{\Ptil \alpha-{\betaw} },
\end{equation}
the $\pm$ denotes $\sgn[(\Ptil \alpha-{\betaw})-b/\jatil]$.

\subsection{Comparison with experiments}

Let us here try to explain the experimental results \cite{Yamaguchi05} assuming the wall there is planar and rigid. 
We first consider the case of regime I-a).
Assuming $\xi\sim \lambda$, we estimate the pinning potential  from the measured depinning field 
$\Bc=0.01-0.1$ T as 
$\Vz= 0.34\times (10^{-2} \sim 10^{-1})$ K
$=4.7\times (10^{-26} \sim 10^{-25})$ J, i.e.,
$\frac{\Vz}{\Kp} = 1.4\times (10^{-3} \sim 10^{-2})$, and so 
${\jc}^{\rm Ia)} = (0.21\sim0.67)\times \jci$. 
This value is still too big to explain the experimental value. 
Velocity jump is estimated as \cite{TTKSNF06} 
$\Delta v^{\rm Ia)} =\frac{\betaw}{\alpha}\times 839$ m/s, so an extremely small $\betaw$ ($\frac{\betaw}{\alpha}\sim 4\times 10^{-3}$) is required to explain the experimental value of $\Delta v\sim 3$ m/s \cite{Yamaguchi04}.
If we assume regime I-b), the threshold is
${\jc}^{\rm Ib)} = \frac{1}{|\betaw|} \times 
 2.8 \times (10^{-3} \sim 10^{-2}) \times \jci$.
The experimental value could be reproduced if $\betaw= 0.1 \sim 1$.
But such a large value of $\betaw$ cannot be explained within the current understanding that $\betaw$ arises from either non-adiabaticity \cite{TK04,Thiaville05} or spin relaxation \cite{Zhang04}.
Instead, $\Delta v$ cannot be explained by use of the above $\Vz$ assuming I-b), as it predicts too large a value of 
$\Delta v^{\rm Ib)}=10^3$ m/s.
Thus, honestly, none of the above predictions based on a rigid 1D wall neglecting the temperature rise due to heating is successful in explaining the experimental result for metals quantitatively.

There are some possibilities to resolve the disagreement. 
The most probable one would be the heating effect of the current, which has been reported to be important \cite{Yamaguchi05,You07}. 
The estimate of $\Vz$ by use of experimental $B_c$ could be an over estimate 
if the effective barrier height $\Vz$ is greatly reduced by heating under current, while such heating does not occur under a static magnetic field.
Let us estimate the pinning potential that gives the experimental 
value of $\jc$.
Assuming regime I-a), the experimental value of $\jc/\jci=0.02$ is reproduced if
$\mu\equiv \frac{\Vz}{\Kp}=1.3\times 10^{-5}$, which corresponds to 
$\Vz=3\times 10^{-5}$ K$=4.5\times 10^{-5}$ T.
This is two orders of magnitude smaller than the value extracted from $B_{\rm c}$.
For I-b), we have $\mu=\betaw\times 10^{-2}$.
From the experiment, 
$\Delta v /(\frac{a^3}{e}\jc) = (3$ m/s)$/(67$ m/s)$=0.05$. 
This value is equal, for regime I, to $\frac{\betaw}{\alpha}$, so
$\betaw=5\times 10^{-4}$ if $\alpha=0.01$.
So in case I-b), $\mu=5\times 10^{-6}$.
Thus, assuming either regime I-a) or I-b), the experimental results could be explained by a pinning potential extremely weakened by heating,
$\frac{\Vz}{\Kp}=10^{-6}\sim 10^{-5}$.

As far as the transverse wall is concerned, the assumption of rigidity seems not essential since the analytical results for a rigid wall \cite{TK04,TKSLL07} and numerical simulation including deformations \cite{Thiaville05,Seo07} predict similar behaviors.
For a vortex or a vortex wall, the behavior is very different from that of rigid wall, as we see next.

\section{Motion of vortex}
\label{SEC:vortex}

\begin{figure}[tbh]
  \begin{center}
  \includegraphics[width=0.4\linewidth]{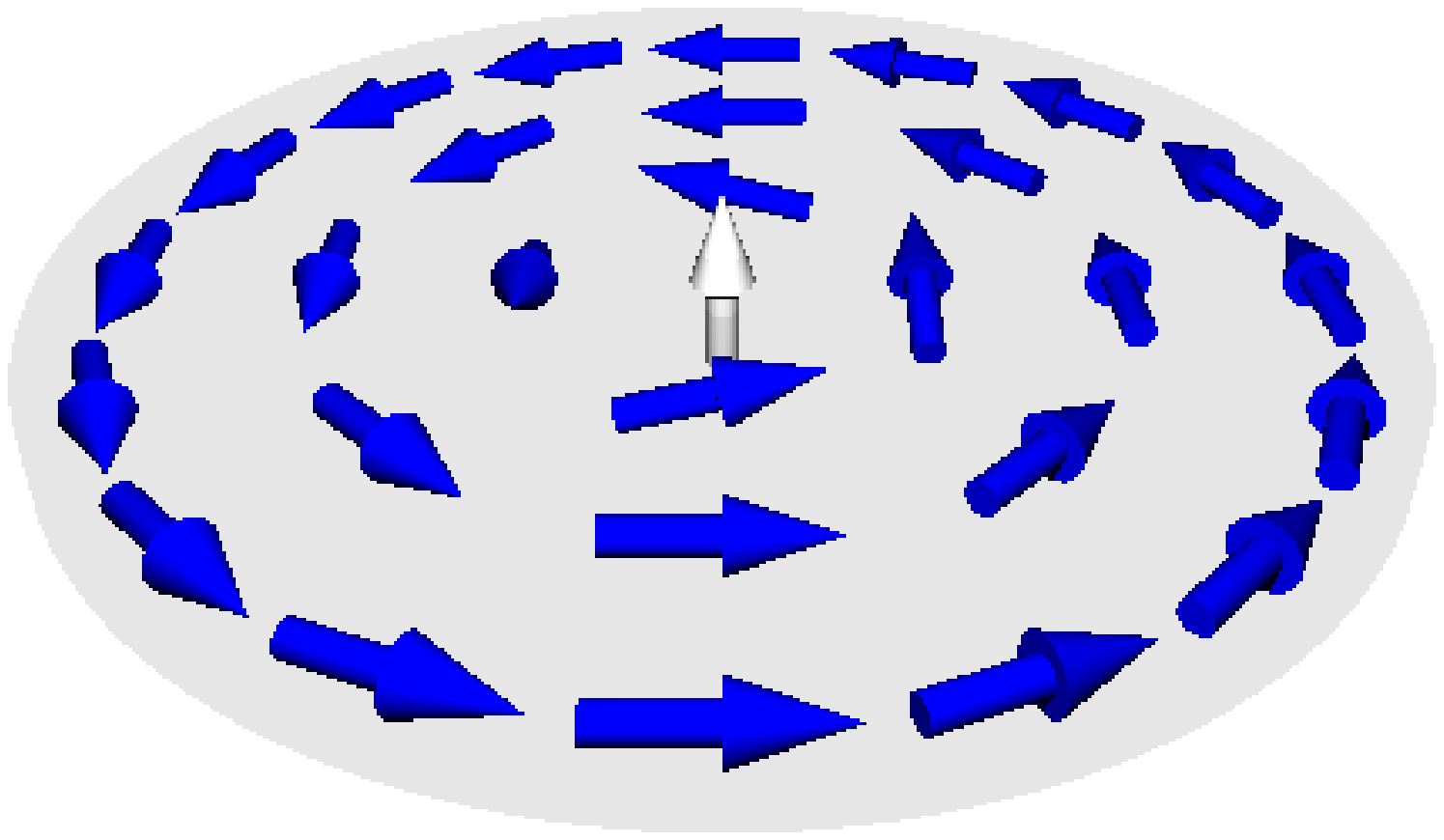}
  \includegraphics[width=0.4\linewidth]{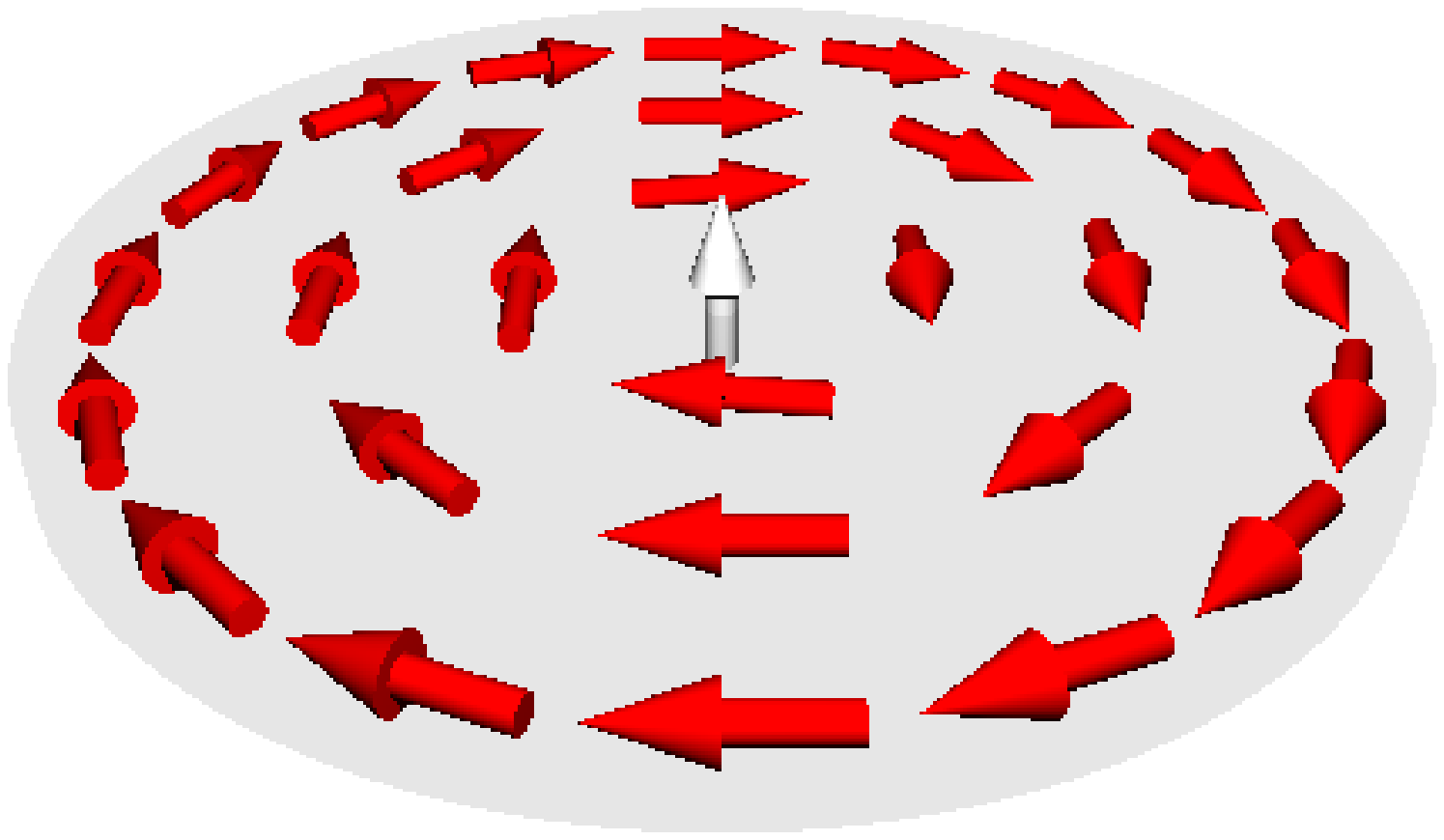}
  \includegraphics[width=0.4\linewidth]{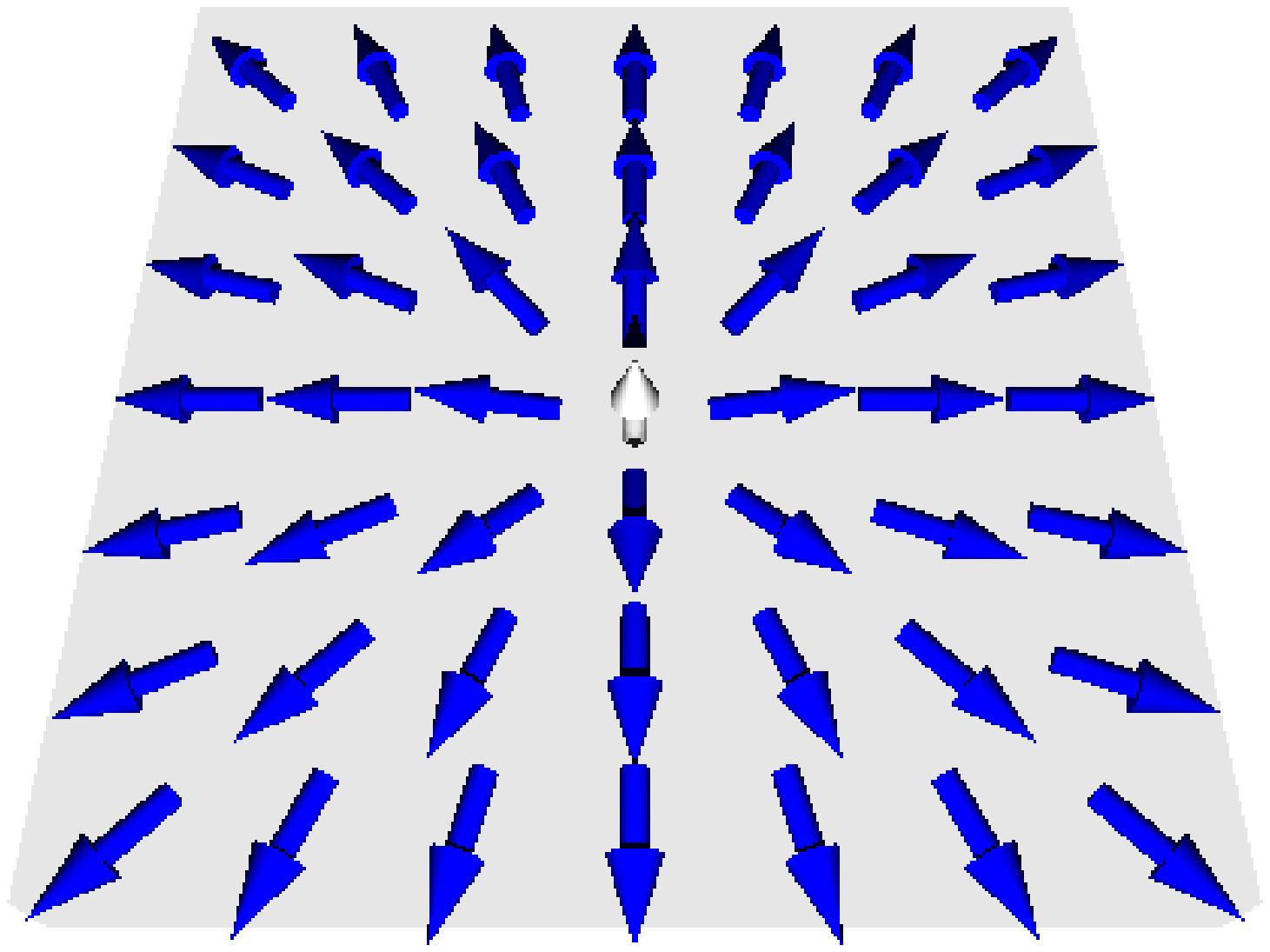}
  \includegraphics[width=0.4\linewidth]{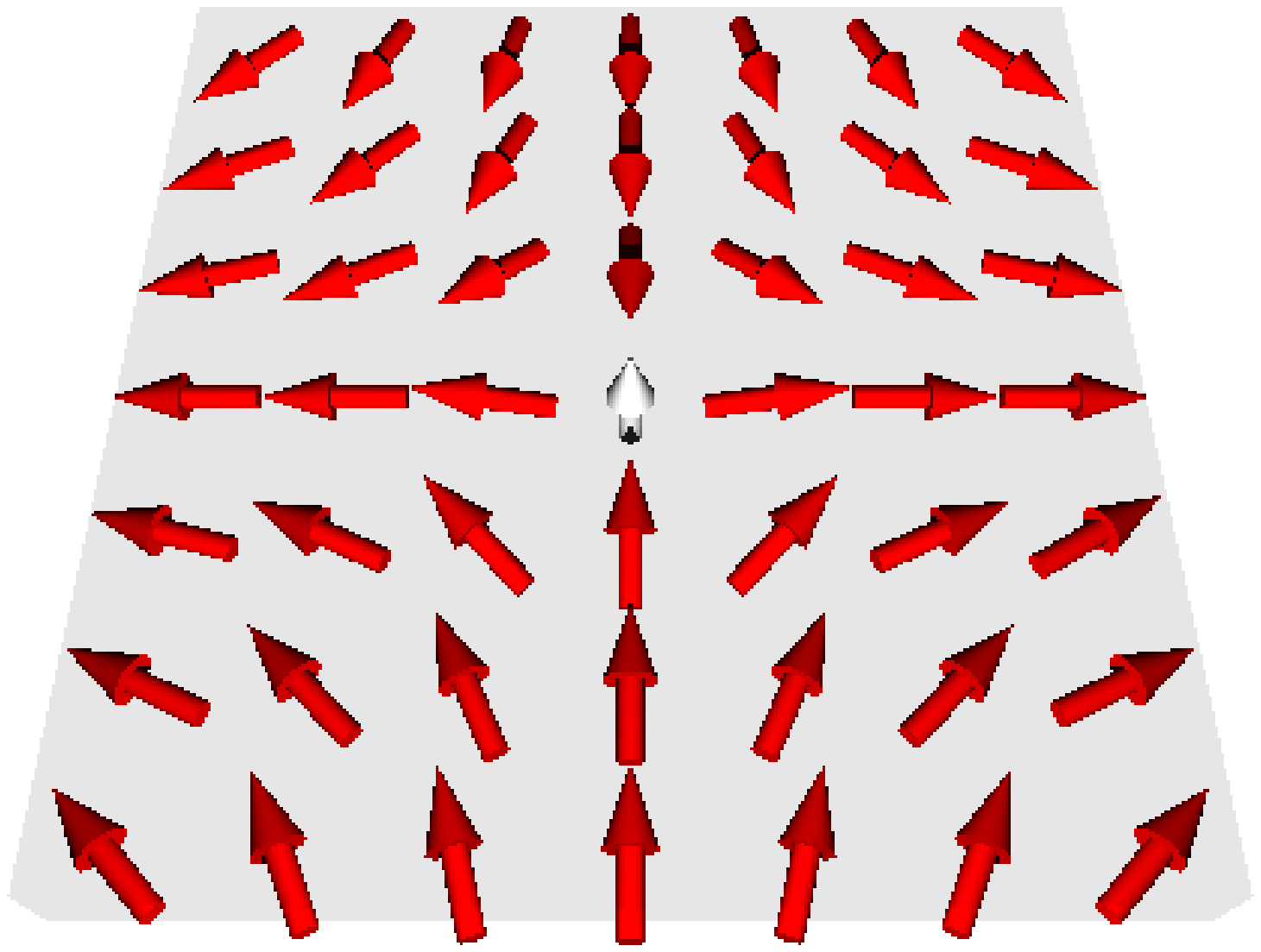}
\caption{
Examples of vortex structure. 
(Core size is very small.)
The first two structures are realized on a small circular disk \cite{Shinjo00,Yamada07}, since the magnetic charge appearing on the edge is zero. 
Structures denoted by blue and red have topological charges of $\nvortex=1$ and $\nvortex=-1$, respectively.
\label{FIGvortex}
}
  \end{center}
\end{figure}

In this section we briefly consider the motion of a vortex in two dimensions (Fig. \ref{FIGvortex}).
The dynamics of a vortex is very different from that of a planar (transverse) domain wall.
Actually, the canonical momentum of vortex position $X$ in the $x$-direction is $Y$, the position in the $y$-direction. 
This leads to vanishing of the intrinsic threshold, which is significantly different from the case of a domain wall, whose canonical momentum is $\phiz$.
This fact may explain why vortices or vortex walls move more easily than (transverse) domain walls \cite{Klaui05,Seo07}.
 
The Lagrangian of a single vortex is given by
\Eqref{LeffST} with a single vortex solution centered at 
$\Xv(t)=(X(t),Y(t))$.
It is given as \cite{SNTKO06}
\begin{equation}
L_{\rm v}=\frac{1}{2}\Gv\cdot(\dot{\Xv}\times\Xv)
     -\Gv\cdot(\vsv\times\Xv) -U(\Xv),
\end{equation}
where $\Gv\equiv G\evz $ 
($G\equiv\frac{4\pi \hbar S}{a^3}\Phi_{xy}$) is the gyrovector of \Eqref{Gvdef}, and the potential energy $U$ is from $\Hs$.
The equation of motion in two dimensions is thus given by \cite{SNTKO06}
\begin{equation}
\dot{\Xv}-\vsv=\frac{1}{G}\evz\times(\Fv-\tilde\alpha \dot{\Xv}),
\label{veq1}
\end{equation}
where 
$\Fv\equiv -\frac{\partial}{\partial \Xv}U(\Xv)$ and
$\tilde\alpha\equiv \alpha\frac{\hbar S}{a^3} \thickness
\int d^2 x [(\nabla\theta)^2+\sin^2\theta (\nabla\phi)^2]$.
As seen, the vortex has a velocity perpendicular to the applied force, $\dot{\Xv}\perp\Fv$, which might sound strange but is an interesting feature of a topological object.
The force due to current is given by Eq. (\ref{totalF}), and so
Eq. (\ref{veq1}) reduces to
\begin{equation}
\dot{\Xv}=\frac{a^3}{2eS}\jsv
+ \evz\times
\left(\frac{f}{e}\jv
-\frac{a^3\tilde\alpha}{4\pi S \Phi_{xy}}
 \dot{\Xv} \right),
\label{veq2}
\end{equation}
where 
$f=\frac{a^3}{8\pi S^2\Phi_{xy}}
( 2Se^2 \Ne\rhos+\Fbetafactor P \beta_{\rm sf})$.
The perpendicular force on the vortex of the adiabatic origin ($\jv\times \Gv$ in Eq. (\ref{totalF}))
is thus simply a spin torque, which induces the vortex to flow in the current direction.
This is easily understood since the spin torque and the time-derivative term of the Landau-Lifshitz equation are combined into a Lagrange derivative along the current flow, 
$\partial_t \rightarrow 
\left( \partial_t-\frac{1}{2eS}\jsv\cdot\nabla \right)$, 
which indicates that spin transfer torque drives any spin structure along the spin current. 
(This is seen from \Eqref{LeffST}.)

A very important consequence of vortex motion described by \Eqref{veq2} is that the vortex (or vortex wall) has no intrinsic threshold \cite{SNTKO06}.
In fact, as soon as any small current is applied, \Eqref{veq2} indicates flow of the vortex, if extrinsic pinning is absent.
This seems to be consistent with experimental and numerical observations \cite{Klaui05,Seo07}.

Let us consider a single vortex in a film in more detail.
The Hamiltonian of localized spin is modeled as
\begin{equation}
H_v=\frac{S^2\thickness}{2}\int  d^2x [
J[(\nabla\theta)^2+\sin^2\theta(\nabla\phi)^2]
+\Kp\cos^2\theta].
\end{equation}
Since we cannot obtain the vortex solution analytically, we approximate it as
\begin{equation}
\phi=\tan^{-1}\frac{y}{x}\pm\frac{\pi}{2}, \;\;\; 
\theta=
\left\{ \begin{array}{cc}
\frac{\pi}{2} & r> \lamv \\
\frac{\pi }{2(1-e^{-1})} (1-e^{-r^2/\lamv^2})  & r \leq \lamv
\end{array} \right.,
\end{equation}
where $r\equiv\sqrt{x^2+y^2}$ and
$\lamv$ is the size of the vortex core.
Then the gauge field is given as
$A_i^\pm\sim \pm i \frac{x_i(x\pm i y)}{\lamv^2 r} e^{-r^2/\lamv^2} \theta(\lamv-r)$, neglecting the small contribution from outside the core.
Choosing the current direction as $x$, we obtain
the Fourier transform as ($A$ is the area of the film)
\begin{equation}
A_x^\pm(\qv)=\pm i \pi\frac{\lamv}{2A} (\sqrt{\pi}(1-2q^2\lamv^2)+2iq\lamv)
 e^{-\frac{\lamv^2 q^2}{4}}.
\end{equation}
The resistivity due to the core is then calculated as
\begin{eqnarray}
\rho_{\rm v} &\simeq&  \frac{\pi^2}{2}
\frac{m^2 \kf \Delta^2 \lamv^2}{e^2 n^2 A}
 e^{-2\zeta^2 \kf^2\lamv^2 },
\end{eqnarray}
where we used the fact that the resistivity is dominated by the contribution from $q\lesssim k_{F+}-k_{F-}=2\kf\zeta$.
The nonadiabatic force is then given as
\begin{equation}
\Fna \simeq \frac{j}{e}\hbar \left(\frac{\Delta}{\eF}\right)^2 
(\kf\lamv)^2 d  e^{-2\zeta^2 \kf^2\lamv^2 }.
\end{equation}
On the other hand, the force due to the topological Hall effect is given by
\begin{equation}
\Fad \simeq \frac{j}{e}\hbar 
\end{equation}
for a vortex with $\nvortex=1$.


Let us estimate the magnitude of the Hall effect.
The Hall resistivity is given as
\begin{equation}
\rhoxy\simeq  \frac{2\pi S^2 P \nvortex}{\hbar e^2 n A}.
\end{equation}
The Hall conductivity is given by
$\sigma_{xy}=\sigma_0
 \frac{\rho_{xy}/\rho_0}{1+(\rho_{xy}/\rho_0)^2}$ 
( $\sigma_0=\rho_0^{-1}$ is the Boltzmann conductivity) and hence
the ratio of the Hall current to applied current 
is obtained as 
\begin{equation}
\frac{j_\perp}{j_0} =  \frac{\rho_{xy}}{\rho_0}
   \sim 2\pi S^2 P\nvortex \frac{\ell}{\kf A}.
\end{equation}
Thus, the deviation of the electric current becomes significant if 
the ratio $ \frac{\rho_{xy}}{\rho_0} $ is of the order of unity.
The deviation of current due to the Hall effect would be large in clean samples.

Current-induced dynamics of a vortex in a disk was observed in Ref.  \cite{Yamada07}.


\section{Spin torques in LLG equation}
\label{SEC:singlespin}

 In this section, we describe the present theoretical status of 
the microscopic derivation of Landau-Lifshitz-Gilbert (LLG) equation in the presence of current. 
 We focus on magnetization dynamics that is slowly varying in 
space and time, as described by the LLG equation with local torques. 
 Here \lq slow' means slow compared to the electronic scales, 
so it is satisfied quite well in most cases for metallic systems. 
One important aim here is to calculate microscopically the $\betasf$ term and damping due to the electron spin relaxation. 
 Throughout this section, we consider general spin configurations.

\subsection{General}

 The LLG equation is given by
\begin{equation}
  \frac{d{\bm M}}{dt} = \gyroz {\bm H}_{\rm eff} \times {\bm M} 
  + \frac{\alphaz}{M}{\bm M} \times \frac{d{\bm M}}{dt} 
  + {\bm T}_{\rm el} , 
\label{eq:LLG00}
\end{equation}
in terms of the magnetization vector ${\bm M}$. 
 The first term on the right-hand side represents the precessional torque 
around the effective field ${\bm H}_{\rm eff}$, 
with $\gyroz(\equiv\frac{g\mub}{\hbar}=-\frac{ge}{2m}>0)$ being the bare gyromagnetic ratio (without modification by conduction electrons). 
 The effective field includes the ferromagnetic exchange 
(gradient energy) field, 
the magnetocrystalline anisotropy field, and the demagnetizing field. 
 The second term represents the Gilbert damping, coming from 
processes that do not involve conduction electrons, 
and is thus present even in insulating ferromagnets. 
 The effects of conduction electrons are contained in the 
third term, ${\bm T}_{\rm el}$, called the spin torque in particular. 
 This term comes from the $s$-$d$ exchange coupling $H_{sd}$ 
to conduction electrons, and is given by 
\begin{equation}
 {\bm T}_{\rm el} 
 = - M \, \nv({\bm r}) \times 
     \langle \sigmav({\bm r}) \rangle_{\rm ne} .
\label{eq:T_el_vec}
\end{equation}
(We wrote here explicitly $ \average{\ }_{\rm ne}$ to indicate that the expectation value is taken in the non-equilibrium state with current flow or dynamical magnetization. 
See \S\ref{SEC:singlespingeneral}.)

 For notational convenience, we introduce a unit vector, $\nv$, 
whose direction is in the $d$-spin direction, 
hence is opposite to magnetization direction, 
\begin{equation}
  {\bm M} = - \gyroz \frac{\hbar S}{a^3} \nv .
\end{equation}
 Note that the magnetization is, by definition, the 
magnetic moment per unit volume, hence 
$|{\bm M}| = \hbar \gyroz S/a^3$. 
 In terms of $\nv$, the LLG equation is written as 
\begin{equation}
  \dot \nv = \gyroz {\bm H}_{\rm eff} \times \nv 
  + \alphaz \dot \nv \times \nv + {\bm t}_{\rm el}' ,
\label{eq:LLG}
\end{equation}
where we have put 
${\bm T}_{\rm el}({\bm M}) = - (\hbar S/a^3) {\bm t}_{\rm el}'(\nv)$. 
 The dot represents the time derivative.

 For long-wavelength, low-frequency dynamics, it may be sufficient 
to consider spin torques that are first order in space/time 
derivatives. 
 Let us call such torques as {\it adiabatic torques}. 
 In the presence of rotational symmetry in spin space, they are expressed as  
\begin{equation}
 {\bm \tau}_{\rm ad}^{0\prime} 
= - ({\bm v}_{\rm s}^0 \!\cdot\! {\bm \nabla})\,  \nv 
  - \betasf \, \nv \times ({\bm v}_{\rm s}^0 \cdot\! {\bm \nabla}) 
    \, \nv 
  - \alphasf \, (\nv \times \dot \nv) 
  - \frac{\delta S}{S} \, \dot \nv . 
\label{eq:tau_ad}
\end{equation}
 The first term on the right-hand side is the celebrated 
spin-transfer torque \cite{Bazaliy98,Ansermet04,Rossier04,STK05,Xiao06}, 
where 
\begin{equation}
 {\bm v}_{\rm s}^0 =  \frac{a^3}{2eS} \, {\bm j}_{\rm s} 
\label{eq:v_s1}
\end{equation}
is the (unrenormalized) \lq\lq spin-transfer velocity,", 
with ${\bm j}_{\rm s}$ being the spin-current density. 
 The second term, sometimes called the \lq $\beta$-term', 
comes from spin-relaxation processes of electrons  
\cite{Zhang04,Thiaville05,Barnes05,Tserkovnyak06,KTS06,Piechon07,Duine07}. 
 Here $\betasf$ is a dimensionless constant. 
 The third term is the Gilbert damping, which also results from 
spin relaxation of electrons. 
 The fourth term contributes as a \lq\lq renormalization'' of spin, 
as seen below. 

 If the magnetization varies rapidly, we have in addition a 
{\it non-adiabatic torque}, ${\bm \tau}_{\rm na}^0$, 
which is oscillatory and nonlocal
(see \S\ref{SEC:singlespinnonadiabatic}). 
 The total torque may thus be given by the sum of the two, 
\begin{equation}
 {\bm t}_{\rm el} '
= {\bm \tau}_{\rm ad}^{0\prime}  + {\bm \tau}_{\rm na}^{0\prime} . 
\label{eq:t_el2}
\end{equation}

 The LLG equation (\ref{eq:LLG}) is then written as 
\begin{equation}
  \left( 1 + \frac{\delta S}{S} \right) \dot \nv 
= \gyroz {\bm H}_{\rm eff} \times \nv 
  - ({\bm v}_{\rm s}^0 \!\cdot\! {\bm \nabla})\,  \nv 
  - \betasf \, \nv \times 
    ({\bm v}_{\rm s}^0 \cdot\! {\bm \nabla}) 
    \, \nv 
  - (\alphaz + \alphasf) \nv \times \dot \nv 
  + {\bm \tau}_{\rm na}^{0\prime} .
\label{eq:LLG1}
\end{equation}
 Here we have transposed the \lq spin renormalization' term 
to the left-hand side. 
 We define the total (\lq\lq renormalized'') spin as 
\begin{eqnarray}
 S_{\rm tot} = S + \delta S ,
\label{eq:S_tot}
\end{eqnarray}
with $\delta S$ being the contribution from conduction electrons (\Eqref{stotdef}), 
and divide both sides of Eq.(\ref{eq:LLG1}) by $S_{\rm tot}/S$. 
 Then we arrive at 
\begin{equation}
  \dot \nv 
= \gamma {\bm H}_{\rm eff} \times \nv 
  - ({\bm v}_{\rm s} \!\cdot\! {\bm \nabla})\,  \nv 
  - \betasf \, \nv \times ({\bm v}_{\rm s} \cdot\! {\bm \nabla}) 
    \, \nv 
  - \alpha \, (\nv \times \dot \nv) 
  + {\bm \tau}_{\rm na}' ,
\label{eq:LLG2}
\end{equation}
where 
$\gyro = (S/S_{\rm tot}) \, \gyroz$, 
$\alpha = (S/S_{\rm tot}) (\alphaz + \alphasf )$, 
${\bm \tau}_{\rm na}' = (S/S_{\rm tot}) \, {\bm \tau}_{\rm na}^{0\prime}$, 
and 
\begin{eqnarray}
  {\bm v}_{\rm s}
= \frac{S}{S_{\rm tot}} \, {\bm v}_{\rm s}^0 
=  \frac{a^3}{2eS_{\rm tot}} {\bm j}_{\rm s} ,
\label{eq:v_s2}
\end{eqnarray}
is the \lq\lq renormalized'' spin-transfer velocity. 
(These torques are schematically shown in Fig. \ref{FIGspintorques}.)


 In the parameter space of the LLG equation, \Eqref{eq:LLG2}, the manifold of 
$\alpha = \betasf$ (with ${\bm \tau}_{\rm na}' = {\bm 0}$) provides 
a very special case for the dynamics, 
and there has been a controversy whether the relation 
$\alpha = \betasf$ holds generally or not. 
 If $\alpha = \betasf$, the following peculiar dynamics are expected. 
 (i) Any static solution, $\nv({\bm r})$, in the absence of spin 
current is used to construct a solution, 
$\nv({\bm r}- {\bm v}_{\rm s} t)$, in the presence of spin current 
${\bm v}_{\rm s}$. 
 (ii) The current-induced spin-wave instability does not occur \cite{Tserkovnyak06}.

 The relation $\alpha = \betasf$ was originally suggested in 
Ref.\cite{Barnes05} based on the Galilean invariance of the system. 
 Although one may argue that the Galilean invariance should be valid 
for the long-wavelength and low-frequency dynamics in which the 
underlying lattice structure is irrelevant, the $\alpha$ and $\betasf$ 
come from spin-relaxation processes \cite{Zhang04}, 
which are usually intimately related to the lattice, 
{\it e.g.}, through the spin-orbit coupling. 
 The problem is thus subtle, and one has to go beyond 
the phenomenological argument such as the one based on the 
Galilean invariance. 
 Instead, a fully microscopic calculation, which starts from a 
definite microscopic model and does not introduce any phenomenological 
assumptions, is desired. 
 The present section is devoted to outline such attempts.

 At present, only a single model, where spin-relaxation processes 
are introduced by magnetic impurities, has been examined, 
with a result that $\alpha \neq \betasf$ 
(even for single-band itinerant ferromagnets) in general. 
 This will be surveyed in \S \ref{SECsmall} and \S \ref{SECspingauge}. 
 Studies on other models, hopefully with more realistic spin-relaxation 
mechanisms, are left to future studies. 
 Readers who are not interested in theoretical details but the results 
can jump to \S\ref{SEC:singleresult}.

\subsection{Small-amplitude method}
\label{SECsmall}

\noindent
\subsubsection{Microscopic model}

 Let us first set up a microscopic model. 
 We take a localized picture for ferromagnetism, and consider the 
so-called $s$-$d$ model. 
 It consists of localized $d$ spins, $\Sv$, and conducting $s$ 
electrons, which are coupled via 
the $s$-$d$ exchange interaction. 
 The total Lagrangian is given by 
$ L_{\rm tot} = L_S + L_{\rm el} - H_{sd}$, 
where $L_S$ is the Lagrangian for $d$ spins (\Eqref{Ls}), 
\begin{equation}
  L_{\rm el} 
= \int d^3x \, c^\dagger \left[ 
     i \hbar \frac{\partial}{\partial t} 
   + \frac{\hbar^2}{2m} \nabla^2 
   + \varepsilon_{\rm F} - V_{\rm imp} -V_{\spinflip} \right] c  
\label{eq:L_el}
\end{equation}
is the Lagrangian for $s$ electrons, and 
\begin{equation}
 H_{sd} 
 = - M \int d^3x \, \nv ({\bm r}) \!\cdot\! 
     \sigmav ({\bm r}) 
\label{eq:H_sd}
\end{equation} 
is the $s$-$d$ exchange coupling. 
 Here, $c^\dagger = (c^\dagger_\uparrow , c^\dagger_\downarrow ) $  
is the spinor of electron creation operators, and 
$\sigmav({\bm r}) 
= c^\dagger ({\bm r}) \sigmav c({\bm r})$ 
represents (twice) the $s$-electron spin density, with 
$\sigmav$ being a vector of Pauli spin matrices. 
 We have put $\Sv=S \nv$ with the magnitude of spin, $S$, 
and a unit vector $\nv$, 
and $M = J_{sd}S$ with $J_{sd}$ being the $s$-$d$ exchange coupling 
constant. 
(We define $\nv$ to be a unit vector in the direction of spin, 
which is opposite to the direction of magnetization. 
)

\begin{figure}[tbh]
\begin{center}
\includegraphics[scale=0.4]{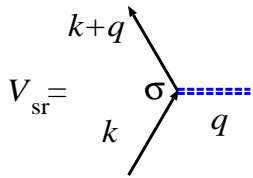}
\caption{Diagrammatic representation of spin relaxation due to spin flip scattering by impurity spin.
\label{FIGVsrl}
}
\end{center}
\end{figure}

 The $s$ electrons are treated as a free electron gas in three 
dimensions subject to the impurity potential 
\begin{equation}
 V_{\rm imp} +V_{\spinflip}
=  u \sum_i \delta ({\bm r} - {\bm R}_i) 
   + u_{\rm s} \sum_j {\bm S}_{{\rm imp},j} \!\cdot\! \sigmav 
               \delta ({\bm r} - {\bm R}_j') . 
\label{eq:Vimp}
\end{equation}
 The first term describes potential scattering. 
 The second term represents quenched magnetic impurities, 
which is aimed at introducing spin relaxation processes
(Fig. \ref{FIGVsrl}). 
 The averaging over the impurity spin direction is taken as 
$\overline{S_{{\rm imp},i}^\alpha} = 0$ and 
\begin{equation}
  \overline{S_{{\rm imp},i}^\alpha S_{{\rm imp},j}^\beta} 
= \frac{1}{3} S_{\rm imp}^2 \delta_{ij} \delta^{\alpha\beta} .
\label{eq:SS}
\end{equation}
 The damping rate of $s$ electrons is then given by 
\begin{equation}
  \gamma_\sigma  
 =  \frac{\hbar}{2\tau_\sigma} 
 =  \pi n_{\rm i}u^2  \nu_\sigma  
  + \frac{\pi}{3}  n_{\rm s} u_{\rm s}^2 S_{\rm imp}^2
    ( 2 \nu_{\bar\sigma} + \nu_\sigma ) . 
\label{eq:gamma}
\end{equation}
 Here $n_{\rm i}$ ($n_{\rm s}$) is the concentration of normal 
(magnetic) impurities, and 
$\nu_\sigma = m \, k_{{\rm F} \sigma}/2\pi^2\hbar^2$ 
(with $\hbar k_{{\rm F} \sigma} = \sqrt{2m\varepsilon_{{\rm F} \sigma}}$) 
is the density of states (per volume) at energy 
$\varepsilon_{{\rm F} \sigma} \equiv \varepsilon_{\rm F} + \sigma M$. 
(The subscript $\sigma = \uparrow, \downarrow$ 
corresponds, respectively, to $\sigma = +1, -1$ in the formula, 
and to $\bar\sigma = \downarrow, \uparrow$ or $-1, +1$.) 
 We assume that $\gamma_\sigma \ll \varepsilon_{{\rm F} \sigma}$ 
and $\gamma_\sigma \ll M$, and calculate the torques 
in the lowest nontrivial order in 
$\gamma_\sigma /\varepsilon_{{\rm F} \sigma}$ and $\gamma_\sigma /M$. 

In this section, the electron lifetime is treated as spin-dependent, since this feature becomes essentially important in calculating the effect of spin relaxation consistently. 
(Determination of $\betasf$ and $\alpha$ terms requires much care,
and so phenomenological argument can easily lose consistency.)

\subsubsection{General framework}
\label{SEC:singlespingeneral}

 The spin torque from $H_{sd}$ is given by 
\begin{equation}
 {\bm t}_{\rm el} ({\bm r}) 
\equiv  M \nv ({\bm r}) \times 
    \langle \sigmav ({\bm r}) \rangle_{\rm ne}. 
\label{eq:torque0}
\end{equation}
 This is related to ${\bm T}_{\rm el}$ of 
Eq.(\ref{eq:T_el_vec}) via 
${\bm t}_{\rm el} [\nv] = - {\bm T}_{\rm el}[{\bm M}]$ ,
and to ${\bm t}_{\rm el}'$ of Eq.(\ref{eq:t_el2}) via 
\begin{equation}
 {\bm t}_{\rm el} = \frac{\hbar S}{a^3} \, {\bm t}_{\rm el}'  
\label{eq:t_t'}
\end{equation}
with $a^3$ being the volume per $d$ spin.

 The calculation of spin torque is thus equivalent to that of 
$s$-electron spin polarization, 
$\ \langle \sigmav ({\bm r}) \rangle_{\rm ne}$, 
or precisely speaking, its orthogonal projection $\langle \sigmav_\perp ({\bm r}) \rangle_{\rm ne}$ 
to $\nv$. 
(We define 
$\hat\sigmav_\perp 
= \hat\sigmav-\nv(\nv \!\cdot\! \hat\sigmav)$ 
and 
$\hat\sigmav_\perp' 
= \hat\sigmav-\hat z (\hat z \!\cdot\! \hat\sigmav)$. 
 Note that $\langle \hat \sigmav \rangle_{\rm ne}$ in 
Eq.(\ref{eq:torque0}) can be replaced by 
$\langle \hat \sigmav_\perp \rangle_{\rm ne}$. )
 The expectation value $\langle \cdots \rangle_{\rm ne}$ 
is taken in the following nonequilibrium states depending on the 
type of the torque. 

(a) Nonequilibrium states under the influence of uniform but 
{\it time-dependent magnetization}. 
  This leads to torques with time derivative of $\nv$,
namely, Gilbert damping and spin renormalization. 

(b) Nonequilibrium states with {\it current flow} under static but 
{\it spatially-varying magnetization}.  
 This leads to current-induced torques, namely, 
spin-transfer torque and the $\beta$-term.

 In the presence of spin rotational symmetry for electrons, 
adiabatic spin torques, which are first order in space/time derivative, 
are expressed as
\begin{equation}
 {\bm \tau}_{\rm ad}^0 
= a_0 \dot \nv + ({\bm a} \!\cdot\! {\bm \nabla})\,  \nv 
  + b_0 \, (\nv \times \dot \nv) 
  + \nv \times ({\bm b} \cdot\! {\bm \nabla}) \, \nv . 
\label{eq:torque1} 
\end{equation}
 The corresponding $s$-electron spin polarization is given by 
\begin{equation}
\langle \sigmav_\perp \rangle_{\rm ne} 
= \frac{1}{M} \left[\, 
    b_0 \dot \nv + ({\bm b} \!\cdot\! {\bm \nabla})\,  \nv 
  - a_0 \, (\nv \times \dot \nv) 
  - \nv \times ({\bm a} \cdot\! {\bm \nabla}) \, \nv \,\right] . 
\label{eq:sigma_perp} 
\end{equation}
 To calculate the coefficients $a_\mu$ and $b_\mu$ microscopically, 
it is sufficient to consider small transverse fluctuations, 
${\bm u} = (u^x,u^y,0)$, $|{\bm u}| \ll 1$, 
around a uniformly magnetized state, $\nv = \hat z$, such that 
$\nv = \hat z + {\bm u} + {\cal O}(u^2)$ \cite{Tserkovnyak06,Tserkovnyak04}. 
 Then, up to ${\cal O}(u)$, Eq.(\ref{eq:sigma_perp}) becomes 
\begin{equation}
\langle \sigmav_\perp \rangle_{\rm ne} 
= \frac{1}{M} \left[\, 
b_0 \dot {\bm u} + ({\bm b} \!\cdot\! {\bm \nabla}) {\bm u} 
   - a_0 (\hat z \times \dot {\bm u}) 
   - \hat z \times ({\bm a} \!\cdot\! {\bm \nabla}) {\bm u}  \,\right] .
\label{eq:sigma_u}
\end{equation}
 This equation can be regarded as a linear response of 
$\sigmav_\perp$ to ${\bm u}$, and the coefficients, 
$a_\mu$ and $b_\mu$, are obtained as linear-response coefficients. 
 (Precisely speaking, $\langle \sigmav_\perp \rangle_{\rm ne}$ 
due to current is calculated as a linear response to the applied 
electric field\cite{Kohno07}.)


\subsubsection{Results}
\label{SEC:singleresult}

 The results are given by 
\begin{eqnarray}
 \delta S &=& \frac{1}{2} \, \rho_{\rm s} a^3 , 
\label{eq:deltaS}
\\
  {\bm v}_{\rm s} 
&=&  \frac{a^3}{2e\, (S+\delta S)} \, {\bm j}_{\rm s} , 
\label{eq:vs}
\\
  \alpha 
&=&  \frac{a^3 (\DOS_++\DOS_-)}{4(S+\delta S)} \!\cdot\! \frac{\hbar}{\tau_{\rm s}} 
   + \frac{S}{S+\delta S} \, \alphaz 
 \equiv \alphasf+\frac{S}{S+\delta S} \, \alphaz, 
\label{eq:alpha1}
\\
  \betasf 
&=&  \frac{\hbar}{2M\tau_{\rm s}} 
\label{eq:beta1}
\end{eqnarray}
where $\rho_{\rm s}  = n_\uparrow - n_\downarrow$, 
and 
${\bm j}_{\rm s} = \sigma_{\rm s}{\bm E}
 = {\bm j}_\uparrow - {\bm j}_\downarrow $ is the spin current, 
with $\sigma_{\rm s} = (e^2/m) 
(n_\uparrow \tau_\uparrow - n_\downarrow \tau_\downarrow)$ 
being the \lq\lq spin conductivity''. 
($n_\sigma$ is the density of spin-$\sigma$ electrons.)
 We have defined the spin-relaxation time $\tau_{\rm s}$ by 
\begin{equation}
  \frac{\hbar}{\tau_{\rm s}} 
 =  \frac{4\pi}{3} \, n_{\rm s}u_{\rm s}^2 \, S_{\rm imp}^2 \, (\DOS_++\DOS_-) 
\label{eq:tau_s}
\end{equation}
 As expected, only the spin scattering ($\sim \tau_{\rm s}^{-1}$) 
contributes to $\alpha$ and $\betasf$, 
and the potential scattering ($\sim n_{\rm i}u^2$) does not. 
 (For $\alpha$, the second term on the right-hand side of 
Eq.(\ref{eq:alpha1}) comes from processes that do not involve $s$-electrons.)

 For a single-band itinerant ferromagnet 
(as described by, {\it e.g.}, the Stoner model), 
the results are obtained by simply putting 
$S=0$ in Eqs.(\ref{eq:deltaS})-(\ref{eq:beta1}), 
and by using the spin polarization of itinerant electrons for $\delta S$.
 Even in this case, we see $\alphasf \neq \betasf$; 
however, it was pointed out \cite{Tserkovnyak06} that the ratio 
\begin{equation}
  \frac{\betasf}{\alphasf} 
=  \frac{\rho_{\rm s}}{M (\DOS_++\DOS_-)} 
\simeq 1 + \frac{1}{12} \left(\frac{M}{\varepsilon_{\rm F}} \right)^2 
\end{equation}
is very close to unity.
 Even in this case, if we make the impurity spins anisotropic 
by generalizing Eq.(\ref{eq:SS}) to 
\begin{eqnarray}
  \overline{S_i^\alpha S_j^\beta} 
&=& \delta_{ij} \delta_{\alpha\beta} \times \left\{ \begin{array}{cc} 
    \overline{S_\perp^2} & (\alpha, \beta = x,y) \\ 
    \overline{S_z^2}     & (\alpha, \beta = z) 
    \end{array} \right.
\end{eqnarray}
we obtain 
\begin{equation}
  \frac{\betasf}{\alphasf} 
=  \frac{3 \overline{S_\perp^2} + \overline{S_z^2} }
        {\, 2 \, (\overline{S_\perp^2} + \overline{S_z^2}) \,} 
\end{equation}
which ranges from $1/2$ (for $\overline{S_\perp^2} \ll \overline{S_z^2}$) 
to $3/2$ (for $\overline{S_\perp^2} \gg \overline{S_z^2}$). 
 From this example, we can learn that the ratio $\betasf / \alphasf$ is very 
sensitive to the details of the spin-relaxation mechanism. 
In reality, a non-electron contribution, $\alphaz$, exists, and so 
$\alpha=\betasf$ predicted in Ref. \cite{Barnes05} 
would never happen.

 The results obtained based on the phenomenological equations \cite{Zhang04} 
can be written in the form 
\begin{eqnarray}
  \alpha_{\rm ZL} 
&=&  \frac{\delta S}{S+\delta S} \!\cdot\! \frac{\hbar}{2 M \tau_{\rm s}} , 
\label{eq:alpha3}
\end{eqnarray}
whereas $\beta_{\rm ZL} = \betasf$ is the same as Eq.(\ref{eq:beta1}). 
 Thus, it predicts $\alpha = \betasf$ for single-band itinerant ferromagnets, 
$S=0$, which is, however, in disagreement with the present microscopic 
calculation, (\ref{eq:deltaS})-(\ref{eq:beta1}), 
predicting $\alpha \neq \betasf$ in general.

\subsection{Gauge field method in the presence of spin relaxation}
\label{SECspingauge}

 In the previous section, we considered small-amplitude fluctuations 
of magnetization, and calculated the torques in the first order 
with respect to these small fluctuations. 
 In this sense, the spin torques calculated there are limited to 
small-amplitude dynamics. 
 (Only for systems with rotational symmetry in spin space, where the 
form of the torque is known as Eq.(\ref{eq:torque1}), 
is this small-amplitude method sufficient to determine the coefficients, 
hence the torque.)
 In this section, we describe a theoretical formalism that is not 
restricted to small-amplitude dynamics, but can treat finite-amplitude 
(arbitrary) dynamics directly \cite{Kohno07}.

\subsubsection{Adiabatic spin frame and gauge field} 

 To treat finite-amplitude dynamics of magnetization, we work 
with a local/instantaneous spin frame 
(called \lq\lq adiabatic frame'' in the following) 
for $s$ electrons whose spin quantization axis is taken 
to be the local/instantaneous $d$-spin direction, $\nv$ 
\cite{Korenman77,TF94,TF97}. 
 The electron spinor $a(x)$ in the new frame is related to the 
original spinor $c(x)$ as $c(x) = U(x) a(x)$, where $U$ is a 
$2\times 2$ unitary matrix satisfying 
$ c^\dagger (\nv \!\cdot\! \sigmav) c 
 = a^\dagger \sigma^z a$. 
 It is convenient to choose $U$ satisfying $U^2=1$. 

 Since 
$\partial_\mu c = U (\partial_\mu + U^\dagger \partial_\mu U) a 
 \equiv U (\partial_\mu +iA_\mu) a$, the Lagrangian for the 
$a$-electrons becomes  
\begin{equation}
 L [a] = \int d^3x a^\dagger \left[ 
 i\hbar \left( \partial_t + i A_{_0} \right) 
 + \frac{\hbar^{2}}{2m}  \left(\nabla_i + iA_i \right)^2 
 + M \sigma_z - \tilde V_{\rm imp} -\tilde V_{\spinflip} \, \right] a .
\label{eq:L_a}
\end{equation}
 The original electrons moving in time-dependent/inhomogeneous magnetization 
are thus mapped to new electrons moving in a uniform 
and static magnetization $M \sigma_z$ but there arises a coupling to 
an SU(2) gauge field 
\begin{equation}
 A_\mu = -i U^\dagger (\partial_\mu U) = A^\alpha_\mu \sigma^\alpha 
\equiv {\bm A}_\mu \!\cdot \sigmav . 
\label{eq:A_UdU}
\end{equation}
 Here ${\bm A}_\mu$ 
is a measure of temporal ($\mu=0$) or spatial ($\mu=1,2,3$) variation 
of magnetization. 
(We use the vector (bold italic) notation for the spin component. 
 The space-time components are indicated by subscripts such as 
$\mu, \nu$ $(=0,1,2,3)$ or $i,j$ $(=1,2,3)$.)

 Let us introduce a $3\times 3$ orthogonal matrix ${\cal R}$, 
representing the same rotation as $U$ 
but in a three-dimensional vector space, 
and satisfying ${\rm det} {\cal R} = 1$.
 Note that ${\cal R} \hat z = \nv$, 
${\cal R} \nv = \hat z$, 
$c^\dagger \sigmav c
= {\cal R} \, (a^\dagger \sigmav a)$,  
and that 
${\cal R} \, ({\bm a} \times {\bm b}) 
= ({\cal R}{\bm a}) \times ({\cal R}{\bm b})$ 
for arbitrary vectors ${\bm a}$ and ${\bm b}$. 
 Then the spin-torque density, Eq.(\ref{eq:torque0}), is written as 
\begin{equation}
 {\bm t}_{\rm el} (x) 
= M {\cal R} \, (\hat z \times \langle \tilde \sigmav (x) 
  \rangle_{\rm ne}) ,
\label{eq:t_el}
\end{equation}
where 
$\tilde \sigmav (x) = (a^\dagger \sigmav a)_x $. 

 Since the gauge field ${\bm A}_\mu$ contains a space/time derivative 
of magnetization, one may naturally formulate a gradient expansion 
in terms of ${\bm A}_\mu$ to calculate, {\it e.g.}, the 
torque (or spin polarization). 
 In particular, the adiabatic torques are obtained as the 
first-order terms in ${\bm A}_\mu$: 
\begin{equation}
 \langle \tilde \sigmav_\perp \rangle_{\rm ne} 
= \frac{2}{M} \left[ 
   a_\mu {\bm A}^\perp_\mu 
  + b_\mu (\hat z \times {\bm A}^\perp_\mu )  \right] .
\label{eq:sigma_perp_A}
\end{equation}
 Here 
$\tilde \sigmav_\perp 
 = \tilde \sigmav 
 - \hat z \, (\hat z \!\cdot\! \tilde \sigmav)$ 
and 
$ {\bm A}^\perp_\mu
= {\bm A}_\mu - \hat z \, (\hat z \!\cdot\! {\bm A}_\mu) $ 
are the respective transverse components, 
and the sums over $\mu = 0,1,2,3$ are understood.  
 From the identities, 
\begin{align}
& {\cal R} {\bm A}^\perp_\mu
= - \frac{1}{2} \nv \times (\partial_\mu \nv) , 
\ \ \ 
{\cal R} (\hat z \times {\bm A}^\perp_\mu) 
= \frac{1}{2} \partial_\mu \nv , 
\label{eq:RA}
\end{align}
together with Eq. (\ref{eq:t_el}), 
we see that Eq. (\ref{eq:sigma_perp_A}) leads to the adiabatic 
torque density 
$ {\bm \tau}_{\rm ad}^0$ of Eq. (\ref{eq:torque1}). 

\begin{figure}[tbh]
\begin{center}
\includegraphics[scale=0.4]{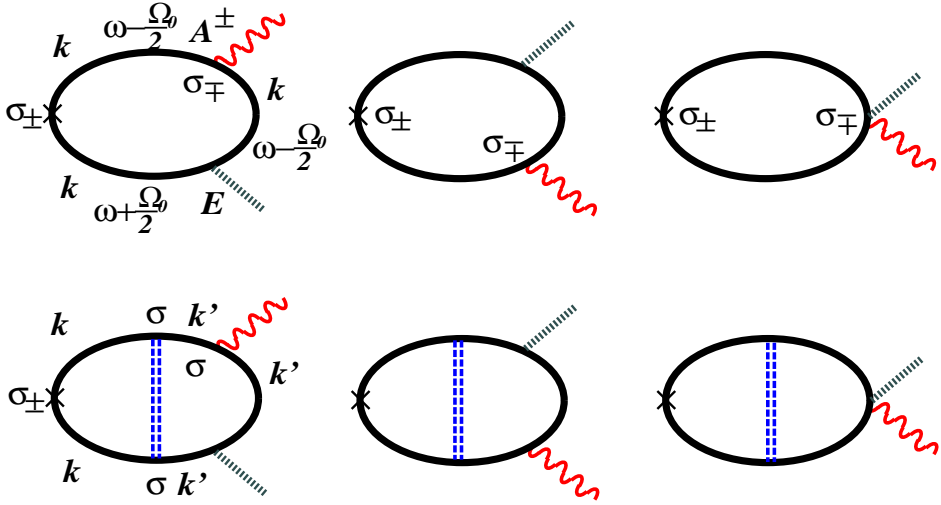}
\caption{Diagrams contributing to the $\beta$ term.
The electron Green's functions here are denoted by thick lines, indicating that they include self-energy from spin flip interaction, $\Hsf$. The last three diagrams are vertex corrections due to spin flip scattering.
Since we are interested in adiabatic limit, the gauge field does not change electron momentum.
\label{FIGbeta}
}
\end{center}
\end{figure}

The  processes contributing to the $\beta$ term are shown in Fig.  \ref{FIGbeta}.\cite{Kohno07}

\subsubsection{Results} 

 If we regard Eq.(\ref{eq:sigma_perp_A}) as a linear response 
to the gauge field ${\bm A}_\mu$ appearing in Eq.(\ref{eq:L_a}), 
the coefficients $a_\mu$ and $b_\mu$ are calculated as 
linear-response coefficients, and  $\delta S, {\bm v}_{\rm s}$, and $\betasf$ 
thus obtained coincide with those obtained by the small-amplitude method, 
Eqs.(\ref{eq:deltaS}), (\ref{eq:vs}), and (\ref{eq:beta1}). 
 However, it predicts $\alphasf = 0$, namely, it
fails to reproduce the Gilbert damping.

\subsubsection{Gilbert damping}

\begin{figure}[tbh]
\begin{center}
\includegraphics[scale=0.4]{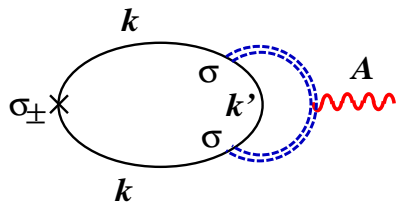}
\caption{Diagrams contributing to damping $\alpha$ in the gauge transformed frame.
The gauge field (wavy line) is induced by dynamical impurity spins (represented by double dotted line). 
\label{FIGalpha}
}
\end{center}
\end{figure}

 The above puzzle with the Gilbert damping can be resolved 
if we note that the impurity spins, 
which are static (quenched) in the original frame, 
become time-dependent in the adiabatic frame. 
 Namely, the spin part of $V_{\spinflip}$ is expressed as 
\begin{equation}
 {\bm S_{{\rm imp},j}} \!\cdot\! c^\dagger \sigmav c
= \tilde {\bm S_j}(t)  \!\cdot\! a^\dagger \sigmav a 
\label{eq:Sc_SSa}
\end{equation}
where 
\begin{equation}
 \tilde {\bm S_j}(t) = {\cal R}(t) {\bm S_{{\rm imp},j}}
\label{eq:SS_RS}
\end{equation}
is the impurity spin in the adiabatic frame, which is time-dependent. 
(This fact is expressed by $\tilde V_{\spinflip}$ in Eq.(\ref{eq:L_a}).) 
 Actually, we can obtain the gauge field from this time dependence as 
\begin{equation}
 [ {\cal R}(t)  \dot {\cal R}(t) ] ^{\alpha \beta} 
= 2 \varepsilon^{\alpha \beta \gamma} A_0^\gamma . 
\label{eq:RdR_A0}
\end{equation}
 Explicit calculation of 
$\langle \tilde \sigmav_\perp \rangle_{\rm ne}$ 
in the second order in $ \tilde {\bm S_j}(t)$ (nonlinear response) 
gives 
\begin{equation}
 \langle \tilde \sigmav_\perp (t) \rangle_{\rm ne} 
= - \frac{2\pi\hbar}{3M} n_{\rm s} u_{\rm s}^2 S_{\rm imp}^2 
  \nu_+^2 \, (\hat z \times {\bm A}_0^\perp (t))  , 
\label{eq:sigma_iso_result_vec}
\end{equation}
leading to the Gilbert damping, with damping constant 
exactly the same as Eq. (\ref{eq:alpha1}). 
The Gilbert damping term is shown diagramatically in Fig. \ref{FIGalpha}.

 The present calculation provides us a new picture of Gilbert damping. 
 While the $s$-electron spin tends to follow the instantaneous 
$d$-spin direction $\nv(t)$, it is at the same time 
pinned by the quenched impurity spins. 
 These two competing effects are expressed by the time dependence 
of $\tilde \Sv_j (t)$ in the adiabatic frame, and this effect 
causes Gilbert damping. 
 Namely, the Gilbert damping arises because spins of $s$-electrons 
are \lq dragged' by impurity spins. 

 Generally, any terms in the Hamiltonian leading to spin relaxation 
break spin rotational symmetry of $s$ electrons, and thus acquire 
time dependence in the adiabatic frame. 
 Therefore, the same scenario as presented here is expected to apply 
to other type of spin-relaxation processes quite generally.

\subsubsection{Non-adiabatic torque} 
\label{SEC:singlespinnonadiabatic}

 The non-adiabatic torques come from large-momentum processes, and 
are expected to be insensitive to impurities, so 
let us put $V_{\rm imp}=V_{\spinflip}=0$ for simplicity. 
 If we retain the full ${\bm q}$-dependence in the response function, 
we have, among others, a contribution 
\begin{equation}
 \langle \tilde\sigmav_\perp ({\bm q}) \rangle_{\rm ne} 
  =  \chi_{ij} ({\bm q}) 
     E_i \, (\hat z \times {\bm A}_{{\bm q},j}^\perp) 
    + \cdots , 
\label{eq:n_na}
\end{equation}
where the function $\chi_{ij} ({\bm q})$ is nonzero only when 
$| k_{{\rm F} \uparrow} - k_{{\rm F} \downarrow} | < q < 
 k_{{\rm F} \uparrow} + k_{{\rm F} \downarrow}$. 
 In particular, $\chi_{ij} ({\bm q}) = 0$ in the vicinity of 
${\bm q} = {\bm 0}$, and hence the long-wavelength approximation 
cannot be applied. 
 The resulting torque 
\begin{equation}
 {\bm \tau}_{\rm na}^0 (x) 
= -ME_i \int d^3x' \, \chi_{ij} ({\bm r}-{\bm r}') 
   {\cal R}(x) {\bm A}_j^\perp (x') , 
\label{eq:t_na}
\end{equation}
is characterized by an oscillatory function 
$\chi_{ij} ({\bm r}-{\bm r}')$ in real space, 
and is essentially nonlocal \cite{Waintal04,TK04,Xiao06,TKSLL07}. 
 This is the non-adiabatic, momentum-transfer torque due to electron 
reflection \cite{TK04}, generalized here to arbitrary magnetization texture. 
 For details, see \S\ref{SEC:LLGequation}, \S\ref{SEC:torque} and Ref. \cite{TKSLL07}.

\subsubsection{Further references}

 The spin-transfer torque in the form of Eq.(\ref{eq:tau_ad}) was 
first derived by Bazaliy {\it et al.} \cite{Bazaliy98}. 
 The $\beta$-term was first derived by Zhang and Li \cite{Zhang04} 
based on a spin diffusion equation with a spin-relaxation term 
included, 
and by Thiaville {\it et al.} \cite{Thiaville05} 
as a continuum limit of a special type of torque known in multilayer systems 
\cite{Heide01}.

 Duine {\it et al.} put the present microscopic (small-amplitude) 
calculation into the Keldysh formalism, and developed a functional 
description of spin torques, which will be used for finite-temperature 
and/or fluctuation dynamics \cite{Duine07}. 
 An attempt with the Boltzmann equation is carried out in \cite{Piechon07}. 
 Some developments in the treatment of Gilbert damping can be seen 
in \cite{Tserkovnyak04,Skadsem07}. 
 The non-adiabatic torque is studied in 
\cite{Waintal04,TK04,Xiao06,TKSLL07}. 
 The domain-wall resistivity is studied in 
\cite{TF97,Simanek05}.

 The effects of spin-orbit coupling is studied on domain-wall 
resistance \cite{Nguyen06} and domain-wall mobility \cite{Nguyen07} 
in a model of ferromagnetic semiconductors.

 Microscopic understanding of each spin torque is also an important 
issue.  
 In particular, the dissipative torques ($\alpha$ and $\beta$ terms), 
coming from spin relaxation of electrons, should be understood 
as material-dependent quantities; 
for each real system, we need to identify the dominant spin-relaxation 
mechanism, and clarify the dependence of $\alpha$ and $\beta$ 
on material parameters. 
 Development of first-principles calculational methods 
for spin torques is also desired for the purpose of material/device 
design.

\section{Domain wall and electron transport}
\label{SEC:dwtransport}
In this section, we briefly review electron transport theory in the presence of a domain wall.  
Electron transport coefficients such as resistivity and Hall coefficients represent the reaction of current-induced forces on magnetic structures.
We calculate these coefficients in linear response theory, and show that the current-induced forces we have discussed in \S\ref{SEC:force} are indeed proportional to them.
For experimental developments, see Ref.  \cite{Marrows05}.

\subsection{Resistivity in magnetic metals}

Since more than a century ago a number of studies has been carried out 
on electric transport properties in ferromagnetic metals. 
They revealed many remarkable 
features that are not seen in non-magnetic metals.
One of the most notable would be the hysteretic and 
anisotropic behavior of 
the resistance in the magnetic field (magnetoresistance) observed at 
small magnetic fields of  $\lesssim 1$ T, which has been already 
noted more than a hundred years ago \cite{Thomson1857,McKeehan30}.
The magnetoresistance in the case of the field $H$ parallel to the 
current $I$
takes a minimum at a finite value of the field ($\sim200$ Oe for 
instance for 
the case of Ni and Fe) .
If the field is applied perpendicular to the current, the curve of 
magnetoresistance is reversed; 
namely the resistivity shows a maximum at a certain field and 
decreases as the field deviates from that value.
The hysteretic behavior of the magnetoresistance is due to the fact 
that the resistivity is mostly governed by 
the total magnetization vector $\Mv$, not the applied magnetic field.
(The total magnetization is given as the average of localized spin vectors, $M=- g\frac{e\hbar}{2m V}\intx{\Sv}$.)
The observed resistivity $\rho$ as a function of the field, $H$, has 
been shown to be well fitted by a phenomenological relation 
$\rho \sim \rho_{0}+\Delta\rho_{\rm ani} <\cos^{2} \theta_M>$,
where $\rho_{0}$ is the field independent part 
and $\Delta\rho_{\rm ani}$ measures the strength of the 
anisotropy in the resistivity \cite{McGuire75,Smit51}.
$\theta_M$ is the mutual angle between the local magnetization and 
the current, which depends on the magnetic field, 
and the brackets denote the average over the sample.
In most ferromagnetic metals, the anisotropy $\Delta\rho_{\rm ani}$ is 
positive.
This anisotropic behavior of the resistivity is called anisotropic 
magnetoresistance (AMR).

\begin{figure}[tbh]
  \begin{center}
  \includegraphics[width=0.6\linewidth]{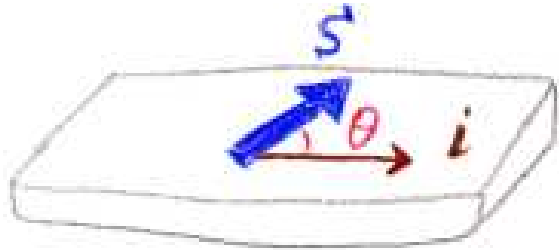}
\caption{
Anisotropic magnetoresistance (AMR) effect. Resistivity depends on the angle $\theta$ between the magnetization $\Mv$ or localized spin $\Sv$ and applied current, $\iv$.
\label{FIGamr}}
  \end{center}
\end{figure}

The mechanism of AMR was argued to be spin-orbit interaction by 
Smit \cite{Smit51} and McGuire and Potter \cite{McGuire75}.
They pointed out that the conduction ($s$-) electron is coupled to the 
magnetization due to the scattering into the
magnetic ($d$-) band and the spin-orbit interaction there, and that this 
process gives rise to a spin asymmetric lifetime, which depends on the 
angle between the current and the field, resulting in an anisotropy.
According to their arguments, the positiveness of $\Delta\rho_{\rm ani}$ is 
explained if the resistivity is dominated by the electron having the minority spin.

\subsection{Domain wall contribution to classical magnetoresistance}
\label{SECclassicalDW}
The magnetoresistance observed in bulk 
materials is mostly understood well in terms of the AMR effect, which 
assumes that the magnetization changes very slowly and hence the 
electron feels only the average magnetization \cite{McGuire75}.
However this assumption may not be good in small magnets that  
contain many domains with different directions of local 
magnetization.
In fact the boundary of these domains is a structure called a domain wall 
where the spins 
rotate spatially within a finite distance.
Such domain walls lead to scattering of the electron, which 
is not taken into account in the standard AMR argument.
In this paper we exclusively consider the effect of domain walls on the electronic resistivity.
Of particular interest would be the case of a small sample less than a 
typical domain size $\simeq 1\sim10\ \mu$m, since there the magnetic 
properties can be described in terms of the domain wall configuration.
In fact in such samples the magnetization process as the field is swept 
will be described by the 
nucleation of one or a few domain walls, motion of the walls by 
depinning followed by the annihilation. 
The effect of domain walls on the resistivity would then appear in the 
magnetoresistance as discrete jumps 
in the measurement with the field is swept.
(Such events of domain walls are faster (e.g., the speed of domain wall motion is 
estimated to be about 182 m/s in submicron wires of NiFe \cite{Ono99}) 
compared to the sweep speed.) 
These effects are
similar to the  Barkhausen noise \cite{Barkhausen19}.
The jump will not be seen very clearly in bulk samples, since there the 
contribution from a macroscopic number of domain walls, whose 
nucleation and pinning energy may differ, will be summed up in the 
observed magnetoresistance and thus the contribution from each domain 
wall will not be visible.
Jumps in the magnetoresistance was first observed in 1994
in a Ni wire of about the diameter of 
300 \AA \cite{Giordano94,Hong95}.
Nowadays resistance of a single domain wall has been measured  
in many systems \cite{Marrows05,Chiba06}.

Let us here give a rough estimation of the effect of the wall on the 
classical electron transport. 
"Classical" here means that the conductivity (or resistivity) is evaluated at the lowest 
order in $\hbar$, which is determined by the 
 probability of reflection by the wall. 
The most important parameter of the wall on the transport is the 
thickness of the wall, $\lambda$.
This quantity is determined by the competition between the 
exchange energy, $J$, which aligns the neighboring spins and 
the 
magnetic anisotropy energy in the easy axis, $K$, which tends to make the wall thinner to minimize the deviation of spins from the easy 
axis,
as $\lambda =\sqrt{J/K}$
(Here $J$ has dimensions of J/m$^2$.)
Thus $\lambda$ depends on the material and also on the sample shape 
since $K$ depends on the shape.
In the case of 3$d$ transition metals such as iron and nickel 
$\lambda\simeq 500\sim1000$ \AA \cite{Hong95}, and $\lambda\sim 150$ \AA\ 
in Co thin films \cite{Gregg96}.
This length scale is very large compared 
with the length scale of the electron, 
$k_{F}^{-1}\sim O(1 $ \AA$)$ ($k_{F}$ being the Fermi wave length of the 
electron).
Thus in such materials the conduction electron can 
adiabatically adjust 
itself to the local magnetization at every point as it passes through 
the wall, resulting in a very small scattering probability by the 
wall. 
The classical resistivity in the Boltzmann sense, which is proportional to 
the reflection probability by the wall, is thus expected to be negligibly 
small. 
This was explicitly shown in 1974 by Cabrera and Falicov \cite{Cabrera74} by 
calculating in the clean limit 
the reflection coefficient based on the one-dimensional 
Schr\"odinger equation for the electron coupled via exchange coupling 
to the magnetization whose configuration is a domain wall. 
They obtained for the case of thick wall, $k_{F}\lambda\gg 1$, the expression 
$\Rw \propto \exp(-2\pi k_{F}\lambda)$ for the wall 
contribution, and showed that the wall can have a large effect only 
if the wall is extremely thin ($k_{F}\lambda \lesssim 1$) and if the 
conduction electron is strongly polarized by the magnetization.
Such a thin domain wall was fictitious at that time, where 3$d$ 
metals in 
the bulk were mostly considered, but can be realized at present \cite{Garcia99,Feigenson07}.

In the past 10 years there has been a renewal of interest in the theory 
of the classical resistivity due to the domain 
wall \cite{Yamanaka96,TF97,Levy97,Brataas98,Brataas99,vanHoof99,TZMG99,Gorkom99,Imamura00,Nakanishi00,GT01}, 
because experimental control of the wall thickness \cite{Garcia99} and 
precise measurements are becoming possible.
(Resistance due to a single domain wall in GaMnAs was experimentally determined to be $\sim 1\ \Omega$ \cite{Chiba06}.)
In Ref. \cite{TF97}, the domain wall resistivity was studied based on a linear response theory. 
There the effective wall-electron 
interaction was derived and treated perturbatively. The calculation 
corresponds to the perturbation expansion with respect to 
$O(k_{F}\lambda)^{-1}$. 
The effect of the finite mean free path due to the impurity scattering 
was taken into account consistently. As far as the classical transport is
concerned, the calculation confirmed the result of Cabrera and Falicov. 
It was pointed out by Levy and Zhang \cite{Levy97}
that in the presence of 
a spin asymmetry in the electron lifetime, which is the case in most 
ferromagnets, the domain wall can have a substantial effect on the 
classical resistivity by mixing the two spin channels with different 
resistivities, in agreement with experiments on Co films at room 
temperature \cite{Gregg96}.
The effect of lifetime asymmetry has been further discussed in detail 
in Refs. \cite{Brataas98,Brataas99}. 
The domain wall resistivity in the ballistic limit has been discussed 
by use of realistic band structures in Ref. \cite{vanHoof99}.
It was shown there that the existence of nearly degenerate bands at 
the Fermi level in real magnets enhances the classical resistivity 
due to the wall.
The calculation of domain wall resistance by use of Landauer's formula 
was compared with that by linear response in Ref. \cite{GT00}.

Of recent particular interest is an atomic scale contact 
of magnetic metals in the ballistic region, as experimentally 
realized in Ref. \cite{Garcia99}.  
In such narrow contacts the profile of the wall is 
determined mostly by the shape rather than the anisotropy energy of the 
bulk magnets, and thus the wall is trapped in a 
contact region, which is typically on a nm 
scale \cite{vanHoof99,TZMG99,Bruno99}. 
The adiabaticity does not hold in such cases of small $k_{F}\lambda$ 
comparable to 1, and the wall can have a large 
effect on resistivity \cite{Cabrera74,vanHoof99,TZMG99}.
Indeed a magnetoresistance of 200\% was observed in 1999 in ballistic Ni 
nanocontacts \cite{Garcia99} where the number of the channels is less than 
$N\lesssim10$, 
and this was interpreted in terms 
of a strong reflection by a nanometer scale domain wall \cite{TZMG99}.
In dirty contacts, the effect is not so large, since 
the wall scattering is smeared out by impurity scattering \cite{ZMTG01}.
More recent studies report magnetoresistance over 1000\%
 \cite{Sullivan05}.
Theoretical studies, however, suggest that such high magnetoresistance is not of electric origin and could be due to mechanical effects in the contact region \cite{Rocha07}.
Whatever the mechanism, such high magnetoresistance would be useful for applications.

\subsection{Quantum electron transport and domain wall}

Most experimental studies have been carried out on low resistivity 
materials, and theoretical studies also have mostly been based on 
classical transport theory ({\it i.e.}, neglecting 
$O(\hbar/\epsilon_{F}\tau)$). 
However, besides the classical transport, there 
is another important aspect of electronic transport at low 
temperature. This is the effect of the quantum coherence among  
electrons, which modifies the low-energy electronic properties significantly.
The effect becomes important in disordered metals with high 
resistivity, where $\ell$ becomes shorter, 
$\ell\equiv (\hbar k_{F}\tau/m)$ being the elastic mean free path.
The electron wave scattered by such normal impurities can interfere 
with the incoming wave, leading to a standing wave. 
This is called weak localization and the resistivity in this case is 
enhanced due to the quantum interference \cite{Bergmann84,Lee85}.
This correction becomes large in small dimensions such as in wires since there 
the interference becomes stronger.
The interesting point of this situation is 
that the electronic properties are very sensitive to a small 
disturbance because of the presence of coherence.
For instance in non-magnetic metals of micron size, 
even the motion of a single impurity atom has been shown to
change the low-temperature conductance ($G\equiv \sigma 
A/L$, $A$ and $L$ being the cross-sectional area and 
length of the system) of the entire 
system by disturbing the coherence \cite{Meisenheimer89}. 
The magnitude of the conductance change turns out in most cases 
to be a universal order of $e^{2}/h$ \cite{Feng86}.

Because of this sensitivity these conductance fluctuations as a consequence of 
quantum interference have already been used
as a probe in studies of various mesoscopic non-magnetic 
metallic or semi-conducting systems. 
For example the telegraph noise due to a two-level oscillation of a 
defect in Bi films has been investigated and it turned out that the 
oscillation at $T\lesssim 1$ K is governed by the quantum tunneling 
subject to the dissipation from the conduction electron \cite{Golding92}. 
Such measurement of the quantum transport properties has been proved to 
be a useful probe also for studies of mesoscopic spin-glass 
systems \cite{Meyer95,Strunk98} and the magnetization flip of 
mesoscopic magnets \cite{Coppinger94,Lee04}.

It would be natural to expect that the domain wall also affects the 
quantum transport.
The effect of domain wall scattering on the quantum correction to the 
conductivity was first studied in Ref.  \cite{TF97}. 
It was shown that a domain wall destroys the 
interference among the electrons and the wall contributes to a 
negative quantum correction to the resistivity. 
In disordered 3$d$ transition metals, this quantum correction
can overcome the classical contribution (i.e., reflection), 
and thus a wall may in total lead to a {\it decrease} of resistivity.
Numerical simulation also supports the negative resistivity 
contribution from 
the wall in disordered thin wires \cite{Jonkers99}. 
It has been also pointed out that the geometric phase attached to 
the electron spin as it passes through the wall can also cause 
an important dephasing effect, which would become important in multiply
 connected geometries \cite{Geller98,Loss99}. 
The effect of dephasing due to the magnetic origin has been 
considered also in a thin film of metal sandwiched by ferromagnetic 
layers \cite{TF00}. There the internal magnetic field at the interface  
causes dephasing in the conduction layer in the 
presence of the spin-orbit scattering, which 
contributes to a positive magnetoresistance.

These theories on the quantum transport in magnetic metals are based 
on the existence of electron coherence in magnetic metals.  
In non-magnetic metals such coherence has been observed as an appearance 
of weak localization, for instance, in thin films of 
Cu \cite{Kobayashi80} and non-magnetic metals with magnetic 
impurities \cite{Raffy87}. 
Magnetic systems have been considered as not suitable for coherent electron transport, since the magnetic field (or magnetization) and magnetic disorder result in decoherence.
However, in ferromagnets, magnetic disorder should be frozen by a strong internal field (magnetization) and would not destroy coherence as theoretically discussed in Refs.  \cite{TB01,TKBB04}. 
The strong internal field in ferromagnets 
($M\sim O(1$ T)) would not affect the coherence in a mesoscopic case like a sufficiently
small wire, since there the magnetic flux penetrating through the wire can be small enough. 
Nevertheless, there have not been so far many observations of electron quantum coherence in ferromagnetic systems.
A negative contribution to resistance in the presence of domain walls was observed in GaMnAs at 4.2 K \cite{Tang04}, which was explained as due to electron decoherence due to the domain wall predicted in Ref.  \cite{TF97} in the weak localization regime.
Semiconductor ferromagnets were found to show further evidence of quantum transport: universal conductance fluctuation \cite{Vila07}.
Observation in metallic systems seems more difficult.
Aharonov-Bohm oscillation was observed in a 500 nm Fe$_{19}$Ni$_{81}$ ring \cite{Saitoh05,Sekiguchi08}. 
Mesoscopic resistance fluctuation due to phase arising from a domain wall was observed in Co nanoparticles \cite{Wei06}.

In the present paper, we will not discuss further the subject of quantum transport.

\subsection{Linear response theory of resistivity and Hall effect}
\label{SECmori}


In this section, we derive the resistivity (both diagonal and off-diagonal (i.e., Hall) components) due to the general spin texture on the 
basis of linear response theory, using the Mori formula \cite{Mori65,Mori65a,Gotze72}.
The Mori formula is valid in the clean limit, $\tau\rightarrow\infty$, where the resistivity is dominated by spin structure.
It relates the resistivity $\rhoS_{ij}$ ($ij$ being spatial directions)
to the correlation of random forces in the weak scattering case as
\begin{equation}
{\rhoS}_{ij}=\left(\frac{e^{2} n}{m}\right)^{-2}\lim_{\omega\rightarrow 0}
\frac{1}{\hbar\omega} 
{\rm Im}[\chi_{\dot{J}_i\dot{J}_j}(\hbar\omega)-\chi_{\dot{J}_i\dot{J}_j}(0)].
\label{Mori}
\end{equation}
Here, $\chi_{\dot{J}_i\dot{J}_j}(i\omega_{\ell})\equiv
-(\hbar/\beta V) <\dot{J}_i(i\omega_{\ell})\dot{J}_j(-i\omega_{\ell})>$
with $\dot{J_i}\equiv dJ_i/dt= \frac{ i}{\hbar}[H,J_i]$,
where $H$ is the total Hamiltonian and 
$J_i \equiv \frac{e\hbar}{m} \sum_\kv k_i \cdag_\kv c_\kv$ is the total current.
The operators here are defined in the imaginary time, $\tau=it$.
The correlation function 
$\chi_{\dot{J}\dot{J}}(\hbar\omega)$ in Eq. (\ref{Mori}) denotes 
an analytical continuation of the correlation function calculated for 
imaginary frequency, 
i.e., $\chi_{\dot{J}\dot{J}}(\hbar\omega)\equiv 
\chi_{\dot{J}\dot{J}}(i\omega_{\ell}\rightarrow \hbar \omega+i0)$.

\begin{figure}[tbh]
  \begin{center}
  \includegraphics[scale=0.3]{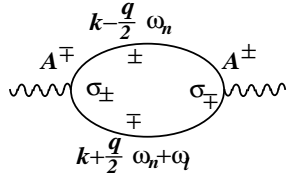}
\caption{Diagram representing the resistivity in the Mori formula at the lowest order in the gauge field.
\label{FIGmori}}
  \end{center}
\end{figure}

The nonconservation of the current (i.e., finite $\dot{J}$) arises from the 
scattering by the spin texture. 
In fact, Eq. (\ref{Le}) leads to 
\begin{eqnarray}
\dot{J_{i}} &=&  i\left(\frac{e}{m}\right)
\sum_{\kv,\qv} \left[ -2\spol  \sum_{\sigma} A_{i}^{-\sigma}  \adag_{\kv+\qv}\sigma_{\sigma} a_{\kv} \right.\nonumber\\
 && \left. +
 \frac{\hbar^2}{2m} 
 \left( 
A_{i}^{\alpha}((2\kv+\qv)\cdot \qv)  - q_i (\Av^{\alpha}\cdot(2\kv+\qv))
-\hbar q_i A_0^{\alpha} \right) 
\adag_{\kv+\qv}\sigma_{\alpha} a_{\kv} \right],
\nonumber\\
&&
\end{eqnarray}
where we neglect higher order terms in $A$.
We consider a static spin texture (i.e., $A_0=0$), 
and then
\begin{equation}
\dot{J_{i}} = -i2\left(\frac{e}{m}\right) \spol
\sum_{\kv,\qv} \sum_{\sigma} A_{i}^{-\sigma}  
\adag_{\kv+\qv}\sigma_{\sigma} a_{\kv} .
\end{equation}
The Fourier transform in imaginary time $\tau$, defined as
$\dot{J_{i}}(i\omega_{\ell})\equiv \int_0^{\beta} e^{-i\omega_\ell \tau}
 \dot{J_{i}}(\tau)$, 
where $\omega_{\ell}\equiv \frac{2\pi\ell}{\beta}$ is 
a bosonic thermal frequency,
is given as
\begin{equation}
\dot{J_{i}}(i\omega_{\ell}) = -i2\left(\frac{e}{m}\right) \spol
\sum_{\kv,\qv,n} \sum_{\sigma} A_{i}^{-\sigma}  
\adag_{\kv+\qv,n+\ell}\sigma_{\sigma} a_{\kv,n} ,
\end{equation}
where 
$a_{\kv,n}\equiv \frac{1}{\sqrt{\beta}} 
 \int_0^{\beta} e^{i\omega_n \tau} a_{\kv}(\tau)d\tau$ and $\omega_n$ 
represents the fermionic frequency $\omega_n\equiv \frac{\pi(2n-1)}{\beta}$.
The thermal correlation function is then obtained as (Fig. \ref{FIGmori})
\begin{equation}
\chi_{\dot{J}_i\dot{J}_j}(i\omega_{\ell}) =
-{\hbar} \left(\frac{2e\spol}{m}\right)^{2} \frac{1}{V}\sum_{kq\sigma} 
 A_i^{\sigma}(\qv) A_j^{-\sigma}(-\qv)
\frac{1}{\beta}\sum_{n}G_{k+q,n+\ell,-\sigma} G_{kn\sigma},
\label{chimori}
\end{equation}
where the imaginary time Green's function is defined as
\begin{equation}
G_{kn\sigma} =\frac{1}{i\omega_n-\ekvs}.
\end{equation}

The summation over $\omega_{n}$ is carried out to obtain
\begin{equation}
\chi_{\dot{J}_i\dot{J}_j}(i\omega_{\ell})=
-\hbar\left(\frac{2e\spol}{m}\right)^{2} \frac{1}{V}\sum_{kq\sigma}
 A_i^{\sigma}(\qv) A_j^{-\sigma}(-\qv)
\frac{
  f(\epsilon_{k+q,-\sigma})- f(\epsilon_{k,\sigma})  }
  { \epsilon_{k+q,-\sigma} -\epsilon_{k,\sigma} -i\omega_{\ell} }.
	\label{chidotcor}
\end{equation}
Thus the resistivity is obtained as
\begin{equation}
{\rhoS}_{ij}=
\lim_{\omega\rightarrow0} \frac{4\spol^2}{e^2n^2} \frac{1}{V}\sum_{\kv\qv\sigma} \Im \left[
 A_i^{\sigma}(\qv) A_j^{-\sigma}(-\qv)
\frac{1}{\omega} \frac{
  f(\epsilon_{k+q,-\sigma})- f(\epsilon_{k,\sigma})  }
  { \epsilon_{k+q,-\sigma} -\epsilon_{k,\sigma} -\omega-i0 }\right].
	\label{rhosij}
\end{equation}
Then, choosing the current direction as $\xw$, we may rewrite the resistivity as 
\begin{equation}
\rhoS \equiv {\rhoS}_{\xw\xw}
= \frac{4\pi \spol^{2}}{e^{2}n^{2}}\frac{1}{V}
 \sum_{\kv q \sigma} |A^\sigma_\xw(\qv)|^{2} 
\delta(\epsilon_{\kv+q,-\sigma} -\epsilon_{\kv,\sigma} ) \delta(\epsilon_{\kv,\sigma} ).
\label{rhocclean}
\end{equation}
Thus, the reflection force of Eq. (\ref{Fna}) is proportional to the resistivity due to spin, as we explained for Eq. (\ref{Fandj}).
The $\kv$-summation can be carried out easily as
\begin{equation}
\Vinv \sum_{\kv} 
\delta(\epsilon_{\kv+q,-\sigma} -\epsilon_{\kv,\sigma} ) \delta(\epsilon_{\kv,\sigma} )
=\frac{\DOS_\sigma m}{2 k_{F\sigma}q}\thetast(q),
\end{equation}
where $q\equiv |\qv|$
($\thetast(q)$ is defined in Eq. (\ref{thetastdef})).

In contrast to the diagonal components of resistivity, which vanish in the adiabatic limit, the off-diagonal (i.e., Hall) components remain finite in this limit.
In fact, we find 
\begin{eqnarray}
{\rhoS}_{ij}^{\rm ad} &=&
\lim_{\omega\rightarrow0} \left(\frac{4\spol^2}{e^2n^2}\right) \frac{1}{V}\sum_{\kv\qv\sigma} \Im \left[
 A_i^{\sigma}(\qv) A_j^{-\sigma}(-\qv)
\frac{1}{\omega} \frac{
  f(\epsilon_{k,-\sigma})- f(\epsilon_{k,\sigma})  }
  { 2\sigma \spol-\omega }\right]
\nonumber\\
&=&
-\frac{1}{\hbar e^2n^2V}\sumqv \Im \left[
 A_i^{\sigma}(\qv) A_j^{-\sigma}(-\qv) \right]
\sum_{\kv\sigma} f(\epsilon_{k,\sigma})
\nonumber\\
&=&
\frac{n_+-n_-}{2\hbar e^2n^2V }\intx 
\evs\cdot(\partial_i\Sv \times \partial_j\Sv )
=\frac{2\pi S^2 P}{\hbar e^2 n V}\Phi_{ij}. \label{rhoxy}
\end{eqnarray}
(We have retained the antisymmetric component, $\half({\rhoS}_{ij}-{\rhoS}_{ji})$, in the second line.)
This is the Hall effect caused by the spin chirality, or the spin Berry phase \cite{Ye99}, demonstrated in the slowly varying case \cite{TK02,OTN04}.
Comparing Eqs. (\ref{Fad}) and (\ref{rhoxy}), we see that the force in the adiabatic limit is exactly due to the Hall effect from spin chirality:
\begin{eqnarray}
\Fad_i 
&=& \frac{2e^2n^2}{n_+-n_-} \sum_j {\js}_j {\rhoS}_{ji} 
\nonumber\\
&=& e^2n \sum_j j_j {\rhoS}_{ji} .
\end{eqnarray}


\subsection{Kubo formula}
\label{SECclassical}

When impurity scattering dominates the resistivity, we need to use the Kubo formula. 
The wall contribution to resistivity is calculated  
perturbatively.  
The conductivity $\sigma$ for the current in the $\xw$-direction is 
calculated from the imaginary-time 
current-current correlation function 
\begin{equation}
Q(\tau)\equiv\frac{\hbar}{V}
 <{\rm T}_\tau J_{\xw}(\tau)J_{\xw}(0)> ,
\end{equation}
where ${\rm T}_\tau$ denotes the time order in the imaginary time and 
the bracket represents the thermal average;
\begin{equation}
 <{\rm T}_\tau J_{\xw}(\tau)J_{\xw}(0)>\equiv 
{\rm Tr}[\rho {\rm T}_\tau 
 J_{\xw}(\tau)J_{\xw}(0)].
\end{equation}
Here $\rho\equiv e^{-\beta H}/Z$, $Z\equiv {\rm Tr} e^{-\beta H}$, 
$\beta\equiv 1/(k_{B}T)$.
The Fourier transform of $Q(\tau)$ is written as 
\begin{eqnarray}
	Q(i\omega_{\ell}) & \equiv & \int_{0}^{\beta} e^{-i 
	\omega_{\ell}\tau} Q(\tau) \nonumber \\
	 & = &  \frac{\hbar}{\beta V} 
\average{ J_{\xw}(i\omega_{\ell})J_{\xw}(-i\omega_{\ell})},
	\label{Qdef}
\end{eqnarray}
where $\omega_{\ell}\equiv 2\pi\ell/\beta$ is the Bosonic Matsubara 
frequency and 
$J_{\xw}(i\omega_\ell)\equiv \int_{0}^{\beta}d\tau 
e^{-i\omega_{\ell}\tau} J_{\xw}(\tau)$.
The Fourier transform of the electron is defined as
$a_n\equiv (1/\sqrt{\beta}) \int_{0}^{\beta}d\tau e^{i\fomega_{n}\tau} 
a(\tau)$, $\fomega_{n}\equiv (2n-1)\pi n/\beta$ being a Fermionic 
Matsubara frequency. 
Here, thermal frequencies are written as 
$\omega_{\ell}, \omega_{\ell'},\cdots$ or $\omega_{n}, 
\omega_{n'},\cdots$, where subscripts $\ell, \ell', \cdots$ are used for 
Bosonic and $n, n', \cdots$ are for Fermionic frequencies.
We consider in this section a static domain wall solution, given by $\thetaz(\xw)$ and $\phi=0$. 
In the imaginary time, the current is then written by use of Eq. (\ref{jdef}) as
\begin{equation}
	J_{\xw}(i\omega_{\ell})=  \frac{e\hbar}{m} \sum_{{\kv}} 
\sum_{\omega_{n}}\left(  k_{\xw}a^{\dagger}_{{\kv},n+\ell}a_{{\kv},n}
- \frac{1}{2} \sum_{q}
\Av_{\xw} (q)a^{\dagger}_{{\kv}+q,n+\ell}\sigmav a_{{\kv},n} \right).
	\label{Jomegadef}
\end{equation}
The electron operator carrying a thermal frequency of 
$\fomega_{n}+\omega_{\ell}$ is denoted as $a_{n+\ell}$. 
The Kubo formula relates the correlation function (\ref{Qdef}) to the 
conductivity as \cite{Mahan90}
\begin{equation}
	\sigma=\lim_{\omega\rightarrow 0}\frac{1}{\omega}{\rm Im}(Q(\hbar\omega+i0)-Q(i0)). 
	\label{sigmadef}
\end{equation}
Here $Q(\hbar \omega+i0)$ is the retarded correlation function 
obtained by analytical continuation, 
$Q(\hbar\omega+i0)\equiv Q(i\omega_{\ell} \rightarrow \hbar\omega+i0)$, 
and ${\rm Im}$ denotes the imaginary part ($i0$ denotes a 
infinitesimal imaginary part).
We estimate the correction to the conductivity due to the wall 
to the second order of gauge field. 
In this section the classical (Boltzmann) contribution (denoted by 
$\sigma_{\rm c}$) is calculated.
The quantum corrections represented by maximally crossed diagrams are 
considered in the next section.

As is well known, the zeroth-order term of $Q$ is obtained as 
\begin{eqnarray}
Q_{0}(i\omega_{\ell}) &=& \left(\frac{e\hbar}{m}\right)^{2} 
\frac{\hbar}{V\beta}\sum_{n\kv \sigma}  k_{z}^{2}
\average{(a^\dagger_{\kv,n+\ell}a_{\kv n})(a^\dagger_{\kv,n}a_{\kv,n+\ell})} 
\nonumber \\
&=& -\left(\frac{e\hbar}{m}\right)^{2} 
\frac{\hbar}{V\beta}\sum_{n\kv \sigma}  
k_{z}^{2}G_{\kv n\sigma}G_{\kv,n+\ell,\sigma},
\end{eqnarray}
and this contribution leads by use of Eq. (\ref{sigmadef}) to the
classical conductivity due to the normal impurity, 
$\sigma_{0}\equiv e^{2}n\tau/m$ ($n$ being the electron 
density) \cite{Mahan90}.
Here the imaginary time electron Green's function 
($G_{\kv n\sigma}\equiv 
 -\average{a_{\kv n\sigma}a^\dagger_{\kv n\sigma}}$) 
includes the effect of the impurity and is given 
by \cite{Mahan90}
\begin{equation}
G_{\kv n\sigma}=
\frac{1}{i(\fomega_{n}+\frac{\hbar}{2\tau}{\rm sgn}(n))-\epsilon_{\kv\sigma} },
	\label{greenfunc}
\end{equation}
where ${\rm sgn}(n)=1$ and $-1$ for $n>0$ and $n<0$, respectively.
(The Green's function carrying a frequency of $\fomega_{n}+\omega_{\ell}$ is 
denoted by $G_{\kv,n+\ell,\sigma}$.)

\begin{figure}[tbhp]
  \begin{center}
  \includegraphics[scale=0.6]{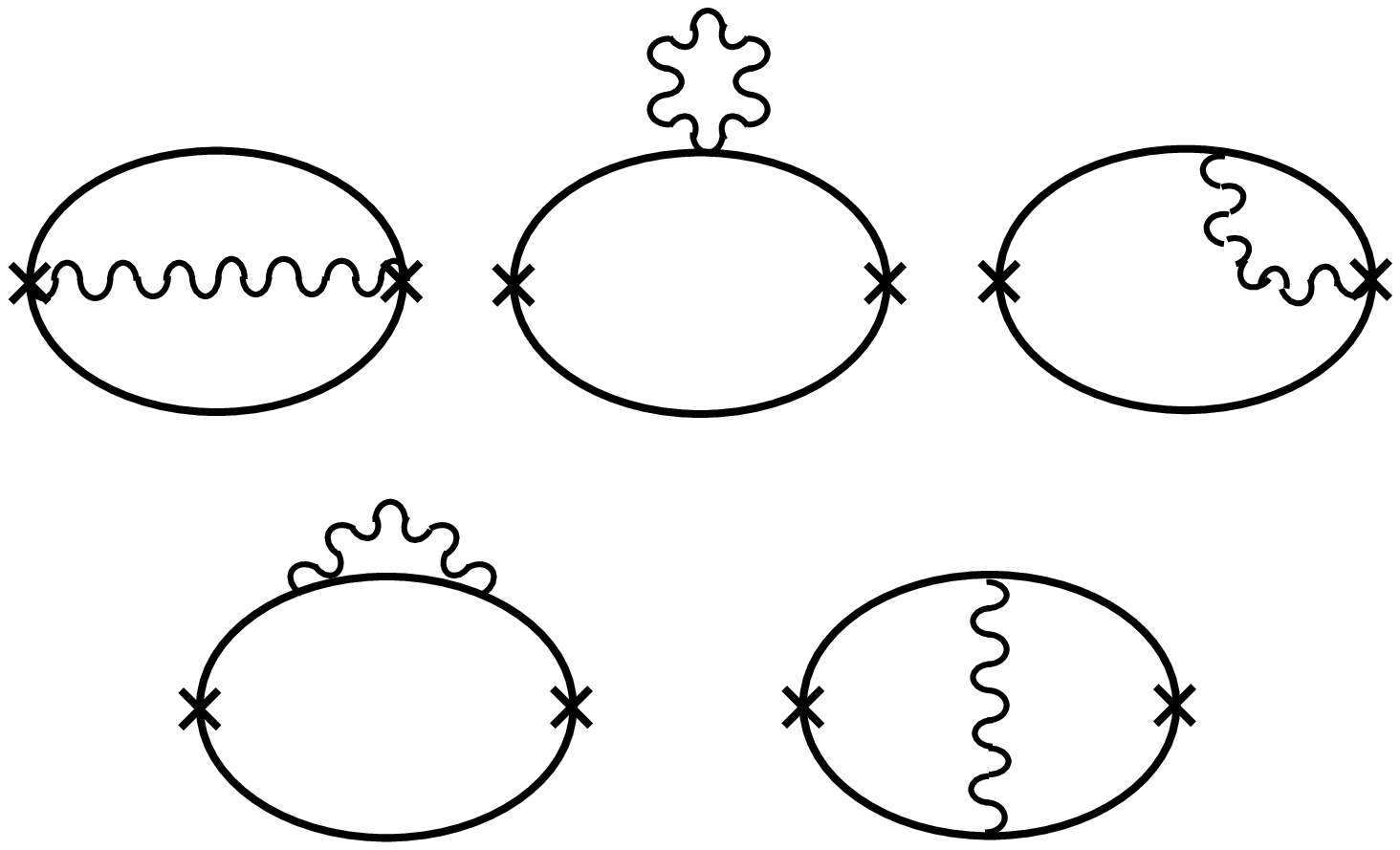}
\caption{The contributions to the Boltzmann conductivity 
in second order with 
respect to the domain wall. The interactions with the wall are 
denoted by wavy 
lines. The current vertex expressed by crosses represents $J_{0}$ and 
the cross with wavy line represents $\delta J$.
\label{FIGdiag_kubo}}
  \end{center}
\end{figure}

Let us include the scattering by the domain wall.
The first-order contribution of $A_{q}$ vanishes.
The second-order contributions to the Boltzmann conductivity
 are shown in Fig. \ref{FIGdiag_kubo}.
The process $Q_{1}$ arises from the correction of both of the two current 
vertices by the wall, $\delta J$, and $Q_{3}$ is due to the 
correction of 
one of the current vertices and a interaction with the wall.  
$Q_{2}$ and $Q_{4}$ are the self-energy due to the wall and $Q_{5}$ 
is the vertex correction.
We here consider a domain wall with $\phi=0$, i.e., 
$\Av_\mu=\hf(\partial_\mu\theta) \evy$.
The contributions in Fig. \ref{FIGdiag_kubo} are written as 
(we write $A^y_{q}\equiv A^y_{\xw}(q)$)
\begin{eqnarray}
Q_{1}&=& 
-\left( \frac{e\hbar}{m} \right)^{2}\frac{1}{4}
\frac{1}{\beta}\sum_{n}\frac{1}{V}\sum_{\kv q\sigma}
|A^y_{q}|^{2}G_{\kv-\frac{q}{2},n,\sigma}
G_{\kv+\frac{q}{2},n+\ell,-\sigma}
\nonumber\\
Q_{2}&=& 
-\left( \frac{e\hbar}{m} \right)^{2}\frac{\hbar^{2}}{8m}\frac{1}{\beta}\sum_{n}
\frac{1}{V}\sum_{\kv q\sigma }
k_{z}^{2}|A^y_{q}|^{2}
[(
G_{\kv,n,\sigma})^{2}G_{\kv,n+\ell,\sigma}+
G_{\kv,n,\sigma}(G_{\kv,n+\ell,\sigma})^{2}
]
\nonumber\\
Q_{3}&=& 
-\left( \frac{e\hbar}{m} \right)^{2}\frac{\hbar^{2}}{2m}\frac{1}{\beta}\sum_{n}
\frac{1}{V}\sum_{\kv q\sigma}
k_{z}\left(k_{z}-\frac{q}{2}\right)|A^y_{q}|^{2}
\nonumber\\
&&
\times
[
G_{\kv-\frac{q}{2},n,\sigma}
G_{\kv-\frac{q}{2},n+\ell,\sigma}
G_{\kv+\frac{q}{2},n,-\sigma}
+
G_{\kv-\frac{q}{2},n,\sigma}
G_{\kv-\frac{q}{2},n+\ell,\sigma}
G_{\kv+\frac{q}{2},n+\ell,-\sigma}
]
\nonumber\\
Q_{4}&=& 
-\left( \frac{e\hbar}{m} \right)^{2}\frac{\hbar^{4}}{4m^{2}}
\frac{1}{\beta}\sum_{n}
\frac{1}{V}\sum_{\kv q\sigma }
k_{z}^{2}\left(k_{z}-\frac{q}{2}\right)^{2} |A^y_{q}|^{2}
\nonumber\\
&&
\times
[(G_{\kv-\frac{q}{2},n,\sigma})^{2}
G_{\kv-\frac{q}{2},n+\ell,\sigma}
G_{\kv+\frac{q}{2},n,-\sigma}  
 +
G_{\kv-\frac{q}{2},n,\sigma}
(G_{\kv-\frac{q}{2},n+\ell,\sigma})^{2}
G_{\kv+\frac{q}{2},n+\ell,-\sigma}]
\nonumber\\
Q_{5}&=& 
-\left( \frac{e\hbar}{m} \right)^{2}\frac{\hbar^{4}}{4m^{2}}\frac{1}{\beta}\sum_{n}
\frac{1}{V}\sum_{\kv q\sigma }
k_{z}^{2}\left(k_{z}^{2}-\frac{q^{2}}{4}\right) |A^y_{q}|^{2}
\nonumber\\
&&
\times
G_{\kv-\frac{q}{2},n,\sigma}
G_{\kv-\frac{q}{2},n+\ell,\sigma}
G_{\kv+\frac{q}{2},n,-\sigma}
G_{\kv+\frac{q}{2},n+\ell,-\sigma} .
\end{eqnarray}

The contribution from the wall needs to vanish in the limit of 
the vanishing Zeeman splitting, $\spol\rightarrow 0$, since no 
scattering occurs there. 
This becomes obvious after summing the contribution 
$ Q_{1-5}\equiv \sum_{i=1,5}Q_{i}$.
Using 
\begin{eqnarray}
G_{\kv,n,\sigma}G_{\kv,n+\ell,\sigma}&=&
-i{\left(
  \omega_{\ell}+\frac{\hbar}{2\tau}({\rm sgn}(n+\ell)-{\rm sgn}(n))
                \right)^{-1}}
	(G_{\kv,n,\sigma}-G_{\kv,n+\ell,\sigma}) 
	\nonumber\\
G_{\kv+\frac{q}{2},n,-\sigma} G_{\kv-\frac{q}{2},n,\sigma}
&=& \left(\frac{\hbar^{2} k_{z}q}{m}+2\sigma \spol \right)^{-1}
 (G_{\kv+\frac{q}{2},n,-\sigma}-G_{\kv-\frac{q}{2},n,\sigma}),	
\end{eqnarray}
we obtain $Q_{1-5}=Q_{\rm c}+Q_{\rm c}'$, where \cite{GT01}
\begin{equation}
Q_{\rm c}(i\omega_{\ell})=
\frac{1}{2}\left(\frac{e\hbar\spol}{m}\right)^{2}
\frac{1}{\beta}\sum_{n}\frac{1}{V}\sum_{\kv q\sigma }
|A^y_{q}|^{2}
G_{\kv-\frac{q}{2},n,\sigma}
G_{\kv-\frac{q}{2},n+\ell,\sigma}
G_{\kv+\frac{q}{2},n,-\sigma}
G_{\kv+\frac{q}{2},n+\ell,-\sigma}
,  \label{QSUMRESULT}
\end{equation}
and 
\begin{equation}
Q_{\rm c}'(i\omega_{\ell})=
-\frac{1}{4}\left(\frac{e\hbar}{m}\right)^{2}
\frac{1}{\beta}\sum_{n}\frac{1}{V}\sum_{\kv q\sigma }
|A^y_{q}|^{2} 
\spol \frac{\spol-\sigma\frac{(k_{z}+q/2)^{2}}{m}} 
{\left[ \spol+\sigma\frac{(k_{z}+q/2)q}{2m} \right]^{2}} 
G_{\kv,n,\sigma} G_{\kv,n+\ell,\sigma}
.  \label{Qp}
\end{equation}
The term $Q_{\rm c}$ is dominant and the 
contribution from the term $Q_{\rm c}'$ turns out to cancel 
with the effect of the shift of the electron density, which is 
calculated later.

The summation over the Matsubara frequency, $\fomega_{n}$, in Eqs. (\ref{QSUMRESULT}) 
and (\ref{Qp}) can be carried 
out by use of contour integration (see \S \ref{APPomegasum})
and the contribution to the Boltzmann conductivity from the five 
classical processes,
$\sigma_{1-5}\equiv 
\lim_{\omega\rightarrow 0}{\rm Im}( Q_{1-5}(\omega+i0)- Q_{\rm 
c}(i0))/\omega$, 
is obtained as $\sigma_{1-5}=\sigma_{\rm c}+\sigma_{\rm c}'$, where
\begin{equation}
\sigma_{\rm c}=
-\frac{\spol^{2}\hbar^{3}}{8\pi\tau^{2}}\left(\frac{e\hbar}{m}\right)^{2}
\frac{1}{V}\sum_{\kv q\sigma }|A^y_{q}|^{2}
\frac{(\epsilon_{\kv-\frac{q}{2},\sigma}
       +\epsilon_{\kv+\frac{q}{2},-\sigma})^{2}}   
   {\left[(\epsilon_{\kv-\frac{q}{2},\sigma})^{2}
         +\left(\frac{\hbar}{2\tau}\right)^{2}\right]^{2}
    \left[(\epsilon_{\kv+\frac{q}{2},-\sigma})^{2}
         +\left(\frac{\hbar}{2\tau}\right)^{2}\right]^{2}},
         \label{sigmacresult}
\end{equation}
and $\sigma_{\rm c}'$ is the contribution from $Q_{\rm c}'$;
\begin{equation}
\sigma_{\rm c}'= \frac{\hbar}{32\pi} \left( 
\frac{e\hbar}{m}\right)^{2} \frac{\spol}{\tau^{2}}
\frac{1}{V}\sum_{\kv q\sigma} 
\frac{|A^y_{q}|^{2}}
 {\left[\epsilon_{\kv\sigma}^{2}+\left(\frac{\hbar}{2\tau}\right)^{2}\right]^{2}}
 \frac{\spol-\sigma\frac{(k_{z}+q/2)^{2}}{m}} 
{ \left[\spol+\sigma\frac{(k_{z}+q/2)q}{2m} \right]^{2}}.
	\label{sigmap}
\end{equation}

Besides the processes in Fig. \ref{FIGdiag_kubo}, 
there is another contribution to the classical conductivity, which is due 
to the change of the electron density in the presence of a domain 
wall \cite{Brataas98,Brataas99}.
(In this paper the chemical potential of the system is fixed as the domain 
wall is introduced, considering a constant voltage. In 
\protect \cite{Brataas98,Brataas99}, on the other hand, the electron 
number is kept constant by shifting the chemical potential, which 
corresponds to a measurement under constant current.
The difference is small if $k_{F}\lambda\gg 1$.) 
The correction to the electron density due to the interaction 
with the gauge field is written as
\begin{equation}
	\delta n=\frac{\hbar^{2}}{4m}\frac{1}{\beta V}\sum_{\kv q n\sigma} 
|A^y_{q}|^{2} \left[ \frac{1}{2}(G_{\kv n\sigma})^{2} 
+\frac{\hbar^{2} k_{z}^{2}}{m}
 (G_{\kv-\frac{q}{2}, n\sigma})^{2} G_{\kv+\frac{q}{2}, n,-\sigma} 
 \right].
	\label{delndef}
\end{equation}
After some calculation it reduces to 
\begin{equation}
\delta n= -\frac{\hbar^{3}}{16\pi m\tau }\frac{1}{V}\sum_{\kv q\sigma} 
\frac{|A^y_{q}|^{2}}
 {\epsilon_{\kv\sigma}^{2}+\left(\frac{\hbar}{2\tau}\right)^{2}}
 \frac{\spol-\sigma\frac{(k_{z}+q/2)k_{z}}{m}} 
{\spol+\sigma\frac{(k_{z}+q/2)q}{2m} }.
	\label{deln}
\end{equation}
This shift of the electron density leads to a correction of the 
zeroth-order conductivity, $\sigma_{0}\rightarrow 
\sigma_{0}+\delta\sigma_{\rm c}$, where $\delta\sigma_{\rm c}=
e^{2}\tau\delta n /m$ 
is obtained from Eq. (\ref{deln}) as
\begin{equation}
\delta \sigma_{\rm c}= -\frac{\hbar}{16\pi} \left( 
\frac{e\hbar}{m}\right)^{2} \frac{1}{V}\sum_{\kv q\sigma} 
\frac{|A^y_{q}|^{2}}
 {\epsilon_{\kv\sigma}^{2}+\left(\frac{\hbar}{2\tau}\right)^{2}}
 \frac{\spol-\sigma\frac{(k_{z}+q/2)k_{z}}{m}} 
{\spol+\sigma\frac{(k_{z}+q/2)q}{2m} }.
	\label{delsigma}
\end{equation}
It turns out after the $\kv$-summation that 
$\sigma_{\rm c}'+\delta\sigma_{\rm c}$ vanishes in the case of 
$k_{F}\lambda\gg1$ \cite{GT01}.
Thus the classical correction to the conductivity due to the wall is given simply by $\sigma_{\rm c}$.

To proceed further we neglect quantities of $O((q/k_{F})^{2})$ and 
approximate $\epsilon_{ \kv\pm q/2,\mp\sigma}\simeq 
\epsilon_{\kv}\pm[(\hbar^{2}  k_{z}q/2m)+\sigma\spol]$.
This is because the momentum transfer, $q$, is limited to a small value of 
$q\lesssim \lambda^{-1}$ due to the 
form factor of the wall, $|A^y_{q}|^{2}\propto [\cosh(\pi q\lambda/2)]^{-2}$, 
and we are considering the 
case of a thick wall, $k_{F}\lambda \gg 1$.
The result of $\kv$-summation is (see \S \ref{APPomegasum}) 
\begin{eqnarray}
	  \sigma_{\rm c}&=&
	-\frac{e^{2}\spol^{2}\tau^{2}}{8\pi\hbar^{3}}n_{\rm w}\sum_{\sigma} 
  \int_{-\infty}^{\infty}\frac{dx}{x}\frac{1}{\cosh^{2}x} 
  \nonumber\\
 && \times
\left[ \tan^{-1} 
\left(\frac{2l_{\sigma}}{\pi \lambda}x+ 2\spol
\frac{\tau}{\hbar}\right)
+
\tan^{-1} 
\left(\frac{2l_{\sigma}}{\pi \lambda}x-2\spol
\frac{\tau}{\hbar}\right)
\right],
	\label{SIGMAC}
\end{eqnarray}
where $x\equiv \pi q \lambda/2$ and
$l_{\sigma}\equiv \hbar k_{F\sigma}\tau/m$ is the mean free path of the 
electron with spin $\sigma$.
We have introduced a density of the wall, $n_{\rm w}\equiv 1/L$ (we have 
one wall in the sample of length $L$).
The extension of the result to the many-wall case is straight-forward 
as long as the distance between walls is larger than $\lambda$.
Using this result the contribution of the wall to the Boltzmann resistivity is 
given  as
\begin{equation}
	\rho_{\rm c}\equiv (\sigma_{0}+\sigma_{\rm c})^{-1} -\sigma_{0}^{-1} \simeq 
	\frac{|\sigma_{\rm c}|}{\sigma_{0}^{2}}.
	\label{rhocdef}
\end{equation}
The last expression is valid if the contribution from the impurity is much 
larger than that from the wall, namely if $|\sigma_{\rm c}|/\sigma_{0} \ll 1$.
(But see also the discussion at the end of this section, before Eq.  (\ref{rhow}).)

Here we consider the case of a ferromagnet with weak disorder, and 
assume two further conditions: 
\begin{equation}
\spol\tau/\hbar \gg1,  \label{condition1}
\end{equation}
which indicates that the effect of the Zeeman splitting is not smeared by 
the width of energy level, and 
\begin{equation}
m\spol\lambda/k_{F}\hbar^{2} \gg 1. \label{condition2}
\end{equation}
The second condition is satisfied if the Zeeman splitting is not too 
small. 
Both inequalities would be satisfied in the case of $d$ electron. 
In this case the classical correction due to the wall is obtained as 
(Eq. (\ref{sigmacferro}))
\begin{eqnarray}
	\sigma_{\rm  c} &\simeq& -\frac{e^{2}}{4\pi^{2}\hbar} n_{\rm w}
	\sum_{\sigma}
	\frac{l_{\sigma}}{\lambda} \int_{0}^{\infty} \frac{dx}{\cosh^{2}x}
	= -\frac{e^{2}}{4\pi^{2}\hbar}n_{\rm w}\sum_{\sigma}
	\frac{l_{\sigma}}{\lambda}
	\nonumber\\
	&&
		(\spol\tau/\hbar\gg 1, m\spol\lambda/k_{F}\hbar^{2} \gg 1).
	\label{SIGMACFERRO0}
\end{eqnarray}

Although the last equality of Eq. (\ref{rhocdef}) is true only when
$|\sigma_{\rm c}|/\sigma_{0}\ll1$, the result of Eq. (\ref{rhocdef})
with $\sigma_{\rm c}$ calculated from the Kubo formula (\ref{SIGMAC})has a 
finite limiting value at $\tau\rightarrow \infty$ of
\begin{equation}
	\frac{|\sigma_{\rm c}|}{\sigma_{0}^{2}}=
\frac{m^{2} \Delta^{2}}{4e^{2}n^{2}\hbar^{3}}n_{\rm w} \sum_{\sigma}
 \int_{\Lambda_{c}(\sigma)}^{\infty} \frac{dx}{x} 
 \frac{1}{\cosh^{2}x} \;\;\; (\tau\rightarrow \infty) ,
\end{equation}
which is identical with the result of the Mori formula, 
(\ref{rhocclean}), and thus is correct in this limit.
Furthermore if we assume a finite lifetime for the electron 
in the Mori formula (\ref{chimori}), which is without justification,
we obtain 
\begin{equation}
\rho_{\rm c}=
\left({e n}\right)^{-2}\frac{\hbar^{3}}{4\pi}
\left(\frac{\hbar \Delta}{\tau}\right)^{2}
\frac{1}{V}\sum_{\kv q\sigma }|A_{q}|^{2}
\frac{1}
{\left[\epsilon_{\kv-\frac{q}{2},\sigma}^{2}
         +\left(\frac{\hbar}{2\tau}\right)^{2}\right]
    \left[\epsilon_{\kv+\frac{q}{2},-\sigma}^{2}
         +\left(\frac{\hbar}{2\tau}\right)^{2}\right]}.
         \label{rhow}
\end{equation}
This is shown to be equivalent to Eqs.  (\ref{SIGMAC}) and 
(\ref{rhocdef}) by a similar calculation as in \S \ref{APPomegasum}.
These results may suggest that the expressions (\ref{SIGMAC}) and 
(\ref{rhocdef}), which are justified only for $|\sigma_{\rm 
c}|/\sigma_{0}\ll1$, are valid for any value of $\tau$.

\subsection{Spin chirality mechanism of Hall effect}

Here we will give an intuitive explanation for the Hall effect due to vorticity in terms of the Josephson effect of 
spin \cite{TK03,TG03,GT04}
.
A persistent current in a metallic ring 
is an equilibrium current that can be 
induced when the time-reversal symmetry is broken. 
 Such a current appears in the presence of a magnetic flux through 
a normal ring.
 The effect is due to a U(1) phase factor attached 
by the flux to the electron wave function. 
 Here we show theoretically that a permanent current is  
induced in a conducting normal ring just by attaching three 
ferromagnets, without magnetic flux through the ring \cite{TK03}. 
 This surprising effect can be seen  in nano-scales at low temperatures. 
 The key here is the non-commutativity of the SU(2) spin algebra, which
breaks the time-reversal symmetry, and leads, in the presence 
of electron coherence,  to a permanent electron current (spin Josephson effect \cite{TG03}).
We also show that this persistent current gives an intuitive explanation of the 
anomalous Hall effect due to the spin Berry phase in frustrated magnets \cite{TK02,TK03}.

\subsubsection{Spin chirality and time reversal symmetry breaking }

 The electron has spin 1/2 (i.e., has two components),  
and the spin obeys SU(2) algebra. 
 The algebra is represented by three 
$2\times 2$ Pauli matrices 
$\sigma_{i}$ 
($i=x,y,z$) satisfying the commutation relation 
\begin{equation}
[\sigma_i,\sigma_j]
= 2i \epsilon_{ijk}\sigma_{k}, \label{sigcom}
\end{equation}
where $\epsilon_{ijk}$ is the totally antisymmetric tensor with 
$\epsilon_{xyz}=1$. 
 When a conduction electron in a conductor is scattered 
by some magnetic object,
the electron wave function is multiplied by an amplitude 
$A(\nv) = \alpha e^{i\beta\nv\cdot\sigmav}
=\alpha(\cos\beta+i(\nv\cdot\sigmav)\sin\beta)$, 
which is generally spin-dependent and 
is represented by 
a 2$\times$2 matrix in spin space. 
Here $\alpha$ and $\beta$ are complex numbers and $\nv$ is a 
three-component unit vector representing the magnetization direction. 
 We consider in this paper only classical, static scattering objects, 
and assume that the $\nv$'s are constant vectors. 

\begin{figure}[tbh]
\begin{center}
\includegraphics[scale=0.5]{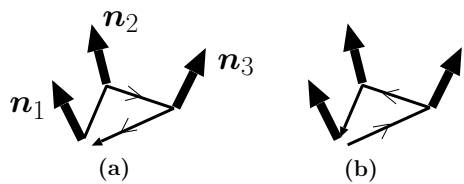} 
\end{center}
\caption{ A closed path contributing to the 
amplitude of the electron propagation from $x$ to $x$.
 At $X_i$, the electron experiences a scattering represented by an 
SU(2) amplitude, $A(\nv_i)$. 
 The contributions from one 
path (a) and the reversed one (b) 
are different in general due to the non-commutativity of $A(\nv_i)$'s.
\label{FIGahe}}
\end{figure}
 Let us consider two successive scattering events 
represented by $A(\nv_1)$ and $A(\nv_2)$ (Fig. \ref{FIGahe}). 
 Due to the non-commutativity of $\sigma_i$, the amplitude depends on the 
order of the scattering events: $A(\nv_1)A(\nv_2) \neq A(\nv_2)A(\nv_1)$ 
in general. 
 Various features in spin transport, which are under intensive pursuit 
recently, arise from this non-commutativity. 
Actually, 
\begin{eqnarray}
A(\nv_1)A(\nv_2) &=& \alpha^2 (\cos^2\beta-(\nv_1\cdot\nv_2)\sin^2\beta
+i\sin\beta\cos\beta(\nv_1+\nv_2)\cdot\sigmav \nonumber\\
&&
 -i\sin^2\beta(\nv_1\times\nv_2)\cdot\sigmav ),
\end{eqnarray}
has an asymmetric part,  
\begin{eqnarray}
A(\nv_1)A(\nv_2)-A(\nv_2)A(\nv_1) &=& 
-2i\alpha^2\sin^2\beta(\nv_1\times\nv_2)\cdot\sigmav .
\end{eqnarray}
This relation indicates that the spin current with polarization $(\nv_1\times \nv_2)$ flows between two spins, i.e., 
\begin{eqnarray}
\jsv^\alpha &\propto& 
(\nv_1\times\nv_2)^\alpha \ev_{12} , \label{js12}
\end{eqnarray}
where $\ev_{12}$ represents a vector connecting two localized spins.
This spin current is spontaneous (equilibrium), and represents an exchange interaction between $\nv_1$ and $\nv_2$ transmitted by the conduction electron \cite{Caroli71}.
In fact, when the spin current given by \Eqref{js12} flows, spin accumulation $\propto(\nv_1\times\nv_2)$ accumulates on $\nv_1$ and $\nv_2$, which rotates $\nv_1$ and $\nv_2$ within the $\nv_1$-$\nv_2$ plane until the two spins become parallel or anti-parallel to each other. 
In the case of smooth spin structures, the equilibrium spin current, \Eqref{js12}, reduces to  \cite{Takeuchi08} 
\begin{eqnarray}
{\js}^\alpha _{\mu} &\propto& 
(\nv \times\nabla_\mu \nv)^\alpha. \label{js12cont}
\end{eqnarray}

This non-commutativity of two spins does not affect the charge transport, 
since the charge is given as a sum of the two spin components 
(denoted by tr), and 
$\tr[A(\nv_1)A(\nv_2)] - \tr[A(\nv_2)A(\nv_1)] = 0$. 
 An anomaly in the charge transport arises at the third order.
 We have, by virtue
of Eq.(\ref{sigcom}) and the relation 
$\tr[\sigma_i \sigma_j] = 2\delta_{ij}$,
\begin{equation}
\tr[A(\nv_1)A(\nv_2)A(\nv_3)] - \tr[A(\nv_3)A(\nv_2)A(\nv_1)]
  = 4 \alpha^3 \sin^3\beta \nv_1 \cdot (\nv_2\times\nv_3) 
  \equiv iC_{123} .  \label{Acom}
\end{equation}
 This relation indicates that in the presence of fixed $\nv_i$'s with
 $ \nv_1 \cdot (\nv_2\times\nv_3) \neq0$, the symmetry under time-reversal 
(more appropriately, reversal of motion) is generally broken in the 
charge transport.  
In fact, the relation (\ref{Acom}) indicates that the 
contribution from one path, $x\ra X_1\ra X_2\ra X_3\ra x$ (Fig.\ref{FIGahe}a), 
and its (time-) reversed one, $x\ra X_3\ra X_2\ra X_1\ra x$ (Fig.\ref{FIGahe}b), 
are not equal, and this difference results in a spontaneous electron 
motion in a direction specified by the sign of $C_{123}$, 
namely, a permanent  current. 
 What is essential here is the non-commutativity of the SU(2) algebra. 
 In fact, $C_{123}$ vanishes if all $\nv_i$'s lie in a plane, 
in which case the algebra is reduced to a commutative U(1) algebra. 
 The degree of the symmetry breaking, $\nv_1 \cdot (\nv_2\times\nv_3)$, 
is given by the non-coplanarity, often called spin chirality.

\subsubsection{Charge current}
The spontaneous current above would be realized 
on a small conducting ring with three ferromagnets or magnetic dots 
with different magnetization directions, 
$\Sv_1$, $\Sv_2$, and $\Sv_3$  \cite{TK03}. 
The electron in a ring feels an effective spin polarization 
when it goes through the region ($F_i$) affected by the 
ferromagnets, and the effect will be modeled by the exchange 
(spin-dependent) potential,  
$V(x) = - \spol \, \nv_i \!\cdot\! \sigmav$ for $x \in F_i$. 
 Here $\spol$ represents the effective exchange field. 
 The equilibrium charge current in the ring is calculated from 
\begin{equation}
 j(x) = \frac{\hbar e}{2m}{\rm Im} (\nabla_{x}-\nabla_{x'}) \tr 
         G(x,x',\tau=0-)|_{x'=x}, 
\end{equation}
where
$G(x,x',\tau)\equiv-<T c(x,\tau)c^\dagger(x',0)>$ is the thermal Green's
function, $e, m, c$ being the charge, mass, and 
annihilation operator of electrons in the imaginary time, 
respectively.
$G(x,x',\tau)$ is calculated perturbatively from the Dyson equation, $G=g+gVG$, 
where $g$ represents free Green function. 
As is seen from Eq. (\ref{Acom}), a possible finite current arises at 
the third order in $V$. 
 By summing the contribution of the two paths, 
$x\ra X_1\ra X_2\ra X_3\ra x$ and the reversed one, we have 
$  j(x) = -\frac{\hbar e}{m} B(x) \, {\rm Re} \, C_{123} $. 
 Here $C_{123}$ is defined by Eq. (\ref{Acom}) with $\alpha=i\spol$, $\beta=\pi/2$,  
and 
\begin{equation} 
B(x) =  \prod_{i=1}^{3}\int_{X_i \in F_i} dX_i
   \int \frac{d\omega}{2\pi} 
   f(\omega) \left. \nabla_{X_0} {\rm Im} 
   [ g_{01} g_{12} g_{23} g_{34}] \right|_{X_4 = X_0=x} \label{B}
\end{equation} 
describes the electron propagation through the ring, 
which is common to both paths. 
 In Eq. (\ref{B}), 
$f(\omega)$ is the Fermi distribution function and 
$g_{ij} = g^r(X_i-X_j,\omega )$ is the retarded Green's function of 
free electrons.  
 Approximating the transport along the ring as one-dimensional and 
neglecting multiple circulation, we have 
$g^r(x, \omega ) \simeq -i\pi(D/L)e^{ik_F|x|}$, 
where 
$k_F$ is the Fermi wavenumber, 
$D$ the density of states ($\sim 1/\epsilon_F$ ;  
$\epsilon_F = \hbar^2 k_F^2/2m$ being the Fermi energy), 
and $L$ the length of the ring perimeter. 
 The final result is given by 
\begin{equation}
  j = -2e \frac{v_F}{L} 
  \cos(k_F L) 
       \left( \frac{J}{\epsilon_F} \right)^3 
       \Sv_1 \cdot (\Sv_2\times\Sv_3), \label{jres}
\end{equation}
at zero temperature. 
 Here $J \equiv \pi W \spol / L$ with $W$ being the width of the ferromagnets,
and $v_F=\hbar k_F/m$ is the Fermi velocity.

 The current is thus induced by the spin chirality 
$\Sv_1\cdot(\Sv_2\times\Sv_3)$ of the ferromagnets. 
 This quantity reduces to the Pontryagin index (density)
for the case of a smoothly varying field $\Sv(x)$,
which is also interpreted as the Berry phase of the 
spin. 
 The effect of the spin Berry phase on electron transport 
has so far been investigated in the limit of strong coupling 
to $\Sv (x)$ where the electron spin adiabatically follows 
$\Sv (x)$  \cite{Loss90}. 
 In contrast, the present result Eq. (\ref{jres}) is obtained in the
 opposite limit; 
we have treated the coupling to $\Sv$ perturbatively (weak 
coupling) and 
made no assumption of smoothness on $\Sv (x)$.

The appearance of the current is due to the symmetry breaking of the
charge (U(1)) sector, as in the case of the current in 
Josephson junction.
But note that here the U(1) symmetry breaking was due to the
non-commutativity of spin (SU(2)) sector
("spin Josephson effect" \cite{TG03}).

The system of a persistent current arising from the spin chirality works as a novel quantum operation gate, where spin Qbits and flux (current) Qbit are combined \cite{TG03}.

\subsubsection{Anomalous Hall effect}

\begin{figure}[tbh]
\begin{center}
\includegraphics[scale=0.4]{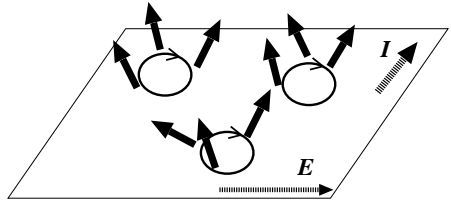} 
\end{center}
\caption{ Schematic picture 
of the chirality-induced Hall effect. 
 Circulating permanent currents are induced local around spin chirality, and this circular current 
drift in the perpendicular direction when an electric field is applied. 
\label{FIGahe_drift}}%
\end{figure}

The phenomenon predicted here is not restricted to artificial nano-structures, 
but will be present rather generally in metallic frustrated spin systems such 
as pyrochlore ferromagnets \cite{Taguchi01,Fujita06} and spin glasses \cite{Kageyama03,Pureur04,Taniguchi04,Fabris06,Taniguchi07}, where finite spin chirality is 
often realized.
 The spin chirality was recently pointed out   
in the adiabatic limit to induce a peculiar anomalous Hall effect \cite{Ye99,Onoda03,Nagaosa06}.
 The present chirality-driven persistent current  \cite{TK03}
affords an intuitive interpretation. 
 The circulating current 
starts to drift when the electric field is applied, in the direction 
perpendicular to the electric field  \cite{Kittel86} 
(Fig. \ref{FIGahe_drift}), 
just as in the normal Hall effect. 
 With the frequency of the circulating motion, read from Eq. (\ref{jres}) as 
\begin{equation}
 \Omega \simeq 
 \frac{2\pi v_F}{L} \left( \frac{J}{\epsilon_F} \right)^3 
 \Sv_1 \cdot ( \Sv_2 \times \Sv_3 ),
\end{equation}
we may estimate the Hall conductivity 
by 
$
\sigma_{xy} = \sigma_0 \Omega\tau.
$
 Here $\sigma_0$ is the classical (Boltzmann) conductivity, 
$\tau$ is the elastic lifetime, 
and the dirty case $\Omega \tau \ll 1$ is assumed. 
 If the spin chirality is located uniformly on every triangle of size of 
inter-atomic distance (i.e., $\Sv_1 \cdot (\Sv_2\times\Sv_3)= \chi_0$ and 
$L\sim 1/k_F$), we have 
$\sigma_{xy}/\sigma_0 \simeq \chi_0 J^3 \tau/\epsilon_F^2$.
 This result agrees with the result based on the linear 
response theory  \cite{TK02,TK03}.

When the spin structure is slowly varying, spin chirality 
can be expanded as  
\begin{equation}
\Sv(\xv) \cdot (\Sv(\xv_1)\times\Sv(\xv_2))
\simeq \sum_{\mu\nu}(x_1-x)_\mu(x_2-x)_\nu
\Sv(\xv) \cdot (\partial_\mu \Sv(\xv)\times\partial_\nu \Sv(\xv)),
\end{equation}
and the spin chirality reduces to the spin Berry phase \cite{OTN04}.

So far, the chirality mechanism are confirmed in spin-glasses  \cite{Kageyama03,Pureur04,Taniguchi04,Fabris06,Taniguchi07}, but the case of Nd$_2$Mo$_2$O$_7$ seems to be under debate
\cite{Nagaosa06,Yasui06}.

\section{Current from magnetization dynamics}
\label{SEC:ishe}
In most parts of this paper, we have studied magnetization dynamics induced by an electric current. 
The opposite effect, conversion of magnetic information into electric information, also exists, and has quite recent developments. 
The electromotive force (voltage) induced as a counter action of spin-transfer torque was in fact pointed out by Berger in 1986 \cite{Berger86}.
Later, Stern showed in a general context that the fictitious field of the Berry phase ($\Phi_{S}$) accumulated by electron spin induces a spin-dependent electromotive force ($V$) via Faraday's law as  $V=-\dot{\Phi}_{S}$ \cite{Stern92}.
This argument was applied to the case of domain wall dynamics by Barnes and Maekawa \cite{Barnes07}.
Duine \cite{Duine08} argued that spin relaxation
($\betasf$) induces a new current contribution, proportional to $\betasf(\dot{\Sv}\cdot\nabla_i \Sv)$, where $\Sv$ is a localized spin and $i$ is the current direction.
The effect of a Rashba-type spin-orbit interaction was studied by
Ohe et al. \cite{OTT07,Takeuchi08} in the weak $s$-$d$ coupling limit.
The current in this case was found to be 
 $j_i \propto \lamso \epsilon_{ijz}\average{\Sv\times\dot{\Sv}}_j$, where
$\lamso$ is the strength of the spin-orbit interaction, $z$ is direction of Rashba field, 
 and $\average{\cdots}$ denotes averaging over electron motion.
The quantity $\average{\Sv\times\dot{\Sv}}$ represents a spin damping and is related phenomenologically to a spin current across the interface in the case of junctions
 \cite{Tserkovnyak02,Tserkovnyak05}.

Such a spin-dynamics-induced current was experimentally observed in systems where the spin-orbit interaction is strong.
Saitoh et al. \cite{SUMT06} demonstrated that when the spin-orbit interaction is strong, an inverse process of the spin Hall effect would occur.
Namely, the spin current pumped by spin dynamnics \cite{Tserkovnyak02,Tserkovnyak05} would be converted into charge current by spin-orbit interaction.
They confirmed this idea by observing charge current in a Pt layer attached to a ferromagnet with dynamical magnetization.
The induced current was perpendicular to the spin current flow and $\average{\Sv\times\dot{\Sv}}$,
consistent with the inverse spin Hall mechanism, and with the  theory of Refs.  \cite{OTT07,Takeuchi08} at least at the qualitative level. 
Experimental confirmation of electromotive force due to the pure spin Berry phase without spin-orbit interaction is challenging and interesting.
Closer studies on these phenomena will lead to a unified picture of the interplay 
between electric/spin current and magnetization.

\section{Summary}

We have reviewed the theoretical aspects of current-driven domain wall motion, 
including microscopic derivation of the equation of motion, wall dynamics, and brief discussion of experimental results. 
The effect of current arises from the $s$-$d$ exchange coupling between the localized spin and conduction electrons. 
 Treating the non-adiabaticity perturbatively, we have derived fully quantum mechanical expressions for torques and forces acting on the wall in terms of Green's functions.
The obtained torques are summarized in Table \ref{table:st}.
With these results, we derived the equation of motion of the wall.
The wall is assumed to be rigid and planar (one-dimensional), described by two collective coordinates, position $X$ and angle $\phiz$ of magnetization out of the easy plane.
Spin-transfer torque reflecting angular-momentum conservation was shown to contribute to wall velocity, and  spin relaxation and non-adiabaticity were shown to produce forces on the wall, which induce $\dot\phiz$. 
Solving the equation of motion, we found that there is a threshold current for driving the wall arising from hard-axis magnetic anisotropy energy $\Kp$ and/or extrinsic pinning potential $\Vz$. 
The threshold current is determined by $\Kp$ in the intrinsic pinning regime.
In the extrinsic pinning regime, it is determined by $\Vz$, $\Kp$, and the force from the current (represented by $\betaw$).

\begin{table}[htb]
\begin{center}
\caption{
Summary of torques induced by electrons.
The solid lines represent the conduction electron Green's functions, 
the wavy lines are the gauge fields representing the domain wall,
the dotted lines are applied electric fields, and 
the double-dotted lines represent spin relaxation due to spin-flip scattering by magnetic impurities.
The Green's functions (thick solid lines) in the contributions to the $\beta$ terms contain lifetimes due to spin-flip impurities.
\label{table:st}}

\vspace{5mm}

{\renewcommand{\arraystretch}{2.0}
\begin{tabular}{ccc} 
\hline\hline 
\multicolumn{3}{c}{local contributions} \\
\hline
   &  torque &  \\
\hline
  \includegraphics[scale=0.15]{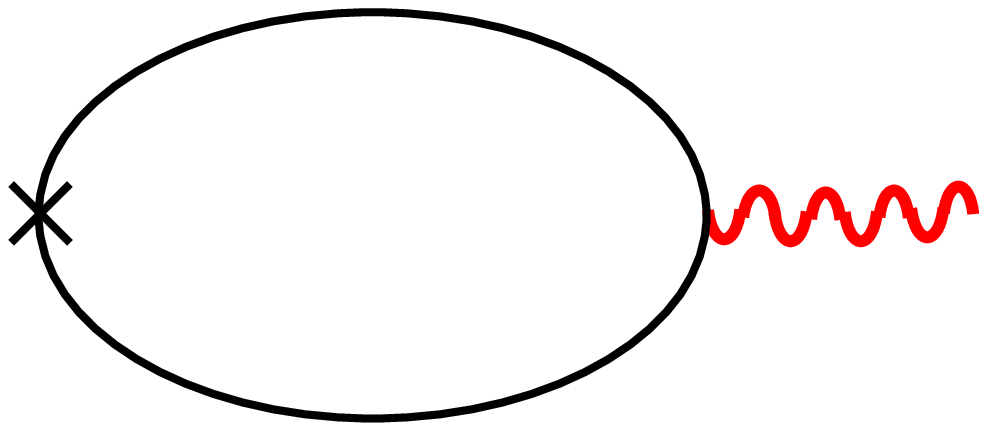} 
 & $\se\dot{\nv}$ & spin renormalization \\
\hline
{\includegraphics[scale=0.15]{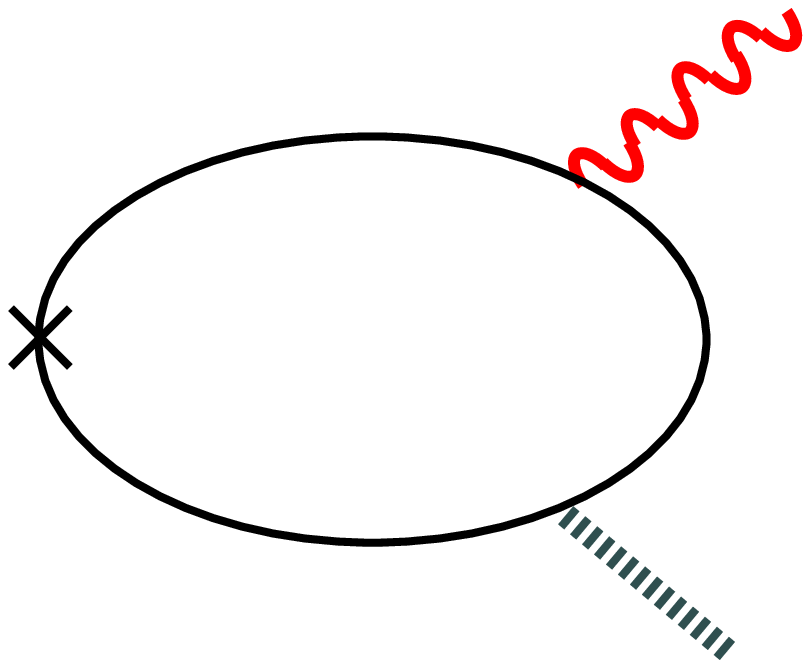} }
\raisebox{1ex}{  \includegraphics[scale=0.15]{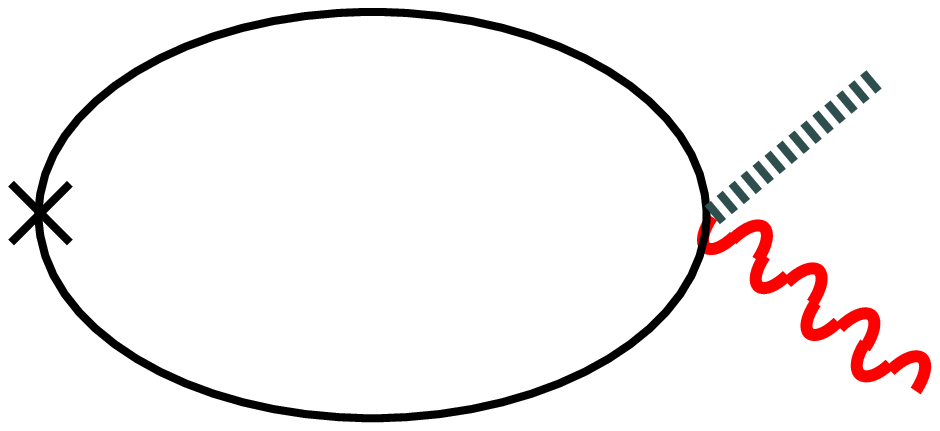}}
  & $\jsv\cdot\nabla \nv$ &  spin-transfer torque \\ 
\hline
  \includegraphics[scale=0.15]{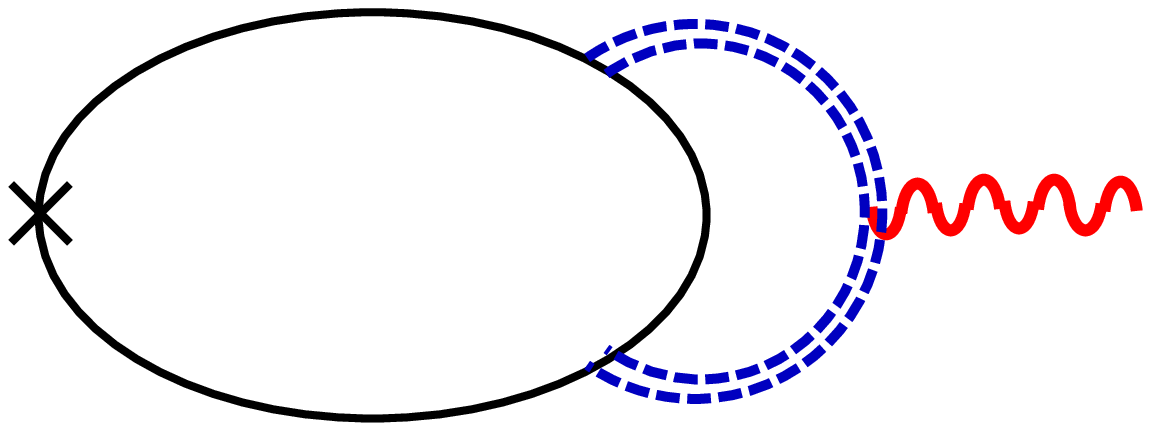}
  & $\alphasf\nv\times\dot\nv$ &  Gilbert damping \\ 
\hline
{\includegraphics[scale=0.15]{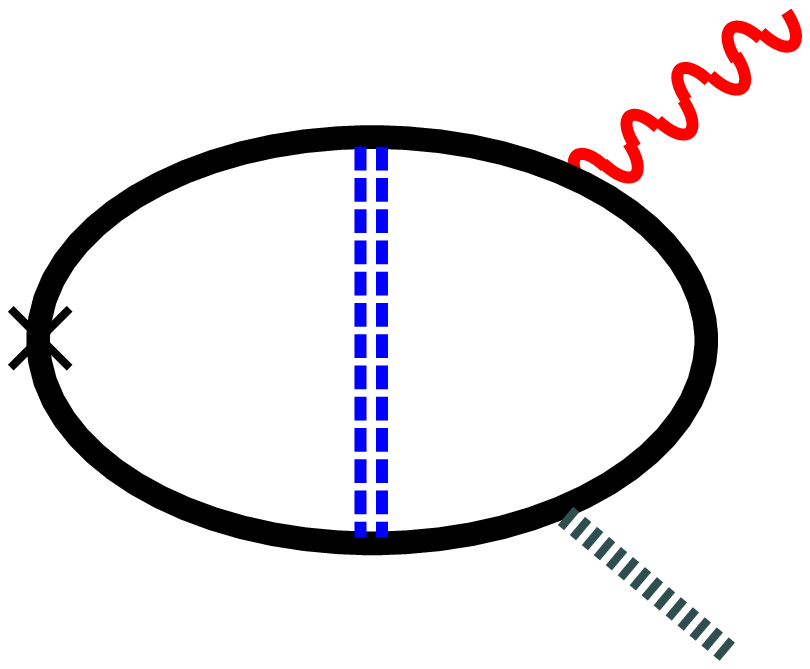}}
\raisebox{1ex}{  \includegraphics[scale=0.15]{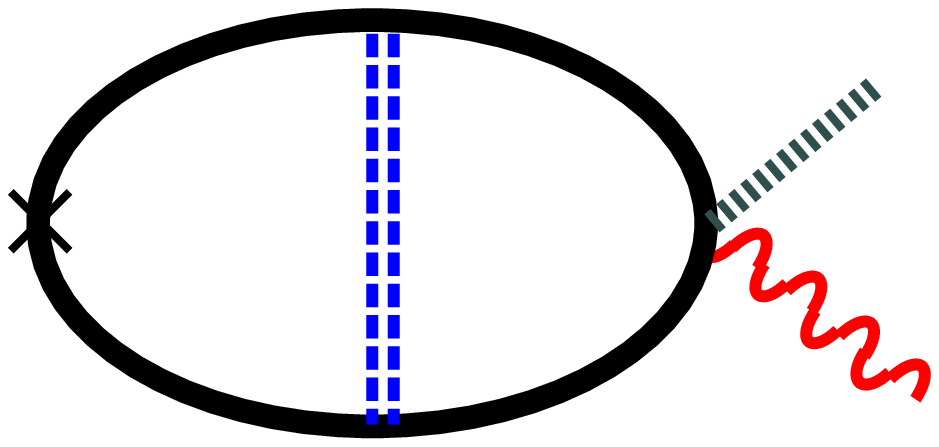}} 
{\includegraphics[scale=0.15]{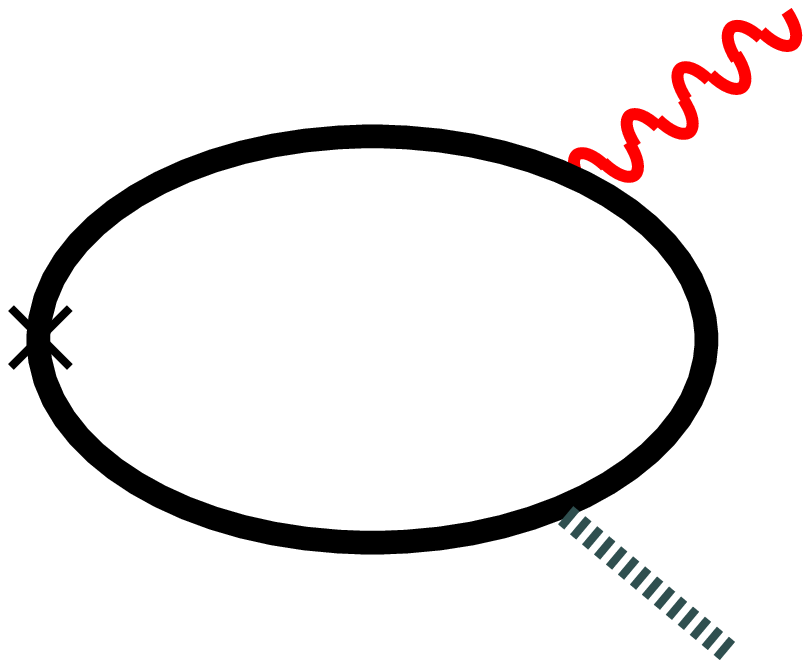}}
\raisebox{1ex}{\includegraphics[scale=0.15]{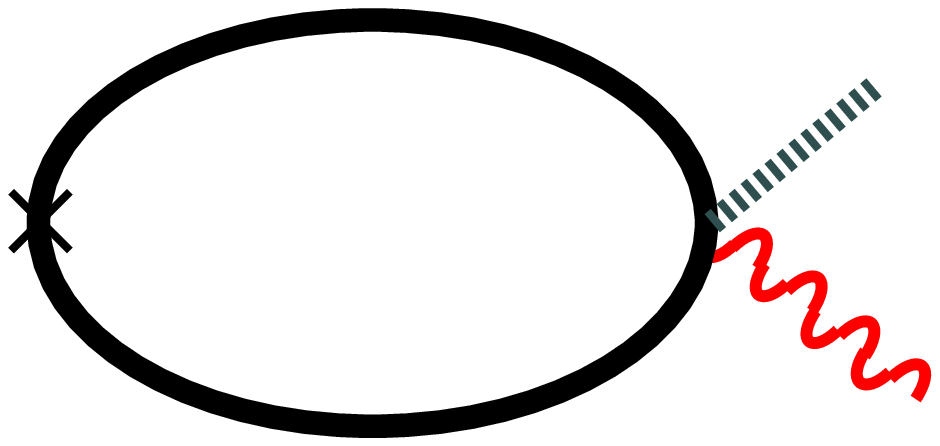}}
  & $\betasf (\nv\times(\jsv\cdot\nabla)\nv)$  & $\beta$ term \\ 
\hline\hline
\multicolumn{3}{c}{non-local contributions} \\
\hline
  \includegraphics[scale=0.15]{se1_1.eps} 
\raisebox{1ex}{  \includegraphics[scale=0.15]{se1_2.eps}}
  & $\Fna$ &  reflection force \\ 
\hline \hline 
\end{tabular}
}

\end{center}
\vspace{5mm}
\end{table}

Current-driven magnetization switching has an advantage in downsizing of the devices compared with the Amp\`ere's mechanism.
As we have seen above, the domain wall motion is governed by the applied current density and material parameters. 
In contrast, the field created by the Amp\`ere's law is 
proportional to the current, which is proportional to the area of the system. 
Therefore, the conventional Amp\`ere's mechanism requires high current density and becomes inefficient in small devices, while the efficienty of the current-induced mechanism is not reduced.
The same applies to the current pumped by the spin dynamics due to the inverse spin Hall effect, discussed in \S\ref{SEC:ishe}.
These two new magnetoelectric effects, based on quantum mechanical material features, have therefore great advantage in high density devices.

Our predictions are briefly summarized as follows.

The wall dynamics depends much on the behavior of $\phiz$.
For $\jtil\ll\jatil$ ($\jatil$ defined in \Eqref{jatildef}), $\phiz$ remains small and there is no intrinsic pinning effect, while
for $\jtil\gtrsim\jatil$, $\phiz$ develops and an intrinsic pinning effect arises due to $\Kp$. 
In most cases, the intrinsic pinning threshold seems rather high, and the extrinsic pinning regime would be suitable for low-current operation. 
This is realized when $\betaw$ is finite if the extrinsic pinning potential is much weaker than $\Kp$ 
(since, then, $\phiz\sim 0$ and the wall does not suffer from the intrinsic pinning effect).
In other words,  wall motion at low current would be realized in
systems with 
\begin{itemize}
\item 
very clean samples (with very few or weak pinning sites) and 
\item 
large $\betaw$ realized by large spin relaxation ($\betasf$) or thin wall ($\betana$)
\end{itemize}
Reduction of thresholds for thin walls seems consistent with experimental results \cite{Ravelosona05,Feigenson07}.

The wall speed is in most cases governed by the spin-transfer rate (determined by angular momentum conservation), and is equal to 
$\vw\sim \vs=\frac{a^3}{2eS}Pj$.
This speed is estimated to be $\vs=100$ m/s for $j=10^{12}\Ams$, 
which is high enough (corresponding to switching at 10 GHz if the device is 10 nm size), but
present experiments on metals are far below this limit.
Faster walls beyond this spin-transfer limit could be realized if $\betaw/\alpha$ is large (\Eqref{vwallslide}).

Although many studies, both theoretical and experimental, 
have been done so far, current-induced magnetization reversal has still exciting problems to be attacked.
The final aim is to achieve fast switching at low current density.
Roughly speaking, domain wall speeds need to be enhanced by a factor of 10 (so that $\vw\gtrsim 10$ m/s) and the threshold current density needs to be reduced by factor of $\frac{1}{10}$ (i.e., below $10^{11}\Ams$).
To do this,  material choice and structural design will be important. 
For instance, spin-orbit interaction is expected to lower the threshold current greatly by inducing effective force ($\betasf$), and this possibility should now be investigated quantitatively taking account of material details.
For this, a microscopic formalism as described in the present paper needs to be combined with first-principle calculations.
In addition, the effects of deformation and fluctuation (spin waves), the case of vortex walls, and details of the extrinsic pinning potential need to be studied.
Systematic studies for realizing efficient domain wall motion are now under way.

\noindent
{Acknowledgments}\\

The authors are grateful  to
T. Ono, Y. Yamaguchi, E. Saitoh, M. Yamanouchi, Y. Otani, 
K. Obata, A. Takeuchi, K. Hosono, K. Taguchi,  
M. Kl\"aui, Y. Nakatani, A. Thiaville, Y. Suzuki, A. H. MacDonald, J. Sinova, R. Duine, Y. Tserkovnyak, 
M. Stiles, S. Parkin,
K.-J. Lee, A. Brataas,  R. Egger, M. Thorwart, J. Ohe, 
J. Ieda, N. Garcia, Y. Le Maho, 
J. Inoue,  S. Barnes, S. Maekawa, and H. Miyajima  
for valuable discussions.
They are in particular grateful to professors H. Fukuyama, H. Ohno, and 
G. E. W. Bauer for continuous advice and encouragement.


\appendix

\section{Spin Lagrangian} 
\label{SEC:spinlag}

In this section, we derive the Lagrangian of a classical spin 
by employing the path integral for quantum spin.

\subsection{Spin coherent state}

Consider a spin $1/2$, which takes two states  
$| \uparrow \,\rangle$ and $| \downarrow \,\rangle$ 
pointing, respectively, parallel and antiparallel to the $z$-axis. 
 The $z$-axis here is not a special one, but could be chosen arbitrary. 
 So one should be able to construct a spin eigen state 
$| \, {\bm n} \, \rangle$ pointing in an arbitrary direction, 
\begin{equation}
 {\bm n} = (\, \sin\theta \cos\phi, \, \sin\theta \sin\phi, \, \cos\theta \,) . 
\label{eq:n}
\end{equation}
 This state is by definition an eigen state, with eigenvalue $+1$, 
of the spin operator projected in the direction of ${\bm n}$, 
and satisfies 
\begin{eqnarray}
 ({\bm n} \!\cdot\! {\bm \sigma}) | \, {\bm n} \, \rangle 
= | \, {\bm n} \, \rangle . 
\label{eq:coherent_eigen1}
\end{eqnarray}
 Here
\begin{eqnarray}
 {\bm n} \!\cdot\! {\bm \sigma}
= \left( \begin{array}{cc} 
           \cos\theta                 & {\rm e}^{-i\phi} \sin\theta \\ 
           {\rm e}^{i\phi} \sin\theta & - \cos\theta   \end{array} \right) 
\end{eqnarray}
is the Pauli spin matrix projected to ${\bm n}$. 
 The solution of the eigenvalue equation (\ref{eq:coherent_eigen1}) 
is given by 
\begin{eqnarray}
| \, {\bm n} \, \rangle 
\equiv | \, \theta, \, \phi \, \rangle 
\equiv {\rm e}^{-i\phi/2} \cos\frac{\theta}{2} \, | \uparrow \,\rangle 
  + {\rm e}^{i\phi/2} \sin\frac{\theta}{2} \, | \downarrow \,\rangle 
\label{eq:coherent_state_1}
\end{eqnarray}
up to an overall phase factor.
\footnote{
 This state is mutivalued as a function of ${\bm n}$ 
(though single-valued as a function of $\theta$ and $\phi$), 
so the notation 
$| \, {\bm n} \, \rangle$ may not be appropriate. 
 However, the quantity, 
$| \, {\bm n} \, \rangle \langle \, {\bm n} \, |$, 
which will be used below, is single-valued. 
 The same applies to the state, 
\begin{eqnarray}
| \, {\bm n} \, \rangle' 
\equiv | \, \theta, \, \phi \, \rangle' 
\equiv  \cos\frac{\theta}{2} \, | \uparrow \,\rangle 
  + {\rm e}^{i\phi} \sin\frac{\theta}{2} \, | \downarrow \,\rangle , 
\label{eq:coherent_state_2}
\end{eqnarray}
which differs from (\ref{eq:coherent_state_1}) by an overall phase factor 
and is used more often.  
 This is due to the spinor nature of the state, and does not mean any 
mathematical difficulty.} 
 The relative magnitude of the two coefficients is related to $\theta$, 
and the relative phase to $\phi$. 
 The state (\ref{eq:coherent_state_1}) is called spin coherent state 
\cite{Auerbach94,Nagaosa99}

Note that it does not satisfy the (three-component) vector identity, 
\begin{eqnarray}
  \hat{\bm \sigma} \, | \, {\bm n} \, \rangle 
\ne {\bm n} \, | \, {\bm n} \, \rangle , 
\label{eq:coherent_eigen2}
\end{eqnarray}
but only the (single-component) scalar identity, 
eq.(\ref{eq:coherent_eigen1}). 
 The expectation value, however, satisfies the vector identity, 
\begin{eqnarray}
  \langle \, {\bm n} \, | \, \hat{\bm \sigma} \, | \, {\bm n} \, \rangle 
= {\bm n} .
\label{eq:coherent_expt}
\end{eqnarray}

The following completeness relation 
(actually, they form an over-complete set) 
is easy to verify, 
\begin{eqnarray}
 \int \frac{d{\bm n}}{2\pi} | \, {\bm n} \, \rangle \langle \, {\bm n} \, | 
&\equiv& 
 \frac{1}{2\pi} 
 \int_0^\pi \sin\theta d\theta \int_0^{2\pi} d\phi \, 
 | \, \theta, \, \phi \, \rangle \langle \, \theta, \, \phi \, | 
\nonumber \\
&=& | \uparrow \, \rangle \langle \, \uparrow | 
  + | \downarrow \, \rangle \langle \, \downarrow | 
\nonumber \\
&=& 1 .
\label{eq:completeness}
\end{eqnarray}

The coherent state (\ref{eq:coherent_state_1}) for $S=1/2$ is obtained 
from the state $| \uparrow \, \rangle$ by two successive rotations as 
\begin{equation}
 | \, \theta, \, \phi \, \rangle 
= {\rm e}^{-i\frac{\phi}{2} \sigma_z} 
  {\rm e}^{-i\frac{\theta}{2} \sigma_y} 
 | \uparrow \, \rangle , 
\end{equation}
namely, first by $\theta$ around the $y$-axis, 
and second by $\phi$ around the $z$-axis.
 For general $S$, the spin coherent state is given by 
\begin{equation}
 | \, \theta, \, \phi \, \rangle_S  
= {\rm e}^{-i\phi \hat S_z} {\rm e}^{-i\theta \hat S_y} |S\, \rangle .
\label{eq:coherent_S}
\end{equation}
 Here, $\hat {\bm S}$ is a spin-$S$ operator, 
and the state $|m\, \rangle$ is defined by 
$\hat S_z \, |m\, \rangle = m\, |m\, \rangle$. 
 One can varify that  
$\langle \, {\bm n} \, | \, \hat{\bm S} \, |\, {\bm n} \, \rangle  = S {\bm n}$.

\noindent
\subsection{Path integral for spin} 

 In the formulation of quantum mechanics in terms of path integral, 
a central role is played by the Lagrangian of the system.

According to Feynman \cite{Feynman65}, the probability amplitude 
that a particle located at $x_0$ at time $t=0$ will be found 
at later time $t=T$ at position $x_{\rm f}$ is given by 
the sum of amplitudes over all possible paths $x(t)$ satisfying 
$x(t=0)=x_0$ and $x(t=T)=x_{\rm f}$ as 
\begin{equation}
 \langle x_{\rm f} | \, {\rm e}^{-iHT/\hbar} | x_0 \rangle 
= \int Dx(t) \, {\rm e}^{i {\cal S}[x(t)] /\hbar } . 
\end{equation}
 Here $H$ is the Hamiltonian of the particle, 
and the ${\cal S}[x(t)]$ is the action 
(time integral of Lagrangian). 
 The same expression holds for spin. 
 The probability amplitude that a spin in a state ${\bm n}_0$ 
at time $t=0$ will be found in a state ${\bm n}_{\rm f}$ 
at some later time $t=T$ is given by the sum of amplitudes 
over all possible paths ${\bm n}(t)$ satisfying 
${\bm n}(t=0)={\bm n}_0$ and ${\bm n}(t=T)={\bm n}_{\rm f}$ as 
\begin{equation}
 \langle {\bm n}_{\rm f} | \, {\rm e}^{-iHT/\hbar} | {\bm n}_0 \rangle 
= \int D{\bm n}(t) \, {\rm e}^{i {\cal S}[{\bm n}(t)] /\hbar } . 
\label{eq:path_spin}
\end{equation}
 Here  $H$ is the Hamiltonian for the spin, 
and ${\cal S}[{\bm n}(t)]$ is the action of spin. 
 Let us utilize this fact to calculate the Lagrangian for spin 
\cite{Auerbach94,Nagaosa99}.

Consider the quantity on the left-hand side of eq. (\ref{eq:path_spin}). 
 By dividing the time interval $T$ into many $(N \gg 1)$ tiny intervals 
$\varepsilon = T/N$, we first write it as 
\begin{eqnarray}
 \langle {\bm n}_{\rm f} \, | \, {\rm e}^{-iHT} 
 | \, {\bm n}_0 \, \rangle 
&=& \langle {\bm n}_{\rm f} \, | \, {\rm e}^{-iH\varepsilon} 
   {\rm e}^{-iH\varepsilon} \cdots {\rm e}^{-iH\varepsilon} \, 
 | \, {\bm n}_0 \, \rangle .
\end{eqnarray}
(For simplicity, we drop $\hbar$ for the time being.) 
 Next, we insert the completeness relation (\ref{eq:completeness}) 
between each neighboring factors ${\rm e}^{-iH\varepsilon}$ as 
\begin{eqnarray}
 \langle {\bm n}_{\rm f} \, | \, {\rm e}^{-iH\varepsilon} 
   {\rm e}^{-iH\varepsilon} \cdots {\rm e}^{-iH\varepsilon} \,
 | \, {\bm n}_0 \, \rangle 
&=& \int \frac{d{\bm n}_{N-1}}{2\pi} \cdots 
    \int \frac{d{\bm n}_2}{2\pi} 
    \int \frac{d{\bm n}_1}{2\pi} 
   \langle {\bm n}_{\rm f} \, | \, {\rm e}^{-iH\varepsilon} \, 
   | \, {\bm n}_{N-1} \, \rangle \cdots 
\nonumber \\ 
&{}&  \hskip 1.5cm \times 
    \langle {\bm n}_2 \, | \, {\rm e}^{-iH\varepsilon} \, 
   | \, {\bm n}_1 \, \rangle 
    \langle {\bm n}_1 \, | \, {\rm e}^{-iH\varepsilon} \, 
 | \, {\bm n}_0 \, \rangle 
\end{eqnarray}
 We have used the variable  ${\bm n}_k$ for the 
$k$-th position counted from the right. 
 We regard $t_k = k \, \varepsilon $ as a discretized time, 
and $|{\bm n}_k \rangle$ as a snapshot $|{\bm n} (t_k) \rangle$ 
of a path $|{\bm n} (t) \rangle$ defined on a continuous time $t$. 
 The $k$-th element is written, 
neglecting terms of ${\cal O}(\varepsilon^2)$ and writing $t_k \equiv t$, 
as 
\begin{eqnarray}
 \langle {\bm n}(t) \, | \, {\rm e}^{-iH\varepsilon} \, 
 | \, {\bm n}(t-\varepsilon ) \, \rangle 
&=&  \langle {\bm n}(t) \, | (1-iH \varepsilon ) 
 \left\{ 
 | \, {\bm n}(t) \, \rangle 
  - \varepsilon \frac{d}{dt} | \, {\bm n}(t) \, \rangle \right\} 
  + {\cal O}(\varepsilon^2) 
\nonumber \\
&=& 1 - \varepsilon \left\{ 
    \langle {\bm n}(t) \, | \frac{d}{dt} | \, {\bm n}(t) \, \rangle 
   +i \langle {\bm n}(t) \, | H | \, {\bm n}(t) \, \rangle 
   \right\} 
  + {\cal O}(\varepsilon^2) 
\nonumber \\
&\equiv& 1 + i \varepsilon  L[{\bm n}(t)] 
  + {\cal O}(\varepsilon^2) 
\nonumber \\
&=& {\rm exp} \left\{ i \varepsilon L[{\bm n}(t)] \right\} 
  + {\cal O}(\varepsilon^2) . 
\end{eqnarray}
 Here we have put
\begin{eqnarray}
  L[{\bm n}(t)] 
&=& i \hbar \langle {\bm n}(t) \, | \frac{d}{dt} | \, {\bm n}(t) \, \rangle 
   - \langle {\bm n}(t) \, | H | \, {\bm n}(t) \, \rangle . 
\label{eq:Lagrangian_1}
\end{eqnarray}
(We have recovered $\hbar$.)
 Therefore, we have 
\begin{equation}
 \langle {\bm n}_{\rm f} \, | \, {\rm e}^{-iHT/\hbar} \, 
 | \, {\bm n}_0 \, \rangle 
= \int \frac{d{\bm n}_{N-1}}{2\pi} \cdots 
    \int \frac{d{\bm n}_2}{2\pi} 
    \int \frac{d{\bm n}_1}{2\pi} \, 
 {\rm exp} \left\{i \, \frac{\varepsilon}{\hbar} 
    \sum_{k=1}^N L[{\bm n}(k\varepsilon)] \right\} 
  + {\cal O}(\varepsilon^2) . 
\end{equation}
 Taking the limit, 
$N \to \infty$, $\varepsilon \to 0$ (with $N\varepsilon = T =$ constant), 
the right-hand side is written as 
\begin{equation}
\int D{\bm n}(t) \, 
 {\rm exp} \left\{\frac{i}{\hbar} 
           \int_0^T dt' L[{\bm n}(t')] \right\} , 
\label{eq:spin_PI}
\end{equation}
expressing the sum over all possible trajectories ${\bm n}(t)$.

 The Lagrangian is thus given by $L$ of eq.(\ref{eq:Lagrangian_1}). 
 The first term on the right-hand side is known as the Berry phase 
\cite{Berry84,Sakurai94}.
 In terms of $\theta$ and $\phi$, we have 
\footnote{
 If we use the spin coherent state $|{\bm n}\rangle'$ 
of (\ref{eq:coherent_state_2}), 
the first term becomes 
$\frac{\hbar}{2} \, \dot\phi\, (\cos\theta -1)$. 
 The difference $\frac{\hbar}{2} \, \dot\phi$ is the total time 
derivative, and does not affect the classical equation of motion. 
}
\begin{eqnarray}
  L[{\bm n}] &\equiv& 
\frac{\hbar}{2} \, \dot\phi\, \cos\theta - H({\bm n}) . 
\label{eq:Lagrangian_2}
\end{eqnarray}
 Here 
$H({\bm n}) \equiv 
 \langle {\bm n} \, | H(\hat {\bm \sigma}) | \, {\bm n} \, \rangle$ 
is the Hamiltonian. 
 Thanks to eq.(10), spin operators in the (quantum) Hamiltonian 
can be replaced by the c-number counterparts. 
 Lagrangian (\ref{eq:Lagrangian_2}) is for $S=1/2$. 
 For general spin $S$, the first term of the Lagrangian 
becomes $\  \hbar S \dot\phi\, \cos\theta$, 
as verified with eq.(\ref{eq:coherent_S}).

\subsection{Lagrangian for classical spin}

In this subsection, we consider a classical spin of magnitude $S$. 
 The Lagrangian is given by 
\begin{eqnarray}
  L_S = \hbar S \dot \phi \, \cos \theta - H ({\bm n}) .
\label{eq:LS0}
\end{eqnarray}
 The first term on the right-hand side is the kinetic term,
\footnote{This has been known as the \lq\lq kinetic potential'' 
\cite{Hubert00}.} 
which is essential to lead to the equation of motion 
for angular momentum. 
 It has the same form as the spin Berry phase in quantum 
mechanics, but the concept of phase does not appear in classical mechanics 
considered here.  
(The appearance of $\hbar$ is due to the fact that we have written the 
angular momentum as $\hbar S$ with $S$ being a dimensionless number.) 
 Comparison with the Lagrangian $L = p \dot q - H$ for ordinary 
particles implies that $\hbar S \cos\theta \equiv \hbar S_z$ and 
$\phi$ are canonically conjugate each other. 
 More precisely, it specifies (in quantum language) the commutation 
relation of spin.
 Let us see this within classical mechanics.

 Let us define the canonical momentum conjugate to $\phi$ by
\begin{equation}
 P_\phi \equiv \frac{\partial L_S}{\partial \dot{\phi}} 
 = \hbar S_z , 
\end{equation}
and the Poisson bracket \cite{Goldstein02} by 
\begin{eqnarray} 
 \{ A , B \}_{\rm PB} 
&=& \frac{\partial A}{\partial \phi} \frac{\partial B}{\partial P_\phi} 
- \frac{\partial B}{\partial \phi} \frac{\partial A}{\partial P_\phi} 
\nonumber \\
&=& \frac{1}{\hbar} \left( 
  \frac{\partial A}{\partial \phi} \frac{\partial B}{\partial S_z} 
- \frac{\partial B}{\partial \phi} \frac{\partial A}{\partial S_z} 
  \right) . 
\end{eqnarray} 
 It satisfies $\{ \phi , \hbar S_z \}_{\rm PB} = 1$. 
 Other components of spin, defined by 
$S^\pm \equiv S_x \pm iS_y = \sqrt{S^2 - S_z^2} \, e^{\pm i\phi}$, 
satisfies 
$\{ \hbar S_z, \hbar S^\pm \}_{\rm PB} = \mp i \hbar S^\pm $,
and thus 
\begin{equation}
 \{ \hbar S_i, \hbar S_j \}_{\rm PB} 
= \sum_k \epsilon_{ijk} \hbar S_k .
\label{spin:Scom}
\end{equation}
 Namely, the SU(2) algebra of spin has been obtained within 
classical mechanics. 
 It is notable that this information is contained in the 
kinetic term of the Lagrangian.

\vskip 3mm
\noindent
{\bf \S A1-4.  Equation of Motion for Classical Spin}
\vskip 2mm

 The equation of motion is obtained by taking the 
variation of the action, 
\begin{equation}
 {\cal S} = \int L ({\bm n}, \dot{\bm n}) \, dt . 
\end{equation}
 Under an infinitesimal variation, 
${\bm n}(t) \to {\bm n}(t) + \delta {\bm n}(t)$, we have 
\begin{equation}
 \delta {\cal S} = \int \left( 
  \frac{\partial L}{\partial {\bm n}} 
- \frac{d}{dt} \frac{\partial L}{\partial \dot {\bm n}} 
  \right) \cdot \delta {\bm n}  \, dt 
= 0 . 
\end{equation}
 Since $|{\bm n}|=1$, all three components of 
$\delta {\bm n}$ are not independent, and we cannot equate their 
coefficients to zero. 
 Since $\delta {\bm n}$ is orthogonal to ${\bm n}$ 
(in the first order in $\delta {\bm n}$), we put 
$\delta {\bm n} = {\bm n} \times \delta{\bm w}$, 
and regard all the components of $\delta{\bm w}$ 
as independent. 
 Then, from 
\begin{equation}
 \delta {\cal S} = \int \left[ \left( 
  \frac{\partial L}{\partial {\bm n}} 
- \frac{d}{dt} \frac{\partial L}{\partial \dot {\bm n}} 
  \right) \, \times {\bm n} \right] \cdot \delta {\bm w} \, dt 
= 0 , 
\end{equation}
we obtain the equation of motion as 
\begin{equation}
 {\bm n} \times \left( 
 \frac{d}{dt} \frac{\partial L}{\partial \dot {\bm n}} 
- \frac{\partial L}{\partial {\bm n}} 
  \right) 
= 0 . 
\label{eq:LLG0}
\end{equation}

 Using 
\begin{eqnarray}
\frac{\delta {\cal S}}{\delta {\bm n}} 
&=& {\bf e}_\theta \frac{\delta {\cal S}}{\delta \theta} 
+ {\bf e}_\phi \frac{1}{\sin\theta} \frac{\delta {\cal S}}{\delta \phi} 
\end{eqnarray}
the differentiation of the kinetic term 
$L_0 \equiv \hbar S \dot \phi \cos \theta$ 
is calculated as 
\begin{eqnarray}
  \frac{d}{dt} \frac{\partial L_0}{\partial \dot {\bm n}}  
- \frac{\partial L_0}{\partial {\bm n}} 
&=& \hbar S \left( {\bf e}_\phi \dot\theta 
                 - {\bf e}_\theta \dot\phi \sin\theta \right) 
\nonumber \\ 
&=& \hbar S (\dot{\bm n} \times {\bm n}) . 
\end{eqnarray}
 The equation of motion (\ref{eq:LLG0}) then becomes 
\begin{eqnarray}
  {\bm n} \times \left( 
  \hbar S \dot {\bm n} \times {\bm n} 
+  \frac{\delta H}{\delta {\bm n}} \right) 
= 0 . 
\label{eq:variation6}
\end{eqnarray}
 Using 
${\bm n} \!\cdot\! \dot{\bm n} = 0$, 
we obtain the well-known expression 
\begin{eqnarray}
   \dot {\bm n} 
&=& \gamma {\bm H}_{{\rm eff}}  \times {\bm n} . 
\label{eq:LLG3}
\end{eqnarray}
 Here we have defined the effective field by 
\begin{eqnarray}
  {\bm H}_{{\rm eff}} 
&\equiv& + \frac{1}{\gamma \hbar}\frac{\delta H}{\delta {\bm S}} 
\ = \  \frac{1}{\gamma \hbar S}\frac{\delta H}{\delta {\bm n}} . 
\label{eq:Heff1}
\end{eqnarray}
with $\gamma$ being the gyromagnetic ratio.

 The equation of motion, (\ref{eq:LLG3}), describes precessional motion 
around the effective field, ${\bm H}_{\rm eff}$, as shown in 
Fig.\ref{fig:precession} (left), but the effect of damping
(energy dissipation) is not included. 
 To treat the damping in the Lagrangian formalism, 
we introduce the dissipation function \cite{Goldstein02}, 
\begin{eqnarray}
  W =  \frac{\alpha}{2} \frac{\hbar}{S} \,\dot {\bm S}^2 
    = \frac{\alpha}{2} \hbar S \, \dot {\bm n}^2 , 
\label{eq:W2}
\end{eqnarray}
where $\alpha$ is a dimensionless damping constant, 
and generalize eq.(\ref{eq:LLG0}) to 
\begin{equation}
 {\bm n} \times \left( 
  \frac{d}{dt} \frac{\partial L}{\partial \dot {\bm n}} 
- \frac{\partial L}{\partial {\bm n}} 
+ \frac{\partial W}{\partial \dot {\bm n}} 
  \right) 
= 0 . 
\label{eq:LLG0_W}
\end{equation}
 The equation of motion then becomes 
\begin{eqnarray}
   \dot {\bm n} 
&=& \gamma {\bm H}_{{\rm eff}}  \times {\bm n} 
     - \alpha \, {\bm n} \times \dot {\bm n} . 
\label{eq:LLG3a}
\end{eqnarray}
 The second term on the right-hand side represents the dapming torque, 
called Gilbert damping.

\subsection{Landau-Lifshitz-Gilbert equation}

So far, we have essentially considered a single spin. 
In this subsection, we consider the dynamics of magnetization 
of ferromagnets consisting of many spins.

For simplicity, we consider a localized model for ferromagnetism, 
and assume that the magnetization is carried by localized spins, 
${\bm S}_i$, located at each site $i$. 
 Writing as ${\bm S}_i = S {\bm n}_i (t)$ where $S$ is the magnitude 
of ${\bm S}_i$ and ${\bm n}_i$ is a unit vector representing its direction,
we assume that $S$ is constant, and only ${\bm n}_i (t)$ can vary. 
 Moreover, we assume that its space/time variation is sufficiently smooth, 
and adopt a continuum approximation, 
${\bm n}_i (t) \to {\bm n}({\bm r},t)$, and write as 
$\theta = \theta ({\bm r},t)$, $\phi = \phi ({\bm r},t)$. 
(Namely, it is a function of continuum space ${\bm r}$ and time $t$.)
 Then the magnetization (magnetic moment per unit volume) is given by 
\begin{equation} 
  {\bm M} ({\bm r},t) = - \gamma \hbar \sum_{{\rm unit \ volume}} {\bm S}_i (t) 
= - \frac{\gamma \hbar S}{a^3} {\bm n} ({\bm r},t) . 
\label{eq:M}
\end{equation} 
 (For simplicity, we consider a simple cubic lattice, and 
put the volume per localized spin to be $a^3$.)

 As a microscopic model, we have the following Hamiltonian in mind, 
\begin{equation} 
  H_S = - \tilde J \sum_{<i,j>} {\bm S}_i \!\cdot\! {\bm S}_j 
 + \frac{1}{2} \sum_i \left( K_\perp S_{i,y}^{\,2} - K S_{i,z}^{\,2} 
\right). 
\label{eq:Hspin}
\end{equation} 
 Here, $\tilde J$ is the exchange coupling constant leading to the 
ferromagnetic state, $K \, (\geq 0)$ and $K_\perp \, ( \geq 0)$ 
are easy-axis and hard-axis 
anisotropy constants, respectively. 
 (In soft magnets such as permalloy, shape magnetic anisotropy due to 
long-range dipole-dipole interaction is important. 
 Here, we assume that such effects can also be modeled by this Hamiltonian. 
 It is of course straightforward to take such long-range interaction 
into the LLG equation, which can be treated numerically, 
but its analytical treatment is almost impossible.) 

 In the continuum approximation, the Hamiltonian is expressed as 
\begin{eqnarray}
 H_S &=& \frac{S^2}{2}  \int \frac{d^3x}{a^3} 
    \Big[\, J (\nabla {\bm n})^2 
          + K_\perp n_y^2 - K n_z^2  \,\Big] . 
\end{eqnarray}
 Here $J = \tilde J a^2$ with $a$ being a lattice constant .
 Similarly, the Lagrangian $L_S$ and the dissipation function $W_S$ 
becomes 
\begin{eqnarray}
 L_S &=& \hbar S \int \frac{d^3x}{a^3} 
         \, \dot\phi \cos\theta - H_S 
\label{eq:LS4}
\\ 
  W_S &=& \frac{\alpha}{2} \hbar S \int \frac{d^3x}{a^3} \, \dot {\bm n}^2 . 
\label{eq:W4}
\end{eqnarray}
 From these, the equation of motion is obtained as 
\begin{equation}
   \dot {\bm n} 
= \gamma {\bm H}_{{\rm eff}}  \times {\bm n} 
     - \alpha \, {\bm n} \times \dot {\bm n} 
\label{eq:LLG4}
\end{equation}
(Usually, this equation is written in terms of magnetization vector, 
${\bm M}$, rather than the spin direction, ${\bm n}$.) 
 Here, 
\begin{equation}
  \gamma {\bm H}_{{\rm eff}} 
\equiv \frac{a^3}{\hbar S} \frac{\delta H_S}{\delta {\bm n}} 
\ =\  \frac{S}{\hbar} 
    \left[ -J \nabla^2 {\bm n} 
           - Kn_z \hat z + K_\perp n_y \hat y  \,\right] 
    + \gamma {\bm H}_{\rm ext} 
\label{eq:Heff2}
\end{equation}
is the effective field coming from $H_S$, 
\footnote{
Here, the derivative means (spatial) functional derivative,
and an additional factor $a^3$ appears compared to eq.(\ref{eq:Heff1}). 
 Mathematically, it comes from the factor $1/a^3$ in   
eqs.(\ref{eq:LS4}) and (\ref{eq:W4}).
} 
and we have added an external field, ${\bm H}_{\rm ext}$. 
 The equation (\ref{eq:LLG4}) is formally the same as eq.(\ref{eq:LLG3a}), 
but here ${\bm n} = {\bm n} ({\bm r},t)$ is a field variable.

 A phenomenological equation describing the space/time variation 
of magnetization vector of ferromagnets was first introduced by 
Landau and Lifshitz in 1935. 
 They used an equation of the form, 
\begin{equation}
   \dot {\bm n} 
= \gamma' {\bm H}_{{\rm eff}}  \times {\bm n} 
     + \alpha_{\rm LL} \, 
       (\gamma' {\bm H}_{{\rm eff}}  \times {\bm n}) 
       \times {\bm n} , 
\label{eq:LL}
\end{equation}
to
study the structure and dynamics of a domain wall, etc. 
 The second tern on the right-hand side represents damping torque, 
called Landau-Lifshitz damping. 
 The Landau-Lifshitz (LL) equation (\ref{eq:LL}) and the 
Landau-Lifshitz-Gilbert (LLG) equation (\ref{eq:LLG4}) are equivalent, 
and their coefficients are related by 
$\alpha_{\rm LL} = \alpha$ and 
$\gamma' = \gamma / (1 + \alpha^2)$.

\vskip 3mm
\noindent
{\bf \S A1-6. Effects of Conduction Electrons}
\vskip 2mm

 To study magnetization dynamics driven by electric/spin current, 
let us introduce conduction electrons. 
 We treat conduction electrons ($s$ electrons) and magetization ($d$ spin) 
as distinct degrees of freedom (the $s$-$d$ model), which are coupled  
by the $s$-$d$ exchange interaction, 
\begin{equation}
 H_{sd} 
 = - M \int d^3x \, {\bm n}({\bm r},t) \!\cdot\! (c^\dagger {\bm \sigma} c) . 
\end{equation}
 Here $c^\dagger$ and $c$ are electron creation/annihilation operators 
(spinors), and $(c^\dagger {\bm \sigma} c)$ represents spin density. 
(We have put $M=J_{sd} S$ with $J_{sd}$ being the $s$-$d$ exchange coupling 
constant.)
 Through $H_{sd}$, the electrons exerts an effective field
\begin{equation}
   \gamma {\bm H}_{{\rm eff}}' 
= \frac{a^3}{\hbar S} 
  \left\langle \frac{\delta H_{sd}}{\delta {\bm n}} \right\rangle 
= - \frac{a^3}{\hbar S} M 
  \langle c^\dagger {\bm \sigma} c \,\rangle 
\label{eq:Heff_sd}
\end{equation}
on magnetization, and thus the torque (times $a^3/\hbar S$)
\begin{equation}
 {\bm t}_{\rm el}' 
\equiv   \gamma {\bm H}_{{\rm eff}}' \times {\bm n} 
= \frac{a^3}{\hbar S} M {\bm n} \times 
  \langle c^\dagger {\bm \sigma} c \,\rangle . 
\label{eq:t_el'}
\end{equation}
 Here $\langle \cdots \rangle$ represents quantum-mechanical and statistical 
average. 
 The LLG equation is then becomes 
\begin{equation}
   \dot {\bm n} 
= \gamma {\bm H}_{{\rm eff}}  \times {\bm n} 
     - \alpha \, {\bm n} \times \dot {\bm n} 
     + {\bm t}_{\rm el}' . 
\label{eq:LLG5}
\end{equation}

\section{Brief introduction to non-equilibrium Green function}
\label{SEC:Keldysh}

We consider the electron system evoluving under the Hamiltonian 
\begin{eqnarray}
\mathcal{H}(t) = H + H^{\rm e}(t),
\label{Hamiltonian1}
\end{eqnarray}
where $H^{\rm e}(t)$ represents the interaction Hamitonian 
driven by an external field which is switchied on at a certain time, $t_{0}$, much earlier than the time we are interested. 
This interaction is treated perturbatively.
The Hamitonian $H$ can contain interaction term which are also  treated perturbatively. 
It is written therefore as
$H\equiv H_{\rm 0}+H^{\rm i}$, 
where $H_{\rm 0}$ is  perturbed part  
and interaction term to be treated perturbatively is 
$H^{\rm i}$. 
(The differece between  $H^{\rm e}(t)$ and $H^{\rm i}$ is that while $H^{\rm i}$ is there at $t<t_{0}$,  $H^{\rm e}(t)$ vanishs.)
For $t < t_{0}$, the system is in the thermal equilibrium state 
determined by $H$ and 
the expectation value of an arbitrary operator ${\cal O}$ is calculated by 
the density matrix 
\begin{eqnarray}
\rho(H)= \frac{e^{-\beta (H-\mu N)}}{{\rm Tr}(e^{-\beta (H-\mu N)})}
\end{eqnarray}
as $\langle {\cal O} \rangle_{H} = {\rm Tr}(\rho(H){\cal O})$. 
Here $\mu$ and $N$ are respectively the chemical potential and the total number operator of the system.

\begin{figure}
\begin{center}
\includegraphics[scale=0.5]{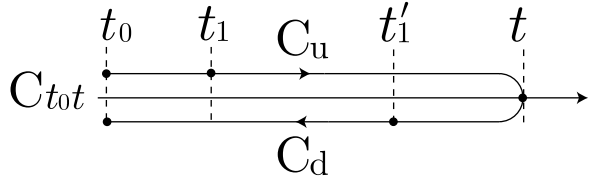}
     \vskip -\lastskip \vskip -8pt
\caption{The contour ${\rm C}_{t_{0}t}$ in the case of $t_{1}<t_{1}'$. 
Times $t_{1}$ and $t_{1}'$ are defined on the contour ${\rm C}_{t_{0}t}$.
The contour runs on the complex-time plane from $t_{0}$ 
on the slightly upper branch (${\rm C}_{\rm u}$) to $t$ on the real axis, 
and back again from $t$ to $t_{0}$ on the slightly downward branch 
$({\rm C}_{\rm d})$. 
}
\label{contour1}
\end{center}
\end{figure}

\begin{figure}
\begin{center}
\includegraphics[scale=0.3]{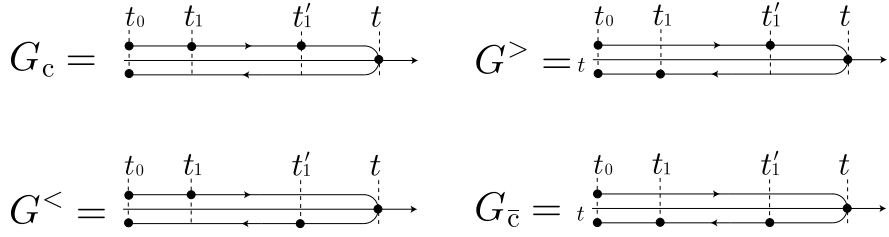}
     \vskip -\lastskip \vskip -8pt
\caption{Four different Green's functions}
\label{contour1-2}
\end{center}
\end{figure}

We introduce the following Green's function \cite{{Rammer86},{Haug98},{Rammer07}}\footnote{In this subsection, we put $\hbar=1$.} defined 
on closed-time pass contour ${\rm C}_{t_{0}t}$ as depicted in Fig. \ref{contour1}. 
\begin{eqnarray}
\label{CPG1}
G(x,x') = -i\langle T_{{\rm C}_{t_{0}t}}
[a_{\mathcal{H}}(x)a^{\dagger}_{\mathcal{H}}(x')]\rangle_{H},
\end{eqnarray}
where $a^{\dagger}_{\mathcal{H}}=(a^{\dagger}_{\mathcal{H}, +},
a^{\dagger}_{\mathcal{H},-})$ is the two-component 
electron operator 
in the Heisenberg picture and 
$T_{{\rm C}_{\rm t_{0}t}}$ is the contour-ordering operator.
We here put $x=({\bm x},t_{1})$ and $x'=({\bm x}' ,t'_{1})$ 
for simplicity. 
Since there is four different combinations 
for the times $t_{1}$ and $t_{1}'$ located on 
either of the two branches of ${\rm C}_{\rm u}$ and ${\rm C}_{\rm d}$, 
the contour-ordered Green's function $G(x,x')$ 
contains four different functions (Fig. \ref{contour1-2}):
\begin{eqnarray}
G(x,x')=
\left\{
\begin{array}{ll}
G_{\rm c}(x,x'), &\quad t_{1},~t_{1}' \in {\rm C}_{\rm u} \\
G^{>}(x,x'), &\quad t_{1}\in {\rm C}_{\rm d},~t_{1}' \in {\rm C}_{\rm u} \\
G^{<}(x,x'), &\quad t_{1}\in {\rm C}_{\rm u},~t_{1}' \in {\rm C}_{\rm d} \\
G_{\rm \bar{c}}(x,x'), &\quad t_{1},~t_{1}' \in {\rm C}_{\rm d} 
\end{array}
\right.
\label{CPG2}
\end{eqnarray}
Here $G_{\rm c}$ is the causal, or time-ordered Green's function 
and $G_{\rm \bar{c}}$ is the antitime-ordered Green's function, 
which are given by 
\begin{eqnarray}
G_{\rm c}(x,x')
&=&-i\langle T[a_{\mathcal{H}}(x)a^{\dagger}_{\mathcal{H}}(x')]
\rangle_{H}\nonumber\\
&=&-i\theta(t_{1}-t_{1}')
\langle a_{\mathcal{H}}(x)a^{\dagger}_{\mathcal{H}}(x')\rangle_{H}+
i\theta(t_{1}'-t_{1})
\langle a^{\dagger}_{\mathcal{H}}(x)a_{\mathcal{H}}(x')\rangle_{H},\\
G_{\rm \bar{c}}(x,x')&=&
-i\langle \bar{T}[a_{\mathcal{H}}(x)a^{\dagger}_{\mathcal{H}}(x')]
\rangle_{H}
\nonumber\\
&=&-i\theta(t_{1}'-t_{1})
\langle a_{\mathcal{H}}(x)a^{\dagger}_{\mathcal{H}}(x')\rangle_{H}
+i\theta(t_{1}-t_{1}')
\langle a^{\dagger}_{\mathcal{H}}(x)a_{\mathcal{H}}(x')\rangle_{H}, 
\nonumber\\
\end{eqnarray}
where $T(\bar{T})$ is the usual (anti)time-ordering operator 
and $\theta(t)$ is the step function. 
The second $G^{>}$ and the third $G^{<}$ in Eq. (\ref{CPG2}) 
are respectively 
the greater and the lesser Green's functions, which are given by 
\begin{eqnarray}
&&G^{>}(x,x')\equiv -i\langle a_{\mathcal{H}}(x)
a^{\dagger}_{\mathcal{H}}(x')\rangle_{H}, \\
&&G^{<}(x,x')\equiv i\langle a^{\dagger}_{\mathcal{H}}(x')a_{\mathcal{H}}(x)\rangle_{H}.
\end{eqnarray}
These four functions are not independent of each other 
and related as 
\begin{eqnarray}
G_{\rm c}(x,x')+G_{\bar{{\rm c}}}(x,x')
=G^{<}(x,x') + G^{>}(x,x'). 
\end{eqnarray}
Furthermore, the retareded and the advanced Green's functions 
are written by using $G^{>}$ and $G^{<}$ as 
\begin{eqnarray}
G^{\rm R}(x,x') &=&-i\theta(t_{1}-t_{1}')\langle
\{a_{\mathcal{H}}(x),a^{\dagger}_{\mathcal{H}}(x')\}\rangle_{H}
\nonumber\\
&=&\theta(t_{1}-t_{1}')[G^{>}(x,x')
-G^{<}(x,x')],\label{RG1}\\
G^{\rm A}(x,x') &=&i\theta(t'_{1}-t_{1})\langle
\{a_{\mathcal{H}}(x),a^{\dagger}_{\mathcal{H}}(x')\}\rangle_{H}
\nonumber\\
&=&\theta(t'_{1}-t_{1})[G^{<}(x,x')
-G^{>}(x,x')],\label{AG1}
\end{eqnarray}

We define the Fourier transform of the contour-ordered Green's function 
as 
\begin{eqnarray}
G_{\sigma,\sigma'}(x,t_{1},x',t'_{1})= \sum_{{\bm k},{\bm k}'}
\int_{-\infty}^{\infty}\frac{d\omega}{2\pi}
\int_{-\infty}^{\infty}\frac{d\omega'}{2\pi}
e^{i{\bm k}\cdot{\bm x}-i{\bm k}'\cdot{\bm x}'-i\omega t_{1} + i\omega' t'_{1}}
G_{{\bm k}\sigma,{\bm k}'\sigma'}(\omega,\omega').
\end{eqnarray}
Note that we here explicitly write down the spin indices $\sigma,\sigma'$. 
In the free Hamiltonian $\mathcal{H} = H_{0}=\sum_{{\bm k},\sigma}
\varepsilon_{{\bm k}\sigma}a^{\dagger}_{{\bm k}\sigma}a_{{\bm k}\sigma}$ and the zero temperature, 
$G^{>}$ and $G^{<}$ are given by 
\begin{eqnarray} 
&&G^{>}_{{\bm k}\sigma,{\bm k}'\sigma'}(\omega,\omega')=2\pi\delta(\omega-\omega')\delta_{{\bm k},{\bm k}'}\delta_{\sigma,\sigma'}g^{>}_{{\bm k}\sigma}(\omega), \\
&&g^{>}_{{\bm k}\sigma}(\omega)= 
-2\pi i (1-f(\omega))\delta(\omega-\xi_{{\bm k}\sigma}), \\
&&G^{<}_{{\bm k}\sigma,{\bm k}'\sigma'}(\omega,\omega')=2\pi\delta(\omega-\omega')\delta_{{\bm k},{\bm k}'}\delta_{\sigma,\sigma'}g^{<}_{{\bm k}\sigma}(\omega), \\
&&g^{<}_{{\bm k}\sigma}(\omega)= 2\pi i f(\omega)\delta(\omega-\xi_{{\bm k}\sigma}),
\end{eqnarray}
where $g^{>}_{{\bm k}\sigma}(\omega)$ and $ g^{<}_{{\bm k}\sigma}(\omega)$ 
are respectively the free greater and lesser Green's functions in 
the momentum space, 
$f(\omega)=(1+e^{\beta\omega})^{-1}$ is the Fermi distribution function 
and $\xi_{{\bm k}\sigma}= \varepsilon_{{\bm k}}-\varepsilon_{{\rm F}\sigma}$ 
with $\varepsilon_{{\rm F}\sigma}= \varepsilon_{\rm F} + \sigma M$. 
Using these and taking account of (\ref{RG1}) and (\ref{AG1}), 
we obtain the free retarded and advanced Green's functions 
in the momentum space as 
\begin{eqnarray}
&&g^{\rm r}_{{\bm k}\sigma}(\omega) = \frac{1}{\omega-\xi_{{\bm k}\sigma} + i0}, 
\label{RG2}\\
&&g^{\rm a}_{{\bm k}\sigma}(\omega) = (g^{\rm r}_{\sigma}({\bm k},\omega))^{*},
\label{AG2}
\end{eqnarray}
where $0$ is the positive infinitesimal 
coming from the step function $\theta$ in Eqs. (\ref{RG1}) and (\ref{AG1}).

\subsection{Perturbation expansion for closed-time ordered Green's function}

In this subsection, we derive the general formula of 
the perturbation expansion for the contour-ordered Green's function. 

We first rewrite $a_{\mathcal{H},{\bm k}}(t)$ in the Heisenberg picture as
\begin{eqnarray}
a_{\mathcal{H},{\bm k}}(t) = U^{\dagger}_{H_{0}}(t,t_{0})
a_{{\bm k}}(t)U_{H_{0}}(t,t_{0}), 
\label{HtoI-1}
\end{eqnarray}
where $a_{{\bm k}}(t)$ is the electron operator 
in the interaction picture 
and $U_{H_{0}}(t,t_{0})$ is the unitary operator defined by 
\begin{eqnarray}
U_{H_{0}}(t,t_{0}) = Te^{-i\int_{t_{0}}^{t}dt'
\{H^{\rm i}_{H_{0}}(t')+H^{\rm e}_{H_{0}}(t')\}
}, 
\end{eqnarray}
with
\begin{eqnarray}
&&H^{\rm i}_{H_{0}}(t) = e^{iH_{0}(t-t_{0})}
H^{\rm i}(t)e^{-iH_{0}(t-t_{0})}\nonumber\\
&&H^{\rm e}_{H_{0}}(t) = e^{iH_{0}(t-t_{0})}H^{\rm e}(t)
e^{-iH_{0}(t-t_{0})} 
\end{eqnarray}
in the interaction picture with respect to the free Hamiltonian $H_{0}$. 
Note that we have chosen $t_{0}$ as reference time so that 
two pictures coincide. 
Using the contour-ordering operator $T_{{\rm C}_{t_{0}t}}$, 
Equation (\ref{HtoI-1}) is rewritten as \cite{Haug98}
\begin{eqnarray}
a_{\mathcal{H},{\bm k}}(t)= T_{{\rm C}_{t_{0}t}}\left(
e^{-i
\int_{{\rm C}_{t_{0}t}}dt' \{H^{\rm i}_{H_{0}}(t')+H^{\rm e}_{H_{0}}(t')\}}
a_{{\bm k}}(t)
\right). 
\label{HtoI-2}
\end{eqnarray}
We thus obtain the contour-ordered Green' function 
\begin{eqnarray}
G_{{\bm k},{\bm k}'}(t_{1},t_{1}') 
&=& -i\langle T_{{\rm C}_{t_{0}t}}
[a_{\mathcal{H},{\bm k}}(t_{1})a_{\mathcal{H},{\bm k}'}(t_{1}')]\rangle_{H}\nonumber\\
&=&-i\bigg\langle
T_{{\rm C}_{t_{0}t}}\bigg[
e^{-i
\int_{{\rm C}_{t_{0}t}}dt' \{H^{\rm i}_{H_{0}}(t')+H^{\rm e}_{H_{0}}(t')\}}
a_{{\bm k}}(t_{1})a^{\dagger}_{{\bm k}'}(t_{1}')
\bigg]
\bigg\rangle_{H}.
\end{eqnarray}

\begin{figure}
\begin{center}
\includegraphics[scale=0.4]{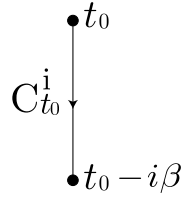}
     \vskip -\lastskip \vskip -8pt
\caption{The imaginary time contour ${\rm C}^{\rm i}_{t_{0}}$.}
\label{contour3}
\end{center}
\end{figure}

\begin{figure}
\begin{center}
\includegraphics[scale=0.5]{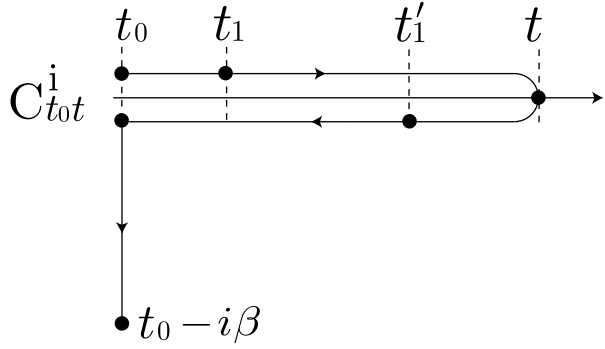}
     \vskip -\lastskip \vskip -8pt
\caption{
The contour $C^{\rm i}_{t_{0}t}$ in the case of $t_{1}<t_{1}'$. 
}
\label{contour4}
\end{center}
\end{figure}

\begin{figure}
\begin{center}
\includegraphics[scale=0.5]{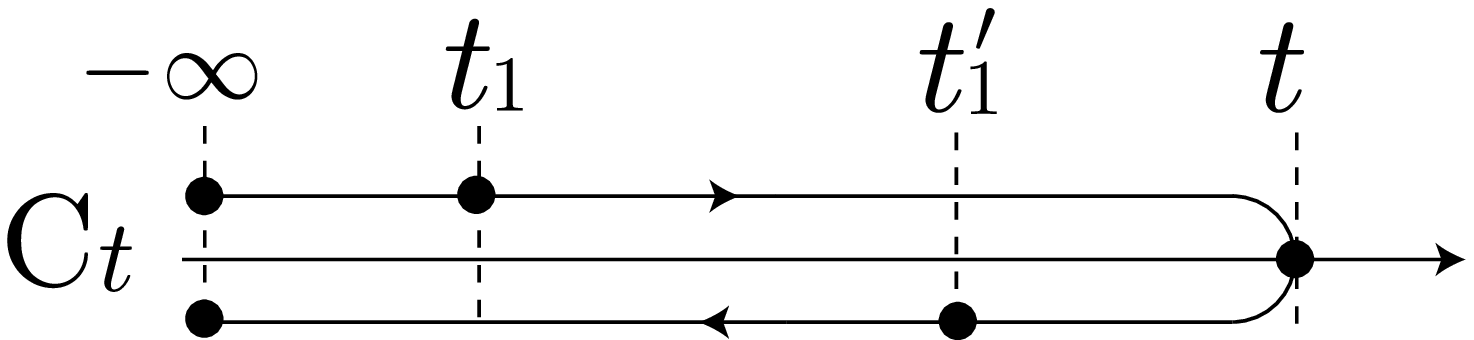}
     \vskip -\lastskip \vskip -8pt
\caption{
The contour ${\rm C}_{t}$. 
}
\label{contour5}
\end{center}
\end{figure}

The next step is to rewrite the density matrix $\rho(H)$ as 
\begin{eqnarray}
\rho(H) &=& e^{-\beta(H_{0}+H^{\rm i}-\mu N)}
/{\rm Tr}(e^{-\beta(H_{0}+H^{\rm i}-\mu N)}) \nonumber\\
&=& e^{-\beta (H_{0}-\mu N)}U_{H_{0}}(t_{0},t_{0}-i\beta)/
{\rm Tr}(e^{-\beta (H_{0}-\mu N)}U_{H_{0}}(t_{0},t_{0}-i\beta)), 
\end{eqnarray}
where 
\begin{eqnarray}
U_{H_{0}}(t_{0},t_{0}-i\beta)=
T_{{\rm C}^{\rm i}_{t_{0}}}
e^{-i
\int_{t_{0}}^{t_{0}-i\beta}
dt' H^{\rm i}_{H_{0}}(t')}
\end{eqnarray}
is defined on the imaginary time contour ${\rm C}^{\rm i}_{t_{0}}$
as depicted in Fig. \ref{contour3}. 
Combining with the following relation 
\begin{eqnarray}
T_{{\rm C}_{t_{0}t}}\left(
e^{-i
\int_{{\rm C}_{t_{0}t}}dt' H^{\rm i}_{H_{0}}(t')}
e^{-i
\int_{{\rm C}_{t_{0}t}}dt' H^{\rm e}_{H_{0}}(t')}
\right)=1,
\label{identity}
\end{eqnarray}
we thus obtain $G_{{\bm k},{\bm k}'}(t_{1},t_{1}')$ in the form 
\begin{eqnarray}
G_{{\bm k},{\bm k}'}(t_{1},t_{1}')&=& \frac{
-i{\rm Tr}\bigg(
e^{-\beta (H_{0}-\mu N)}
T_{C^{\rm i}_{t_{0}t}}\bigg[
e^{-i\int_{C^{\rm i}_{t_{0}t}}dt'
H^{\rm i}_{H_{0}}(t')
}e^{-i\int_{{\rm C}_{t_{0}t}}dt'
H^{\rm e}_{H_{0}}(t')
}a_{{\bm k}}(t_{1})a^{\dagger}_{{\bm k}'}(t_{1}')
\bigg]
\bigg)}{{\rm Tr}\bigg(
e^{-\beta (H_{0}-\mu N)}
T_{{\rm C}^{i}_{t_{0}t}}\bigg(
e^{-\frac{i}{\hbar}\int_{{\rm C}^{\rm i}_{t_{0}t}}dt'
H^{\rm i}_{H_{0}}(t')
}e^{-\frac{i}{\hbar}\int_{{\rm C}_{t_{0}t}}dt'
H^{\rm e}_{H_{0}}(t')
}\bigg)
\bigg)}\nonumber\\
&=&
-i\left\langle
T_{{\rm C}^{\rm i}_{t_{0}t}}\left[
e^{-i\int_{{\rm C}^{\rm i}_{t_{0}t}}dt'
H^{\rm i}_{H_{0}}(t')
}e^{-i\int_{{\rm C}_{t_{0}t}}dt'
H^{\rm e}_{H_{0}}(t')
}a_{{\bm k}}(t_{1})a^{\dagger}_{{\bm k}'}(t_{1}')
\right]
\right\rangle_{H_{0}}
\label{pG1}
\end{eqnarray}
where the contour $C^{\rm i}_{t_{0}t}$ is depicted in Fig. \ref{contour4}. 
In this paper, we do not consider initial correlations and 
therefore put $t_{0}=-\infty$ to ignore the contribution from 
the imaginary part of contour $C^{\rm i}_{t_{0}t}$, namely,
 $C^{\rm i}_{t_{0}t}\rightarrow  C_{t_{0}t}$ \cite{Rammer07}. 
In this prescription, we can expand the exponents 
of Eq. (\ref{pG1}) with respect to 
$H^{\rm i}_{H_{0}}(t)$and $H^{\rm e}_{H_{0}}(t)$ on equal footing 
and can use the Langreth method, which is provided in the next subsection. 
Finally we rewrite Eq. (\ref{pG1}) as 
\begin{eqnarray}
G_{{\bm k},{\bm k}'}(t_{1},t_{1}')
&=&
-i\left\langle
T_{{\rm C}_{t}}\left[
e^{-i\int_{{\rm C}_{t}}dt'
H^{\rm i}_{H_{0}}(t')
}e^{-i\int_{{\rm C}_{t}}dt'
H^{\rm e}_{H_{0}}(t')
}a_{{\bm k}}(t_{1})a^{\dagger}_{{\bm k}}(t_{1}')
\right]
\right\rangle_{H_{0}},
\label{pG2}
\end{eqnarray}
where the contour ${\rm C}_{\rm t}$ is depicted in Fig. \ref{contour5}. 

\begin{figure}[h]
\begin{center}
\includegraphics[scale=0.5]{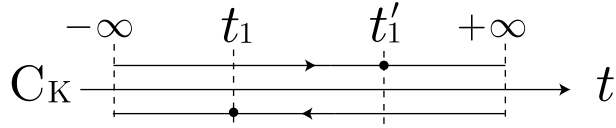}
     \vskip -\lastskip \vskip -8pt
\caption{The Keldysh contour $C^{\rm K}$.}
\label{contour6}
\end{center}
\end{figure}

If we furthermore put $t=\infty$ in the contour ${\rm C}_{t}$, 
this is often called the Keldysh contour $C^{\rm K}$ as depicted in Fig. \ref{contour6} 
and the corresponding Green's function is called 
the Keldysh Green's function \cite{Rammer86,Keldysh65}. 
The Keldysh Green's function consists of $G^{\rm r}$, $G^{\rm a}$ and 
$G^{\rm K} = G^{<} + G^{<}$, which is fundamentally equivalent 
to the contour-ordered Green's function. 
In this paper, since our main evaluation is to calculate 
the lesser Green's function $G^{<}$, we do not use the Keldysh 
representation. 

\subsection{Langreth method}

In this subsection, 
we briefly review the Langreth method 
\cite{Haug98,Rammer07,Langreth76}, which is useful to 
evaluate perterbation expansion of 
the contour-ordered Green's function 
in Eq. (\ref{pG2}). 

\begin{figure}
\begin{center}
\includegraphics[scale=0.4]{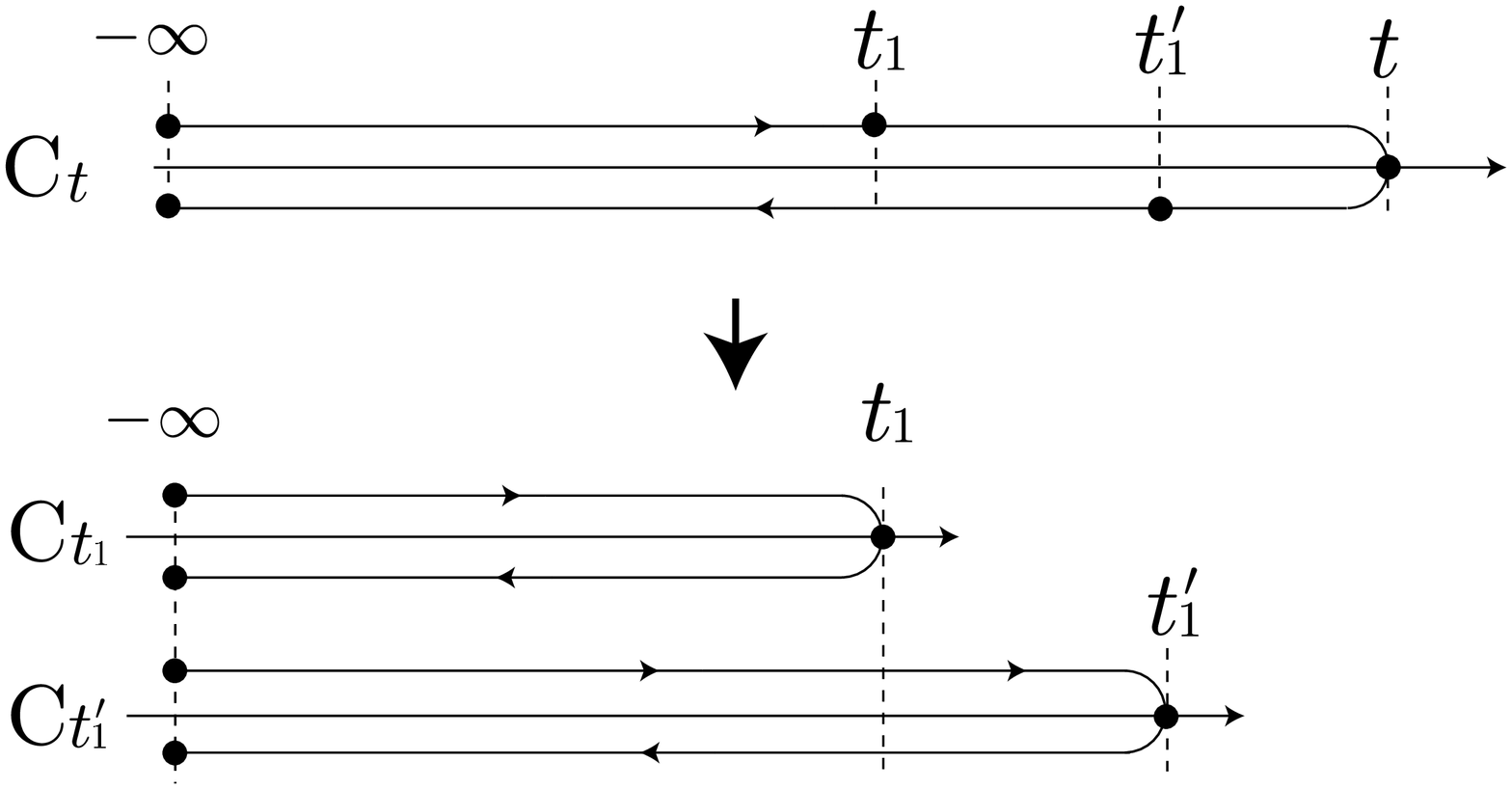}
     \vskip -\lastskip \vskip -8pt
\caption{
The contour ${\rm C}_{t}$ changed into 
${\rm C}_{t_{1}}+
{\rm C}_{t'_{1}}$. The both contour are equivalent.  
}
\label{contour7}
\end{center}
\end{figure}

In a concrete calculation for perturbative expansion 
of the contour-ordered Green's function in Eq. (\ref{pG2}), 
we need to take the lesser component of the following integration 
\begin{eqnarray}
C(t_{1},t_{1}') = \int_{{\rm C}_{t}}d\tau~A(t_{1},\tau)B(\tau,t_{1}'), 
\label{C-Langreth}
\end{eqnarray}
where $C, A$, and $B$ is the contour-ordered Green's functions, which are 
defined on the contour ${\rm C}_{t}$ in Fig. \ref{contour5}. 
Taking the lesser component of $C(t_{1},t_{1}')$ and changing the contour 
${\rm C}_{t}$ into ${\rm C}_{t_{1}}+
{\rm C}_{t'_{1}}$ as depicted in Fig. \ref{contour7}, 
the lesser component $C^{<}$ is given in the form 
\begin{eqnarray}
C^{<}(t_{1},t_{1}') &=& \int_{{\rm C}_{t_{1}}}d\tau~
A(t_{1},\tau)B^{<}(\tau,t_{1}')+
\int_{{\rm C}_{t_{1}'}}d\tau~A^{<}(t_{1},\tau)B(\tau,t_{1}'). \label{C-Langreth2}
\end{eqnarray}
Futhermore the first term in the right hand side of the above expression 
is written as
\begin{eqnarray}
\int_{{\rm C}_{t_{1}}}d\tau~
A(t_{1},\tau)B^{<}(\tau,t_{1}') &=& 
\int_{-\infty}^{t_{1}}d\tau~A^{>}(t_{1},\tau)B^{<}(\tau,t_{1}')+
\int_{t_{1}}^{-\infty}d\tau~A^{<}(t_{1},\tau)B^{<}(\tau,t_{1}')\nonumber\\
&=&\int_{-\infty}^{\infty}d\tau~\theta(t_{1}-\tau)
\left\{
A^{>}(t_{1},\tau)B^{<}(\tau,t_{1}')-A^{<}(t_{1},\tau)B^{<}(\tau,t_{1}')
\right\}\nonumber\\
&=&\int_{-\infty}^{\infty}d\tau~A^{\rm r}(t_{1},\tau)B^{<}(\tau,t_{1}'). 
\end{eqnarray}
In a similar manner for the second term in Eq. (\ref{C-Langreth2}), 
we thus obtain 
\begin{eqnarray}
\label{Langreth-l}
C^{<}(t_{1},t_{1}') = \int_{-\infty}^{\infty} d\tau~
(A^{\rm r}(t_{1},\tau)B^{<}(\tau,t_{1}')+
A^{<}(t_{1},\tau)B^{\rm a}(\tau,t_{1}')
). 
\end{eqnarray}
For the retarded and the advanced Green's functions, 
we obtain 
\begin{eqnarray}
\label{Langreth-r}
C^{\rm r}(t_{1},t_{1}') &=& \int_{-\infty}^{\infty} d\tau~
A^{\rm r}(t_{1},\tau)B^{\rm r}(\tau,t_{1}')\nonumber\\
C^{\rm a}(t_{1},t_{1}') &=& \int_{-\infty}^{\infty} d\tau~
A^{\rm a}(t_{1},\tau)B^{\rm a}(\tau,t_{1}').
\end{eqnarray}
These formula are abbreviated as 
\begin{eqnarray}
[AB]^{<} = A^{\rm r}B^{<} + A^{<}B^{\rm a},~~
[AB]^{\rm r} = A^{\rm r}B^{\rm r}\;\;\;
[AB]^{\rm a} = A^{\rm a}B^{\rm a}. 
\label{Langreth-formula1}
\end{eqnarray}
By using these, 
equations (\ref{Langreth-l}) and (\ref{Langreth-r}), 
and the Fourier components $C^{\eta}(\omega,\omega')$ 
are written as 
\begin{eqnarray}
\label{Langreth-t}
&&C^{\eta}(t_{1},t_{1}') = \int_{-\infty}^{\infty} d\tau
[A(t_{1},\tau)B(\tau,t_{1}')]^{\eta},\\
&&C^{\eta}(\omega,\omega) = 
\int_{-\infty}^{\infty}\frac{d\omega''}{2\pi}
[A(\omega,\omega'')B(\omega'',\omega')]^{\eta}, 
\label{Langreth-omega}
\end{eqnarray}
where $\eta$ is denoted by $\eta = <, {\rm r}, {\rm a}$. 
The result (\ref{Langreth-formula1}) can be extended for 
a product of three or more functions as 
\begin{eqnarray}
[ABC]^{<}  &=& [AB]^{\rm r}C^{<} + [AB]^{<}C^{\rm a} \nonumber\\
           &=& A^{\rm r}B^{\rm r}C^{\rm <} 
             + A^{\rm r}B^{<}C^{\rm a} 
             + A^{<}B^{\rm a}C^{\rm a}, \\
~[ABC]^{\rm r} &=& A^{\rm r}B^{\rm r}C^{\rm r}\nonumber\\
~[ABC]^{\rm a} &=& A^{\rm a}B^{\rm a}C^{\rm a}.
\end{eqnarray}
\subsection{Impurity averaged Green's function}

\begin{figure}
\begin{center}
\includegraphics[scale=0.25]{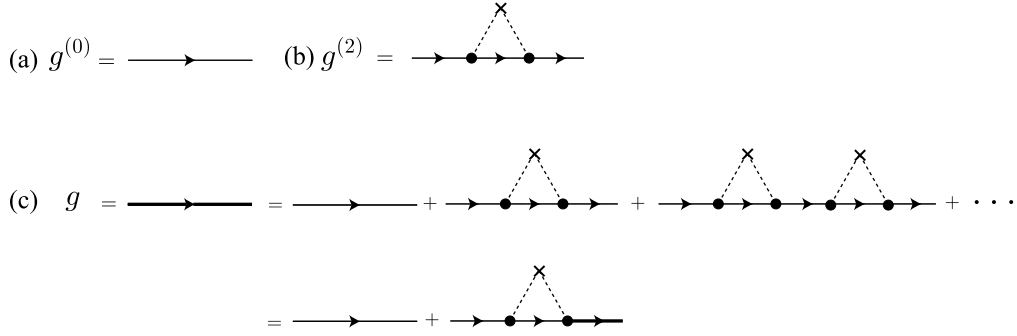}
     \vskip -\lastskip \vskip -8pt
\caption{
Diagrams of the Green's functions. 
(a)The free Green's function. 
(b)The second order Green's function. 
A dot line represents the impurity potential and 
two dot lines with one vertex represents 
averaging over impurity positions. 
(c)The Green's function averaging over the rondomly 
distributed impurities. 
}
\label{green1}
\end{center}
\end{figure}

In this subsection, using 
the paturbation expansion of Eq. (\ref{pG2}) and 
the Langreth method, we calculate the contour-ordered Green's function 
averaging over the rondomly distributed impurities 
without spin-flip scattering. 
The Hamiltonian is given by 
\begin{eqnarray}
&&H_{0} = \sum_{{\bm k},\sigma}\varepsilon_{{\bm k}\sigma}a^{\dagger}_{{\bm k}\sigma}a_{{\bm k}\sigma},\\
&&H^{\rm i} = H_{\rm imp} = \sum_{{\bm k}\sigma}V_{{\bm q}}a^{\dagger}_{{\bm k}+{\bm q}\sigma}a_{{\bm k}\sigma}, ~~V_{{\bm q}}= u_{\rm i} \sum_{i=1}^{n_{\rm i}}e^{-i{\bm q}\cdot{\bm R}_{i}},\\
&&H^{\rm e}(t) = 0, 
\end{eqnarray}
where we here consider the short ranged potential $V({\bm r})= u_{\rm i}\delta({\bm r}-{\bm R}_{i})$ 
with ${\bm R}_{i}$ being position of impurity. 
Since we can regard the averaged $\overline{V_{{\bm q}}}$ 
in terms of position of impurity as zero, 
the first order Green's function 
$g^{(1)}_{{\bm k}\sigma,{\bm k}'\sigma'}(t_{1},t_{1}')$ with respect to $H^{\rm i}$ 
is zero. The next is the second order Green's function 
$g^{(2)}_{{\bm k}\sigma,{\bm k}'\sigma'}(t_{1},t_{1}')$, which is given by 
\begin{eqnarray}
{g^{(2)}_{{\bm k}\sigma ,{\bm k}'\sigma'}(t_{1},t_{1}')}&=&
\delta_{\sigma,\sigma'}
\sum_{{\bm p}}\overline{V_{{\bm k}-{\bm p}}V_{{\bm p}-{\bm k}'}}
\int_{{\rm C}_{t}}dt_{2}\int_{C_{t}}dt'_{2}g^{(0)}_{{\bm k}\sigma}(t_{1}-t_{2})
g^{(0)}_{{\bm p}\sigma}(t_{2}-t_{2}')g^{(0)}_{{\bm k}'\sigma}(t_{2}-t_{1}')
\nonumber\\
&=& \delta_{{\bm k},{\bm k}'} \delta_{\sigma,\sigma'} 
\int_{C_{t}}dt_{2}
\int_{C_{t}}dt'_{2}g^{(0)}_{{\bm k}\sigma}(t_{1}-t_{2})
\Pi_{\sigma}(t_{2}-t_{2}')g^{(0)}_{{\bm k}'\sigma}(t_{2}-t_{1}'),
\end{eqnarray}
where $g^{(0)}_{{\bm k}\sigma}$ is the free contour-ordered Green's function 
and 
\begin{eqnarray}
\Pi_{\sigma}(t) = n_{\rm i}u^{2}_{\rm i}\sum_{{\bm k}}
 g^{(0)}_{{\bm k}\sigma}(t) 
\end{eqnarray}
is the self energy in the first Born approximation. 
We here have used the relation 
$\overline{V_{{\bm k}-{\bm p}}V_{{\bm p}-{\bm q}'}}
=n_{\rm i}u^{2}_{\rm i}\delta_{{\bm k},{\bm k}'}$, 
where $n_{\rm i}$ is the concentration of impurity. 
The diagram representing $g^{(0)}_{{\bm k}\sigma}(\omega)$ 
is shown in Fig. \ref{green1}(a). 

Using the Langreth method (\ref{Langreth-t}), we can write 
\begin{eqnarray}
[{g^{(2)}_{{\bm k}\sigma ,{\bm k}'\sigma'}(t_{1},t_{1}')}]^{\tau}
&=& \delta_{{\bm k},{\bm k}'} \delta_{\sigma,\sigma'} 
\int_{-\infty}^{\infty}dt_{2}
\int_{-\infty}^{\infty}dt'_{2}\left[
g^{(0)}_{{\bm k}\sigma}(t_{1}-t_{2})
\Pi_{\sigma}(t_{2}-t_{2}')
g^{(0)}_{{\bm k}'\sigma}(t_{2}-t_{1}')\right]^{\tau}, 
\nonumber\\
&&
\end{eqnarray}
and this Fourier component is 
\begin{eqnarray}
[{g^{(2)}_{{\bm k}\sigma ,{\bm k}'\sigma'}(\omega,\omega')}]^{\tau}
&=&2\pi \delta(\omega-\omega')\delta_{{\bm k},{\bm k}'}\delta_{\sigma,\sigma'}
g^{(2),\tau}_{{\bm k}\sigma}(\omega), \\
g^{(2),\tau}_{{\bm k}\sigma}(\omega)&=& 
 [g^{(0)}_{{\bm k}\sigma}(\omega)\Pi_{\sigma}(\omega)
  g^{(0)}_{{\bm k}\sigma}(\omega)]^{\tau}. 
\end{eqnarray}
The diagram representing $g^{(2)}_{{\bm k}\sigma}(\omega)$ is shown 
in Fig. \ref{green1}(b). 
In a simmilar manner, we can calculate the higher order Green's functions and 
sum the diagrams as dipicted in Fig. \ref{green1}(c), 
$g_{{\bm k}\sigma}(\omega)$ satisfies the Dyson equation
\begin{eqnarray}
g_{{\bm k}\sigma}(\omega) = g^{(0)}_{{\bm k}\sigma}(\omega) + g^{(0)}_{{\bm k}\sigma}(\omega)
\Pi_{\sigma}(\omega)g_{{\bm k}\sigma}(\omega).
\end{eqnarray}
Using the Langreth method, we obtain 
\begin{eqnarray}
g^{\rm r}(\omega) &=& \frac{1}{(g^{(0){\rm r}}_{{\bm k}\sigma}(\omega))^{-1}-\Pi^{\rm r}
_{\sigma}(\omega)} = \frac{1}{\omega -\varepsilon_{{\bm k}}+\varepsilon_{{\rm F}\sigma} + i\gamma_{\sigma}}, 
~~g^{\rm a}(\omega)=(g^{\rm r}(\omega))^{*}, 
\nonumber\\&&
\label{Green-r-a-imp}
\end{eqnarray}
where $2\gamma_{\sigma} = 1/\tau_{\sigma}$ is the damping rate, 
which is coming from
\begin{eqnarray}
\Pi_{\sigma}^{\rm r}(\omega) \simeq -i \pi n_{\rm i}u^{2} \nu_{\sigma} 
= -i\gamma_{\sigma}
\end{eqnarray}
with $\nu_{\sigma}$ being the density of state at $\varepsilon_{{\rm F}\sigma}$. 

Similarly, 
taking account of (\ref{Green-r-a-imp}), 
the lesser component of $g_{{\bm k}\sigma}(\omega)$ is given by 
\begin{eqnarray}
g^{<}_{{\bm k}\sigma}(\omega)&=& g^{(0)<}_{{\bm k}\sigma} +[g^{(0)}_{{\bm k}\sigma}
\Pi_{\sigma}g_{{\bm k}\sigma}]^{<}\nonumber\\
&=& g^{(0)<}_{{\bm k}\sigma} +g^{(0){\rm r}}_{{\bm k}\sigma}
\Pi^{\rm r}_{\sigma}g^{<}_{{\bm k}\sigma}
+g^{(0){\rm r}}_{{\bm k}\sigma}
\Pi^{<}_{\sigma}g^{\rm a}_{{\bm k}\sigma}
+g^{(0)<}_{{\bm k}\sigma}
\Pi^{\rm a}_{\sigma}g^{\rm a}_{{\bm k}\sigma}\nonumber\\
&=&   f(\omega) 
(g^{\rm a}_{{\bm k}\sigma}(\omega)-g^{\rm r}_{{\bm k}\sigma}(\omega)).
\end{eqnarray}


\section{ Details of linear response calculation}
\label{APPomegasum}
\subsection{Summation over Matsubara frequency}
\begin{figure}[htbp]
  \begin{center}
  \includegraphics[scale=0.6]{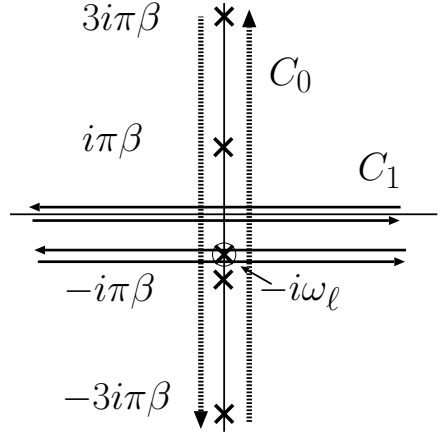}
\caption{ The contour $C_{0}$  (denoted by two dotted lines)
of integration in complex $z$-plane 
which appears in rewriting the summation over the Matsubara frequencies 
(Eq. (\ref{I1})), which surrounds the imaginary axis. 
Because of the function $f(z)$ in the integrand, 
there are poles on the imaginary axis at $z=(2n-1)\pi i/\beta$  where 
$n=0,\pm1,\pm2\cdots$.
$C_{1}$ denoted by four solid lines is a deformed contour parallel to the real axis.
The imaginary part of two lines are $z=\pm i 0$ and 
the other two are $z=-i\omega_\ell \pm i 0$, where $\omega_\ell$ (denoted by a thick cross with circle) is a pole and $i0$ represents small imaginary part. 
\label{FIGcontour_omega}}
  \end{center}
\end{figure}

The summation over Matsubara frequency, $\omega_{n}$, in the 
expression of classical contribution to the conductivity, 
\Eqref{QSUMRESULT}, is carried out in a standard manner by use of contour 
integration as follows.
The quantity we consider is 
\begin{equation}
	Q_{\rm c}(i\omega_{\ell})=-B\frac{1}{V}\sum_{\kv q\sigma } 
|A^y_{\xw}{(q)}|^{2} I_{\kv q\sigma }(i\omega_{\ell}),
	\label{Qc1}
\end{equation}
where $B=((e\hbar)^{2}/2m^{2})$ and 
\begin{equation}
I_{\kv q\sigma }(i\omega_{\ell})\equiv 
\frac{1}{\beta}\sum_{n}
G_{\kv-\frac{q}{2},n,\sigma}
G_{\kv-\frac{q}{2},n+\ell,\sigma}
G_{\kv+\frac{q}{2},n,-\sigma}
G_{\kv+\frac{q}{2},n+\ell,-\sigma}.
	\label{Idef}
\end{equation}
This function is written by use of contour integration with respect 
to $z\equiv i\omega_{n}$ as
\begin{equation}
	I_{\kv q\sigma }(i\omega_{\ell})=
-\frac{1}{2\pi i}\oint_{C_{0}} dz f(z)
G_{\kv-\frac{q}{2},\sigma}(z)
G_{\kv-\frac{q}{2},\sigma}(z+i\omega_{\ell})
G_{\kv+\frac{q}{2},-\sigma}(z)
G_{\kv+\frac{q}{2},-\sigma}(z+i\omega_{\ell}),
	\label{I1}
\end{equation}
where $f(z)\equiv 1/(1+e^{\beta z})$ and Green's function $G_{\kv\sigma n}$ is 
here denoted by $G_{\kv\sigma}(i\omega_{n})$, and the contour $C_{0}$ goes 
around the imaginary axis in complex $z$-plane (Fig. \ref{FIGcontour_omega}).
Noting that 
$G_{\kv\sigma}(z)=[z+(1/2\tau){\rm sgn}({\rm Im}[z])-\epsilon_{\kv\sigma}]^{-1}$
has a cut along ${\rm Im}[z]=0$, the contour can be changed to four 
paths $C_{1}$ parallel to the real axis, namely 
$z\equiv \pm(\omegap+i0)$ and $z\equiv 
-i\omega_{\ell}\pm(\omegap+i0)$,
where $\omegap$ runs from $-\infty$ to $\infty$ (Fig. \ref{FIGcontour_omega}).                                                                                    
We then obtain
\begin{eqnarray}
I_{\kv q\sigma }(i\omega_{\ell})&=&
-\frac{1}{\pi}\int_{-\infty}^{\infty} d\omegap 
\left[ f(\omegap) 
 \frac{1}{(\omegap+i\omega_{\ell}+\frac{i\hbar}{2\tau}-\epsilon_{\kvmq,\sigma})}
 \frac{1}{(\omegap+i\omega_{\ell}+\frac{i\hbar}{2\tau}-\epsilon_{\kvpq,-\sigma})}
\right.
    \nonumber\\
   && \times {\rm Im} \left(
      \frac{1}{(\omegap+\frac{i\hbar}{2\tau}-\epsilon_{\kvmq,\sigma})}
     \frac{1}{(\omegap+\frac{i\hbar}{2\tau}-\epsilon_{\kvpq,-\sigma})}
     \right)
     \nonumber\\
  &&  +f(\omegap-i\omega_{\ell}) 
 \frac{1}{(\omegap-i\omega_{\ell}-\frac{i\hbar}{2\tau}-\epsilon_{\kvmq,\sigma})}
\frac{1}{(\omegap-i\omega_{\ell}-\frac{i\hbar}{2\tau}-\epsilon_{\kvpq,-\sigma})}
 \nonumber\\
   && \left.
     \times {\rm Im} \left(
      \frac{1}{(\omegap+\frac{i\hbar}{2\tau}-\epsilon_{\kvmq,\sigma})}
     \frac{1}{(\omegap+\frac{i\hbar}{2\tau}-\epsilon_{\kvpq,-\sigma})}
     \right)
     \right]
 . \label{I2}
\end{eqnarray}
The classical conductivity is expressed by taking the imaginary part of the 
correlation function analytically continued to 
$\omega+i0\equiv i\omega_{\ell}$ as
\begin{equation}
	\sigma_{\rm c}=\lim_{\omega\rightarrow 0}{\rm Im} 
	\frac{Q_{\rm c}(\omega+i0)-Q_{\rm c}(i0)}{\omega}.
	\label{sigmacdef}
\end{equation}
The imaginary part of $I_{\kv q\sigma }(i\omega_{\ell}=\omega+i0)$ is 
obtained from 
(\ref{I2}) as 
\begin{eqnarray}
{\rm Im}I_{\kv q\sigma }(\omega+i0)&=&
-\frac{1}{\pi}\int_{-\infty}^{\infty} d\omegap 
   {\rm Im} \left(
       \frac{1}{(\omegap+\frac{i\hbar}{2\tau}-\epsilon_{\kvmq,\sigma})}
      \frac{1}{(\omegap+\frac{i\hbar}{2\tau}-\epsilon_{\kvpq,-\sigma})}
      \right)
      \nonumber\\
&&\times \left[ 
   f(\omegap) {\rm Im} \left( 
\frac{1}{(\omegap+\omega+\frac{i\hbar}{2\tau}-\epsilon_{\kvmq,\sigma})}
\frac{1}{(\omegap+\omega+\frac{i\hbar}{2\tau}-\epsilon_{\kvpq,-\sigma})}
\right.\right.
 \nonumber\\
&&
- \left.\left.
f(\omegap-\omega)
\frac{1}{(\omegap-\omega+\frac{i\hbar}{2\tau}-\epsilon_{\kvmq,\sigma})}
\frac{1}{(\omegap-\omega+\frac{i\hbar}{2\tau}-\epsilon_{\kvpq,-\sigma})}
      \right)
      \right]
    \nonumber\\
 &=& 
 -\frac{1}{\pi}\int_{-\infty}^{\infty} d\omegap 
\left[ f(\omegap) -f(\omegap+\omega) \right]
\nonumber\\
&&\times
        {\rm Im} \left( 
\frac{1}{\omegap+\omega+\frac{i\hbar}{2\tau}-\epsilon_{\kvmq,\sigma}}
\frac{1}{\omegap+\omega+\frac{i\hbar}{2\tau}-\epsilon_{\kvpq,-\sigma}}
            \right)
\nonumber\\
&&\times
        {\rm Im} \left(
       \frac{1}{\omegap+\frac{i\hbar}{2\tau}-\epsilon_{\kvmq,\sigma}}
      \frac{1}{\omegap+\frac{i\hbar}{2\tau}-\epsilon_{\kvpq,-\sigma}}
      \right)
 . \label{I3}
\end{eqnarray}
By use of 
\begin{equation}
	f(\omegap)-f(\omegap+\omega)= -\omega\frac{d f(\omegap)}{d\omegap} 
	        +O(\omega^{2})
	     \sim \omega\delta(\omegap)   ,
	\label{delfrel}
\end{equation}
which holds at small $\omega$ and low temperature, we obtain 
\begin{equation}
	{\rm Im}I_{\kv q\sigma }(\omega+i0) \sim
	-\frac{\omega}{\pi}
        \left[  {\rm Im} \left( 
       \frac{1}{\frac{i\hbar}{2\tau}-\epsilon_{\kvmq,\sigma}}
      \frac{1}{\frac{i\hbar}{2\tau}-\epsilon_{\kvpq,-\sigma}} 
      \right) \right]^{2}.	
      \label{ImI}
\end{equation}
The classical contribution to the conductivity, 
(\ref{sigmacdef}),  is then obtained as 
\begin{equation}
	\sigma_{\rm c}=
-\frac{\Delta^{2}\hbar^{3}}{8\pi\tau^{2}}\left(\frac{e\hbar}{m}\right)^{2}
\frac{1}{V}\sum_{\kv q\sigma }|A_{q}|^{2}
\frac{(\epsilon_{\kv-\frac{q}{2},\sigma}
       +\epsilon_{\kv+\frac{q}{2},-\sigma})^{2}}   
   {\left[(\epsilon_{\kv-\frac{q}{2},\sigma})^{2}
         +\left(\frac{\hbar}{2\tau}\right)^{2}\right]^{2}
    \left[(\epsilon_{\kv+\frac{q}{2},-\sigma})^{2}
         +\left(\frac{\hbar}{2\tau}\right)^{2}\right]^{2}}.
\label{sigmacresult2}
\end{equation}

\subsection{ Summation over wave vector $\kv$
}
\label{APPksum}
\begin{figure}[htbp]
  \begin{center}
  \includegraphics[scale=0.6]{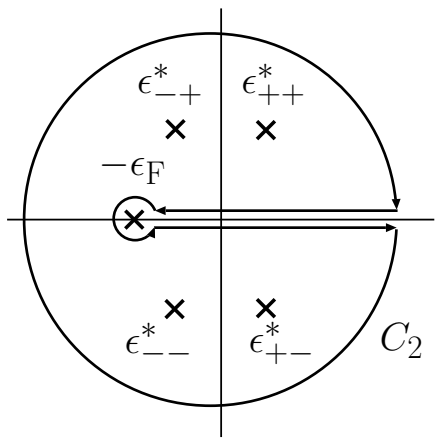}
\caption{The contour of integration $C_{2}$ in complex $\epsilon$-plane. 
There is a cut on the real axis from $\epsilon=-\epsilon_{F}$ to $+\infty$,
because of the behavior of the density of states, 
$N(\epsilon)\propto(\epsilon+\epsilon_F)^{1/2} $ in 
three-dimensions. 
\label{FIGContour2}}
  \end{center}
\end{figure}

The $\kv$-summation in (\ref{sigmacresult}) is carried out as follows.
We neglect quantities of $O((q/k_{F})^{2})$ and 
approximate $\epsilon_{\kv\pm q/2,\mp\sigma}\simeq 
\epsilon_{\kv}\pm[(k_{z}q/2m)+\sigma\Delta]$.
This is because the momentum transfer, $q$, is limited to a small value of 
$q\lesssim \lambda^{-1}$ due to the 
form factor of the wall, $|A_{q}|^{2}\propto [\cosh(\pi q\lambda/2)]^{-2}$, and we are considering the 
case of a thick wall, $k_{F}\lambda \gg 1$.
Then $\sigma_{\rm w}$  is written as
\begin{equation}
       \sigma_{\rm c} \simeq 
	-\frac{\Delta^{2}\hbar^{3}}{8\pi\tau^{2}}\left(\frac{e\hbar}{m}\right)^{2}
	\frac{1}{V}\sum_{q}|A_{q}|^{2}J_{q},
	\label{sigamcJ}
\end{equation}
where
\begin{equation}
	J_{q} \equiv \sum_{{\kv\sigma}} 
	\frac{4(\epsilon_{\kv})^{2}}
	        {   \left[   \left\{
	\epsilon_{\kv}-\left(\frac{\hbar^{2}k_{z}q}{2m}+\sigma\Delta\right)
	    \right\}^{2} +\left(\frac{\hbar}{2\tau}\right)^{2}  \right]^{2}
                      \left[   \left\{
	\epsilon_{\kv}+\left(\frac{\hbar^{2}k_{z}q}{2m}+\sigma\Delta\right)
	    \right\}^{2} +\left(\frac{\hbar}{2\tau}\right)^{2}  \right]^{2}
	    }.
	\label{Jqdef}
\end{equation}
This function is written as
\begin{eqnarray}
J_{q}&\equiv & 2\int_{-1}^{1}\frac{d\cos\theta}{2} 
\int_{-\epsilon_{F}}^{\infty}d\epsilon N(\epsilon) \nonumber\\
&&
\frac{4\epsilon^{2}}
	        {   \left[   \left\{
	  \epsilon-\left(\frac{\hbar^{2}kq\cos\theta}{2m}+\Delta\right)
	    \right\}^{2} +\left(\frac{\hbar}{2\tau}\right)^{2}  \right]^{2}
                      \left[   \left\{
	  \epsilon+\left(\frac{\hbar^{2}kq\cos\theta}{2m}+\Delta\right)
	    \right\}^{2} +\left(\frac{\hbar}{2\tau}\right)^{2}  \right]^{2}
	    }
	    \nonumber\\
&\equiv& \int_{-1}^{1}{d\cos\theta}
\int_{-\epsilon_{F}}^{\infty}d\epsilon N(\epsilon) F_{\theta}(\epsilon),
\label{Jq1}
\end{eqnarray}
where $N(\epsilon)\equiv 
(Vm^{3/2}/\pi^{2}\sqrt{2}\hbar^{3})\sqrt{\epsilon+\epsilon_{F}}$ is the 
density of states, and  
$k\equiv\sqrt{2m(\epsilon+\epsilon_{F})}/\hbar$, $\theta$ being the 
angle between $\kv$ and the $z$-axis, and $F_{\theta}(\epsilon)$ is 
defined by the last line.
In the first line we have included the factor of two due to the 
summation over the spin.
The integration over $\epsilon$ can be carried out by use of contour 
integration. 
In doing this we need to be a little bit careful 
since the density of states in three-dimensions 
$N(\epsilon)\propto\sqrt{\epsilon+\epsilon_{F}}$ has a cut on the real 
axis running from $\epsilon=-\epsilon_{F}$ to $\epsilon={\infty}$.
Choosing a closed path $C_{2}$ in the $\epsilon$-plane as in Fig. 
\ref{FIGContour2}, 
the $\epsilon$-integral in (\ref{Jq1}) is written as
\begin{equation}
\int_{-\epsilon_{F}}^{\infty}d\epsilon N(\epsilon) F_{\theta}(\epsilon)
=\frac{1}{2} \oint_{C_{2}} d\epsilon N(\epsilon) F_{\theta}(\epsilon).
	\label{Jq2}
\end{equation}
The contour $C_{2}$ contains four poles at 
$\epsilon=\epsilon^{*}_{\sigma,\pm}\equiv 
\sigma\left(\Delta+\frac{\hbar^{2}k(\epsilon^{*}_{\sigma})q}{2m} 
\cos\theta\right)\pm i\frac{\hbar}{2\tau}$.
In ferromagnets with strong polarization we are interested in, 
$\Delta\sim O(\epsilon_{F})$, and then neglecting $O(q/k_{F})$ 
contributions we may approximate 
$\epsilon^{*}_{\sigma,\pm}\sim \sigma 
\left(\Delta+\frac{\hbar^{2}k_{F\sigma}q}{2m}\cos\theta\right)
\pm  i\frac{\hbar}{2\tau}$,  where 
$\hbar k_{F\sigma}\equiv \sqrt{2m(\epsilon_{F}+\sigma\Delta)}$ is 
the Fermi momentum of the polarised electron.
The density of states estimated at the pole in the lower half plane 
(i.e., at
$\epsilon^{*}_{\sigma,-}\simeq e^{2\pi i}\epsilon^{*}_{\sigma,+}$) 
has a opposite sign as the upper half 
plane, i.e., $N(\epsilon^{*}_{\sigma,-}) \sim e^{\pi 
i}N(\epsilon^{*}_{\sigma,+})=-N_{\sigma}$
$(N_{\sigma}\equiv N(\epsilon^{*}_{\sigma,+})=N(\sigma\Delta))$. 
We therefore obtain
\begin{equation}
\int_{-\epsilon_{F}}^{\infty}d\epsilon N(\epsilon) F_{\theta}(\epsilon)
\simeq 2\pi\left(\frac{\tau}{\hbar}\right)^{3}\sum_{\sigma}N_{\sigma} 
\frac{1}{\left(\frac{\hbar^{2}k_{F\sigma}q}{2m}\cos\theta 
+\sigma\Delta\right)^{2}+\left(\frac{\hbar}{2\tau}\right)^{2} },
	\label{Jq3}
\end{equation}
where $N_{\sigma}\equiv(Vmk_{F\sigma}/2\pi^{2}\hbar^{2})$.
The integration over $\cos\theta$ in eq. (\ref{Jq1}) is then easily carried 
out to obtain
\begin{equation}
J_{q}=\frac{2m^{2}V\tau^{4}}{\pi\hbar^{8}} \sum_{\sigma} \frac{1}{q} 
\left[ 
\tan^{-1}\frac{2\tau}{\hbar}
       \left(\frac{\hbar^{2}k_{F\sigma}q}{2m}+\Delta\right)
+
\tan^{-1}\frac{2\tau}{\hbar}
       \left(\frac{\hbar^{2}k_{F\sigma}q}{2m}-\Delta\right)
\right].
	\label{Jq4}
\end{equation}
The conductivity (\ref{sigamcJ}) is then obtained by use of the expression of 
$A^y_\xw({q})$ 
($\sum_{q}|A^y_\xw{(q)}|^{2}\cdots=(\pi/L)\int dq [\cosh^{2}(\pi q\lambda/2)]^{-1}\cdots$) 
as
\begin{eqnarray}
  \sigma_{\rm c} &=& -\frac{e^{2}\Delta^{2}\tau^{2}}{8\pi\hbar^{3}L}\sum_{\sigma} 
  \int_{-\infty}^{\infty}\frac{dq}{q}\frac{1}{\cosh^{2}\frac{\pi}{2}q\lambda}
 \nonumber\\
 && \times
  \left[ 
\tan^{-1}\frac{2\tau}{\hbar}
       \left(\frac{\hbar^{2}k_{F\sigma}q}{2m}+\Delta\right)
+
\tan^{-1}\frac{2\tau}{\hbar}
       \left(\frac{\hbar^{2}k_{F\sigma}q}{2m}-\Delta\right)
\right].
	\label{sigmacJ1}
\end{eqnarray}
Changing the variable $x\equiv\pi\lambda q/2$ we obtain (\ref{SIGMAC}).

To look into the asymptotic behaviors at large $\Delta$ it is useful to rewrite the result
by use of $\tan^{-1}x= \pi/2-\tan^{-1}(1/x)$ for $x>0$ and
$\tan^{-1}x= -\pi/2-\tan^{-1}(1/x)$ for $x<0$as
\begin{equation}
\sigma_{\rm c}=-\frac{e^{2}}{4\pi\hbar}\Deltatil^{2} n_{\rm w}\sum_{\sigma} 
 \left[
 \pi \int_{\Lambda_\sigma}^{\infty}\frac{dx}{x}\frac{1}{\cosh^{2}x}
+
\int_{0}^{\infty}\frac{dx}{x}\frac{1}{\cosh^{2}x}
\tan^{-1} 
\frac{2 \lstil x}{(2\Deltatil)^{2}-(\lstil x)^{2}+1} \right],
	\label{sigmac2}
\end{equation}
where $\Deltatil\equiv \Delta\tau/\hbar$, 
$\lstil\equiv (2 l_{\sigma}/\pi\lambda)$, and $\Lambda_\sigma\equiv 2\Deltatil/\lstil 
=(\pi m \Delta\lambda/k_{F\sigma}\hbar^{2})$.
We consider the case $\Deltatil\gg1$ and $\Lambda_{\sigma}\gg1$, 
which would be satisfied for $d$ electrons in 3$d$ transition metals. 
Then the conductivity correction is approximated as
\begin{eqnarray}
\sigma_{\rm c}&=&-\frac{e^{2}}{4\pi\hbar}\Deltatil^{2} n_{\rm w}\sum_{\sigma} 
 \left[ 4\pi \int_{\Lambda_\sigma}^{\infty}\frac{dx}{x} e^{-2x}
+
\frac{\lstil}{2\Deltatil^{2}} \right]\int_{0}^{\infty}\frac{dx}{\cosh^{2}x}
 \nonumber\\
&=& -\frac{e^{2}}{4\pi^{2}\hbar} \frac{\Delta\tau^{2}}{m\lambda} n_{\rm w}
\sum_{\sigma} k_{F\sigma}
 \left[ 2\pi  e^{-2\Lambda_{\sigma}} +\frac{\hbar}{\Delta\tau} \right].
	\label{sigmac3}
\end{eqnarray}
In the case of a thick wall ($\lambda\gg k_{F}^{-1}$) with a finite 
$\tau$, the first term 
in (\ref{sigmac3}) is exponentially small and thus neglected.
The conductivity in this case is
\begin{equation}
	\sigma_{\rm  c} \simeq -\frac{e^{2}}{8\pi\hbar}n_{\rm w}\sum_{\sigma}
	\lstil\int_{0}^{\infty} \frac{dx}{\cosh^{2}x}
	= -\frac{e^{2}}{8\pi\hbar}n_{\rm w}\sum_{\sigma}
	\lstil   \;\;\;\;\;
	(\Delta\tau/\hbar, m\Delta\lambda/k_{F\sigma} \hbar^{2} \gg1).
	\label{sigmacferro}
\end{equation}

On the other hand if we take the limit of $\tau\rightarrow\infty$ 
first, the first term in (\ref{sigmac3}) becomes dominant and the 
result of the Mori formula is obtained\cite{GT01}. 
The result of Cabrera and Falicov\cite{Cabrera74}, obtained by
calculating the reflection coefficient in the absence of impurities, 
corresponds to this limit.



\bibliographystyle{elsart-num}
\bibliography{/home/tatara/References/dw08}

\begin{thebibliography}{100}
\expandafter\ifx\csname url\endcsname\relax
  \def\url#1{\texttt{#1}}\fi
\expandafter\ifx\csname urlprefix\endcsname\relax\def\urlprefix{URL }\fi

\bibitem{Thomson1857}
W.~Thomson, On the electrodynamic qualities of metals, Proc, Royal. Soc. London
  8 (1857) 546.

\bibitem{McGuire75}
T.~R. McGuire, R.~I. Potter, Anisotropic magnetoresistance in ferromagnetic 3d
  alloys, IEEE Trans. Magn. MAG-11 (1975) 1018.

\bibitem{Baibich88}
M.~N. Baibich, J.~M. Broto, A.~Fert, F.~N. Van~Dau, F.~Petroff, P.~Eitenne,
  G.~Creuzet, A.~Friederich, J.~Chazelas, Giant magnetoresistance of
  (001)Fe/(001)Cr magnetic superlattices, Phys. Rev. Lett. 61~(21) (1988)
  2472--2475.

\bibitem{Binasch89}
G. Binasch, P. Gr\"unberg, F. Saurenbach, W. Zinn, 
Enhanced magnetoresistance in layered magnetic structures with antiferromagneticinterlayer exchange,
Phys. Rev. B 39 (7) (1989) 4828--4830.

\bibitem{Miyazaki95}
T.~Miyazaki, N.~Tezuka, Giant magnetic tunneling effect in fe/al$_2$o$_3$/fe
  junction, J. Magn. Magn. Mater. 139 (1995) L231.

\bibitem{Yuasa04}
S.~Yuasa, T.~Nagahama, A.~Fukushima, Y.~Suzuki, K.~Ando, Giant room-temperature
  magnetoresistance in single-crystal Fe/MgO/Fe magnetic tunnel junctions,
  Nature Mater. 3 (2004) 868.

\bibitem{Parkin04}
S.~S.~P. Parkin, C.~Kaiser, A.~Panchula, P.~M. Rice, B.~Hughes, M.~Samant,
  S.-H. Yang, Giant tunnelling magnetoresistance at room temperature with mgo
  (100) tunnel barriers, Nat. Mater. 3 (2004) 862.

\bibitem{Slonczewski96}
J.~C. Slonczewski, Current-driven excitation of magnetic multilayers, J. Magn
  Magn Mater. 159 (1996) L1.

\bibitem{Berger96}
L.~Berger, Emission of spin waves by a magnetic multilayer traversed by a
  current, Phys. Rev. B 54~(13) (1996) 9353--9358.

\bibitem{Parkin08}
S.~S.~P. Parkin, M.~Hayashi, L.~Thomas, Magnetic domain-wall racetrack memory,
  Science 320 (2008) 190--194.

\bibitem{Hayashi08}
M.~Hayashi, L.~Thomas, R.~Moriya, C.~Rettner, S.~S.~P. Parkin,
  Current-controlled magnetic domain-wall nanowire shift register, Science 320
  (2008) 209--211.

\bibitem{Berger78}
L.~Berger, Low-field magnetoresistance and domain drag in ferromagnets, J.
  Appl. Phys. 49 (1978) 2156.

\bibitem{Kittel49}
C.~Kittel, Physical theory of ferromagnetic domains, Rev. Mod. Phys. 21~(4)
  (1949) 541--583.

\bibitem{Malozemoff79}
A.~P. Malozemoff, J.~C. Slonczewski, Magnetic Domain Walls in Bubble Materials,
  Academic press, 1979.

\bibitem{Chikazumi97}
S.~Chikazumi, Physic of ferromagnetism, Oxford University Press.

\bibitem{Hubert98}
A.~Hubert, R.~Schafer, Magnetic Domains, Springer-Verlag, 1998.

\bibitem{Marrows05}
C.~H. Marrows, Spin-polarised currents and magnetic domain walls, Advances in
  Physics 54 (2005) 585.

\bibitem{Klaui08}
M.~Kl\"aui, Head-to-head domain walls in magnetic nanostructures,
J. Phys.: Condens. Matter 20 (2008) 313001.

\bibitem{Hong95}
K.~Hong, N.~Giordano, Approach to mesoscopic magnetic measurements, Phys. Rev.
  B. 51 (1995) 9855.

\bibitem{SMYT04}
E.~Saitoh, H.~Miyajima, T.~Yamaoka, G.~Tatara, Current-induced resonance and
  mass determination of a single magnetic domain wall, Nature 432 (2004) 203.

\bibitem{Gregg96}
J.~F. Gregg, W.~Allen, K.~Ounadjela, M.~Viret, M.~Hehn, S.~M. Thompson,
  J.~M.~D. Coey, Giant magnetoresistive effects in a single element magnetic
  thin film, Phys. Rev. Lett. 77 (1996) 1580.

\bibitem{Feigenson07}
M.~Feigenson, J.~W. Reiner, L.~Klein, Efficient current-induced domain-wall
  displacement in SrRuO$_3$, Phys. Rev. Lett. 98~(24) (2007) 247204.

\bibitem{Garcia99}
N.~Garc\'ia, M.~Mu\~noz, Y.~W. Zhao, Magnetoresistance in excess of 200\% in
  ballistic ni nanocontacts at room temperature and 100 Oe, Phys. Rev. Lett 82
  (1999) 2923.

\bibitem{Berger92}
L.~Berger, Motion of a magnetic domain wall traversed by fast-rising current
  pulses, J. Appl. Phys. 71 (1992) 2721.

\bibitem{Kittel86}
C.~Kittel, Introduction to Solid State Physics, John Wiley \& Sons, 1986.

\bibitem{TK04}
G.~Tatara, H.~Kohno, Theory of current-driven domain wall motion: Spin transfer
  versus momentum transfer, Phys. Rev. Lett. 92~(8) (2004) 086601.

\bibitem{KT06}
H.~Kohno, G.~Tatara, Introduction to a theory of current-driven domain wall
  motion, Spintronic Materials and Technology (Taylor \& Francis Group, 2006).

\bibitem{Tserkovnyak06}
Y.~Tserkovnyak, H.~J. Skadsem, A.~Brataas, G.~E.~W. Bauer, Current-induced
  magnetization dynamics in disordered itinerant ferromagnets, Phys. Rev. B
  74~(14) (2006) 144405.

\bibitem{KTS06}
H.~Kohno, G.~Tatara, J.~Shibata, Microscopic calculation of spin torques in
  disordered ferromagnets, J. Phys. Soc. Jpn. 75 (2006) 113706.

\bibitem{Braun96}
H.-B. Braun, D.~Loss, Berry's phase and quantum dynamics of ferromagnetic
  solitons, Phys. Rev. B 53~(6) (1996) 3237--3255.

\bibitem{TT96}
S.~Takagi, G.~Tatara, Macroscopic quantum coherence of chirality of a domain
  wall in ferromagnets, Phys. Rev. B 54~(14) (1996) 9920--9923.

\bibitem{Dietl01}
T.~Dietl, H.~Ohno, F.~Matsukura, Hole-mediated ferromagnetism in tetrahedrally
  coordinated semiconductors, Phys. Rev. B 63~(19) (2001) 195205.

\bibitem{Berger84}
L.~Berger, Exchange interaction between ferromagnetic domain wall and electric
  current in very thin metallic films, J. Appl. Phys. 55 (1984) 1954.

\bibitem{Hung88}
C.-Y. Hung, L.~Berger, Exchange forces between domain wall and electric current
  in permalloy films of variable thickness, J. Appl. Phys. 63~(8) (1988)
  4276--4278.

\bibitem{Berger86}
L.~Berger, Possible existence of a Josephson effect in ferromagnets, Phys. Rev.
  B 33~(3) (1986) 1572--1578.

\bibitem{Salhi93}
E.~Salhi, L.~Berger, Current-induced displacements and precession of a bloch
  wall in ni-fe thin films, J. Appl. Phys. 73 (1993) 6405.

\bibitem{Grollier02}
J.~Grollier, D.~Lacour, V.~Cros, A.~Hamzic, A.~Vaures, A.~Fert, D.~Adam,
  G.~Faini, Switching the magnetic configuration of a spin valve by
  current-induced domain wall motion, J. Appl. Phys 92 (2002) 4825.

\bibitem{Grollier03}
J.~Grollier, P.~Boulenc, V.~Cros, A.~Vaures, A.~Fert, G.~Faini, Switching a
  spin valve back and forth by current-induced domain wall motion, Appl. Phys.
  Lett. 83 (2003) 509.

\bibitem{Tsoi03}
M.~Tsoi, R.~E. Fontana, S.~S.~P. Parkin, Magnetic domain wall motion triggered
  by an electric current, Appl. Phys. Lett 83 (2003) 2617.

\bibitem{Klaui03}
M.~Kl\"{a}ui, C.~A.~F. Vanz, J.~A.~C. Bland, W.~Wernsdorfer, G.~Faini,
  E.~Cambril, L.~J. Heyderman, Domain wall motion induced by spin polarized
  currents in ferromagnetic ring structure, Appl. Phys. Lett 83 (2003) 105.

\bibitem{Yamaguchi04}
A.~Yamaguchi, T.~Ono, S.~Nasu, K.~Miyake, K.~Mibu, T.~Shinjo, Real-space
  observation of current-driven domain wall motion in submicron magnetic wires,
  Phys. Rev. Lett. 92~(7) (2004) 077205.

\bibitem{Klaui05}
M.~Kl\"aui, P.-O. Jubert, R.~Allenspach, A.~Bischof, J.~A.~C. Bland, G.~Faini,
  U.~Rudiger, C.~A.~F. Vaz, L.~Vila, C.~Vouille, Direct observation of
  domain-wall configurations transformed by spin currents, Phys. Rev. Lett.
  95~(2) (2005) 026601.

\bibitem{TKS08}
G.~Tatara, H.~Kohno, J.~Shibata, Theory of domain wall dynamics under current,
  J. Phys. Soc. Jpn. 77 (2008) 031003.

\bibitem{Slonczewski72}
J.~C. Slonczewski, Dynamics of magnetic domain walls, Int. J. Magn. 2 (1972)
  85.

\bibitem{Thiaville04}
A.~Thiaville, Y.~Nakatani, J.~Miltat, N.~Vernier, Domain wall motion by
  spin-polarized current: a micromagnetic study, J. Appl. Phys. 95 (2004) 7049.

\bibitem{Li04st}
Z.~Li, S.~Zhang, Domain-wall dynamics driven by adiabatic spin-transfer
  torques, Phys. Rev. B 69 (2004) 134416.

\bibitem{Zhang04}
S.~Zhang, Z.~Li, Roles of nonequilibrium conduction electrons on the
  magnetization dynamics of ferromagnets, Phys. Rev. Lett. 93~(12) (2004)
  127204.

\bibitem{Thiaville05}
A.~Thiaville, Y.~Nakatani, J.~Miltat, Y.~Suzuki, Micromagnetic understanding of
  current-driven domain wall motion in patterned nanowires, Europhys. Lett. 69
  (2005) 990.

\bibitem{Xiao06}
J.~Xiao, A.~Zangwill, M.~D. Stiles, Spin-transfer torque for continuously
  variable magnetization, Phys. Rev. B 73 (2006) 054428.

\bibitem{TKSLL07}
G.~Tatara, H.~Kohno, J.~Shibata, Y.~Lemaho, K.-J. Lee, Spin torque and force
  due to current for general spin textures, J. Phys. Soc. Jpn. 76 (2007)
  054707.

\bibitem{TTKSNF06}
G.~Tatara, T.~Takayama, H.~Kohno, J.~Shibata, Y.~Nakatani, H.~Fukuyama,
  Threshold current of domain wall motion under extrinsic pinning, $\beta$-term
  and non-adiabaticity, J. Phys. Soc. Jpn. 75 (2006) 64708.

\bibitem{Kohno07}
H.~Kohno, J.~Shibata, Gauge field formulation of adiabatic spin torques, J.
  Phys. Soc. Jpn. 76 (2007) 063710.

\bibitem{Duine07}
R.~A. Duine, A.~S.~Nunez, J.~Sinova, A.~H. MacDonald, Functional keldysh
  theory of spin torques, Phys. Rev. B 75~(21) (2007) 214420.

\bibitem{Barnes05}
S.~E. Barnes, S.~Maekawa, Current-spin coupling for ferromagnetic domain walls
  in fine wires, Phys. Rev. Lett. 95~(10) (2005) 107204.

\bibitem{Barnes06}
S.~E. Barnes, Comment on ``Theory of current-driven domain wall motion: Spin
  transfer versus momentum transfer'', Phys. Rev. Lett. 96~(18) (2006) 189701.

\bibitem{TK06}
G.~Tatara, H.~Kohno, Tatara and Kohno reply:, Phys. Rev. Lett. 96~(18) (2006)
  189702.

\bibitem{Edwards05}
D.~M. Edwards, F.~Federici, J.~Mathon, A.~Umerski, Self-consistent theory of
  current-induced switching of magnetization, Phys. Rev. B 71~(5) (2005)
  054407.

\bibitem{Thomas06}
L.~Thomas, M.~Hayashi, X.~Jiang, R.~Moriya, C.~Rettner, S.~S.~P. Parkin,
  Oscillatory dependence of current-driven magnetic domain wall motion on
  current pulse length, Nature 443 (2006) 197--200.

\bibitem{Heyne08}
L.~Heyne, M.~Kl\"aui, D.~Backes, T.~A. Moore, S.~Krzyk, U.~Rudiger, L.~J.
  Heyderman, A.~F. Rodriguez, F.~Nolting, T.~O. Mentes, M.~A. Nino,
  A.~Locatelli, K.~Kirsch, R.~Mattheis, Relationship between nonadiabaticity
  and damping in permalloy studied by current induced spin structure
  transformations, Phys. Rev. Lett. 100~(6) (2008) 066603.

\bibitem{Oogane06}
M.~Oogane, T.~Wakitani, S.~Yakata, R.~Yilgin, Y.~Ando, A.~Sakuma, T.~Miyazaki,
  Magnetic damping in ferromagnetic thin films, Jpn. J. Appl. Phys. 45 (2006)
  3889.

\bibitem{Elliott54}
R.~J. Elliott, Theory of the effect of spin-orbit coupling on magnetic
  resonance in some semiconductors, Phys. Rev. 96~(2) (1954) 266--279.

\bibitem{Waintal04}
X.~Waintal, M.~Viret, Current-induced distortion of a magnetic domain wall,
  Europhys. Lett. 65 (2004) 427.

\bibitem{Ohe06}
J.-I. Ohe, B.~Kramer, Dynamics of a domain wall and spin-wave excitations
  driven by a mesoscopic current, Phys. Rev. Lett. 96~(2) (2006) 027204.

\bibitem{Piechon07}
F.~Piechon, A.~Thiaville, Spin transfer torque in continuous textures:
  Semiclassical boltzmann approach, Phys. Rev. B 75~(17) (2007) 174414.

\bibitem{Thorwart07}
M.~Thorwart, R.~Egger, Current-induced nonadiabatic spin torques and
  domain-wall motion with spin relaxation in a ferromagnetic metallic wire,
  Phys. Rev. B 76~(21) (2007) 214418.

\bibitem{Nguyen07}
A.~K. Nguyen, H.~J. Skadsem, A.~Brataas, Giant current-driven domain wall
  mobility in (Ga,Mn)As, Phys. Rev. Lett. 98~(14) (2007) 146602.

\bibitem{Obata08}
K.~Obata, G.~Tatara, Current-induced domain wall motion in Rashba spin-orbit
  system, Phys. Rev. B 77  (2008) 214429.

\bibitem{Nakagawa07}
T.~Nakagawa, O.~Ohgami, Y.~Saito, H.~Okuyama, M.~Nishijima, T.~Aruga,
  Transition between tetramer and monomer phases driven by vacancy
  configuration entropy on Bi/Ag(001), Phys. Rev. B 75~(15) (2007) 155409.
  
\bibitem{Ast07}
C.~R. Ast, J.~Henk, A.~Ernst, L.~Moreschini, M.~C. Falub, D.~Pacil\'{e},
  P.~Bruno, K.~Kern, M.~Grioni, Giant spin splitting through surface alloying,
  Phys. Rev. Lett. 98~(18) (2007) 186807.

\bibitem{Yamaguchi05}
A.~Yamaguchi, S.~Nasu, H.~Tanigawa, T.~Ono, K.~Miyake, K.~Mibu, T.~Shinjo,
  Effect of joule heating in current-driven domain wall motion, Appl. Phys.
  Lett. 86~(1) (2005) 012511.

\bibitem{Yamaguchi06err}
A.~Yamaguchi, T.~Ono, S.~Nasu, K.~Miyake, K.~Mibu, T.~Shinjo, Erratum:
  Real-space observation of current-driven domain wall motion in submicron
  magnetic wires [phys. rev. lett. 92, 077205 (2004)], Phys. Rev. Lett. 96~(17)
  (2006) 179904.

\bibitem{Yamaguchi06}
A.~Yamaguchi, K.~Yano, H.~Tanigawa, S.~Kasai, T.~Ono, Jap. J. Appl. Phys. 45
  (2006) 3850.

\bibitem{Togawa06}
Y.~Togawa, T.~Kimura, K.~Harada, T.~Akashi, T.~Matsuda, A.~Tonomura, Y.~Otani,
  Current-excited magnetization dynamics in narrow ferromagnetic wires, Jpn. J.
  Appl. Phys. 45 (2006) L683.

\bibitem{Togawa06a}
Y.~Togawa, T.~Kimura, K.~Harada, T.~Akashi, T.~Matsuda, A.~Tonomura, Y.~Otani,
  Domain nucleation and annihilation in uniformly magnetized state under
  current pulses in narrow ferromagnetic wires, Jpn. J. Appl. Phys. 45 (2006)
  L1322.

\bibitem{Biehler07}
A.~Biehler, M.~Kl\"{a}ui, M.~Fonin, C.~K\"{o}nig, G.~G\"{u}ntherodt,
  U.~R\"{u}diger, Domain structures and the influence of current on domains and
  domain walls in highly spin-polarized CrO$_2$ wire elements, Phys. Rev. B
  75~(18) (2007) 184427.

\bibitem{Seo07}
S.-M. Seo, K.-J. Lee, W.~Kim, T.-D. Lee, Effect of shape anisotropy on
  threshold current density for current-induced domain wall motion, Appl. Phys.
  Lett. 90~(25) (2007) 252508.

\bibitem{You07}
C.-Y. You, S.-S. Ha, Temperature increment in a current-heated nanowire for
  current-induced domain wall motion with finite thickness insulator layer,
  Appl. Phys. Lett. 91 (2007) 022507.

\bibitem{Lim04}
C.~K. Lim, T.~Devolder, C.~Chappert, J.~Grollier, V.~Cros, A.~Vaures, A.~Fert,
  G.~Faini, Domain wall displacement induced by subnanosecond pulsed current,
  Appl. Phys. Lett 84 (2004) 2820.

\bibitem{Laufenberg06}
M.~Laufenberg, W.~B\"{u}hrer, D.~Bedau, P.-E. Melchy, M.~Kl\"{a}ui, L.~Vila,
  G.~Faini, C.~A.~F. Vaz, J.~A.~C. Bland, U.~R\"{u}diger, Temperature
  dependence of the spin torque effect in current-induced domain wall motion,
  Phys. Rev. Lett. 97~(4) (2006) 046602.

\bibitem{Ravelosona05}
D.~Ravelosona, D.~Lacour, J.~A. Katine, B.~D. Terris, C.~Chappert, Nanometer
  scale observation of high efficiency thermally assisted current-driven domain
  wall depinning, Phys. Rev. Lett. 95~(11) (2005) 117203.

\bibitem{Fukami08}
S.~Fukami, T.~Suzuki, N.~Ohshima, K.~Nagahara, N.~Ishiwata, Micromagnetic
  analysis of current driven domain wall motion in nanostrips with
  perpendicular magnetic anisotropy, 
J. Appl. Phys. 103~(7) (2008) 07E718.

\bibitem{Suzuki08}
T.~Suzuki, S.~Fukami,  N.~Ohshima, K.~Nagahara, N.~Ishiwata, Analysis of current-driven domain wall motion from pinning sites in nanostrips with perpendicular magnetic anisotropy,
J. Appl. Phys. 103 (2008) 113913.

\bibitem{Yamanouchi04}
M.~Yamanouchi, D.~Chiba, F.~Matsukura, H.~Ohno, Current-induced domain-wall
  switching in a ferromagnetic semiconductor structure, Nature 428 (2004) 539.

\bibitem{Yamanouchi06}
M.~Yamanouchi, D.~Chiba, F.~Matsukura, T.~Dietl, H.~Ohno, Velocity of
  domain-wall motion induced by electrical current in the ferromagnetic
  semiconductor (Ga,Mn)As, Phys. Rev. Lett. 96~(9) (2006) 096601.

\bibitem{Chiba06}
D.~Chiba, M.~Yamanouchi, F.~Matsukura, T.~Dietl, H.~Ohno, Domain-wall
  resistance in ferromagnetic (Ga,Mn)As, Phys. Rev. Lett. 96~(9) (2006) 096602.

\bibitem{Lemerle98}
S.~Lemerle, J.~Ferr\`e,  C.~Chappert, V.~Mathet, T.~Giamarchi, P.~Le~Doussal,
  Domain wall creep in an ising ultrathin magnetic film, Phys. Rev. Lett.
  80~(4) (1998) 849.

\bibitem{TVF05}
G.~Tatara, N.~Vernier, J.~Ferr\`e, Universality of thermally assisted magnetic
  domain-wall motion under spin torque, Appl. Phys. Lett. 86~(25) (2005)
  252509.

\bibitem{Yamanouchi07}
M.~Yamanouchi, J.~Ieda, F.~Matsukura, S.~E. Barnes, S.~Maekawa, H.~Ohno,
  Universality classes for domain wall motion in the ferromagnetic
  semiconductor (ga,mn)as, Science 317 (2007) 1726--1729.

\bibitem{Duine07ta}
R.~A. Duine, A.~S. N\'{u}\~{n}ez, A.~H. MacDonald, Thermally assisted
  current-driven domain-wall motion, Phys. Rev. Lett. 98~(5) (2007) 056605.

\bibitem{Thomas07}
L.~Thomas, M.~Hayashi, X.~Jiang, R.~Moriya, C.~Rettner, S.~Parkin, Resonant
  amplification of magnetic domain-wall motion by a train of current pulses,
  Science 315 (2007) 1553--1556.

\bibitem{Hayashi07}
M.~Hayashi, L.~Thomas, C.~Rettner, R.~Moriya, S.~S.~P. Parkin, Direct
  observation of the coherent precession of magnetic domain walls propagating
  along permalloy nanowires, Nat. Phys. 3 (2007) 21.

\bibitem{Auerbach94}
A.~Auerbach, Intracting electrons and quantum magnetism (1994) Chap. 10.

\bibitem{Landau35}
L.~Landau, E.~Lifshitz, Theory of the dispersion of magnetic permeability in
  ferromagnetic bodies, Phys. Z. Sowietunion 8 (1935) 153.

\bibitem{Goldstein02}
H.~Goldstein, C.~Poole, J.~Safko, Classical mechanics, Third Edition (2002)
  Chap. 1. Sec. 5.

\bibitem{Rajaraman82}
R.~Rajaraman, Solitons and Instantons, North-Holland, 1982.

\bibitem{Sakita85}
B.~Sakita, Quantum theory of many-variable systems and fields, World
  Scientific, 1985.

\bibitem{Hubert00}
A.~Hubert, R.~Sch\"afer, Magnetic Domains, Springer-Verlag, Berlin, 2000.

\bibitem{Landau77}
L.~D. Landau, E.~M. Lifshitz, Quantum Mechanics, Pergamon, 1977.

\bibitem{Doring48}
W.~D\"oring, Uber die tragheit der wande zwischen weisschen bezirken, Z.
  Naturforsch 3A (1948) 373.

\bibitem{Rammer86}
J.~Rammer, H.~Smith, Quantum field-theoretical methods in transport theory of
  metals, Rev. Mod. Phys. 58~(2) (1986) 323--359.

\bibitem{Stern92}
A.~Stern, Berry's phase, motive forces, and mesoscopic conductivity, Phys. Rev.
  Lett. 68~(7) (1992) 1022--1025.

\bibitem{Popp03}
M.~Popp, D.~Frustaglia, K.~Richter, Conditions for adiabatic spin transport in
  disordered systems, Phys. Rev. B 68 (2003) 041303.

\bibitem{Feynman65}
R.~P. Feynman, A.~R. Hibbs, Quantum mechanics and path integrals, McGraw-Hill,
  1965.

\bibitem{TE08}
G.~Tatara, P.~Entel, private communication.

\bibitem{Haug07}
H.~Haug, A.-P. Jauho, Quantum Kinetics in Transport and Optics of
  Semiconductors, Springer-Verlag, 2007.

\bibitem{Mahan90}
G.~D. Mahan, Many-Particle Physics, Plenum press, New York, 1990.

\bibitem{Buttiker85}
M.~B\"uttiker, Y.~Imry, R.~Landauer, S.~Pinhas, Generalized many-channel
  conductance formula with application to small rings, Phys. Rev. B 31 (1985)
  6207.

\bibitem{Ye99}
J.~Ye, Y.~B. Kim, A.~J. Millis, B.~I. Shraiman, P.~Majumdar, Z.~Tesanovic,
  Berry phase theory of the anomalous Hall effect: Application to colossal
  magnetoresistance manganites, Phys. Rev. Lett. 83 (1999) 3737.

\bibitem{TK02}
G.~Tatara, H.~Kawamura, Chirality-driven anomalous Hall effect in weak coupling
  regime, J. Phys. Soc. Jpn 71 (2002) 2613.

\bibitem{OTN04}
M.~Onoda, G.~Tatara, N.~Nagaosa, Anomalous Hall effect and skyrmion number in
  real and momentum spaces, J. Phys. Soc. Jpn. 73 (2004) 2624.

\bibitem{Thiele73}
A.~A. Thiele, Steady-state motion of magnetic domains, Phys. Rev. Lett. 30~(6)
  (1973) 230--233.

\bibitem{KTSS07}
H.~Kohno, G.~Tatara, J.~Shibata, Y.~Suzuki, Microscopic calculation of spin
  torques and forces, J. Magn. Magn. Mater. 310 (2007) 2020.

\bibitem{SNTKO06}
J.~Shibata, Y.~Nakatani, G.~Tatara, H.~Kohno, Y.~Otani, Current-induced
  magnetic vortex motion by spin-transfer torque, Phys. Rev. B 73~(2) (2006)
  020403.

\bibitem{Bazaliy98}
Y.~B. Bazaliy, B.~A. Jones, S.-C. Zhang, Modification of the landau-lifshitz
  equation in the presence of a spin-polarized current in colossal- and
  giant-magnetoresistive materials, Phys. Rev. B 57~(6) (1998) R3213--R3216.

\bibitem{Rossier04}
J.~Fern\'andez-Rossier, M.~Braun, A.~S. N\'u\~nez, A.~H. MacDonald, Influence
  of a uniform current on collective magnetization dynamics in a ferromagnetic
  metal, Phys. Rev. B 69~(17) (2004) 174412.

\bibitem{Li04}
Z.~Li, S.~Zhang, Domain-wall dynamics and spin-wave excitations with
  spin-transfer torques, Phys. Rev. Lett. 92 (2004) 207203.

\bibitem{STK05}
J.~Shibata, G.~Tatara, H.~Kohno, Effect of spin current on uniform
  ferromagnetism: Domain nucleation, Phys. Rev. Lett. 94~(7) (2005) 076601.

\bibitem{NSTKTM08}
Y.~Nakatani, J.~Shibata, G.~Tatara, H.~Kohno, A.~Thiaville, J.~Miltat,
  Nucleation and dynamics of magnetic vortices under spin-polarized current,
  Phys. Rev. B 77~(1) (2008) 014439.

\bibitem{TSIK05}
G.~Tatara, E.~Saitoh, M.~Ichimura, H.~Kohno, Domain-wall displacement triggered
  by an ac current below threshold, Appl. Phys. Lett. 86~(23) (2005) 232504.

\bibitem{Schryer74}
N.~L. Schryer, L.~R. Walker, The motion of 180бы domain walls in uniform dc
  magnetic fields, J. Appl. Phys. 45 (1974) 5406.

\bibitem{Shinjo00}
T.~Shinjo, T.~Okuno, R.~Hassdorf, K.~Shigeto, T.~Ono, Magnetic vortex core
  observation in circular dots of permalloy, Science 289 (2000) 930--932.

\bibitem{Yamada07}
K.~Yamada, S.~Kasai, Y.~Nakatani, K.~Kobayashi, H.~Kohno, A.~Thiaville, T.~Ono,
  Electrical switching of the vortex core in a magnetic disk, Nat. Mater. 6
  (2007) 269.

\bibitem{Ansermet04}
J.~P. Ansermet, IEEE Trans.~Magn. 40 (2004) 358.

\bibitem{Tserkovnyak04}
Y.~Tserkovnyak, G.~A. Fiete, B.~I. Halperin, Mean-field magnetization
  relaxation in conducting ferromagnets, Appl. Phys. Lett. 84 (2004) 5234.

\bibitem{Korenman77}
V.~Korenman, J.~L. Murray, R.~E. Prange, Local-band theory of itinerant
  ferromagnetism. i. fermi-liquid theory, Phys. Rev. B 16~(9) (1977)
  4032--4047.

\bibitem{TF94}
G.~Tatara, H.~Fukuyama, Macroscopic quantum tunneling of a domain wall in a
  ferromagnetic metal, Phys. Rev. Lett. 72~(5) (1994) 772--775.

\bibitem{TF97}
G.~Tatara, H.~Fukuyama, Resistivity due to a domain wall in ferromagnetic
  metal, Phys. Rev. Lett. 78~(19) (1997) 3773--3776.

\bibitem{Heide01}
C.~Heide, P.~E. Zilberman, R.~J. Elliott, Current-driven switching of magnetic
  layers, Phys. Rev. B 63~(6) (2001) 064424.

\bibitem{Skadsem07}
H.~J. Skadsem, Y.~Tserkovnyak, A.~Brataas, G.~E.~W. Bauer, Magnetization
  damping in a local-density approximation, Phys. Rev. B 75~(9) (2007) 094416.

\bibitem{Simanek05}
E.~Sim\'{a}nek, A.~Rebei, Spin transport and resistance due to a bloch wall,
  Phys. Rev. B 71~(17) (2005) 172405.

\bibitem{Nguyen06}
A.~K. Nguyen, R.~V. Shchelushkin, A.~Brataas, Intrinsic domain-wall resistance
  in ferromagnetic semiconductors, Phys. Rev. Lett. 97~(13) (2006) 136603.


\bibitem{McKeehan30}
L.~W. McKeehan, Electrical resistance of nickel and permalloy wires as affected
  by longitudinal magnetization and tension, Phys. Rev. 36~(5) (1930) 948--977.

\bibitem{Smit51}
J.~Smit, Physica XVI (1951) 612.

\bibitem{Ono99}
T.~Ono, H.~Miyajima, K.~Shigeto, K.~Mibu, N.~Hosoito, T.~Shinjo, Propagation of
  a magnetic domain wall in a submicrometer magnetic wire, Science 284 (1999)
  468.

\bibitem{Barkhausen19}
H.~Barkhausen, Zwei mit hilfe der neuen verstarker entdeckte erscheinungen,
  Phys. Zeitschrift 20 (1919) 401.

\bibitem{Giordano94}
N.~Giordano, J.~D. Monnier, Physica. B. 194-196 (1994) 1009.

\bibitem{Cabrera74}
G.~G. Cabrera, L.~M. Falicov, Theory of the residual resistivity of bloch
  walls. pt. 1. paramagnetic effects, Phys. Stat. Sol. (b) 61 (1974) 539.

\bibitem{Yamanaka96}
M.~Yamanaka, N.~Nagaosa, J. Phys. Soc. Jpn. 65 (1996) 3088.

\bibitem{Levy97}
P.~M. Levy, S.~Zhang, Resistivity due to domain wall scattering, Phys. Rev.
  Lett. 79~(25) (1997) 5110--5113.

\bibitem{Brataas98}
A.~Brataas, G.~Tatara, G.~Bauer, Phil. Mag. B. 78 (1998) 545.

\bibitem{Brataas99}
A.~Brataas, G.~Tatara, G.~Bauer, Ballistic and diffuse transport through a
  ferromagnetic domain wall, Phys. Rev. B. 60 (1999) 3406.

\bibitem{vanHoof99}
J.~B. A.~N. van Hoof, K.~M. Schep, A.~Brataas, G.~E.~W. Bauer, P.~J. Kelly,
  Ballistic electron transport through magnetic domain walls, Phys. Rev. B
  59~(1) (1999) 138--141.

\bibitem{TZMG99}
G.~Tatara, Y.-W. Zhao, M.~Mu\~noz, N.~Garc\'ia, Domain wall scattering explains
  300
  (1999) 2030--2033.

\bibitem{Gorkom99}
R.~P. Gorkom, A.~Brataas, G.~E.~W. Bauer, Negative domain wall resistance in
  ferromagnets, Phys. Rev. Lett. 83 (1999) 4401.

\bibitem{Imamura00}
H.~Imamura, N.~Kobayashi, S.~Takahashi, S.~Maekawa, Conductance quantization
  and magnetoresistance in magnetic point contacts, Phys. Rev. Lett. 84~(5)
  (2000) 1003--1006.

\bibitem{Nakanishi00}
K.~Nakanishi, Y.~Nakamura, Phys. Rev. B. 69 (2000) 2969.

\bibitem{GT01}
G.~Tatara, Effect of domain-wall on electronic transport properties of metallic
  ferromagnet, Int. J. Mod. Phys. B. 15 (2001) 321.

\bibitem{GT00}
G.~Tatara, Domain wall resistance based on landauer's formula, J. Phys. Soc.
  Jpn. 69 (2000) 2969.

\bibitem{Bruno99}
P.~Bruno, Geometrically constrained magnetic wall, Phys. Rev. Lett. 83 (1999)
  2425.

\bibitem{ZMTG01}
Y.-W. Zhao, M.~Munoz, G.~Tatara, N.~Garc\'ia, From ballistic to non-ballistic
  magnetoresistance in nanocontacts: Theory and experiment, J. Magn. Magn.
  Mater. 223 (2001) 169--174.

\bibitem{Sullivan05}
M.~R. Sullivan, D.~A. Boehm, D.~A. Ateya, S.~Z. Hua, H.~D. Chopra, Ballistic
  magnetoresistance in nickel single-atom conductors without magnetostriction,
  Phys. Rev. B 71~(2) (2005) 024412.

\bibitem{Rocha07}
A.~R. Rocha, T.~Archer, S.~Sanvito, Search for magnetoresistance in excess of
  1000
  76~(5) (2007) 054435.

\bibitem{Bergmann84}
G.~Bergmann, Weak localization in thin films a time-of-flight experiment with
  conduction electrons, Phys. Reports 107 (1984) 1.

\bibitem{Lee85}
P.~A. Lee, T.~V. Ramakrishnan, Disordered electronic systems, Rev. Mod. Phys.
  57~(2) (1985) 287--337.

\bibitem{Meisenheimer89}
T.~L. Meisenheimer, N.~Giordano, Conductance fluctuations in thin silver films,
  Phys. Rev. B 39~(14) (1989) 9929--9936.

\bibitem{Feng86}
S.~Feng, P.~A. Lee, A.~D. Stone, Sensitivity of the conductance of a disordered
  metal to the motion of a single atom: Implications for $1/f$ noise, Phys. Rev.
  Lett. 56~(18) (1986) 1960--1963.

\bibitem{Golding92}
B.~Golding, N.~M. Zimmerman, S.~N. Coppersmith, Dissipative quantum tunneling
  of a single microscopic defect in a mesoscopic metal, Phys. Rev. Lett. 68
  (1992) 998.

\bibitem{Meyer95}
K.~A. Meyer, M.~B. Weissman, Mesoscopic electrical noise from spins in
  au1-xfex, Phys. Rev. B. 51 (1995) 8221.

\bibitem{Strunk98}
C.~Strunk, M.~Henny, C.~Schoenenberger, G.~Neuttiens, C.~V. Haesendonck, Size
  dependent thermopower in mesoscopic aufe wires, Phys. Rev. Lett. 81 (1998)
  2982.

\bibitem{Coppinger94}
F.~Coppinger, J.~Genoe, D.~K. Maude, U.~Gennser, J.~C. Portal, K.~Singer,
  P.~Rutter, T.~Taskin, A.~R. Peaker, A.~C. Wright, Single domain switching
  investigated using telegraph noise spectroscopy: Possible evidence for
  macroscopic quantum tunneling, Phys. Rev. Lett. 75 (1994) 3513.

\bibitem{Lee04}
W.~L. Lee, S.~Watauchi, V.~L. Miller, R.~J. Cava, N.~P. Ong, Dissipationless
  anomalous Hall current in the ferromagnetic spinel CuCr$_2$Se$_{4-x}$Br, Science 303
  (2004) 1647.

\bibitem{Jonkers99}
P.~A.~E. Jonkers, S.~J. Pickering, H.~De~Raedt, G.~Tatara, Quantum transport in
  disordered mesoscopic ferromagnetic films, Phys. Rev. B 60~(23) (1999)
  15970--15974.

\bibitem{Geller98}
Y.~Lyanda-Geller, I.~L. Aleiner, P.~M. Goldbart, Domain walls and conductivity
  of mesoscopic ferromagnets, Phys. Rev. Lett. 81~(15) (1998) 3215--3218.

\bibitem{Loss99}
D.~Loss, H.~Schoeller, P.~M. Goldbart, Observing the berry phase in diffusive
  conductors: Necessary conditions for adiabaticity, Phys. Rev. B 59~(20)
  (1999) 13328--13337.

\bibitem{TF00}
G.~Tatara, H.~Fukuyama, Anomalous magnetoresistance by dephasing in a
  disordered layer with ferromagnetic boundary, J. Phys. Soc. Jpn. 69 (2000)
  2407.

\bibitem{Kobayashi80}
S.~Kobayashi, J. Phys. Soc. Jpn. 49 (1980) 1635.

\bibitem{Raffy87}
H.~Raffy, L.~Dumoulin, J.~P. Burger, Phys. Rev. B 36 (1987) 2158.

\bibitem{TB01}
G.~Tatara, B.~Barbara, Ferromagnetism's affect on the aharonov-bohm effect,
  Phys. Rev. B 64~(17) (2001) 172408.

\bibitem{TKBB04}
G.~Tatara, H.~Kohno, E.~Bonet, B.~Barbara, Aharonov-bohm oscillation in a
  ferromagnetic ring, Phys. Rev. B 69~(5) (2004) 054420.

\bibitem{Tang04}
H.~X. Tang, S.~Masmanidis, R.~K. Kawakami, D.~D. Awschalom, M.~L. Roukes,
  Negative intrinsic resistivity of an individual domain wall in epitaxial (ga,
  mn) as microdevices, Nature 431 (2004) 52.

\bibitem{Vila07}
L.~Vila, R.~Giraud, L.~Thevenard, A.~Lemaitre, F.~Pierre, J.~Dufouleur,
  D.~Mailly, B.~Barbara, G.~Faini, Universal conductance fluctuations in
  epitaxial gamnas ferromagnets: Dephasing by structural and spin disorder,
  Phys. Rev. Lett. 98~(2) (2007) 027204.

\bibitem{Saitoh05}
E.~Saitoh, S.~Kasai, H.~Miyajima, T.~Yamaoka, Electron coherence and magnetic
  structure in a nanostructured ferromagnetic ring, J. Appl. Phys. 97 (2005)
  10J709.

\bibitem{Sekiguchi08}
K.~Sekiguchi, A.~Yamaguchi, H.~Miyajima, A.~Hirohata, Effect of ferromagnetism
  on ab oscillations in a normal-metal ring, Phys. Rev. B 77~(14) (2008)
  140401.

\bibitem{Wei06}
Y.~G. Wei, X.~Y. Liu, L.~Y. Zhang, D.~Davidovi\'{c}, Mesoscopic resistance
  fluctuations in cobalt nanoparticles, Phys. Rev. Lett. 96~(14) (2006) 146803.

\bibitem{Mori65}
H.~Mori, Transport, collective motion, and brownian motion, Prog. Theor. Phys.
  33 (1965) 423.

\bibitem{Mori65a}
H.~Mori, A continued-fraction representation of the time-correlation, Prog.
  Theor. Phys. 34 (1965) 399.

\bibitem{Gotze72}
W.~G\"otze, P.~W\"olfle, Homogeneous dynamical conductivity of simple metals,
  Phys. Rev. B 6 (1972) 1226.

\bibitem{TK03}
G.~Tatara, H.~Kohno, Permanent current from noncommutative spin algebra, Phys.
  Rev. B 67~(11) (2003) 113316.

\bibitem{TG03}
G.~Tatara, N.~Garcia, Quantum toys for quantum computing: Persistent currents
  controlled by the spin josephson effect, Phys. Rev. Lett. 91~(7) (2003)
  076806.

\bibitem{GT04}
G.~Tatara, Persistent current and Hall effect due to spin chirality, phys. sta.
  sol. (b) 241 (2004) 1174--1179.

\bibitem{Caroli71}
R.~Caroli, C.and~Combescot, P.~Nozieres, D.~Saint-James, Direct calculation of
  the tunneling current, J. Phys. C: Solid St. Phys. 4 (1971) 916.

\bibitem{Takeuchi08}
A.~Takeuchi, G.~Tatara, Charge and spin currents generated by dynamical spins,
  J. Phys. Soc. Jpn. 77 (2008) 074701.

\bibitem{Loss90}
D.~Loss, P.~Goldbart, A.~V. Balatsky, Berry's phase and persistent charge and
  spin currents in textured mesoscopic rings, Phys. Rev. Lett. 65~(13) (1990)
  1655--1658.

\bibitem{Taguchi01}
Y.~Taguchi, Y.~Oohara, H.~Yoshizawa, N.~Nagaosa, Y.~Tokura, Spin chirality,
  berry phase, and anomalous Hall effect in a frustrated ferromagnet, Science
  291 (2001) 2573.

\bibitem{Fujita06}
T.~Fujita, M.~Soda, M.~Sato, Anomalous Hall resistivity of tbbaco2o5.53 with
  nontrivial magnetic structure, J. Phys. Soc. Jpn. 75 (2006) 114710.

\bibitem{Kageyama03}
T.~Kageyama, N.~Aito, S.~Iikubo, M.~Sato, Anomalous Hall effect of reentrant
  spin glass system Fe$_{1-x}$Al$_{x}$ ($x\sim$0.3), J. Phys. Soc. Jpn. 72
  (2003) 1491--1494.

\bibitem{Pureur04}
P.~Pureur, F.~W. Fabris, J.~Schaf, I.~A. Campbell, Chiral susceptibility in
  canonical spin glass and re-entrant alloys from Hall effect measurements,
  Europhys. Lett. 67 (2004) 123.

\bibitem{Taniguchi04}
T.~Taniguchi, K.~Yamanaka, H.~Sumioka, T.~Yamazaki, Y.~Tabata, S.~Kawarazaki,
  Direct observation of chiral susceptibility in the canonical spin glass aufe,
  Phys. Rev. Lett. 93~(24) (2004) 246605.

\bibitem{Fabris06}
F.~W. Fabris, P.~Pureur, J.~Schaf, V.~N. Vieira, I.~A. Campbell, Chiral
  anomalous Hall effect in reentrant aufe alloys, Phys. Rev. B 74~(21) (2006)
  214201.
  
\bibitem{Taniguchi07}
T.~Taniguchi, Chiral susceptibility of canonical spin glasses from Hall effect
  measurements, J. Phys.: Condens. Matter 19 (2007) 145213.

\bibitem{Onoda03}
S.~Onoda, N.~Nagaosa, Spin chirality fluctuations and anomalous Hall effect in
  itinerant ferromagnets, Phys. Rev. Lett. 90~(19) (2003) 196602.

\bibitem{Nagaosa06}
N.~Nagaosa, Anomalous Hall effect -a new perspective-, J. Phys. Soc. Jpn. 75
  (2006) 042001.

\bibitem{Yasui06}
Y.~Yasui, T.~Kageyama, T.~Moyoshi, M.~Soda, M.~Sato, K.~Kakurai, Studies of
  anomalous Hall effect and magnetic structure of Nd$_2$Mo$_2$O$_7$ -test of
  chirality mechanism-, J. Phys. Soc. Jpn. 75 (2006) 084711.

\bibitem{Barnes07}
S.~E. Barnes, S.~Maekawa, Generalization of faraday's law to include
  nonconservative spin forces, Phys. Rev. Lett. 98~(24) (2007) 246601.

\bibitem{Duine08}
R.~A. Duine, Spin pumping by a field-driven domain wall, Phys. Rev. B 77~(1)
  (2008) 014409.

\bibitem{OTT07}
J.-I. Ohe, A.~Takeuchi, G.~Tatara, Charge current driven by spin dynamics
  in disordered rashba spin-orbit system, Phys. Rev. Lett. 99~(26) (2007)
  266603.

\bibitem{Tserkovnyak02}
Y.~Tserkovnyak, A.~Brataas, G.~E.~W. Bauer, Enhanced gilbert damping in thin
  ferromagnetic films, Phys. Rev. Lett. 88~(11) (2002) 117601.
  
\bibitem{Tserkovnyak05}
Y.~Tserkovnyak, A.~Brataas, G.~E.~W. Bauer, B.~I. Halperin, Nonlocal
  magnetization dynamics in ferromagnetic heterostructures, Rev. Mod. Phys.
  77~(4) (2005) 1375.

\bibitem{SUMT06}
E.~Saitoh, M.~Ueda, H.~Miyajima, G.~Tatara, Conversion of spin current into
  charge current at room temperature: Inverse spin-Hall effect, Appl. Phys.
  Lett. 88 (2006) 182509.

\bibitem{Nagaosa99}
N.~Nagaosa, Quantum Field Theory in Condensed Matter Physics, Springer-Verlag,
  1999.

\bibitem{Berry84}
M.~V. Berry, Quantal phase factors accompanying adiabatic changes, Proc. Roy.
  Soc. London A392 (1984) 45--57.

\bibitem{Sakurai94}
J.~J. Sakurai, Modern Quantum Mechanics, Addison Wesley, 1994.

\bibitem{Haug98}
H.~Haug, A.~P. Jauho, Quantum Kinetics in Transport and Optics of
  Semi-conductors, Springer-Verlag, 1998.

\bibitem{Rammer07}
J.~Rammer, Quantum Field Theory of Non-equilibrium States, Cambridge, 2007.

\bibitem{Keldysh65}
L.~P. Keldysh, Sov. Phys. JETP 20 (1965) 1018.

\bibitem{Langreth76}
D.~C. Langreth, Linear and Nonlinear Electron Transport in Solids, NATO
  Advanced Study Institute Series B, Vol. 17, Plenum, NewYork/London, 1976.

\end{thebibliography}
\end{document}